\documentclass[preprint,prd,aps,floatfix,preprintnumbers,superscriptaddress,bibnotes,nofootinbib]{revtex4-1}
\usepackage{latexsym}
\usepackage[psamsfonts]{amssymb}
\usepackage{graphicx}
\usepackage{pstricks}

\usepackage{subfigure}

\usepackage{version}    
\usepackage{amsmath}    
\usepackage{hyperref}
\usepackage{longtable}
\usepackage{ulem}
\usepackage{bm}  
\usepackage{color}
\usepackage{enumerate}  
\usepackage{multirow}

\usepackage{comment}

\newcommand{\be}{\begin{equation}}
\newcommand{\ee}{\end{equation}}
\newcommand{\bea}{\begin{eqnarray}}
\newcommand{\eea}{\end{eqnarray}}
\newcommand{\bi}{\begin{itemize}}
\newcommand{\ei}{\end{itemize}}

\begin{document}
\preprint{UTHEP-783, UTCCS-P-149, HUPD-2307, YITP-23-143}
\title{Nucleon form factors in $N_f=2+1$ lattice QCD \\
at the physical point : finite lattice spacing effect \\ on the root-mean-square radii
}
%
\author{Ryutaro Tsuji\:}
\email[E-mail: ]{tsuji@nucl.phys.tohoku.ac.jp}
\affiliation{Department of Physics, Tohoku University, Sendai 980-8578, Japan}
\affiliation{RIKEN Center for Computational Science, Kobe 650-0047, Japan}
\author{Yasumichi Aoki\:}
\affiliation{RIKEN Center for Computational Science, Kobe 650-0047, Japan}
\author{Ken-Ichi~Ishikawa\:}
\affiliation{Core of Research for the Energetic Universe, Graduate School of Advanced Science and Engineering, Hiroshima University, Higashi-Hiroshima 739-8526, Japan}
\author{Yoshinobu~Kuramashi\:}
\affiliation{Center for Computational Sciences, University of Tsukuba, Tsukuba, Ibaraki 305-8577, Japan}
\author{Shoichi~Sasaki\:}
\email[E-mail: ]{ssasaki@nucl.phys.tohoku.ac.jp}
\affiliation{Department of Physics, Tohoku University, Sendai 980-8578, Japan}
\author{Kohei~Sato\:}
\affiliation{Degree Programs in Pure and Applied Sciences, Graduate School of Science and Technology, University of Tsukuba, Ibaraki 305-8571, Japan}
\author{Eigo~Shintani\:}
\affiliation{Center for Computational Sciences, University of Tsukuba, Tsukuba, Ibaraki 305-8577, Japan}
\author{Hiromasa~Watanabe\:}
\affiliation{Yukawa Institute for Theoretical Physics, Kyoto University, Kyoto 606-8502, Japan}
\author{Takeshi~Yamazaki\:}
\affiliation{Institute of Pure and Applied Sciences, University of Tsukuba, Tsukuba, Ibaraki, 305-8571, Japan}
\affiliation{Center for Computational Sciences, University of Tsukuba, Tsukuba, Ibaraki 305-8577, Japan}
\collaboration{PACS Collaboration}

\begin{abstract}

We present results for the nucleon form factors: electric ($G_E$), magnetic ($G_M$), axial ($F_A$), induced pseudoscalar ($F_P$) and pseudoscalar ($G_P$) form factors, using the second PACS10 ensemble that is one of three sets of $2+1$ flavor lattice QCD configurations at physical quark
masses in large spatial volumes (exceeding $(10\ \mathrm{fm})^3$).
The second PACS10 gauge configurations are generated by the PACS Collaboration with the six stout-smeared $O(a)$ improved Wilson quark action and Iwasaki gauge action
at the second gauge coupling $\beta=2.00$ corresponding to the lattice spacing of $a=0.063$ fm.
We determine the isovector electric, magnetic and axial radii and magnetic moment from the corresponding form factors, as well as the axial-vector coupling $g_A$.
Combining our previous results for the coarser lattice spacing
[E. Shintani {\it et al.}, Phys.~Rev.~{\bf D}99 (2019) 014510; Phys.~Rev.~{\bf D}102 (2020) 019902 (erratum)],
the finite lattice spacing effects on the isovector radii, magnetic moment and axial-vector coupling are 
investigated using the difference between the two results.
It was found that the effect on $g_A$ is kept 
smaller than the statistical error of 2\%
while the effect on the isovector radii was observed as a possible discretization error of about 10\%, regardless of the channel. 
We also report the partially conserved axial vector current (PCAC) relation using a set of nucleon three-point correlation functions in order to verify the effect by $O(a)$-improvement of the axial-vector current.

\end{abstract}

\maketitle

\section{Introduction}
In the standard model of the particle physics,
the proton and neutron, in short nucleon, which are the building blocks of nuclei, are composite particles of quarks and gluons,
and the interaction among them are formulated by the Quantum Chromodynamics (QCD).
This indicates that the structure of nucleon is itself a nontrivial consequence of quark-gluon dynamics.
The nucleon form factors are very good probes to investigate the nucleon structure~\cite{Thomas:2001kw}.
Although great theoretical and experimental efforts for the form factors have been devoted to improving our knowledge of the nucleon structure,
there are several unsolved problems and tensions associated with the fundamental properties of nucleons.

The proton radius puzzle~\cite{Pohl:2010zza}, which has become well known as the discrepancy
in experimental measurements of electric root-mean-square (RMS) radius of the proton,
has not been solved. 
In this puzzle, 
high-precision measurements of 
the proton's charge radius using the muonic hydrogen spectroscopy disagree with
its long-established value measured from both elastic electron-proton scattering and hydrogen spectroscopy.
In order to solve the puzzle, recent perspectives have focused primarily on systematic uncertainties~\cite{Bernauer:2020ont}.
Furthermore,
there is a significant tension in empirical parameterizations of the proton magnetic form factor obtained by experiments~\cite{Bradford:2006yz, Borah:2020gte}.
A percent-level measurement is needed to resolve these issues and should be performed in future experiments.

Not only the electric and magnetic form factors,
but also the axial form factor and axial radius are important inputs for the weak process associated with the neutrino-nucleus scattering~\cite{Ruso:2022qes, Kronfeld:2019nfb, Meyer:2022mix}.
The axial-vector coupling ($g_A$), 
which can be determined from the axial form factor at zero-momentum transfer, 
is associated with the neutron lifetime puzzle~\cite{Czarnecki:2018okw}. 
Since the discrepancy between the results of beam experiments and storage experiments remains unsolved, 
it is still an open question that deserves further investigation.
Furthermore,
the $q^2$-dependence of the axial form factor can be used as an important input
~\cite{CLAS:2021neh,Kuzmin:2007kr, Meyer:2016oeg, MINERvA:2023avz}
for the current neutrino oscillation experiments such as T2K, NOvA and so on~\cite{Meyer:2022muy}.

Lattice QCD is the only known way to compute rigorously the nucleon form factors and their corresponding radii as the first principles of QCD.
Recent developments in computational technology and a tremendous increase in computational resources have made it possible
to perform realistic lattice QCD with light quark (degenerate up and down quarks) and strange quark flavors even in baryon physics,
which has more complex systematic uncertainties than meson physics.
Indeed, 
lattice QCD successfully reproduces high accuracy the experimental values of  $g_A$~\cite{FlavourLatticeAveragingGroupFLAG:2021npn, Tsuji:2022ric, Bali:2023sdi, Park:2021ypf} that are precisely measured by the current precision measurements of neutron $\beta$-decay,
and its reproduction is an important benchmark study for the structure of the nucleon based on lattice QCD. 
This success finally reveals the major sources of the systematic uncertainty:
the chiral extrapolation to the physical point, 
the finite volume effect, 
the finite lattice spacing effect, 
and the excited-state contamination~\cite{Djukanovic:2021qxp}.
Furthermore,
it motivates the current efforts to improve precision of theoretical predictions for the nucleon structure and solve related puzzles and tensions~\cite{Ohta:2022csu, QCDSF:2017ssq, Alexandrou:2017hac, Alexandrou:2017ypw, Alexandrou:2020okk, Alexandrou:2021jok, Yoon:2016dij, Gupta:2017dwj, Gupta:2018qil, Jang:2019jkn, Jang:2019vkm, Jang:2023zts, Djukanovic:2021qxp, Djukanovic:2022wru, Djukanovic:2023beb, Djukanovic:2023jag, Shintani:2018ozy}.

For the sake of the high-precision determination with a few percent level,
we perform \textit{fully dynamical lattice QCD simulations} at the physical point with lattice volume larger than $(10\;{\mathrm{fm}})^4$,
which can eliminate the systematic uncertainties due to chiral and infinite-volume extrapolations (called as ``PACS10'' project). 
In the PACS10 project, the PACS Collaboration plans to generate three sets of the PACS10 gauge configurations at three different lattice spacings.
We have reported the first result obtained with the PACS10 gauge configurations generated at the lattice spacing of 0.085 fm (denoted as coarse lattice).~\cite{Shintani:2018ozy}.

This work uses 
the second ensemble of the PACS10 gauge configurations generated at
the lattice spacing of 0.063 fm (denoted as fine lattice) in order to 
investigate the finite lattice spacing effects on the nucleon form factors
toward the continuum limit. 
In a series of our studies,
we have retained some essential features, carried over from our earlier work:
(i) We perform \textit{fully dynamical lattice QCD simulation} with the stout-smeared $O(a)$-improved Wilson-clover quark action and the Iwasaki gauge action~\cite{Iwasaki:1983iya}.
(ii) The physical spatial volume is kept at about $(10\ \mathrm{fm})^4$ where the finite volume effect is sufficiently suppressed and furthermore the nonzero minimum value of the momentum transfer reaches about 
$q^2\sim 0.01\ \mathrm{GeV}^2$.
(iii) The quark masses are carefully tuned to the physical point, which indicates that our simulations are free from the chiral extrapolation.
(iv) For high statistics analysis, the all-mode-averaging (AMA) method~\cite{Bali:2009hu, Shintani:2014vja, Blum:2012uh, vonHippel:2016wid}, which is optimized by the deflation technique~\cite{Luscher:2007se} and implemented by multigrid bias correction by truncated solver method~\cite{Bali:2009hu}, is utilized to significantly reduce the computational cost.

These specific features enable us to overcome the systematic uncertainties due to chiral and infinite-volume extrapolations and approach the low $q^2$ region, which is essential to evaluate
the nucleon RMS radii from the nucleon form factors. 
However, since our previous study~\cite{Shintani:2018ozy} had been performed at a single lattice spacing, the uncertainty associated with the finite lattice spacing still remains.
Therefore, the main objective of this study is to investigate the remaining uncertainties associated with the finite lattice
spacing with respect to the nucleon form factors and associated RMS radii.

This paper is organized as follows.
In Sec.~\ref{sec:calculation_method},
we describe our method to calculate 
the nucleon form factors and their RMS radii 
from the nucleon two- and three-point correlation functions in lattice QCD simulation.
Definitions and notations for the nucleon form factors and their general properties are also summarized in this section.
In addition, we will explain the methodology to determine the RMS radii and magnetic moment, and the strategy to examine the systematic uncertainties of the excited-state contamination.
In Sec.~\ref{sec:simulation_details},
we present a brief description of our gauge configurations, which are a partial set of the PACS10 gauge configurations generated by the PACS Collaboration~\cite{Shintani:2019wai}.
The basic results obtained from the nucleon two-point function including the dispersion relation are also summarized in this section. 
In Sec.~\ref{sec:numerical_results_I}, 
the results for five form factors, $G_E$, $G_M$, $F_A$, $F_P$ and $G_P$ are presented. 
We then investigate the source-sink separation dependence of these form factors and three RMS radii (electric, magnetic and axial RMS radii).
Section \ref{sec:numerical_results_II} is devoted to a discussion of the
results of three form factors, $F_A$, $F_P$ and $G_P$ obtained in the axial-vector and pseudoscalar channels, which
are related to the axial Ward--Takahashi identity.
Finally, we close with summary and outlook with Sec.~\ref{sec:summary}.

In this paper,
the matrix elements are given in the Euclidean metric convention.
$\gamma_5$ is defined by $\gamma_5 \equiv \gamma_1\gamma_2\gamma_3\gamma_4=-\gamma_5^M$, which has the opposite sign relative to that in the Minkowski convention ($\vec{\gamma}^M=i\vec{\gamma}$ and $\gamma_0^M=\gamma_4$) adopted in the particle data group.
The sign of all the form factors is chosen to be positive. The Euclidean four-momentum squared $q^2$ corresponds to the spacelike momentum squared as $q^2_M=-q^2<0$ in Minkowski space.

\clearpage
\section{Calculation method}
\label{sec:calculation_method}

\subsection{General properties of nucleon form factors}
\label{sec:properties_of_ff}

In this paper, we would like to focus on
five target form factors: the electric ($G_E$), magnetic ($G_M$), axial ($F_A$), induced pseudoscalar ($F_P$) and pseudoscalar ($G_P$) form factors, which appear in the nucleon matrix elements of the vector , axial-vector and pseudoscalar currents
as below.

Let us consider, for example, the nucleon matrix elements of the weak current that can describe the neutron $\beta$ decay.
In addition to the standard beta-decay transition, which 
is described by the vector and axial-vector currents, we also include the nonstandard one as the pseudoscalar current as below.
The nucleon matrix element of a given quark bilinear operator as $J^{\mathcal{O}}_\alpha = \bar{u}\Gamma^\mathcal{O}_\alpha d$
with $\Gamma^\mathcal{O}_\alpha = \gamma_\alpha, \gamma_\alpha\gamma_5, \gamma_5$ for the vector $(V_\alpha)$, axial-vector $(A_\alpha)$ and pseudoscalar $(P)$ currents
have the following relativistically covariant decomposition in terms of the five different form factors: the vector ($F_V$), induced tensor
($F_T$), axial-vector ($F_A$), induced pseudoscalar ($F_P$) and pseudoscalar ($G_P$) as
\begin{align}
    \label{eq:nme_ff}
    \langle p(\bm{p})| V_\alpha(x) | n(\bm{p}^{\prime}) \rangle
    &=
    \bar{u}_p(\bm{p})
    \left(
    \gamma_\alpha F_V(q^2) + \sigma_{\alpha\beta}q_\beta F_T(q^2)
    \right)
    u_n(\bm{p}^{\prime}) e^{iq\cdot x},\\
    \label{eq:nme_fap}
    \langle p(\bm{p})| A_\alpha(x) | n(\bm{p}^{\prime}) \rangle 
    &=
    \bar{u}_p(\bm{p})
    \left(
    \gamma_\alpha\gamma_5 F_A(q^2) + iq_\alpha\gamma_5 F_P(q^2)
    \right)
    u_n(\bm{p}^{\prime}) e^{iq\cdot x},\\
    \label{eq:nme_gp}
    \langle p(\bm{p})| P(x) | n(\bm{p}^{\prime}) \rangle
    &=
    \bar{u}_p(\bm{p})
    \left(
    \gamma_5 G_P(q^2)
    \right)
    u_n(\bm{p}^{\prime}) e^{iq\cdot x},
\end{align}
where $|p(\bm{p})\rangle$ and $|n(\bm{p})\rangle$ are the proton ($p$) and neutron ($n$) ground state with the three-dimensional momentum $\bm{p}$.
In the above equation, the four-dimensional momentum transfer $q$ between the proton and neutron is given by $q=P^\prime-P$ with
$P=(E_p(\bm{p}), \bm{p})$ and $P^\prime=(E_n(\bm{p}^\prime), \bm{p}^\prime)$.

The vector part of the weak matrix elements
of neutron $\beta$ decay described by the $F_V$ and $F_T$ form factors is
related to the nucleon's electromagnetic matrix element via an isospin rotation, as long as the heavy-flavor contributions can be neglected under {\it the exact isospin symmetry}.
Suppose that the electromagnetic current can be expressed in terms of the \textit{up} and \textit{down} quark currents as $j^{\mathrm{e.m.}}_{\mu}=\frac{2}{3}\bar{u}\gamma_\alpha u - \frac{1}{3}\bar{d}\gamma_\alpha d$, neglecting the strange and heavier quarks.
Then the electromagnetic matrix elements of the proton and neutron 
are written by the proton's matrix elements of the up and down quark currents as
\begin{align}
    \langle p | 
    j^{\mathrm{e.m.}}_{\mu}
    | p \rangle
    & =
    \frac{2}{3}\langle p | \bar{u}\gamma_\alpha u | p \rangle - \frac{1}{3} \langle p | \bar{d}\gamma_\alpha d | p \rangle, \\
    \langle n | 
    j^{\mathrm{e.m.}}_{\mu}
    | n \rangle
    & =
     -\frac{1}{3}\langle p |\bar{u}\gamma_\alpha u | p \rangle + \frac{2}{3} \langle p | \bar{d}\gamma_\alpha d | p \rangle,
\end{align}
which lead to the following relation
\begin{align}
\langle p | 
    j^{\mathrm{e.m.}}_{\mu}
    | p \rangle - \langle n | 
    j^{\mathrm{e.m.}}_{\mu}
    | n \rangle
& = \langle p | \bar{u}\gamma_\alpha u - \bar{d}\gamma_\alpha d | p \rangle \\
& = \langle p |\bar{u}\gamma_\alpha d|n\rangle,
\end{align}
where in the second equality an isospin rotation is used to show a connection to the vector part of the weak matrix elements of the neutron $\beta$ decay. 
Therefore, the $F_V$ and $F_T$ form factors in neutron $\beta$ decay are related to the isovector part of the electromagnetic Dirac ($F_1$) and Pauli ($F_2$) form factors
\begin{align}
    F^v_1(q^2) & = F_V(q^2), \\
    F^v_2(q^2) & = 2M_NF_T(q^2),
\end{align}
where the nucleon mass $M_N$ is defined by the average of the proton and neutron masses.
The {\it isovector} form factor $F_1^v$ ($F_2^v$) is given by the difference between the Dirac (Pauli) form factors
of the proton and neutron as
\begin{align}
    F^v_l(q^2) = F^p_l(q^2) - F^n_l(q^2),
    \quad l=\{1,2\},
\end{align}
where the individual form factors $F_l^N$ ($N=p, n$)
are defined by
\begin{align}
\langle N(\bm{p})|j^{\mathrm{e.m.}}_{\alpha}(x)
|N(\bm{p}^\prime)\rangle
= \bar{u}_N(\bm{p})
    \left(
    \gamma_\alpha F_1^N(q^2) + \sigma_{\alpha\beta}\frac{q_\beta}{2M_N}F^N_2(q^2)
    \right)
    u_N(\bm{p}^{\prime}) e^{iq\cdot x},
\end{align}
where $q=P^\prime - P$ with $P=(E_N(\bm{p}),\bm{p})$
and $P^\prime=(E_N(\bm{p}^\prime),\bm{p}^\prime)$.

The electric ($G_E$) and magnetic ($G_M$) Sachs form factors are related to the Dirac ($F_1$)
and Pauli ($F_2$) form factors of the proton and neutron, individually as below
\begin{align}
    G^N_E(q^2)
    &=
    F^N_1(q^2) - \frac{q^2}{4M_N}F^N_2(q^2),\\
    G^N_M(q^2)    &=
    F^N_1(q^2) + F^N_2(q^2),
\end{align}
with $N=p, n$. 
$G_E(q^2)$ and $G_M(q^2)$ are relevant quantities to describe experimental data obtained from elastic electron-nucleon scattering experiments.
Even for the electric $G_E(q^2)$ and magnetic $G_M(q^2)$ Sachs form factors,
the isovector part is also given as
\begin{align}
    G^v_l(q^2) = G^p_l(q^2) - G^n_l(q^2),
    \quad l=\{E,M\}.
\end{align}

In this study, we primarily calculate $G^v_E(q^2)$ and $G^v_M(q^2)$ which can be evaluated only by the connected-type contribution as explained previously
and hence can be used for comparison with experiments,
while the electromagnetic form factors of the proton and neutron
are also evaluated separately without the disconnected-type contribution. 

The normalization of the electromagnetic form factors of the proton and neutron is given by the electric charge and the magnetic moment, which are defined as the electromagnetic form factors at the zero momentum transfer, $q^2=0$.
The electric charge and the magnetic moment are given as $G^p_E(0)=1$ and $G^p_M(0)=\mu_p=+2.7928473446(8)$ for the proton and $G^n_E(0)=0$ and $G^n_M(0)=\mu_n
=-1.9130427(5)$ for the neutron~\cite{ParticleDataGroup:2022pth}.

The axial-vector form factor $F_A(q^2)$ and the induced pseudoscalar form factor $F_P(q^2)$ can be extracted from the axial-vector part of the weak matrix elements of the neutron $\beta$ decay, which is associated with
the nucleon matrix element with the isovector axial-vector current $A_\alpha=
\bar{u}\gamma_\alpha\gamma_5 u - \bar{d}\gamma_\alpha\gamma_5 d$.
Especially the axial-vector coupling defined by $g_A=F_A(q^2=0)$ is experimentally well determined as $g_A=1.2754(13)$~\cite{ParticleDataGroup:2022pth}.
Therefore, we can use this quantity as a good reference for verifying the accuracy and reliability of our calculations.
The $q^2$-dependence of the axial form factor
can be directly compared with phenomenological values provided by the neutron $\beta$ decay. 
This indicates that first-principles calculations of the nucleon form factors provide useful information to understand the neutrino-nucleus interactions~\cite{Garvey:2014exa, NuSTEC:2017hzk, Meyer:2022mix}.
On the other hand, the induced pseudoscalar form factor has much less information in experiments~\cite{Bernard:1994pk, Gorringe:2002xx}.

In addition to the two form factors $F_A(q^2)$ and $F_P(q^2)$,
we also calculate the pseudoscalar form factor $G_P(q^2)$, which is also associated with
the nucleon matrix element with the isovector pseudoscalar current $P=
\bar{u}\gamma_5 u - \bar{d}\gamma_5 d$.
Recall that the $q^2$-dependence of these three form factors, $F_A(q^2)$, $F_P(q^2)$ and $G_P(q^2)$,
should be constrained by the generalized Goldberger--Treiman (GGT) relation~\cite{Weisberger:1966ip, Sasaki:2007gw}
\begin{align}
\label{eq:ggt}
    2M_NF_A(q^2) - q^2 F_P(q^2)
    =
    2mG_P(q^2)
\end{align}
with a degenerate up and down quark mass $m=m_u=m_d$, since
the GGT relation can be derived as a consequence of the axial Ward--Takahashi identity (AWTI): 
\begin{align}
\label{eq:awti}
\partial_\alpha A_\alpha(x)=2mP(x),
\end{align}
which is phenomenologically referred to as the partial conservation of the axial-vector current (PCAC). 
Therefore, it is important to evaluate each of the three form factors individually and then verify whether the GGT relation is satisfied among them. This is really a nontrivial check of 
the PCAC relation in terms of the nucleon form factors.

\subsection{Root-mean-square radius of the nucleon}
\label{sec:def_rms}

The root-mean-square (RMS) radius $R_l=\sqrt{\langle r^2_l \rangle}$,
which measures a typical size in the coordinate space is defined from the expansion of the normalized form factor $G_l(q^2)$ for $l=\{E,M,A\}$ 
in the powers of $q^2$:
\begin{align}
    G_l(q^2)
    &=G_l(0)
    \left(
    1-\frac{1}{6}\langle r^2_l\rangle q^2+
    \frac{1}{120}\langle r^4_l\rangle q^4
    +\cdots
    \right),
\end{align}
where the first coefficient determines the mean squared radius $\langle r^2_l \rangle$ that can be read off the slope of $G_l(q^2)$ at $q^2=0$ as
\begin{align}
    \langle r^2_{l} \rangle
    &=
    -\frac{6}{G_{l}(0)}
    \left.
        \frac{dG_{l}(q^2)}{d q^2}
    \right|_{q^{2}=0}.
\end{align}
Here we use the notation of $G_A\equiv F_A$ for the axial-vector form factor.

The $z$-expansion method,
which is known as a model independent analysis and has been widely used in the analyses of the form factors in both experiments and lattice calculations~\cite{Hill:2010yb, T2K:2023smv, Simons:2022ltq},
is mainly employed in this study~\footnote
{In Appendix~\ref{app:model-dep_anal},
we additionally present results obtained from the model dependent analyses with
the dipole form and the polynomial (linear or quadratic) forms, for comparisons.}.
In the $z$-expansion method, 
the given form factor $G(q^2)$ is fitted by the following functional form
\begin{align}
    \label{eq:z-expansion}
    G(q^2)
    &=
    \sum_{k=0}^{k_{\mathrm{max}}}c_kz(q^2)^k=
    {c_0+c_1z(q^2)+c_2z(q^2)^2+c_3z(q^2)^3}+\dots,
\end{align}
where a new variable $z$ is defined by a 
conformal mapping from $q^2$ as
\begin{align}
    \label{eq:conformal-map}
    z(q^2)
    =\frac{\sqrt{t_{\rm cut}+q^2}-\sqrt{t_{\rm cut}-t_0}}
    {\sqrt{t_{\rm cut}+q^2}+\sqrt{t_{\rm cut}-t_0}}
\end{align}
with $t_{\rm cut}=4m_\pi^2$ for $G=G_E$ and $G_M$, or with $t_{\rm cut}=9m_\pi^2$ 
for $G=F_A$. 
Since respective values of $t_{\rm cut}$ are associated with the two-pion continuum
or the three-pion continuum, the value of $m_\pi$ is set to be the simulated pion mass. 
A parameter $t_0$ can be taken arbitrarily within the range of $t_{\rm cut}>t_0$.
For a simplicity, $t_0=0$ is chosen in this study~\footnote{
The optimal choice of $t_0$ is given by
$t_0^{\rm opt}
=
t_\mathrm{cut}
\left(
1-\sqrt{1+q^2_{\mathrm{max}}/t_{\mathrm{cut}}}
\right)$ for minimizing the maximum size of $|z|^2$ when
the value of $q^2$ ranges from 0 to $q_{\rm max}^2$~\cite{Meyer:2016oeg}. 
However, the maximum of the momentum transfer $q^2_\mathrm{max}\approx 0.1\ \mathrm{GeV}^2$ used in this study is so small that the fit result is insensitive to the choice of either $t_0=0$ or $t_0=t_0^{\rm opt}$.
}.
The transformation~(\ref{eq:conformal-map}) maps the analytic domain inside a unit circle $|z|<1$ in the $z$ plane so that Eq.~(\ref{eq:z-expansion}) is supposed to be
a convergent Taylor series in terms of $z$. 
To achieve a model independent fit,
$k_{\mathrm{max}}$ that truncates an infinite series expansion in $z$ should be chosen to ensure
that terms $c_k z^k$ become numerically negligible for $k>k_{\mathrm{max}}$.

To check the stability of the fit results with a given $k_{\mathrm{max}}$, 
we use the singular value decomposition (SVD) algorithm to solve the least squares problem for high degree polynomials. 
We determine the optimal value of $k_{\mathrm{max}}$ in each channel so that the fitting is stable against the variation 
of $k_{\mathrm{max}}$ with a reasonable $\chi^2/$d.o.f. and does not change the value of the RMS radius,
which is given by $R_l=\sqrt{-6\left(c_1/c_0\right)/\left(4t_{\mathrm{cut}}\right)}$.

Recall that the normalization of the neutron electric form factor is $G^n_E(0)=0$ and $\langle r_E^2 \rangle$ is negative.
Therefore,
the constant term in Eq.~(\ref{eq:z-expansion}) is fixed to zero as $c_0=0$ 
during the fitting, 
and the neutron's mean-square radius is determined as  $\langle (r^n_E)^2 \rangle = -6(c_1)/(4t_{\mathrm{cut}})$.

\subsection{Correlation functions with momentum}
\label{sec:corrlation_functions}
In this study,
the exponentially smeared quark operator $q_S(t, \bm{x})$ with the Coulomb gauge fixing is
used for the construction of the nucleon interpolating operator as well as a local quark
operator $q(t, \bm{x})$. 
The smeared quark operator is given by a
convolution of the local quark operator with a smearing function $\phi(\bm{x},\bm{y})$
as
\begin{align}
    \label{eq:smear_qua}
    q_{S}(t,\bm{x})
    =
    \sum_{\bm{y}}
    \phi(\bm{x},\bm{y})
    q(t,\bm{y}),
\end{align}
where the color and Dirac indices are omitted. 

A smearing function 
$\phi(\bm{x},\bm{y})$ is given by an isotropic function of $r=|\bm{x}-\bm{y}|$
in a linear spatial extent of $L$ as the following form
\begin{align}
    \label{eq:exp_smear_nuc}
    \phi(\bm{x},\bm{y})=\phi(r)=
\begin{cases}
    1 & (r=0) \\
    A\mathrm{e}^{-Br} & (r<L/2), \\
    0 & (r\ge L/2)
\end{cases}
\end{align}
with two smearing parameters $A$ and $B$. 
This procedure does not preserve the full gauge invariance of the 
hadron two-point correlation functions consisting of the spatially smeared quark operators, so that 
the Coulomb gauge fixing is necessary. 
Let us define the nucleon two-point function with the local nucleon sink operator $N_L(t,\bm{p})$ 
located at $t=t_{\rm sink}$ 
and the nucleon source operator $\bar{N}_X(t,\bm{p})$
located at $t=t_{\rm src}$ for either smeared ($X=S$) or local ($X=L$) cases as
\begin{align}
    \label{eq:two_pt_func}
    C_{XS}(t_{\rm sink}-t_{\rm src}; \bm{p})
    =
    \frac{1}{4}\mathrm{Tr}
    \left\{
        \mathcal{P_{+}}\langle N_X(t_{\mathrm{sink}}; \bm{p})
        \overline{N}_S(t_{\mathrm{src}}; -\bm{p})  \rangle
    \right\}\ \mathrm{with}\ X=\{S,L\},
\end{align}
\noindent
where the nucleon operator with a three-dimensional momentum $\bm{p}$ is given for the proton state by
\begin{align}
    \label{eq:interpolating_op}
N_L(t,\bm{p})&
    =\sum_{\bm{x}}e^{-i\bm{p}\cdot\bm{x}}\varepsilon_{abc}
    \left[
        u^{T}_{a}(t,\bm{x})C\gamma_5d_b(t,\bm{x})
    \right]
    u_c(t,\bm{x})
\end{align}
with the charge conjugation matrix, $C=\gamma_4\gamma_2$. The superscript $T$ denotes a transposition, while the indices $a$, $b$, $c$ and
$u$, $d$ label the color and the flavor, respectively.
The smeared source operator $N_S(t,\bm{p})$ is the same
as the local one $N_L(t,\bm{p})$, but all the quark operator
$u$, $d$ are replaced by the smeared ones defined in Eq.~(\ref{eq:smear_qua}). 
The lattice momentum is defined as $\bm{p}=2\pi/(La)\times \bm{n}$ with a vector of integers $\bm{n}\in Z^3$ and $L$
 the number of the spatial lattice sites. A projection operator $\mathcal{P_{+}}=(1+\gamma_4)/2$ can eliminate the unwanted contributions from the opposite-parity state for $|\bm{p}|=0$~\cite{Sasaki:2001nf, Sasaki:2005ug}.

In order to calculate the isovector nucleon form factors, 
we evaluate the nucleon three-point functions, which are constructed with the spatially smeared sources and
sink operators of the nucleon as 
\begin{align}
    \label{eq:three_pt_func}
    C^{k}_{\mathcal{O}_\alpha}(t; \bm{p}^{\prime}, \bm{p})
    =
    \frac{1}{4} \mathrm{Tr}
    \left\{
        \mathcal{P}_{k}\langle N_S(t_{\mathrm{sink}}; \bm{p}^{\prime})
        J^{\mathcal{O}}_{\alpha}(t; \bm{p}-\bm{p}^{\prime})
        \overline{N}_S(t_{\mathrm{src}}; -\bm{p})  \rangle
    \right\}
\end{align}
\noindent
with a given isovector bilinear current operator $J^{\mathcal{O}}_\alpha$ defined at Eqs.~(\ref{eq:nme_ff})-(\ref{eq:nme_gp}).
In the above equation, 
$\mathcal{P}_k$ denotes the projection operator to
extract the form factors for unpolarized case $\mathcal{P}_k=\mathcal{P}_t\equiv\mathcal{P}_{+}\gamma_4$ 
and polarized case (in $z$ direction) $\mathcal{P}_k=\mathcal{P}_{5z}\equiv\mathcal{P}_{+}\gamma_5\gamma_z$.

Recall that in the case of the exact SU(2) isospin symmetry ($m_u=m_d$),
the nucleon three-point functions with the isovector currents do not 
receive any contributions from
the disconnected diagrams of all quark flavors thanks to their mutual cancellations.
In this paper, we present results for the isovector nucleon form factors 
that can be determined solely by the connected-type contribution in 2+1 flavor QCD,
while the isoscalar ones require computation of both connected and disconnected-type contributions.
Since the connected parts are in general precisely computed rather than the disconnected parts,
we still have a good opportunity for accurate prediction of the nucleon-neutrino elastic scattering, which is governed by the isovector interaction.

\subsection{Extraction of nucleon form factors}
\label{sec:ext_of_ff}
In a conventional way to extract the nucleon form factors, 
we introduce the following ratio constructed by
an appropriate combination of two-point functions~(\ref{eq:two_pt_func}) and three-point function~(\ref{eq:three_pt_func})~\cite{Hagler:2003jd, Gockeler:2003ay} with a fixed source-sink separation ($t_{\mathrm{sep}}\equiv t_{\mathrm{sink}}-t_{\mathrm{src}}$)
as
\begin{align}
\mathcal{R}_{\mathcal{O}_\alpha}^{k}(t; \bm{p}^{\prime}, \bm{p}) = \frac{C_{\mathcal{O}_\alpha}^{k}(t; \bm{p}^{\prime}, \bm{p})}{C_{SS}(t_{\mathrm{sink}}-t_{\mathrm{src}}; \bm{p}^{\prime})}
\sqrt{\frac{C_{LS}(t_{\mathrm{sink}}-t; \bm{p}) C_{SS}(t-t_{\mathrm{src}}; \bm{p}^{\prime}) C_{L S}(t_{\mathrm{sink}}-t_{\mathrm{src}}; \bm{p}^{\prime})}{C_{L S}(t_{\mathrm{sink}}-t; \bm{p}^{\prime}) C_{SS}(t-t_{\mathrm{src}}; \bm{p}) C_{LS}(t_{\mathrm{sink}}-t_{\mathrm{src}}; \boldsymbol{p})}},
\label{eq:ratio_3pt_2pt}
\end{align}
which is a function of the current operator insertion time $t$ at the given values of momenta $\bm{p}$ and $\bm{p}^{\prime}$
for the initial and final states of the nucleon.
In this work, we consider only the rest frame of the final state with $\bm{p}^\prime=\bm{0}$,
which leads to the condition of 
$\bm{q}=\bm{p}-\bm{p}^{\prime}=\bm{p}$. 
Therefore, 
the squared four-momentum transfer is given by $q^2=2M_N(E_N(\bm{q})-M_N)$ where $M_N$ and $E_N(\bm{q})$ represent the nucleon mass and energy with the momentum $\bm{q}$. 
In this kinematics, $\mathcal{R}_{\mathcal{O}_\alpha}^{k}(t; \bm{p}^{\prime},
\bm{p})$ is rewritten by a simple notation  $\mathcal{R}_{\mathcal{O}_\alpha}^{k}(t; \bm{q})$.

The ratio $\mathcal{R}_{\mathcal{O}_\alpha}^{k}(t; \bm{q})$  with appropriate combinations of the projection operator
$\mathcal{P}_k$ ($k=t$, $5z$) and the $\alpha$ component of the
isovector bilinear bare current $\widetilde{J}^{\mathcal{O}}_\alpha$
gives the following asymptotic values including the respective form factors
in the asymptotic region 
($t_{\mathrm{sep}}/a\gg (t-t_{\mathrm{src}})/a \gg 1$):
\begin{align}
    \label{eq:ge_def}
    \mathcal{R}^t_{V_4}(t;\bm{q})
    &=
    \sqrt{\frac{E_N+M_N}{2E_N}}\widetilde{G}^{v}_E(q^2),\\
    \label{eq:gm_def}
    \mathcal{R}^{5z}_{V_i}(t;\bm{q})
    &=
    \frac{-i\varepsilon_{ij3}q_j}{\sqrt{2E_N(E_N+M_N)}}\widetilde{G}^{v}_M(q^2),\\
    \label{eq:fa_def}
    \mathcal{R}^{5z}_{A_i}(t;\bm{q})
    &=
    \sqrt{\frac{E_N+M_N}{2E_N}}
    \left[
        \widetilde{F}_A(q^2)\delta_{i3}-\frac{q_iq_3}{E_N+M_N}\widetilde{F}_P(q^2)
    \right],\\
    \label{eq:fa4_def}
    \mathcal{R}^{5z}_{A_4}(t;\bm{q})
    &=
    \frac{iq_3}{\sqrt{2E_N(E_N+M_N)}}
    \left[
        \widetilde{F}_A(q^2)-(E_N-M_N)\widetilde{F}_P(q^2)
    \right],\\
    \label{eq:gp_def}
    \mathcal{R}_{P}^{5z}(t; \boldsymbol{q}) 
    & =
    \frac{iq_3}{\sqrt{2 E_{N}(E_N+M_N)}} \widetilde{G}_{P}\left(q^{2}\right),
\end{align}
where the nucleon energy $E_N(\bm{q})$ is simply abbreviated as $E_N$, and the indices $i$ and $j$ run the three spatial directions.
If the condition $t_{\mathrm{sep}}/a\gg (t-t_{\mathrm{src}})/a \gg 1$ is satisfied, 
the five target quantities: the electric ($G_E$), magnetic ($G_M$), axial ($F_A$), induced pseudoscalar ($F_P$) and pseudoscalar ($G_P$) form factors can be read off from an asymptotic plateau
of the ratio $\mathcal{R}_{\mathcal{O}_\alpha}^{k}(t; \bm{q})$,
being independent of the choice of $t_{\mathrm{sep}}$. 
This approach is hereafter referred to as the standard plateau method.
Following our previous work~\cite{Shintani:2018ozy},
we simply use the uncorrelated constant fit to evaluate the plateau values from the ratio, since a fixed fit range can be maintained for all seven momentum transfers.

Finally,
we recall that the quark local currents on the lattice receive finite renormalizations relative to their continuum counterparts in general.
The renormalized values of the form factors require the renormalization factors $Z_O\ (O=V,A,P)$:
\begin{align}
  G^v_E(q^2) &= Z_V\widetilde{G}^v_E(q^2),\\
  G^v_M(q^2) &= Z_V\widetilde{G}^v_M(q^2),\\
  F_A(q^2) &= Z_A\widetilde{F}_A(q^2),\\
  F_P(q^2) &= Z_A\widetilde{F}_P(q^2),\\
  G_P(q^2) &= Z_P\widetilde{G}_P(q^2),
\end{align}
where the renormalization factors are defined through the renormalization of the quark currents
$J^{\mathcal{O}}_\alpha=Z_O\widetilde{J}^{\mathcal{O}}_\alpha$~\footnote{
Hereafter, the form factors and currents with and without tilde indicate bare and renormalized ones. 
}.
The renormalization factors $Z_V$ and $Z_A$ are scale independent, while $Z_P$ depends on the renormalization scale. 
In order to compare the experimental values, 
four form factors, $G_E$, $G_M$, $F_A$ and $F_P$, 
will be properly renormalized with $Z_V$ and $Z_A$,
which are determined by the Schr\"odinger functional method as given in Appendix~\ref{app:sf_scheme}, while the pseudoscalar form factor presented in this study is only bare quantity as indicated by $\widetilde{G}_P$.

\subsection{Test for the PCAC relation using the nucleon}
\label{sec:pcac_test}

As described in Sec.~\ref{sec:properties_of_ff}, 
the three form factors $F_A(q^2)$, $F_P(q^2)$ and $G_P(q^2)$ 
are related to each other through the GGT relation~(\ref{eq:ggt})
and are not independent. Therefore, if 
the GGT relation is well satisfied in our simulations, 
it offers a way to define a {\it bare} quark mass~\footnote{
The axial Ward--Takahashi identity
on the lattice may be represented by
$\partial_\alpha A_{\alpha}(x)=2mP(x)$ with 
the renormalized currents $A_{\alpha}=Z_A \widetilde{A}_{\alpha}$ and $P=Z_P \widetilde{P}$. Thus, $m$ represents the bare value unless $P$ is renormalized.}
by the following 
specific ratio
\begin{align}
\label{eq:m_awti_ggt}
m_{\mathrm{GGT}}^{\mathrm{nucl}}
=\frac{2M_NF_A(q^2) - q^2 F_P(q^2)}{2\widetilde{G}_P(q^2)},
\end{align}
which may have no apparent $q^2$-dependence~\cite{Sasaki:2007gw}.
As reported in our previous studies at the coarse lattice spacing~\cite{{Ishikawa:2018rew},{Shintani:2018ozy}}, the quark mass
$m_{\mathrm{GGT}}^{\mathrm{nucl}}$ defined by Eq.~(\ref{eq:m_awti_ggt}) is roughly 3 times larger
than a (bare) quark mass (hereafter denoted as $m_{\rm PCAC}^{\mathrm{pion}}$) obtained from the pion two-point correlation
functions with the PCAC relation. We concluded that both of $F_P(q^2)$ and $\widetilde{G}_P(q^2)$ significantly suffer from the excited-state contamination, that induces the distortion of their pion-pole structures~\cite{{Ishikawa:2018rew},{Shintani:2018ozy}}.  

To verify the PCAC relation using the nucleon, we may directly use the nucleon three-point correlation functions
instead of the three form factors, following Ref.~\cite{Bali:2018qus} 
\begin{align}
\label{eq:m_awti_pcac}
m_{\mathrm{PCAC}}^{\mathrm{nucl}}
=
\frac{
Z_A\partial_{\alpha} C^{5z}_{A_{\alpha}}(t;\bm{q})
}{
2 C^{5z}_{P}(t;\bm{q})
}
=
\frac{1}{2}
\frac{
\langle N_S(t_{\mathrm{sink}}) \partial_{\alpha} A_{\alpha}(t)
\overline{N}_S(t_{\mathrm{src}}) \rangle
}{
\langle N_S(t_{\mathrm{sink}}) \widetilde{P}(t) \overline{N}_S(t_{\mathrm{src}}) \rangle
}
,
\end{align}
which does not require the spectral decomposition.
In other words, the value of $m_{\mathrm{PCAC}}^{\mathrm{nucl}}$ can be evaluated at the level of the nucleon three-point correlation function without isolating the ground-state contribution
from the excited-state contributions, similar to the determination of $m_{\rm PCAC}^{\mathrm{pion}}$ from the pion two-point correlation functions.

In Eq.~(\ref{eq:m_awti_pcac}), the derivatives of the nucleon three-point function 
with respect to the coordinate are evaluated by
\begin{align}
\label{eq:del_three-pt_time}
\partial_4 C^{5z}_{A_4}(t;\bm{q})
& =
\frac{1}{2a}
\left\{
C^{5z}_{A_4}(t+a;\bm{q})
-C^{5z}_{A_4}(t-a;\bm{q})
\right\}
\end{align}
for the time component and 
\begin{align}
\label{eq:del_three-pt_space}
\partial_k C^{5z}_{A_k}(t;\bm{q})
& =
iq_k C^{5z}_{A_k}(t;\bm{q})
\end{align}
for the spatial components ($k=1,2,3$). Here, we adopt the naive discrete momentum $q_k=\frac{2\pi}{aL}n_k$ ($n_k=0, 1, 2, \cdots, (L-1)$), while the lattice discrete momentum $q_k=\frac{1}{a}\sin[\frac{2\pi}{L}n_k]$ is adopted in the original proposal~\cite{Bali:2018qus}. This is simply because we would like to
treat the momentum in a manner equivalent to the analysis for the nucleon 
form factors which are extracted from the common three-point functions.
Indeed, there is no difference in either case at low momenta.~\footnote{
The momentum region used in study is a significantly lower region thanks to the large spatial extent of $La\sim 10$ fm.}

Both $m_{\mathrm{GGT}}^{\mathrm{nucl}}$ and $m_{\mathrm{PCAC}}^{\mathrm{nucl}}$
can be regarded as a bare quark mass defined through the axial Ward--Takahashi identity as long as they exhibit
$q^2$-independent behavior as a function of $q^2$. 
Therefore, it is worth comparing the quark masses by these two definitions. 
If a difference is observed between $m_{\mathrm{GGT}}^{\mathrm{nucl}}$ and $m_{\mathrm{PCAC}}^{\mathrm{nucl}}$, 
it would confirm that the main reason why $m_{\mathrm{GGT}}^{\mathrm{nucl}}$ is overestimated compared to $m_{\mathrm{PCAC}}^{\mathrm{pion}}$ is due to the excited-state contamination. On the other hand, if $m_{\mathrm{PCAC}}^{\mathrm{nucl}}$ coincides
with $m_{\mathrm{PCAC}}^{\mathrm{pion}}$, we justify that 
${\cal O}(a)$ improvement of the axial-vector current 
$\widetilde{A}^{\mathrm{imp}}_\alpha = \widetilde{A}_\alpha+c_Aa\partial_\alpha \widetilde{P}$
does not help to solve the discrepancy between $m_{\mathrm{GGT}}^{\mathrm{nucl}}$ and $m_{\mathrm{PCAC}}^{\mathrm{nucl}}$. This is simply because the second term
of $\partial_\alpha \widetilde{P}$ causes a momentum dependence on $m_{\mathrm{PCAC}}^{\mathrm{nucl}}$, though
it is not the case for $m_{\mathrm{PCAC}}^{\mathrm{pion}}$ determined by the {\it zero-momentum projected} two-point functions of the pion.

\clearpage
\section{Simulation details}
\label{sec:simulation_details}
\subsection{PACS10 configurations on a $160^4$ lattice}

In this paper, we use the second PACS10 ensemble, which is
a set of gauge configurations generated by the PACS Collaboration 
with $L^3 \times T=160^3 \times 160$ lattice
and physical light quark masses at the second gauge coupling $\beta=2.00$ corresponding to the lattice spacing of
$a=0.06343(14)$ fm [$a^{-1} = 3.1108(70)$ GeV]~\cite{Shintani:2019wai, Tsuji:2022kjt} 
using the six stout-smeared $O(a)$-improved Wilson quark action and the Iwasaki gauge action~\cite{Iwasaki:1983iya}. 
The stout-smearing parameter is set to $\rho=0.1$~\cite{Morningstar:2003gk}.
The improved coefficient, $c_{\mathrm{SW}}=1.02$, is nonperturbatively
determined using the Schr\"odinger functional (SF) scheme~\cite{Taniguchi:2012gew}. 
The hopping parameters of $(\kappa_{ud},\kappa_{s})=(0.125814, 0.124925)$ are 
carefully chosen to be almost at the physical point. 
The scale is determined from the $\Xi$ baryon mass input $M_\Xi = 1.3148$ GeV ~\cite{Shintani:2019wai}.
A brief summary of the simulation parameters is given in Table~\ref{tab:simulation_details}.

The previous works by the PACS Collaboration showed that 
the finite-size effects on both the nucleon mass~\cite{PACS:2019ofv} 
and the nucleon matrix elements~\cite{Ishikawa:2021eut}
are negligible on two lattice volumes (linear spatial extents of 10.9 fm and 5.5 fm) at the coarse lattice spacing of $a=0.08520(16)$ fm. Therefore,
lattice QCD simulations with the spatial size more than 10 fm using the PACS10 configurations provide us very unique opportunity to explore the nucleon structure
without any serious finite-size effect. Especially, the large spatial volume of 
$(10~\mathrm{fm})^3$ allows us to investigate the form factors in the small
 momentum transfer region. The lowest nonzero momentum transfer 
 reaches the value of $q^2=0.015\ \mathrm{(GeV)}^2$, which is smaller than
 $m_\pi^2$ even at almost physical pion mass ($m_\pi\approx 138$ MeV). 
 
We use 19 gauge configurations separated by 5 molecular dynamics trajectories. Since there are four choices for a temporal direction on a $160^4$ lattice, we rotate the temporal direction using hypercubic symmetry of each gauge configuration and then increase the total number of measurements by a factor of four to treat them as 76 gauge configurations in total~\footnote{
Alternatively, the averages of the four data sets can be combined into one and analyzed as 19 statistics. We have confirmed that there is no significant difference between the two analyses. 
}.
The statistical errors are estimated by the single elimination jackknife method~\footnote{There is no significant difference in bin size ranging from 1 to 5 in the jackknife analysis of the nucleon two-point functions.
}.

%
%
\begin{table*}[ht]
\caption{
Parameters of the second PACS10 ensemble. See Refs.~\cite{Shintani:2019wai} for further details.
\label{tab:simulation_details}}
\begin{ruledtabular}
\begin{tabular}{cccccccc}
 $\beta$ &$L^3\times T$ & $a^{-1}$ [GeV] &  $\kappa_{ud}$ & $\kappa_{s}$ &$c_{\mathrm{SW}}$ &  $m_\pi$ [GeV] \\
          \hline
     2.00& $160^3\times 160$ &3.1108(70)& 0.125814 & 0.124925 & 1.02 & 0.138\\
\end{tabular}
\end{ruledtabular}
\end{table*}

\subsection{Computational and technical details}

In this study,
we compute the correlation functions multiple times 
with respect to the geometrical symmetries of the lattice at a given configuration.
The computational cost is significantly reduced by adopting
the all-mode-averaging (AMA) method~\cite{Blum:2012uh,Shintani:2014vja,vonHippel:2016wid, Bali:2009hu} with the deflated Schwarz Alternating Procedure  (SAP)~\cite{Luscher:2003qa} and 
Generalized Conjugate Residual (GCR) \cite{Luscher:2007se}
for the measurements as shown in our previous works~\cite{{Shintani:2018ozy},{Ishikawa:2021eut}}.
We compute the combination of the correlation function with high-precision $O^{(\mathrm{org})}$ 
and low-precision $O^{(\mathrm{approx})}$ as
\begin{align}
    O^{(\mathrm{AMA})} 
    =
    \frac{1}{N_{\mathrm{org}}}\sum^{N_{\mathrm{org}}}_{f\in G}
    \left(O^{(\mathrm{org})f} - O^{(\mathrm{approx})f} \right)
    +
    \frac{1}{N_{G}}\sum^{N_{G}}_{g\in G}O^{(\mathrm{approx})g},
\end{align}
where the superscripts $f, g$ denote the transformation under the lattice symmetry $G$. 
In our calculations, it is translational symmetry, e.g., changing the position of the source operator
as in Refs.~\cite{Ishikawa:2018rew,  Ishikawa:2018jee, PACS:2019ofv, PACS:2019hxd}. $N_{\mathrm{org}}$ and $N_{G}$ are the numbers for $O^{(\mathrm{org})}$ and $O^{(\mathrm{approx})}$, respectively. The numbers and the stopping conditions of the quark propagator for the high- and low-precision measurements are summarized in Table~\ref{tab:measurements}. 

In calculation of the nucleon two- and three-point functions, 
we use the same quark action 
as in the gauge configuration generation with the hopping parameter 
$\kappa=\kappa_{ud}=0.125814$ for the degenerated up-down quarks, 
the improved coefficient, $c_{\mathrm{SW}}=1.02$ and 
six steps of stout-smearing to the link variables.
The periodic boundary condition in all the temporal 
and spatial directions is adopted in the quark propagator calculation.

The quark propagator is calculated using the exponential smeared source (sink) with the Coulomb gauge fixing. 
The smearing parameters for the quark propagator 
defined in Eq.~(\ref{eq:exp_smear_nuc}) are chosen 
as $(A,B)=(1.2, 0.11)$, which optimize
the effective mass plateau for the smear-local case.

The nucleon three-point functions are calculated using the sequential source method with a fixed
source-sink separation~\cite{Martinelli:1988rr, Sasaki:2003jh}. 
This method requires the sequential quark propagator for each choice of a projection operator
$\mathcal{P}_k$ regardless of the types of current $J^{O}_\alpha$. 

As for the source-sink separation of $t_{\mathrm{sep}}$ (denoted as 
$t_{\mathrm{sep}}=t_{\mathrm{sink}}-t_{\mathrm{src}}$), we use three variations of $t_{\mathrm{sep}}/a=\{13,16,19\}$ as summarized in Table~\ref{tab:measurements}.
We investigate the effects of the excited-state contamination by varying $t_{\mathrm{sep}}$ from 0.82
to 1.20 fm in the standard plateau method that was explained in Sec.~\ref{sec:ext_of_ff}.
In this study, for nonzero spatial momentum, we choose the seven variations of $q^2\neq 0$ as listed in Table~\ref{tab:qsq_list}.

The renormalization constants for vector and axial-vector currents, $Z_{O}(O=V,A)$ are obtained with the Schr\"odinger functional (SF) scheme at vanishing quark mass, 
which is described in Appendix~\ref{app:sf_scheme}.
The resultant values are $Z_{V}=0.96677(41)(316)$ and $Z_{A}=0.9783(21)(81)$, where the first error represents the statistical one and the second error represents the systematic one that is evaluated from the difference between the results given by two volumes, $12^3$ and $16^3$.
However,
the second errors are simply ignored in the later analysis,
since we choose the larger volume to set the physical scale.

%
%
\begin{table*}[ht]
    \caption{
    Details of the measurements: the spatial extent ($L$), 
    time separation ($t_{\mathrm{sep}}$),
    the stopping condition of quark propagator in the high- and low-precision calculations ($\epsilon_{\mathrm{high}}$ and $\epsilon_{\mathrm{low}}$), the number of measurements for 
    the high- and low-precision calculations ($N_{\mathrm{org}}$ and $N_{G}$), the number of configurations ($N_{\mathrm{conf}}$) and the total number of the measurements ($N_{\mathrm{meas}}=N_{G}\times N_{\mathrm{conf}}$), respectively. 
    \label{tab:measurements}}
\begin{ruledtabular}
\begin{tabular}{cccccccccc}
$L$ &  $t_{\mathrm{sep}}$ & $\epsilon_\mathrm{high}$ & $\epsilon_\mathrm{low}$& $N_{\mathrm{org}}$
    & $N_{G}$ & $N_{\mathrm{conf}}$ & $N_{\mathrm{meas}}$ & Fit range\\
\hline
160 & 13& $10^{-8}$& ---\footnotemark[1]\footnotetext[1]{The low-precision calculations use a fixed number of iterations for the stopping condition as $6$ GCR iterations using $10^4$ SAP domain size with $40$ deflation fields.}& 1& 64& 76& 4,864 & [4:8]\\
    & 16& $10^{-8}$& ---\footnotemark[1] & 3& 192& 76& 14,592& [6:10]\\
    & 19& $10^{-8}$& ---\footnotemark[1] & 4& 768& 76& 58,368& [7:11]\\
\end{tabular}
\end{ruledtabular}
\end{table*}
%

%
%
\begin{table*}[ht]
    \caption{Choices for the nonzero spatial momenta: $\bm{q}=\frac{2\pi}{160a}\times \bm{n}$. The bottom row shows the degeneracy of $\bm{n}$ due to the permutation symmetry between $\pm x,\pm y,\pm z$ directions.
\label{tab:qsq_list}}
\begin{ruledtabular}
\begin{tabular}{ccccccccc}
    Label& Q0& Q1& Q2& Q3& Q4& Q5& Q6& Q7 \\
          \hline
    $\bm{n}$& (0,0,0)& (1,0,0)& (1,1,0)& (1,1,1)& (2,0,0)& (2,1,0)& (2,1,1)& (2,2,0)\\
    $|\bm{n}|^2$ & 0& 1& 2& 3& 4& 5& 6& 8 \\
    Degeneracy& 1& 6& 12& 8& 6& 24& 24& 12
\end{tabular}
\end{ruledtabular}
\end{table*}

\subsection{Nucleon spectra and the dispersion relation}
\label{sec:disp}

\begin{figure*}
\centering
\includegraphics[width=0.8\textwidth,bb=0 0 792 612,clip]{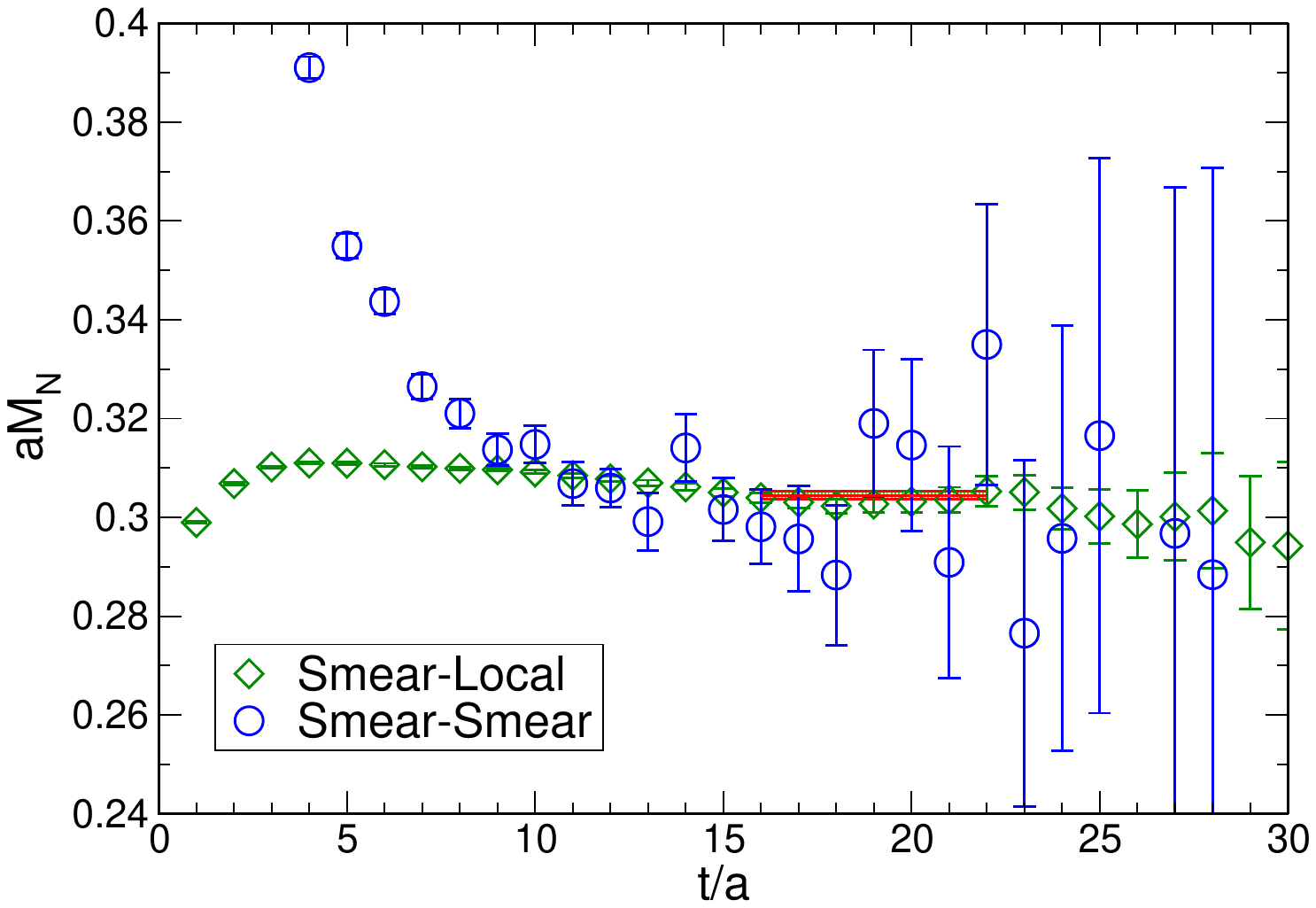}
\caption{Nucleon effective mass plot for 
the smear-local (diamond symbols, left panel) and the smear-smear (circle symbols, right panel) operators 
in the nucleon two-point functions. The three red lines represent the fit result of the smear-local operators in the range of $t/a=16-22$ with one-standard error band.}
\label{fig:effm_Apr22}
\end{figure*}

Figure~\ref{fig:effm_Apr22} shows the nucleon effective mass plot with $|\bm{p}|=0$.
We compute two types of nucleon two-point functions.
The smear-smear denotes that both of the source and sink operators 
in the nucleon two-point function are exponentially smeared.
On the other hand, the smear-local means that only the source operator is smeared, while the local one is used for the sink operator.
We observe the good plateau for $t/a\ge16$ in the smear-local effective mass plot
with our choice of smearing parameters.
Thus, 
the correlated single-exponential fit
with the range of $t/a=16-22$ is used for the smear-local nucleon two-point function in order to measure the nucleon rest mass $M_N$.
The nucleon mass is obtained as
\begin{align}
    aM_N = 0.3045(8), 
    \quad
    M_N = 0.9472(34)\ \mathrm{GeV,}
\end{align}
where the error is statistical only.
This value is slightly heavier but very close to
the experimental value of
$M_N^{\mathrm{exp}}=0.938918754(5)\ \mathrm{GeV}$, {which is given by the average of} the proton and neutron masses.

%
%

\begin{figure*}
\centering
\includegraphics[width=0.8\textwidth,bb=0 0 792 612,clip]{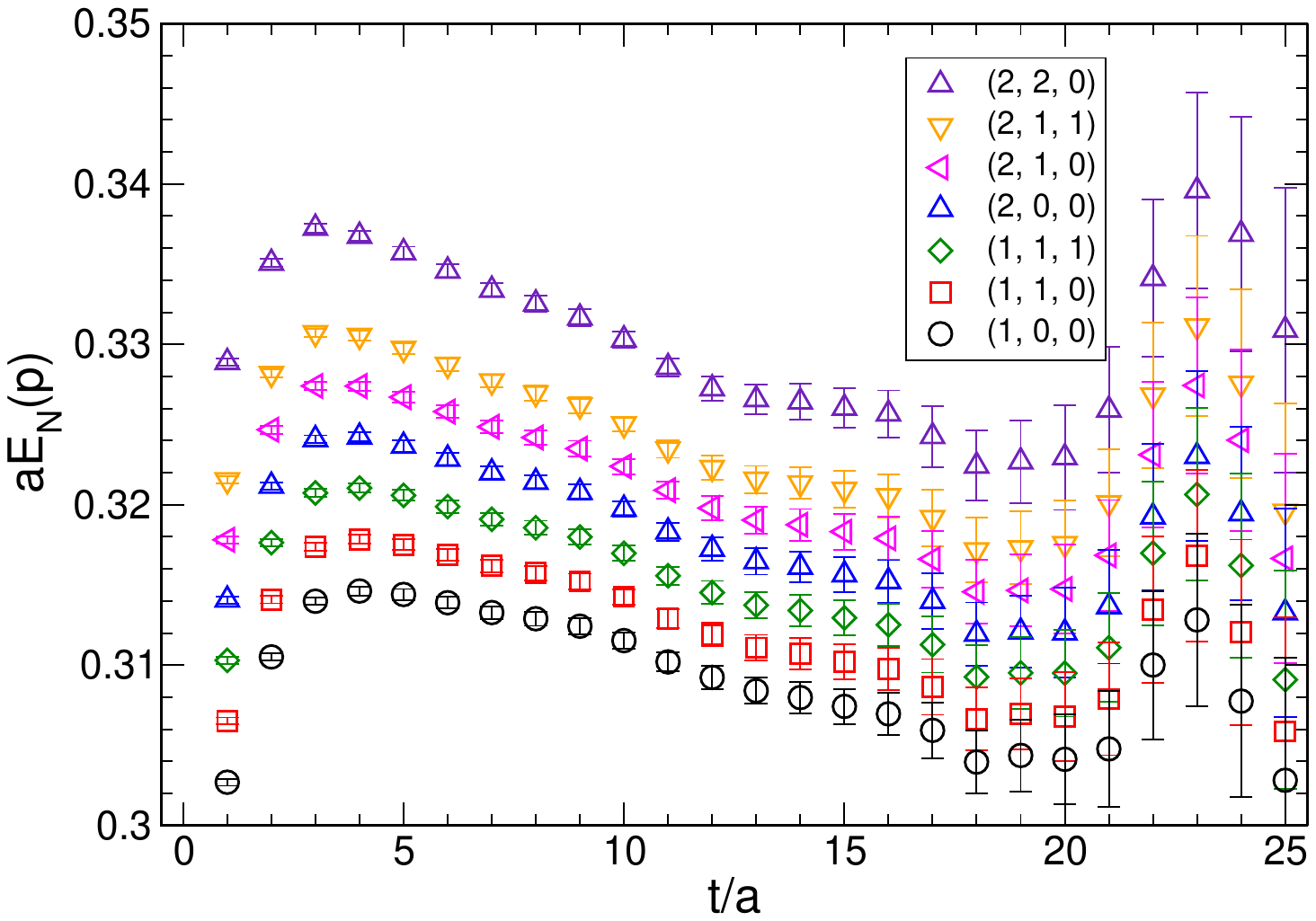}
\caption{
Nucleon effective energy plot for the momentum projected two-point function with the smear-local operators. 
}
\label{fig:NaiveE_N}
\end{figure*}

We also measure the nucleon energies, $E_N(\bm{p})$, from the smear-local 
nucleon two-point functions for all finite momenta $\bm{p}$.
In Fig.~\ref{fig:NaiveE_N}, we show the effective energy plot
for the momentum projected nucleon two-point function in the smear-local case. 
Since the smearing parameters are optimized for the zero-momentum case,
the suppression of the excited-state contamination near the source operator 
gets worse with higher momentum. However, the plateau behaviors are commonly shown at least in the region of $t/a\ge 16$ as well as the zero momentum. Therefore, 
the nucleon energies $E_N(\bm{p})$ are evaluated in the same way as the rest mass of the nucleon.

%
%

\begin{figure*}
\centering
\includegraphics[width=0.8\textwidth,bb=0 0 792 612,clip]{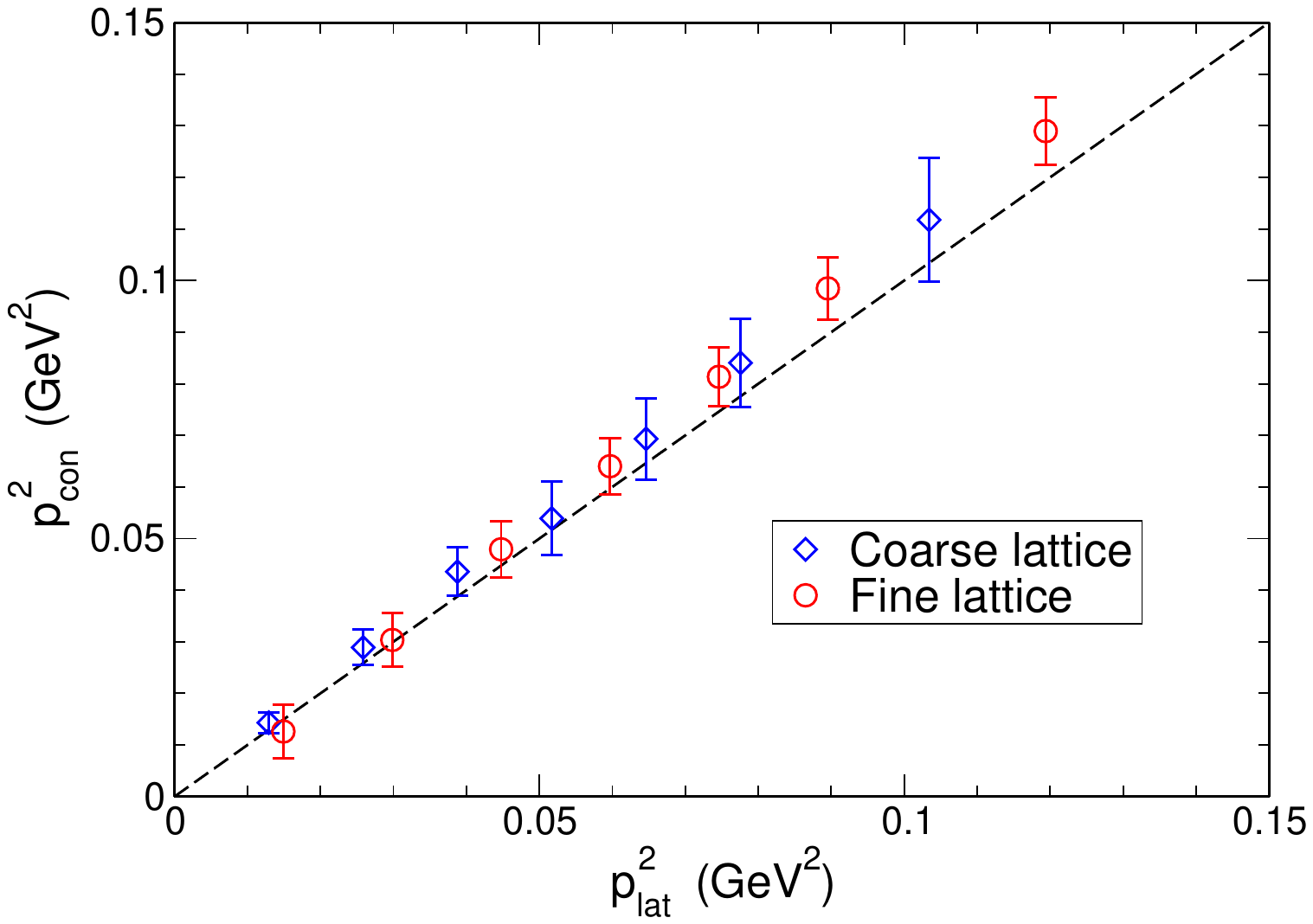}
\caption{
Check of the dispersion relation 
for the nucleon by using the measured values of $E_N(\bm{p})$.
The variables $p^2_{\rm con}$ and $p^2_{\rm lat}$ appearing on the $x$ axis and $y$ axis are defined in the text. For comparison, the relativistic continuum dispersion relation is denoted as a dashed line.}
\label{fig:Disp_from_NaiveE_N}
\end{figure*}

Next, the fitted values of $E_N(\bm{p})$ are used to verify the nucleon dispersion relation as shown in Fig.~\ref{fig:Disp_from_NaiveE_N}. 
To discuss the ${\cal O}(ap)$ lattice discretization artifacts on the dispersion relation, 
we plot the results obtained from the present and previous calculations carried out on the $160^4$ lattice (denoted as the fine lattice) and 
the $128^4$ lattice (denoted as the coarse lattice), respectively.
The horizontal axis shows the momentum squared given by lattice momentum as $p_{\mathrm{lat}}^2=\left(\frac{2\pi}{aL}\right)^2\times |\bm{n}|^2$, while the vertical axis represents the momentum squared obtained from $p_{\mathrm{con}}^2=E_N^2(\bm{p})-M_N^2$ in physical units.
As can be seen in Fig.~\ref{fig:Disp_from_NaiveE_N}, both fine and coarse lattice results satisfy the relativistic continuum dispersion relation at least up to $|\bm{n}|^2=8$ within statistical precision. Since there is no clear systematic difference between the two results, the size of possible ${\cal O}(a^2)$ corrections 
is not evident.

%
%

For a more accurate check of the dispersion relation, we evaluate the energy
splittings, $\Delta E_N(\bm{p}) \equiv E_N(\bm{p}) - M_N$, 
from the ratio of the nonzero and zero momentum
two point functions of the nucleon 
\begin{align}
R^{\mathrm{2pt.}}(t;\bm{p})=
        \frac{C_{LS}(t;\bm{p})}{C_{LS}(t;\bm{0})},
\end{align}
where the smear-local combination is used for both nonzero and zero-momentum nucleon two-point functions. 

\begin{figure*}
\centering
\includegraphics[width=0.8\textwidth,bb=0 0 792 612,clip]{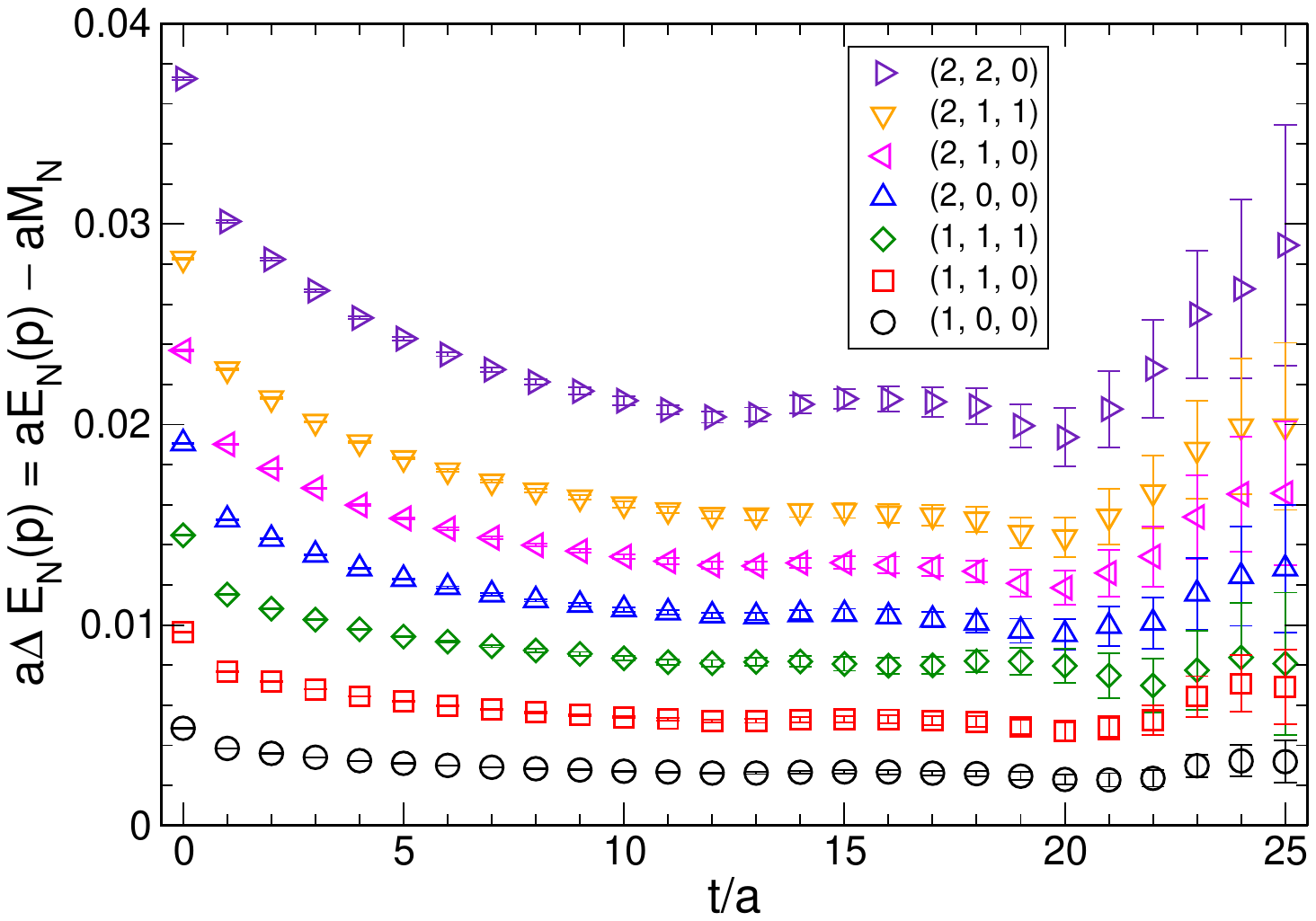}
\caption{
Effective energy plot for the energy splittings $\Delta E_N({\bm p})$ 
calculated by the ratio 
correlator $R^{\mathrm{2pt.}}(t;\bm{p})$.
}
\label{fig:DeltaE_N}
\end{figure*}

As shown in Fig.~\ref{fig:DeltaE_N}, large statistical fluctuations at large $t$ region are suppressed, while the excited-state contamination at small $t$ region 
is significantly reduced. It is observed that the energy splittings 
provide more convincing plateaus than the cases of the nucleon energies. 
This is because that the statistical fluctuations in $R^{\mathrm{2pt.}}(t;\bm{p})$ are eliminated by the strong correlation between zero and non-zero momentum two-point functions. Furthermore, the excited-state contributions at small $t$ region seem to be canceled out from the denominator and numerator of the ratio.

\begin{figure*}
\centering
\includegraphics[width=0.8\textwidth,bb=0 0 792 612,clip]{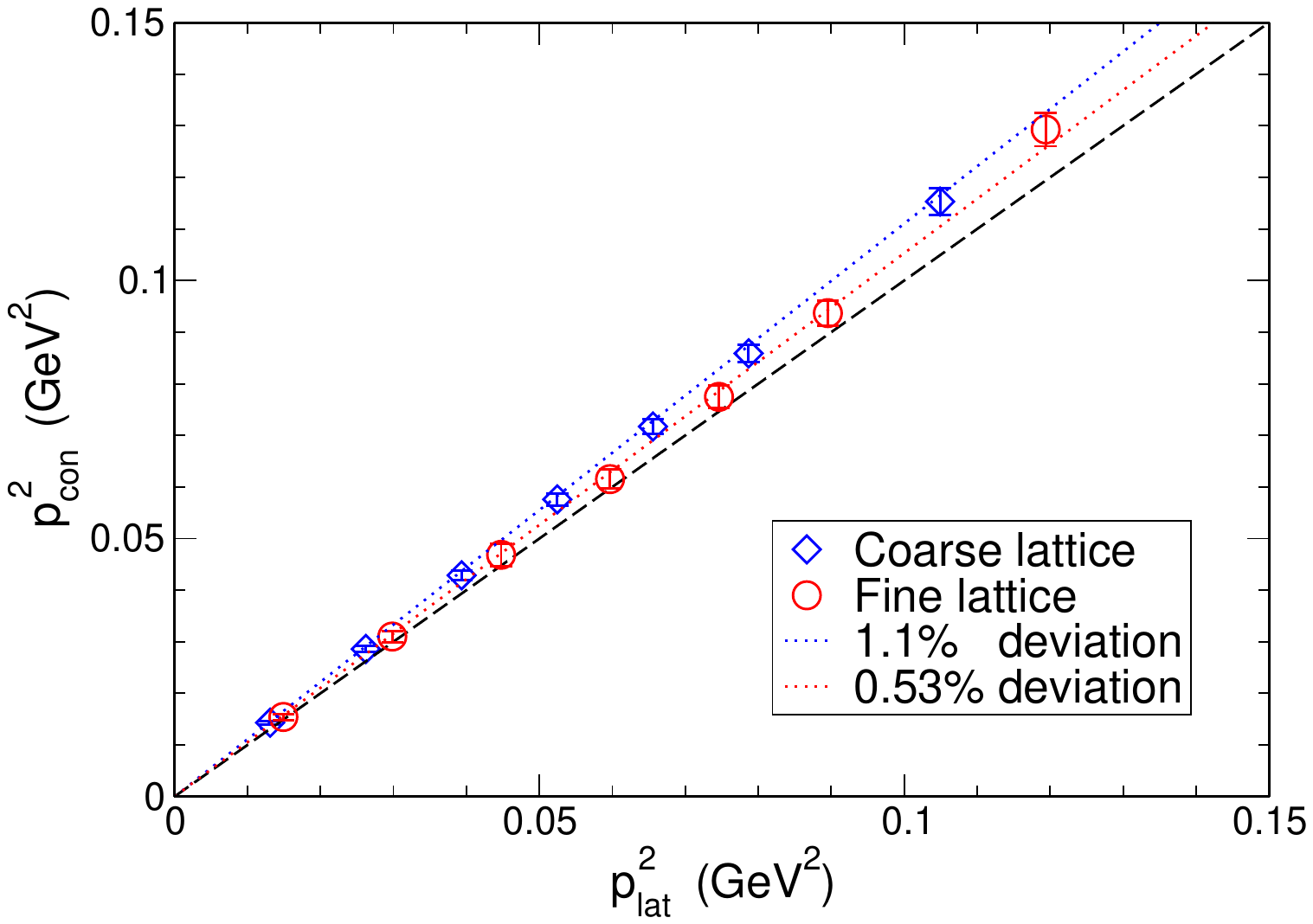}
\caption{
Check of the dispersion relation for the nucleon in an improved method, where the values of $p^2_{\mathrm{con}}$ are precisely evaluated
with the values of $\Delta E_N(q^2)$ instead
of $E_N(q^2)$.
A dashed line represents the relativistic continuum dispersion relation, while red and blue dotted lines
are given by the linear fit of each data set.
The discrepancies from the relativistic 
continuum dispersion relation become visible, but each is very tiny as a deviation of 1.1\%
for the coarse lattice and 0.53\% for the fine lattice from a slope of the continuum one.
}
\label{fig:Disp_from_DeltaE_N}
\end{figure*}

The values of $\Delta E_N(\bm{p})$ are evaluated by the correlated fit with the same fit range of $t/a=16-20$ with high accuracy, though the long plateau start at much earlier $t$ than the case of nucleon energy. The fitted values of $\Delta E_N(\bm{p})$ are summarized in Table~\ref{tab:deltaE_Apr22}.
These values of $\Delta E_N(\bm{p})$ are useful to verify the nucleon dispersion relation more accurately, since the values of $p^2_{\rm con}$ can be alternatively evaluated as $p^2_{\rm con}=\Delta E_N(\bm{p})(\Delta E_N(\bm{p})+2 M_N)$. 
In Fig.~\ref{fig:Disp_from_DeltaE_N}, the checks of the nucleon dispersion relation for the fine ($160^4$) and coarse ($128^4$) lattices are displayed by using $\Delta E_N(\bm{p})$ instead of $E_N(\bm{p})$. 

Figure~\ref{fig:Disp_from_DeltaE_N} reveals a slight deviation from the continuum dispersion relation thanks
to the accurate estimations of $\Delta E_N(\bm{p})$. 
This new way to evaluate $p_{\mathrm{con}}^2$ exposes
each size of the lattice discretization uncertainties 
at the fine ($160^4$) and coarse ($128^4$) lattices 
through the check of the dispersion relation. 
A linear fit applied to the data points results in a deviation of 0.53(3)\% for the fine lattice and 1.1(2)\% 
for the coarse lattice from the dashed line whose slope corresponds to the continuum dispersion relation.

%
%

These sizes are roughly consistent with the $O(a^2)$ 
corrections on the speed of light, which are expected from our usage of nonperturbatively $O(a)$ improved Wilson fermions. 
The observed $O(a^2)$ correction to the continuum dispersion relation at each lattice spacing does not affect the
analysis to evaluate the nucleon form factors from Eqs.~(\ref{eq:ge_def})-(\ref{eq:gp_def}). Therefore,
we simply use the continuum dispersion relation to 
evaluate the values of $E_N(\bm{q})$, which appears 
in Eqs.~(\ref{eq:ge_def})-(\ref{eq:gp_def}), 
with the measured value of $M_N$. However, the values of
$q^2$ are slightly influenced by choosing one of two methods
to evaluate the momentum transfer. One is to use the continuum
dispersion relation similar to the analysis for the
nucleon form factors as  $q^2_{\mathrm{disp}}=2M_N\left(\sqrt{M_N^2+(2\pi\bm{n}/(La))^2}-M_N\right)$, while the other is to use the measured values
of $\Delta E_N(\bm{q}^2)$ as $q^2_{\mathrm{meas}}=2M_N\Delta E_N(\bm{q})$.
Both values of $q^2_{\mathrm{meas}}$ and $q^2_{\mathrm{disp}}$
are tabulated in Table~\ref{tab:deltaE_Apr22}.
The discrepancy between $q^2_{\mathrm{meas}}$ and $q^2_{\mathrm{disp}}$ should be taken into account in examining $q^2$-dependence of the nucleon form factors. 
We primarily use the definition of $q^2_{\mathrm{disp}}$ 
for studying $q^2$-dependence of the nucleon form factors
and then evaluate the systematic uncertainties by a difference associated with the choice of $q^2_{\mathrm{disp}}$
or $q^2_{\mathrm{meas}}$.

%
%
%

%
%
\begin{table*}[ht]
    \caption{Fitted nucleon energy splitting $\Delta E_N(\bm{q})$ obtained from the smear-local nucleon two-point function in lattice units. Results for $\Delta E_N(\bm{q})$ with nonzero momentum $\bm{q}=\frac{2\pi}{aL}\times \bm{n}$ are averaged over all possible permutations of $\bm{n}=(n_x, n_y, n_z)$. 
    In addition, two types of the corresponding momentum transfers: $q^2_{\mathrm{disp}}=2M_N\left(\sqrt{M_N^2+\bm{q}^2}-M_N\right)$
    and
    $q^2_{\mathrm{meas}}=2M_N\Delta E_N(\bm{q})$ 
    with and without the assumption of the continuum dispersion relation
    for $E_N(\bm{q})$ 
    are also summarized 
    for each momentum $\bm{q}$.
\label{tab:deltaE_Apr22}}
\begin{ruledtabular}
\begin{tabular}{ccccc}
    Label& $a\Delta E_N(\bm{q})$ & Fit range  & $q^2_{\mathrm{disp}}\ \left[\mathrm{GeV}^2\right]$ &  $q^2_{\mathrm{meas}}\ \left[\mathrm{GeV}^2\right]$ \\
          \hline
    Q1& 0.00263(10) & [16:20] & 0.0149 & 0.0157(6)  \\
    Q2& 0.00528(19) & [16:20] & 0.0296 & 0.0311(11) \\
    Q3& 0.00801(38) & [16:20] & 0.0442 & 0.0472(22) \\
    Q4& 0.01039(32) & [16:20] & 0.0587 & 0.0612(19) \\
    Q5& 0.01291(37) & [16:20] & 0.0731 & 0.0760(22) \\
    Q6& 0.01546(42) & [16:20] & 0.0874 & 0.0911(25) \\
    Q7& 0.02111(54) & [16:20] & 0.1157 & 0.1245(32) \\
\end{tabular}
\end{ruledtabular}
\end{table*}

\clearpage

\section{Numerical results I: Electromagnetic form factors and axial form factor}
\label{sec:numerical_results_I}
\subsection{Electric form factor and electric charge radius}
\label{sec:ge_re}
\subsubsection{Isovector sector}
\begin{figure*}
\centering
\includegraphics[width=.48\textwidth,bb=0 0 864 720,clip]{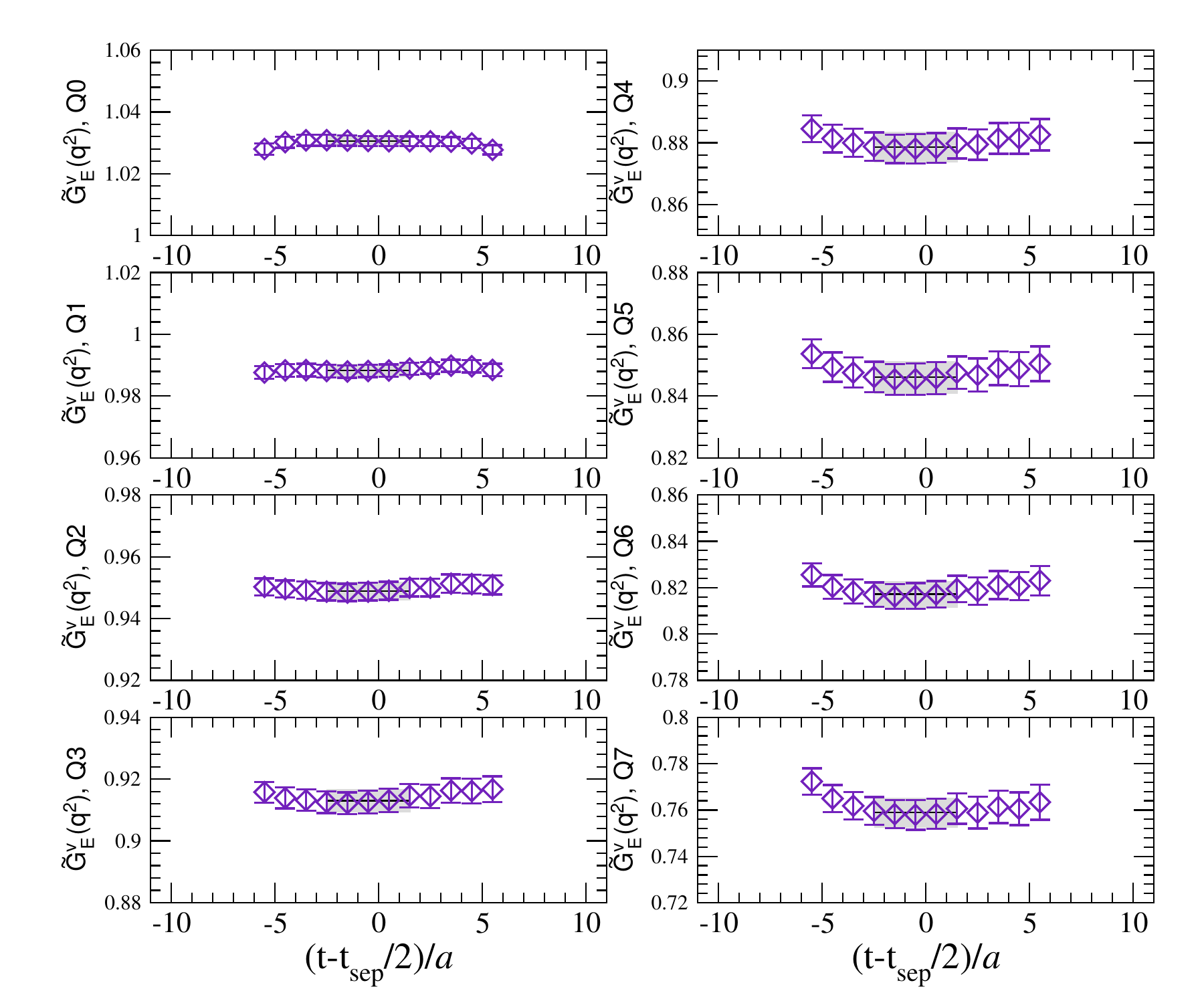}
\includegraphics[width=.48\textwidth,bb=0 0 864 720,clip]{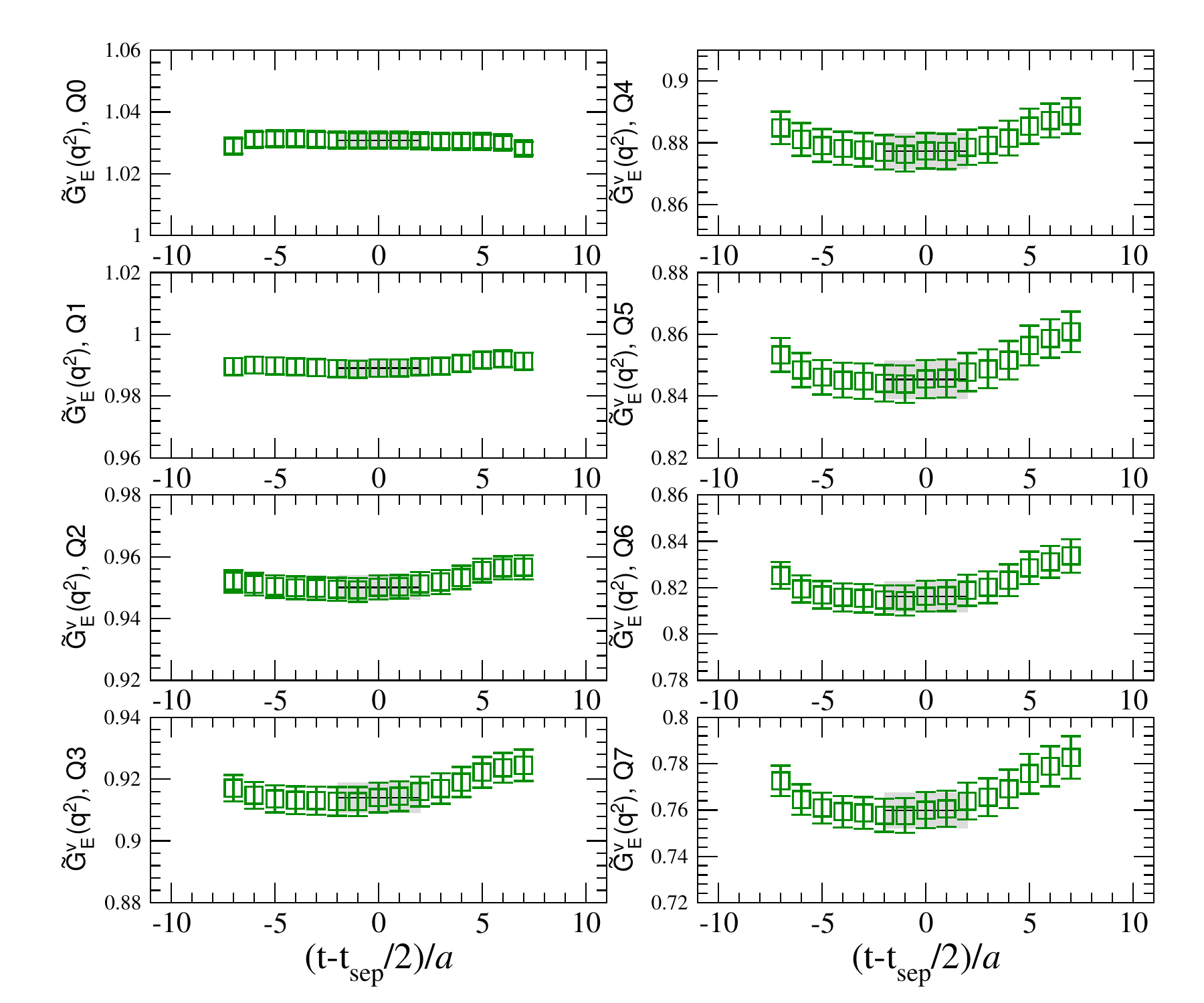}
\includegraphics[width=.48\textwidth,bb=0 0 864 720,clip]{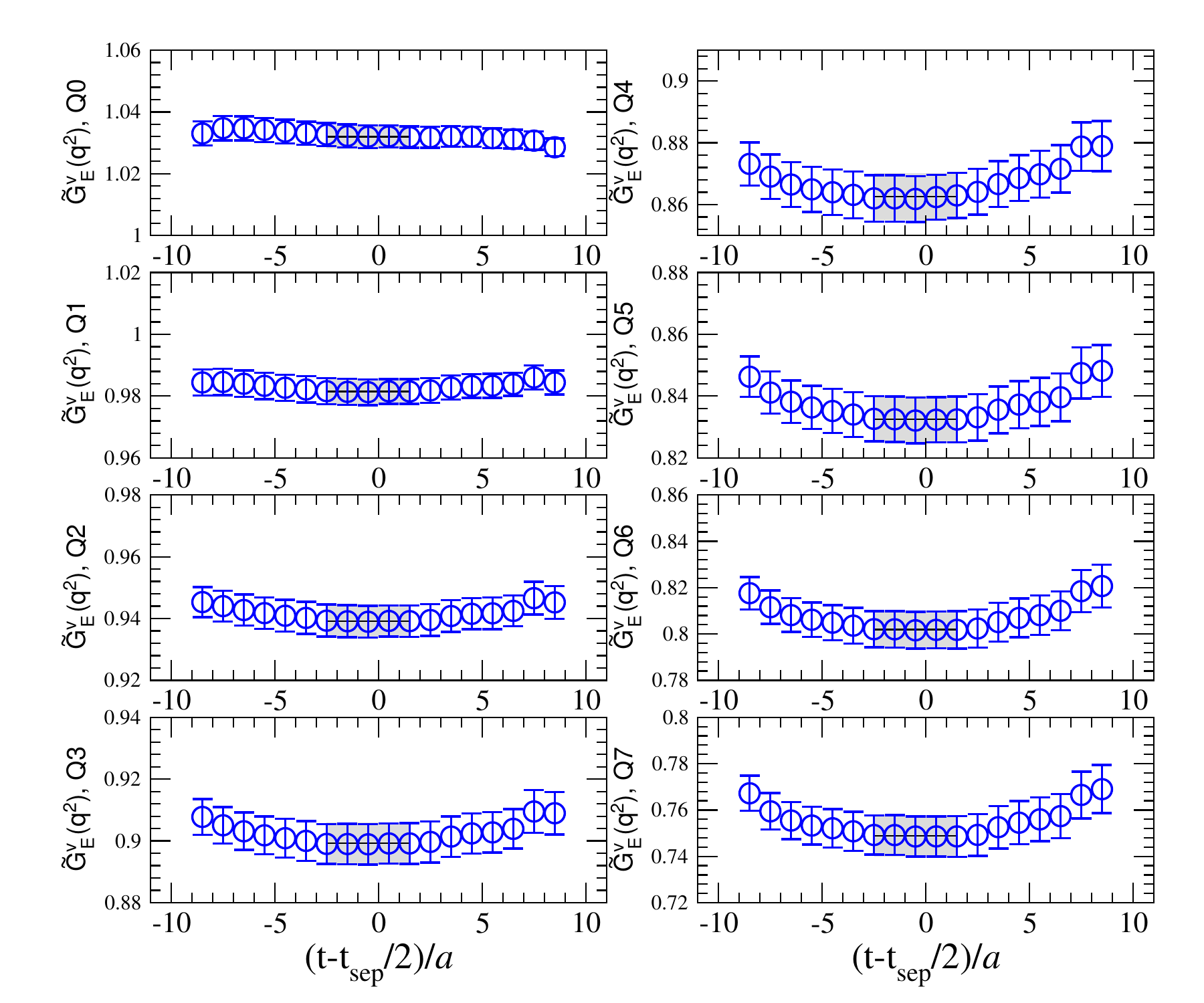}
\caption{Isovector electric form factor $\widetilde{G}^v_E(q^2)$ obtained from the ratio of Eq.~(\ref{eq:ge_def}) as a function of the current operator insertion time
$t$ for $t_{\mathrm{sep}}/a=13$ (purple symbols), $16$ (green symbols), $19$ (blue symbols) with all eight momentum transfers labeled from Q0 to Q7.
The gray bands display the fit range and one standard deviation in each panel.
}
\label{fig:ge_qdep_p-n_ts1X}
\end{figure*}

\begin{figure*}
\centering
\includegraphics[width=1\textwidth,bb=0 0 864 720,clip]{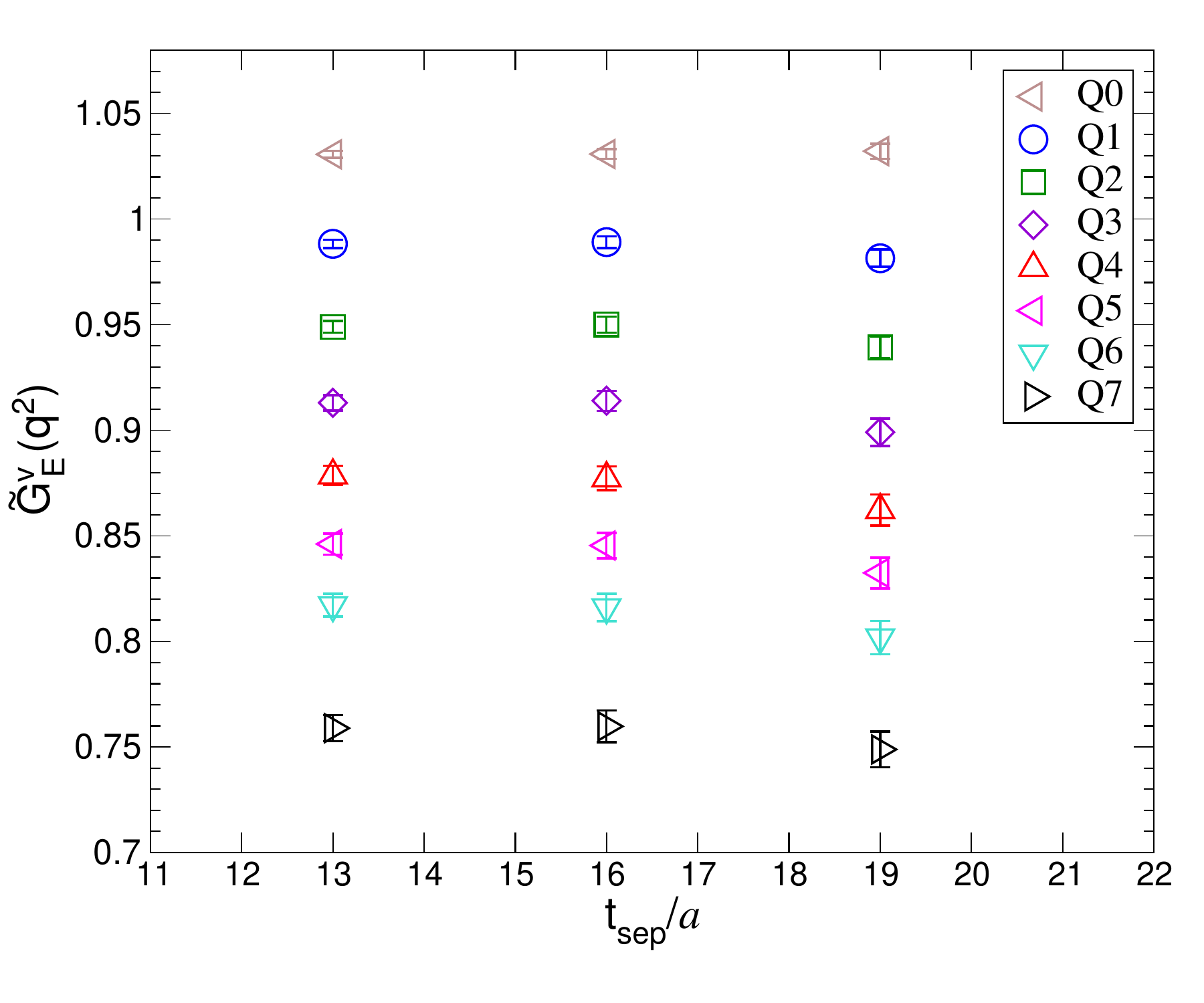}
\caption{
The source-sink separation ($t_{\mathrm{sep}}$) dependence of the isovector electric form factor $\widetilde{G}^v_E(q^2)$ with all eight momentum transfers.
}
\label{fig:ge_tsdep_p-n}
\end{figure*}

The electric form factor is extracted from the ratio $R^t_{V_4}(t;\bm{p})$ defined in Eq.~(\ref{eq:ge_def}).
In Fig.~\ref{fig:ge_qdep_p-n_ts1X},
$t$-dependence of the isovector (bare) electric form factor $\widetilde{G}^v_E(q^2)$
for all eight variations of $q^2$ including $q^2=0$ with $t_{\mathrm{sep}}/a=\{13, 16, 19\}$ is displayed.
Since the excited-state contamination could not be completely eliminated 
even by fine tuning of the smearing parameters in practice,
one should calculate $\widetilde{G}^v_E(q^2)$ with several choices of $t_{\mathrm{sep}}$ and confirm that the evaluated values do not show distinct $t_{\mathrm{sep}}$-dependence during the variations of $t_{\mathrm{sep}}$ with a certain statistical precision for every $q^2$.
Indeed, we observe the good plateaus for all choices of 
$t_{\mathrm{sep}}$ and all variation of $q^2$.
The values of $\widetilde{G}_E(q^2)$ are extracted by the standard plateau method 
using the uncorrelated constant fit. 
In Fig.~\ref{fig:ge_qdep_p-n_ts1X}, 
the solid lines represent the fit results and the gray-shaded
bands display their statistical uncertainties and fit ranges. 
Figure \ref{fig:ge_tsdep_p-n} shows that the $t_{\mathrm{sep}}$-dependence of 
the extracted values of $\widetilde{G}^v_E(q^2)$, which are summarized in Appendix~\ref{app:table_of_ff}.
The results given with the different choices of $t_{\mathrm{sep}}$ are mutually  
consistent with each other within the statistical uncertainties
for all eight variations of $q^2$.

We hereafter make the best estimates of the RMS radius, which  includes the statistical error and two systematic errors in the following way.
First, we perform the simultaneous fit with two data sets of $t_{\mathrm{sep}}/a=\{16,19\}$ as our final estimate for the central value with one standard
deviation given by the jackknife analysis. We also use a single data set
of $t_{\mathrm{sep}}/a=19$ for comparison and quote a difference between two results as the first systematic error. 
In addition, another possible source of the systematic uncertainties for determination of
the RMS radius is the slight deviation from the continuum dispersion relation observed in the measured nucleon energies as previously discussed in Sec.~\ref{sec:disp}.
Therefore, the second systematic error
associated to the choice of $q^2$ definitions is also quoted
as the difference between the results obtained from either $q^2_{\mathrm{disp}}$ or $q^2_{\mathrm{meas}}$.

\begin{figure*}
\centering
\includegraphics[width=1\textwidth,bb=0 0 792 612,clip]{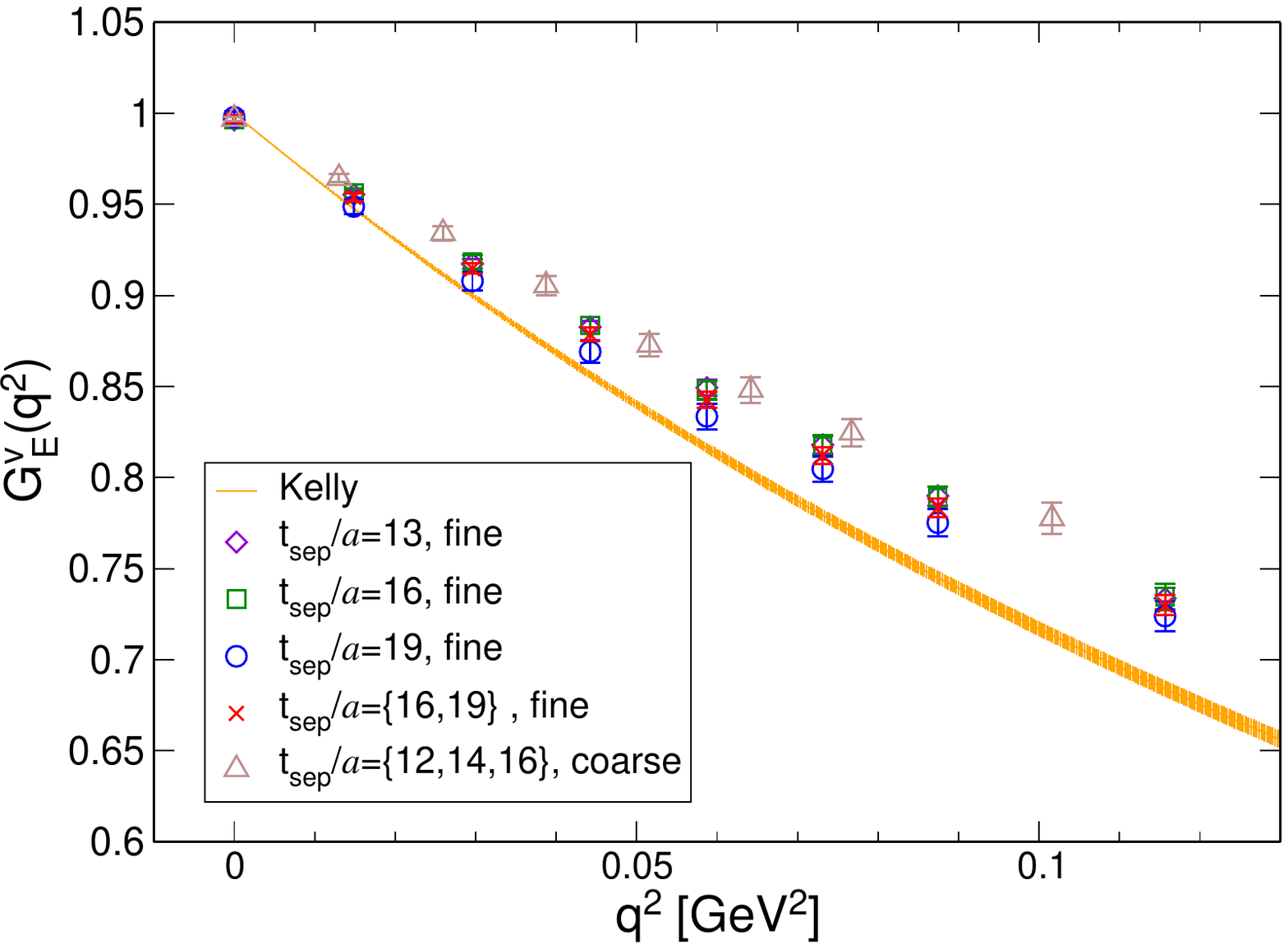}
\caption{
Results of the isovector (renormalized) electric form factor $G^v_E(q^2)$
as a function of four-momentum squared $q^2$ for 
each data set of $t_{\mathrm{sep}}/a=13$ (diamond symbols), $t_{\mathrm{sep}}/a=16$ (square symbols) and $t_{\mathrm{sep}}/a=19$ (circle symbols), and a combined data of $t_{\mathrm{sep}}/a=\{16,19\}$ (cross symbols).
The orange band represents Kelly's fit~\cite{Kelly:2004hm} as the experimental data. 
Triangle symbols, which are obtained from the
coarse ($128^4$) lattice, 
are also plotted for comparison.
}
\label{fig:ge_oct22}
\end{figure*}

Figure~\ref{fig:ge_oct22} shows the $q^2$-dependence of $G_E^v(q^2)=Z_V\widetilde{G}_E^v(q^2)$ with a choice of $q^2_{\mathrm{disp}}$ for the horizontal axis together with the Kelly's fit~\cite{Kelly:2004hm} as the experimental data.
In addition, the coarse lattice results 
are also plotted for comparison~\footnote{
The coarse lattice results presented in this paper are obtained with our slightly improved analyses compared to those used in Ref.~\cite{Shintani:2018ozy}.
The present results are statistically 
consistent with those of Ref.~\cite{Shintani:2018ozy}, 
but is more stable with respect to the fitting performed in the analysis of the $q^2$-dependence of each form factor.}.
One can easily see that the results obtained from the fine lattice locate slightly 
above the Kelly's fit, but appear systematically lower than the coarse lattice results.

Next, let us evaluate the isovector electric RMS radius by the $z$-expansion method.
The analyses with other model-dependent 
functional forms are performed in Appendix~\ref{app:model-dep_anal}.
Here it should be reminded that the size of the linear spatial extent $L$ limits the nonzero minimum value of $q^2$ that can be accessed on the lattice. 
This situation causes the uncertainty in estimating the RMS radius,
which is determined from the slope of the corresponding form factors at $q^2=0$.
In this sense, the large spatial volume ($L=160$) used in our study is an advantage to considerably reduce this particular uncertainty. 
In fact, we can access $q^2=0.015\ (\mathrm{GeV})^2$ for the nonzero momentum transfer, by using the PACS10 configurations at the lattice spacing of 0.063 fm.

\begin{figure*}
\centering
\includegraphics[width=1\textwidth,bb=0 0 792 612,clip]{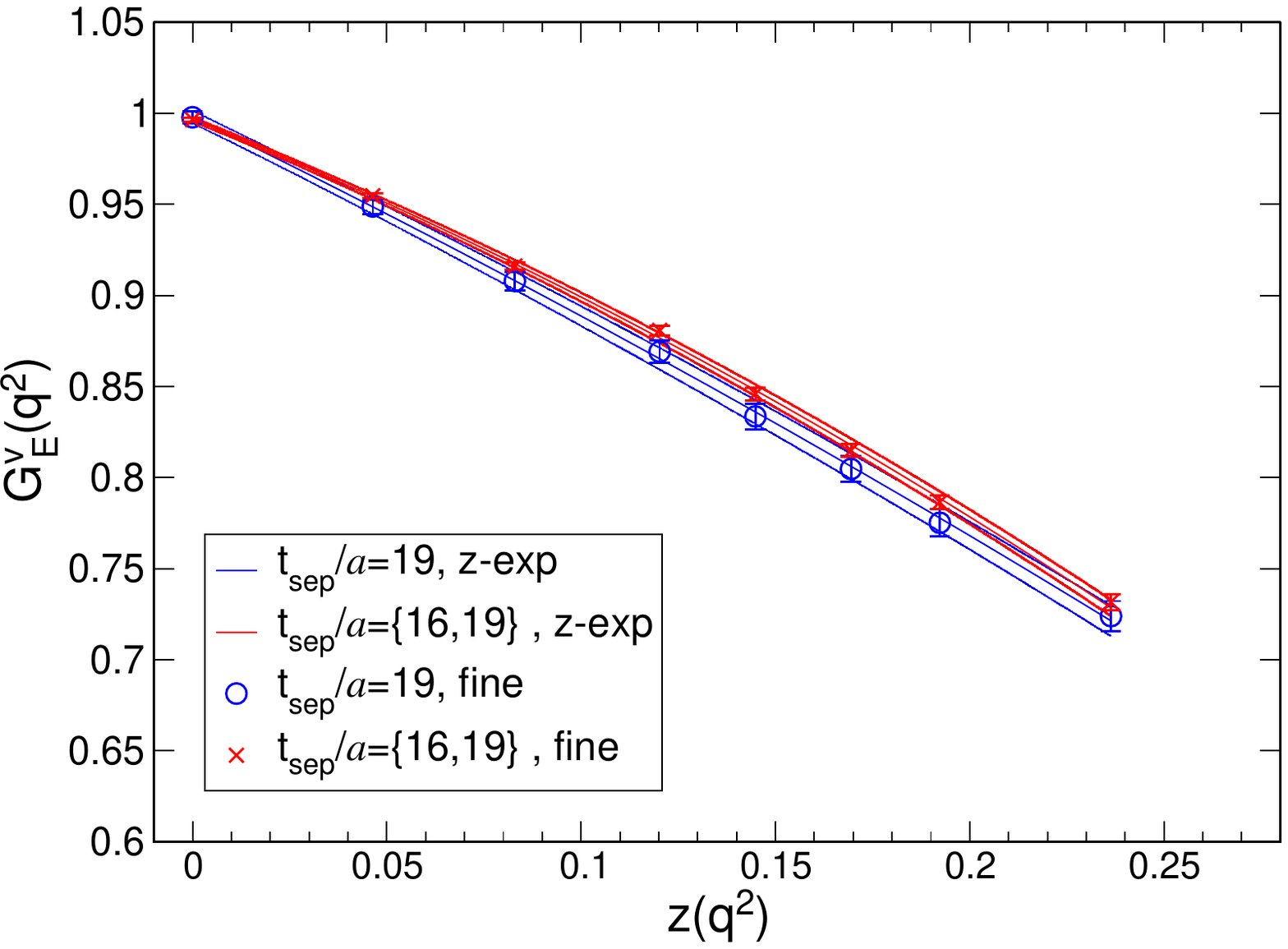}
\caption{
Results of the isovector electric form factor $G^v_E(q^2)$ as a function of $z(q^2)$ for $t_{\mathrm{sep}}/a=19$ data (circle symbols) and a combined data of $t_{\mathrm{sep}}/a=\{16,19\}$ (cross symbols).
On each data set, the inner curve of the band represents the central value obtained from the $z$-expansion analyses,
while the outer curves represent the one standard deviation.}
\label{fig:ge_oct22_zexp}
\end{figure*}

\begin{figure*}
\centering
\includegraphics[width=1\textwidth,bb=0 0 792 612,clip]{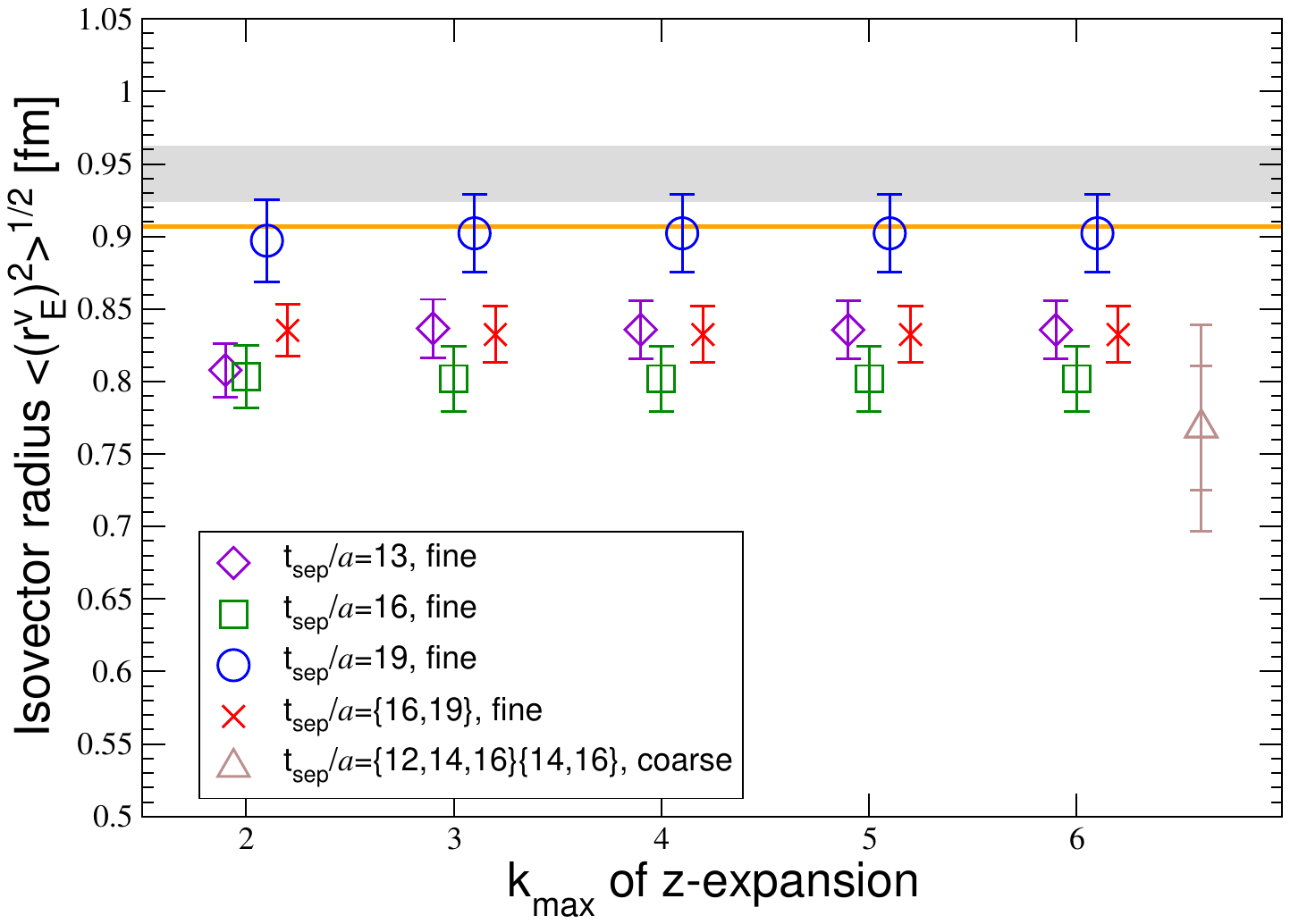}
\caption{
Stability of the variation of $k_{\mathrm{max}}$ for isovector electric charge radius obtained by the  $z$-expansion method. The horizontal axis represents the truncation number of the infinite series $k_{\mathrm{max}}$ in the $z$-expansion. Results from each data set of 
$t_{\mathrm{sep}}/a=13$ (diamond symbols), $t_{\mathrm{sep}}/a=16$ (square symbols) and $t_{\mathrm{sep}}/a=19$ (circle symbols), and also a combined data of $t_{\mathrm{sep}}/a=\{16,19\}$ (cross symbols) 
are displayed. Each result is slightly shifted along the horizontal axis for visibility.
Triangle symbols correspond to
our previous results obtained from the coarse ($128^4$) lattice.
Horizontal bands represent the experimental results
from the $ep$ scattering (gray band) and $\mu$H atom spectroscopy (orange band).}
\label{fig:ge_zexp_rmscomp}
\end{figure*}

Figure~\ref{fig:ge_oct22_zexp} shows $G^v_E(q^2)$
as a function of $z(q^2)$ together with the $z$-expansion
fitting results. 
The circle symbols
are plotted for $t_{\rm sep}/a=19$ data, while the cross symbols
are plotted for a combined data of $t_{\rm sep}/a=\{16,19\}$. 
On each data set, the inner curve of the band represents
the central value obtained from the $z$-expansion fit, while
the outer curves represent the one standard deviation. 
In Fig.~\ref{fig:ge_zexp_rmscomp},
we show stability of the variation of 
$k_{\mathrm{max}}$ in extracting $\sqrt{\langle(r^v_E)^2\rangle}$ for 
each $t_{\mathrm{sep}}$ data and a combined data of $t_{\mathrm{sep}}/a=\{16,19\}$.
It is clearly seen that the resultant values given by the $z$-expansion are stable under the variation of $k_{\mathrm{max}}$ for all cases if $k_{\mathrm{max}}\ge 3$.
Therefore, in this study, we choose 
$k_{\mathrm{max}}=4$ for the evaluation of the RMS radius in the $z$-expansion analysis
and then quote the fit result with $k_{\mathrm{max}}=4$ as our best estimate. 
The obtained values of $\sqrt{\langle (r^v_E)^2 \rangle}$ are summarized in Table~\ref{tab:re_zexp}.

Next, let us discuss $t_{\mathrm{sep}}$-dependence of $\sqrt{\langle (r^v_E)^2 \rangle}$. 
It is observed that the results obtained from $t_{\mathrm{sep}}/a=13$ and 16 are in good agreement with each other within their statistical errors,
while they are relatively underestimated 
in comparison to the corresponding experimental values.
On the other hand, the result of $t_{\mathrm{sep}}/a=19$ seems to be consistent with
the experimental values. 

In this situation, one might consider
that discrepancy between $t_{\mathrm{sep}}/a=\{13,16\}$ and $t_{\mathrm{sep}}/a=19$ is associated with the systematic uncertainties stemming from the excited-state contamination.
Although there is no
significant $t_{\mathrm{sep}}$-dependence of $\widetilde{G}^v_E(q^2)$ at every $q^2$ as shown in Fig.~\ref{fig:ge_tsdep_p-n},
a more careful look at the data reveals that the result of 
$t_{\mathrm{sep}}/a=19$ at zero momentum transfer is slightly larger than the other two data of $t_{\mathrm{sep}}/a=\{13,16\}$,
while the results at nonzero momentum transfers show an opposite trend.
This slight difference between $t_{\mathrm{sep}}/a=19$ and  $t_{\mathrm{sep}}/a=\{13,16\}$ sensitively affects the determination of the RMS radius that is determined as the slope of the form factor with respect to $q^2$ at the zero momentum transfer using the $z$-expansion method.

What we observe here may suggest that there is some
strong correlation among the values of 
$\widetilde{G}^v_E(q^2)$ evaluated at different $q^2$ in data set of $t_{\mathrm{sep}}/a=19$. 
In general, such data correlation can be addressed by performing the correlated fit using a covariance matrix.
However, there is no significant change in fit results obtained from the $z$-expansion method, 
regardless of correlated or uncorrelated fits.
In that sense, we do not have a firm conclusion on this point. 
We simply use the uncorrelated fits to examine the $q^2$-dependence of the form factors in the $z$-expansion method, hereafter.

Comparing with our previous  results obtained with the $128^4$ (coarse) lattice, 
the statistical precision achieved in this study is slightly better due to the increased number of measurements. 
Therefore, it can be clearly seen that the central value of our best estimate of the electric RMS radius obtained from the fine lattice deviates from
that of the coarse lattice by about 8.3\%, which is beyond the statistical uncertainty.
Contrary to expectations from checking the nucleon dispersion relation, the quantity of
$\sqrt{\langle (r^v_E)^2 \rangle}$ is subject to fairly large systematic uncertainties associated with the finite lattice spacing.
To remove this lattice discretization artifact, it is necessary to take the continuum limit.

%
%
\begin{table*}[ht!]
     \renewcommand{\arraystretch}{1.1}
    {\scriptsize
\begin{ruledtabular}
\caption{
    Results for the electric RMS charge radius $\sqrt{\langle (r_E)^2\rangle}$ in the isovector, proton and neutron channels. In the row of ``This work'' we present our best estimates. The first error is statistical error, while the second and third errors are systematic ones explained in the text.
    Our previous results obtained from the $128^4$ (coarse) lattice~\cite{Shintani:2018ozy} are also included.     
    Results for the proton and neutron are obtained without the disconnected diagram. 
\label{tab:re_zexp}}
\begin{tabular}{ccccccccc}
  & & & \multicolumn{2}{c}{Isovector} & \multicolumn{2}{c}{Proton} & \multicolumn{2}{c}{Neutron}\\
  \cline{4-5}\cline{6-7}\cline{8-9}
  Fit type & $q^2$ [GeV$^2$] & $t_{\rm sep}/a$ &
  $\sqrt{\langle (r^v_E)^2\rangle}$ [fm] & $\chi^2$/d.o.f. &
  $\sqrt{\langle (r^p_E)^2\rangle}$ [fm] & $\chi^2$/d.o.f. &
  $\langle (r^n_E)^2\rangle$ [fm$^2$] & $\chi^2$/d.o.f.\\
  \hline
  \multicolumn{2}{ c }{$160^4$ (fine) lattice} \\
  \multirow{2}{*}{$k_{\rm max}=4$}& \multirow{2}{*}{$q^2_{\mathrm{disp}}\le 0.116 $} & $\{16,19\}$&  0.832(19)  &  1.5&  0.804(14)  &  1.4&  $-$0.054(23) & 0.84\\
                    & & $19$&  0.902(27)  &  0.04&  0.853(20)  &  0.04& $-$0.100(40) &0.12 \\
  \multirow{2}{*}{$k_{\rm max}=3$}& \multirow{2}{*}{$q^2_{\mathrm{meas}}\le 0.091 $} & $\{16,19\}$&  0.810(19)  &  1.6&  0.786(14)  &  1.5&  $-$0.050(32) & 0.89\\
                    & & $19$&  0.874(26)  &  0.07&  0.832(19)  &  0.05& $-$0.091(43) &0.12\\
  \multicolumn{2}{ c }{This work} & &
  $0.832(19)(70)(22)$ & &
  $0.804(14)(49)(18)$ & &
  $-0.054(23)(46)(4)$\\
  \multicolumn{2}{ c }{$128^4$ (coarse) lattice} \\
    \multirow{2}{*}{$k_{\rm max}=4$}& \multirow{2}{*}{$q^2_{\mathrm{disp}}\le 0.102 $} & $\{12,14,16\}$&  0.768(43)  &  0.9&  0.767(14)  &  0.6&  $-$0.027(22) & 0.9\\
                    & & $\{14,16\}$&  0.813(48)  &  1.0&  0.802(40)  &  0.6& $-$0.060(29) &1.1 \\
  \multirow{2}{*}{$k_{\rm max}=4$}& \multirow{2}{*}{$q^2_{\mathrm{meas}}\le 0.112 $} & $\{12,14,16\}$&  0.734(42)  &  0.8&  0.734(31)  &  0.6&  $-$0.026(20) & 0.9\\
                    & & $\{14,16\}$&  0.780(47)  &  1.0&  0.768(39)  &  0.6& $-$0.056(27) &1.1 \\
  \multicolumn{2}{ c }{PACS10 $128^4$ lattice} & &
  $0.768(43)(45)(34)$ & &
  $0.767(14)(35)(33)$ & &
  $-0.027(23)(33)(1)$\\
  \multicolumn{2}{ c }{Experimental value~\cite{ParticleDataGroup:2022pth, Tiesinga:2021myr}}  &  &  \\
  \multicolumn{2}{ c }{$ep$ scattering} & & 0.943(19) & & 0.880(20) & & $-$0.1155(17) &\\
  \multicolumn{2}{ c }{$\mu$H atom} & & 0.907(1) & & 0.8409(4) & & $-$ &\\
\end{tabular}
\end{ruledtabular}
}
\end{table*}

%
%
%
%
%
%
%
%
%
%

\subsubsection{Proton and Neutron sector}
\begin{figure*}
\centering
\includegraphics[width=0.48\textwidth,bb=0 0 864 720,clip]{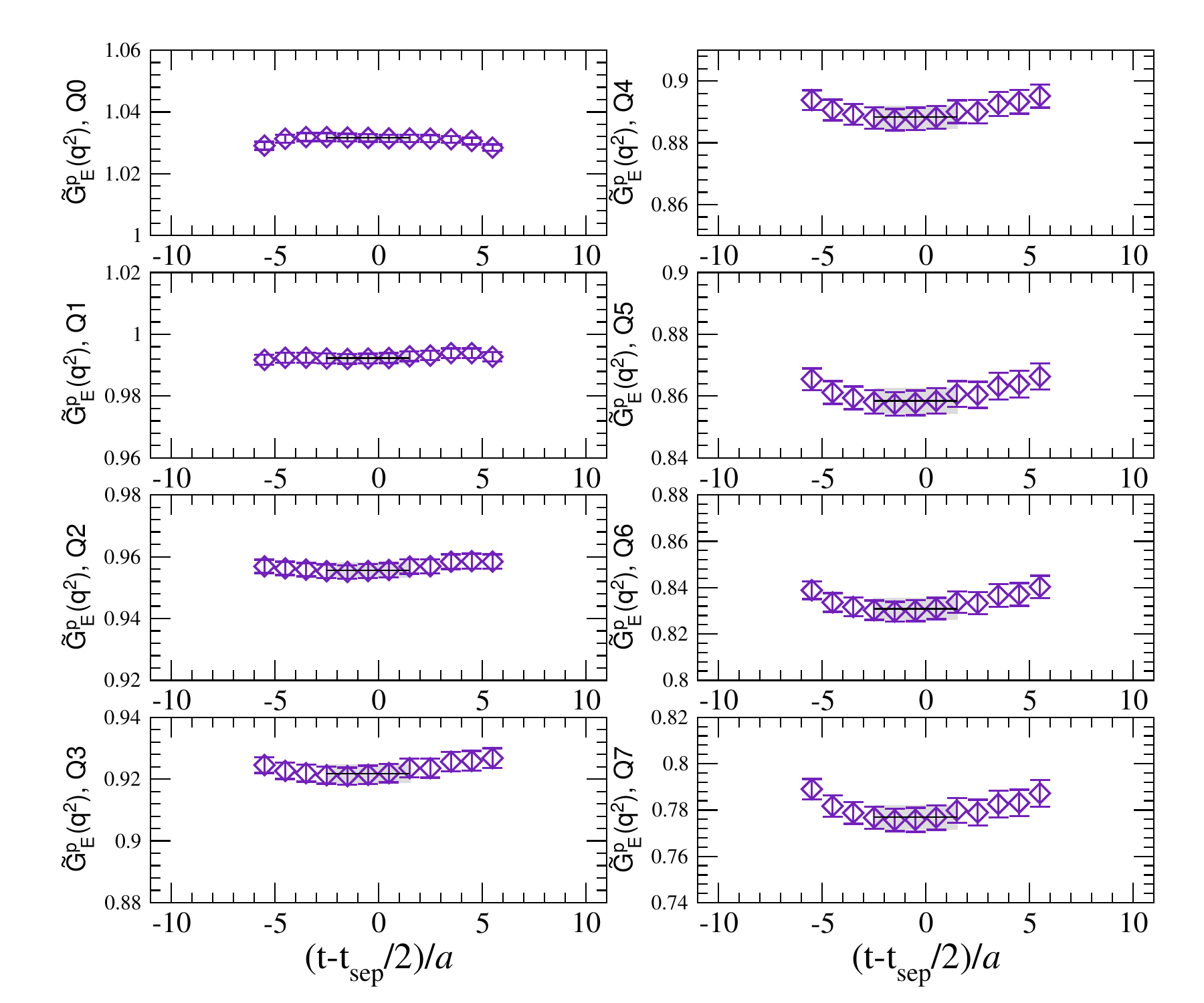}
\includegraphics[width=0.48\textwidth,bb=0 0 864 720,clip]{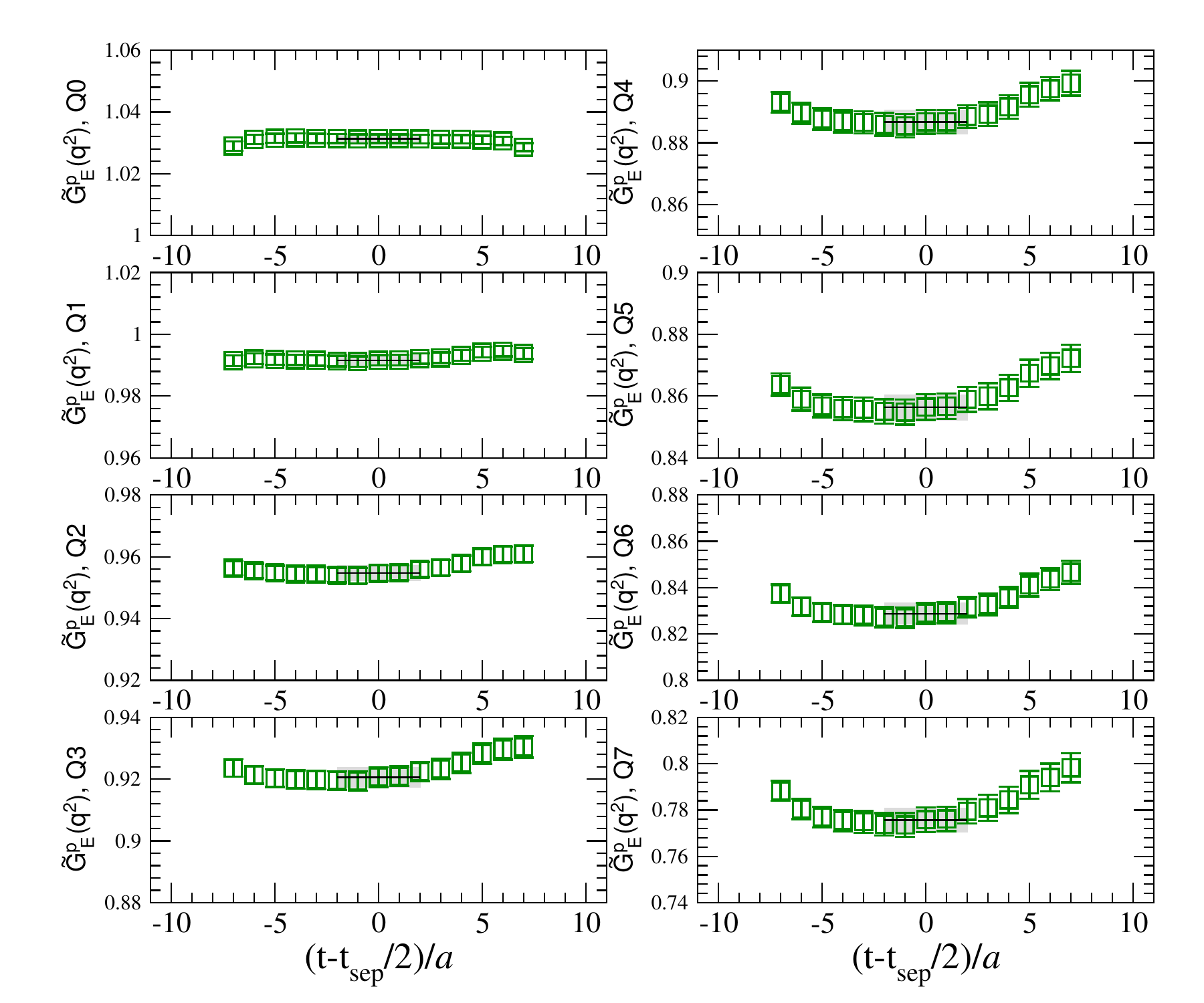}
\includegraphics[width=0.48\textwidth,bb=0 0 864 720,clip]{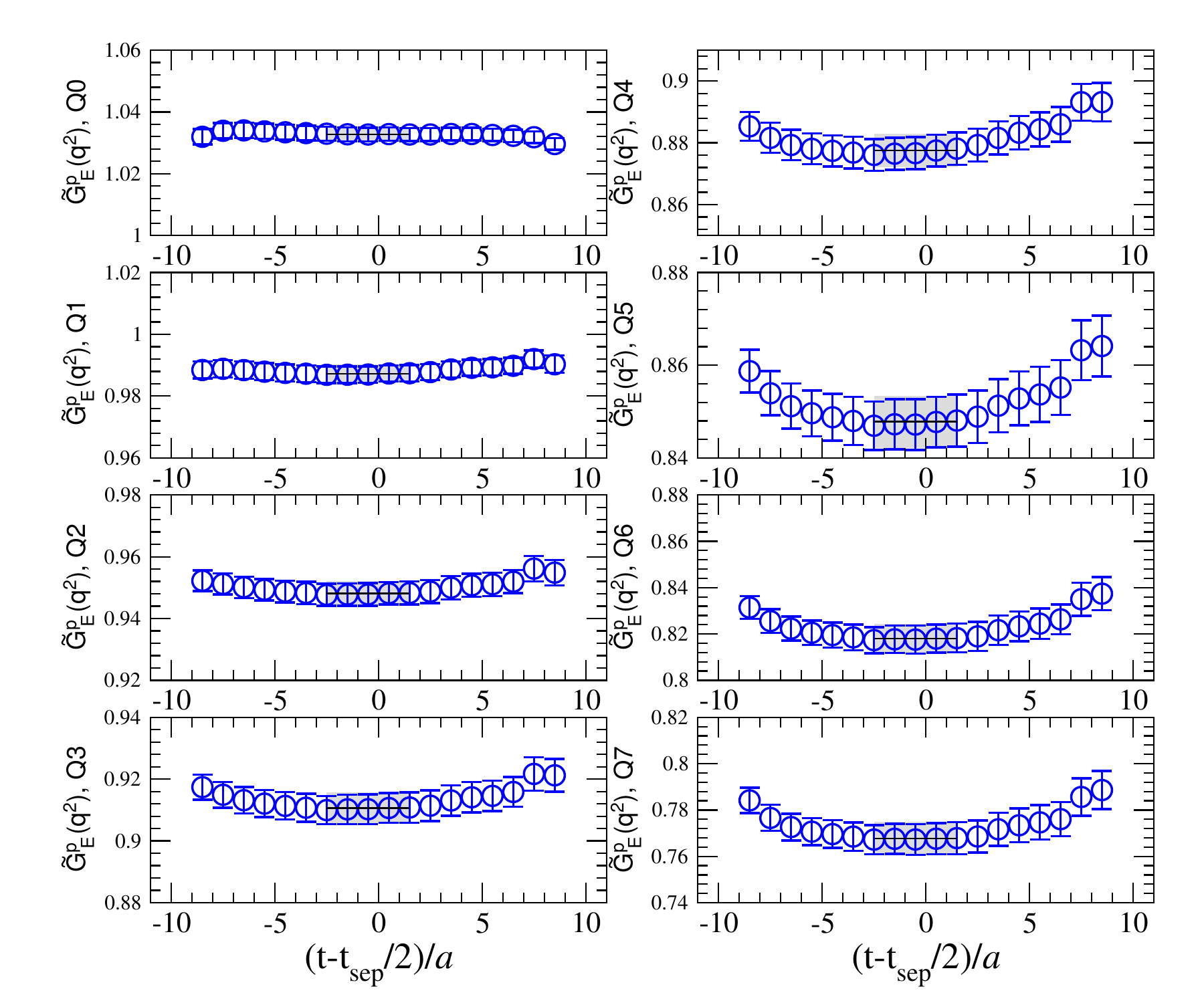}
\caption{Same as Fig.~\ref{fig:ge_qdep_p-n_ts1X} for the proton.}
\label{fig:gep_qdep_p-n_ts1X}
\end{figure*}

\begin{figure*}
\centering
\includegraphics[width=0.48\textwidth,bb=0 0 864 720,clip]{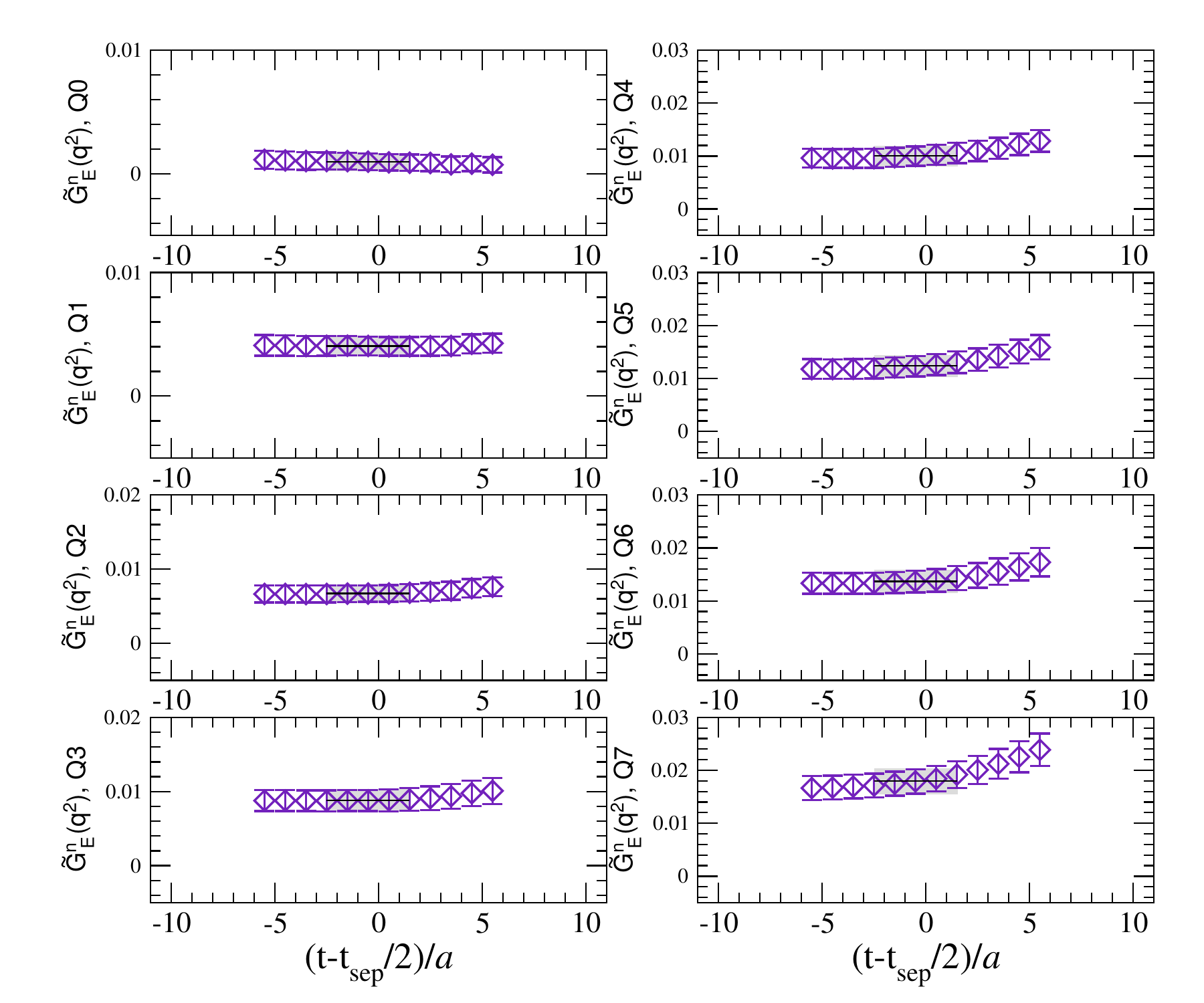}
\includegraphics[width=0.48\textwidth,bb=0 0 864 720,clip]{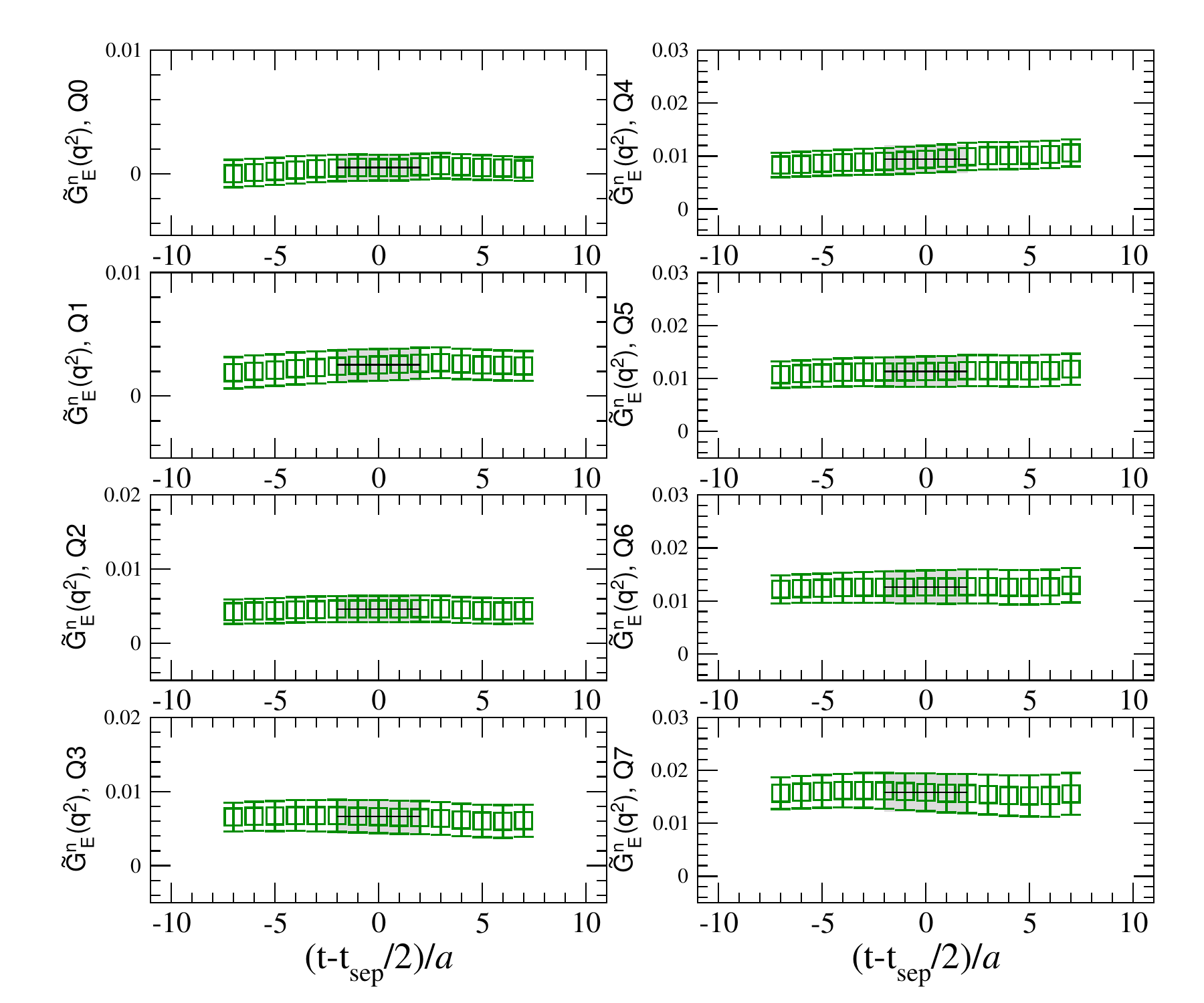}
\includegraphics[width=0.48\textwidth,bb=0 0 864 720,clip]{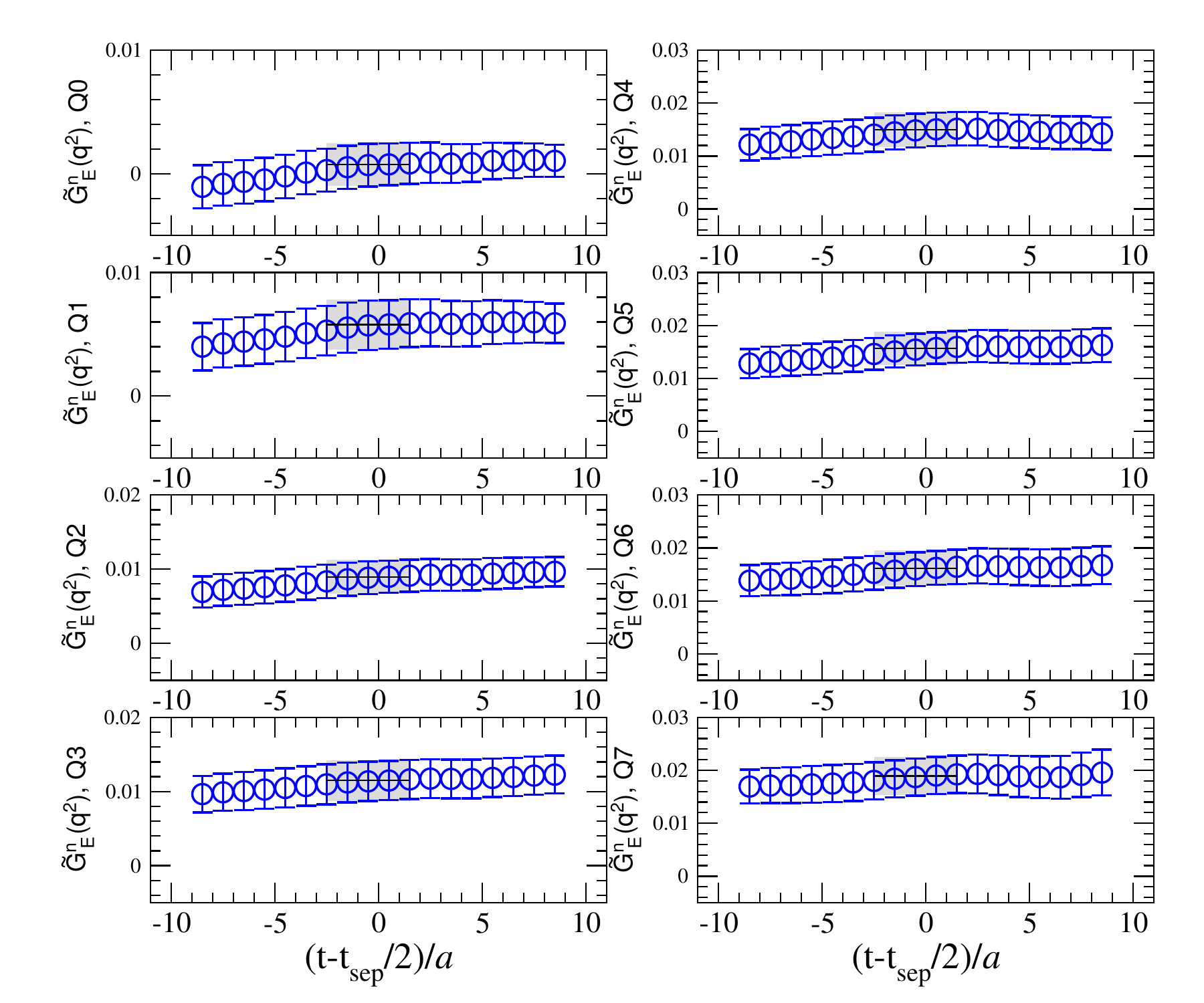}
\caption{Same as Fig.~\ref{fig:ge_qdep_p-n_ts1X} for the neutron.}
\label{fig:gen_qdep_p-n_ts1X}
\end{figure*}

\begin{figure*}
\centering
\includegraphics[width=0.49\textwidth,bb=0 0 864 720,clip]{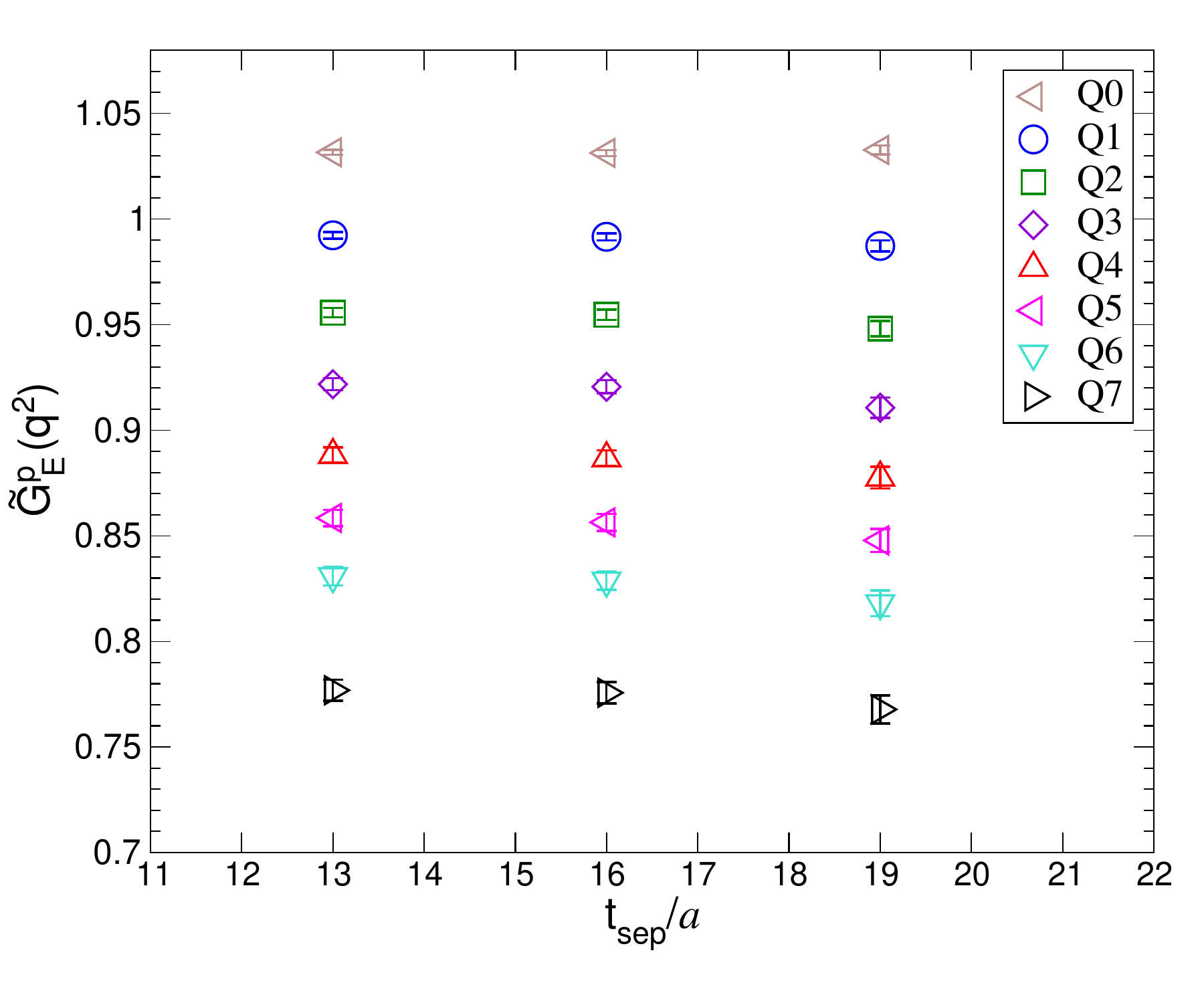}
\includegraphics[width=0.49\textwidth,bb=0 0 864 720,clip]{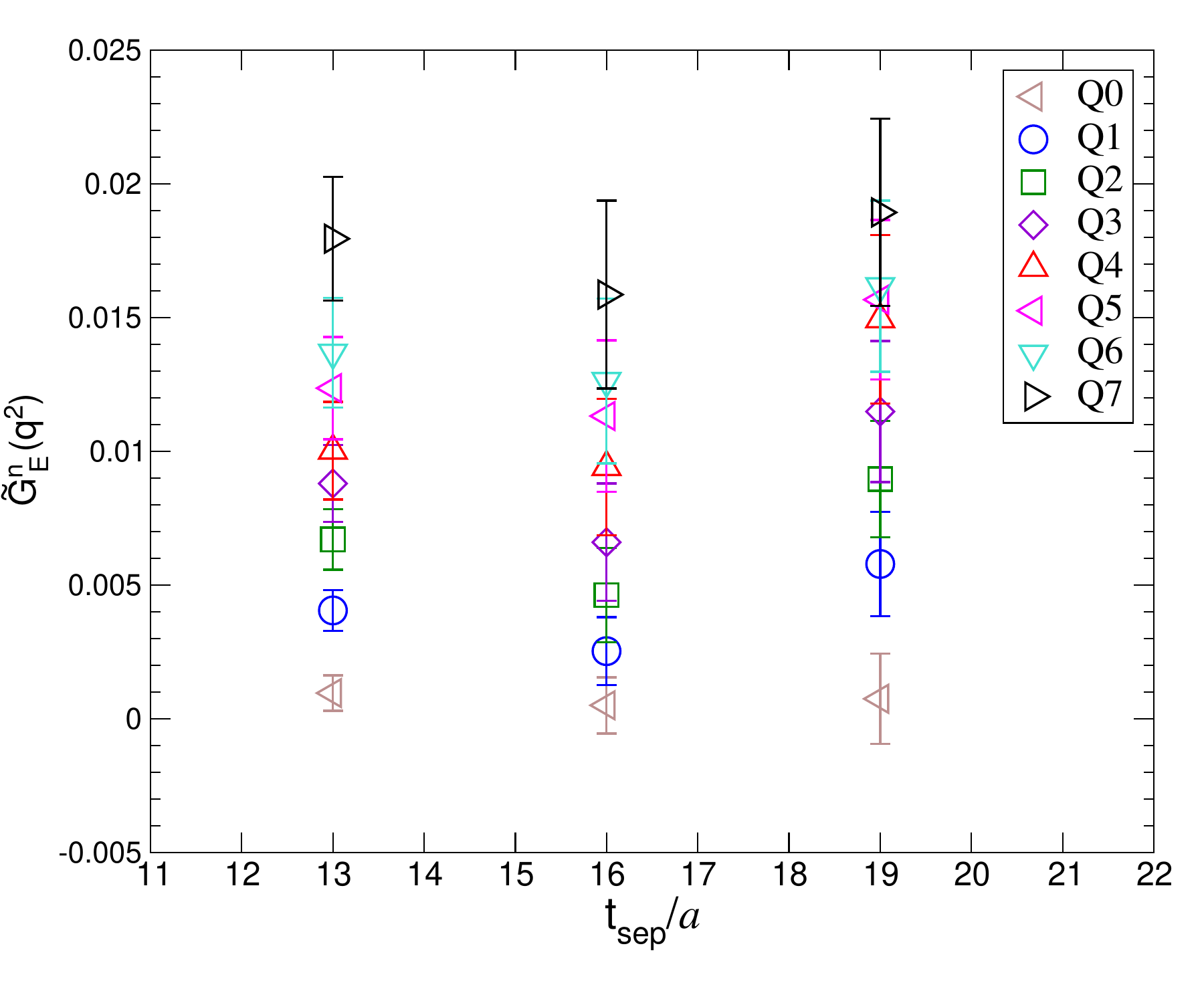}
\caption{Same as Fig.~\ref{fig:ge_tsdep_p-n} for the proton (left) and neutron (right).}
\label{fig:gepn_tsdep_p-n}
\end{figure*}

We also calculate the electric form factors, $\widetilde{G}^p_E(q^2)$ and $\widetilde{G}^n_E(q^2)$, separately from the
ratio $\mathcal{R}^{t}_{V_4}(t;\bm{q})$ for each proton and neutron, where we omit the disconnected contributions.
Similar to the isovector case, 
Fig.~\ref{fig:gep_qdep_p-n_ts1X} for the proton
and Fig.~\ref{fig:gen_qdep_p-n_ts1X} for the neutron
show the good plateaus can be observed in each 
case of $t_{\mathrm{sep}}$.
The $t_{\mathrm{sep}}$-dependence 
of $\widetilde{G}^p_E(q^2)$ and $\widetilde{G}^n_E(q^2)$ are also 
examined in Fig.~\ref{fig:gepn_tsdep_p-n}.
The $t_{\mathrm{sep}}$-dependence of the proton's electric form factor reveals the similar tendency found in the isovector case as shown in Fig.~\ref{fig:ge_tsdep_p-n}.
Although there is no significant $t_{\mathrm{sep}}$-dependence observed within the statistical errors,
the results of $t_{\mathrm{sep}}/a=19$ at nonzero momentum transfers seem to be slightly smaller than the other two data of $t_{\mathrm{sep}}/a=\{13, 16\}$
with a closer look at the data.
On the other hand, there is no $t_{\mathrm{sep}}$-dependence observed in the electric form factor of the neutron within the statistical errors.
The whole results of $G^p_E(q^2)$ and $G^n_E(q^2)$ obtained by the standard plateau method are displayed in Fig.~\ref{fig:gepn_oct22}, and their values are summarized 
in Appendix~\ref{app:table_of_ff}
together with the results of $G_E^v(q^2)$.

\begin{figure*}
\centering
\includegraphics[width=0.49\textwidth,bb=0 0 792 612,clip]{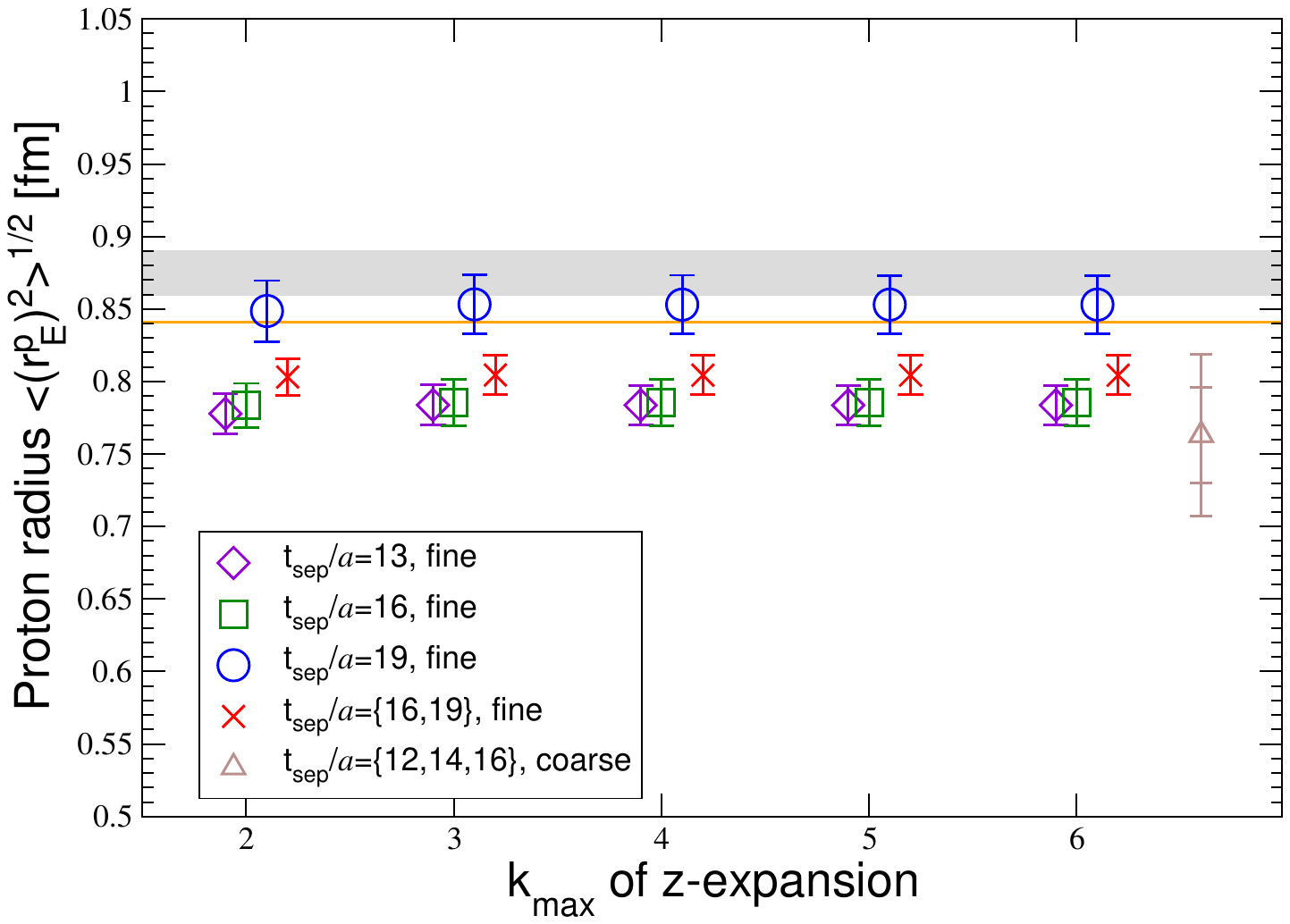}
\includegraphics[width=0.49\textwidth,bb=0 0 792 612,clip]{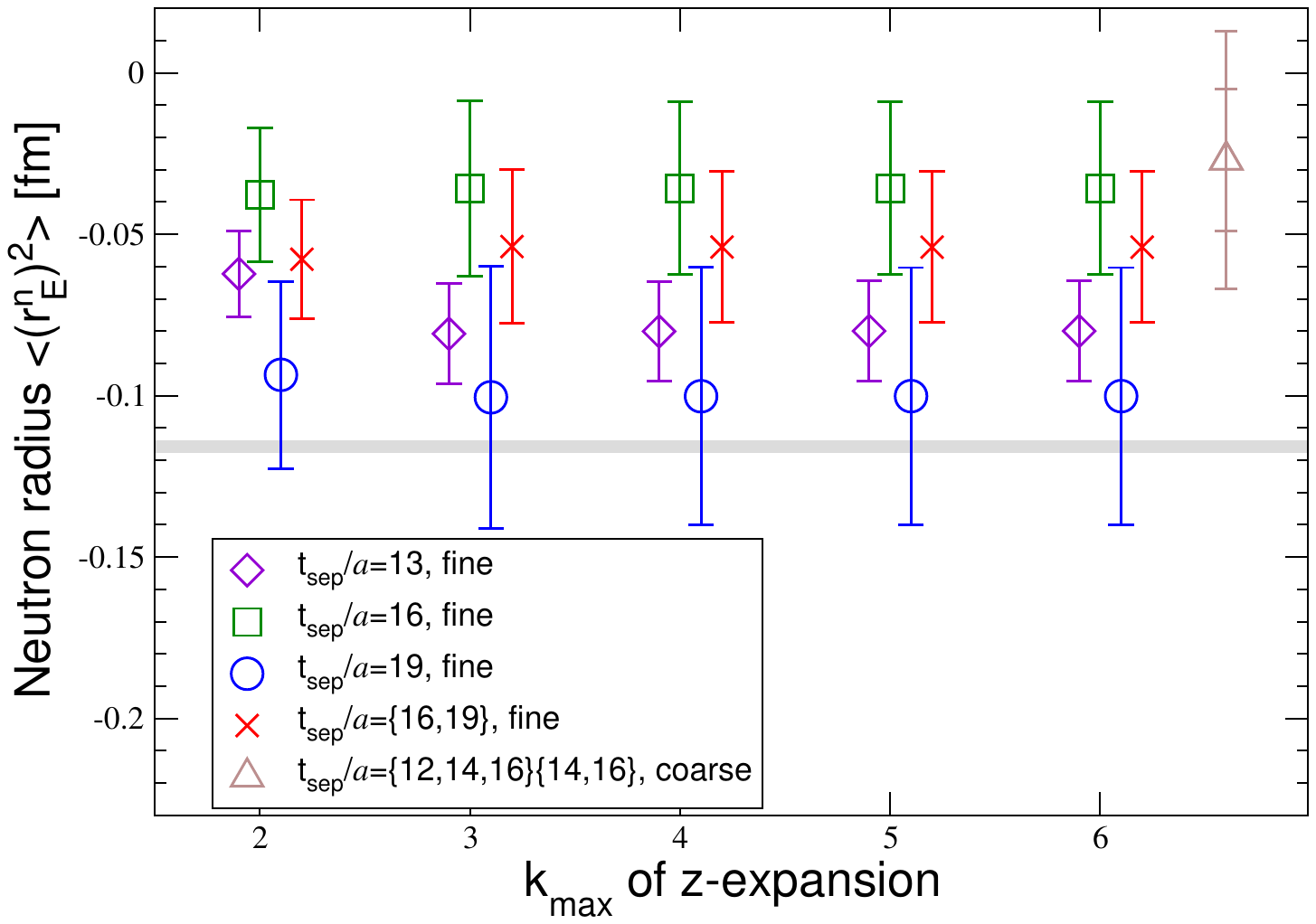}
\caption{
Same as Fig.~\ref{fig:ge_zexp_rmscomp} for the proton (left) and neutron (right).}
\label{fig:gepn_zexp_rmscomp}
\end{figure*}

The proton's and neutron's electric charge radii are determined by the $z$-expansion method,
though the analyses with other model-dependent functional forms are discussed in Appendix~\ref{app:model-dep_anal}.
Figures~\ref{fig:gepn_oct22_zexp} and \ref{fig:gepn_zexp_rmscomp} show the results obtained from the $z$-expansion fit.
The former represents $z(q^2)$-dependence of $G^v_E(q^2)$, while the latter shows the stability of the variation of 
$k_{\mathrm{max}}$ in extracting the radii.
In Fig.~\ref{fig:ge_zexp_rmscomp},
The results of $\sqrt{\langle (r^p_E)^2 \rangle}$ and $\langle (r^n_E)^2 \rangle$ obtained from the $z$-expansion method are summarized 
in Table~\ref{tab:re_zexp}, where 
the two systematic errors are quoted in the similar manner to the isovector case.

It is observed that the differences between the fine and coarse lattice results are not as large as the $isovector$ case.
However, more accurate calculations including the disconnected-type contributions are need to make a firm conclusion.

%
%
%
%
%
%
%
%
\begin{figure*}
\centering
\includegraphics[width=0.49\textwidth,bb=0 0 792 612,clip]{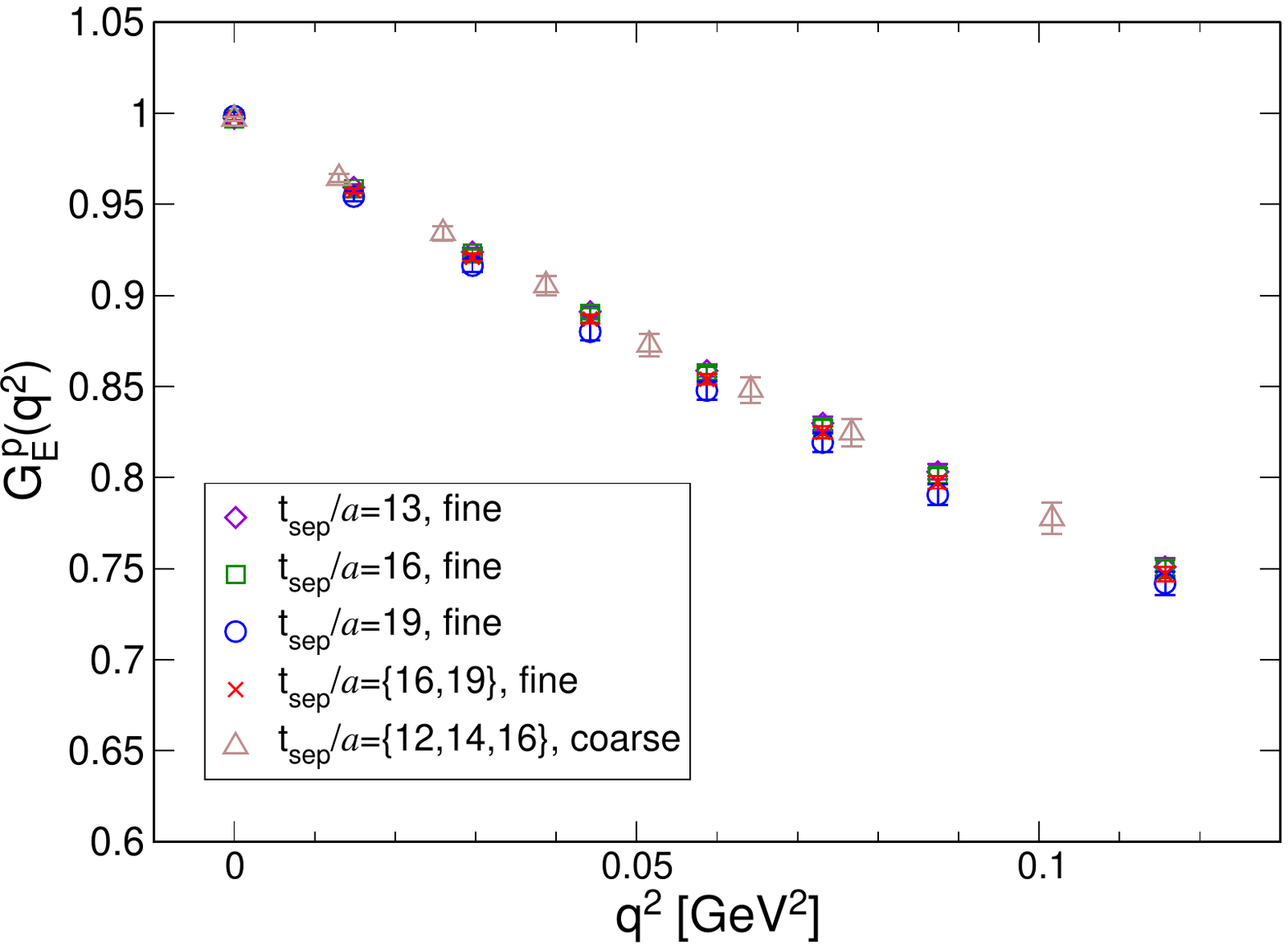}
\includegraphics[width=0.49\textwidth,bb=0 0 792 612,clip]{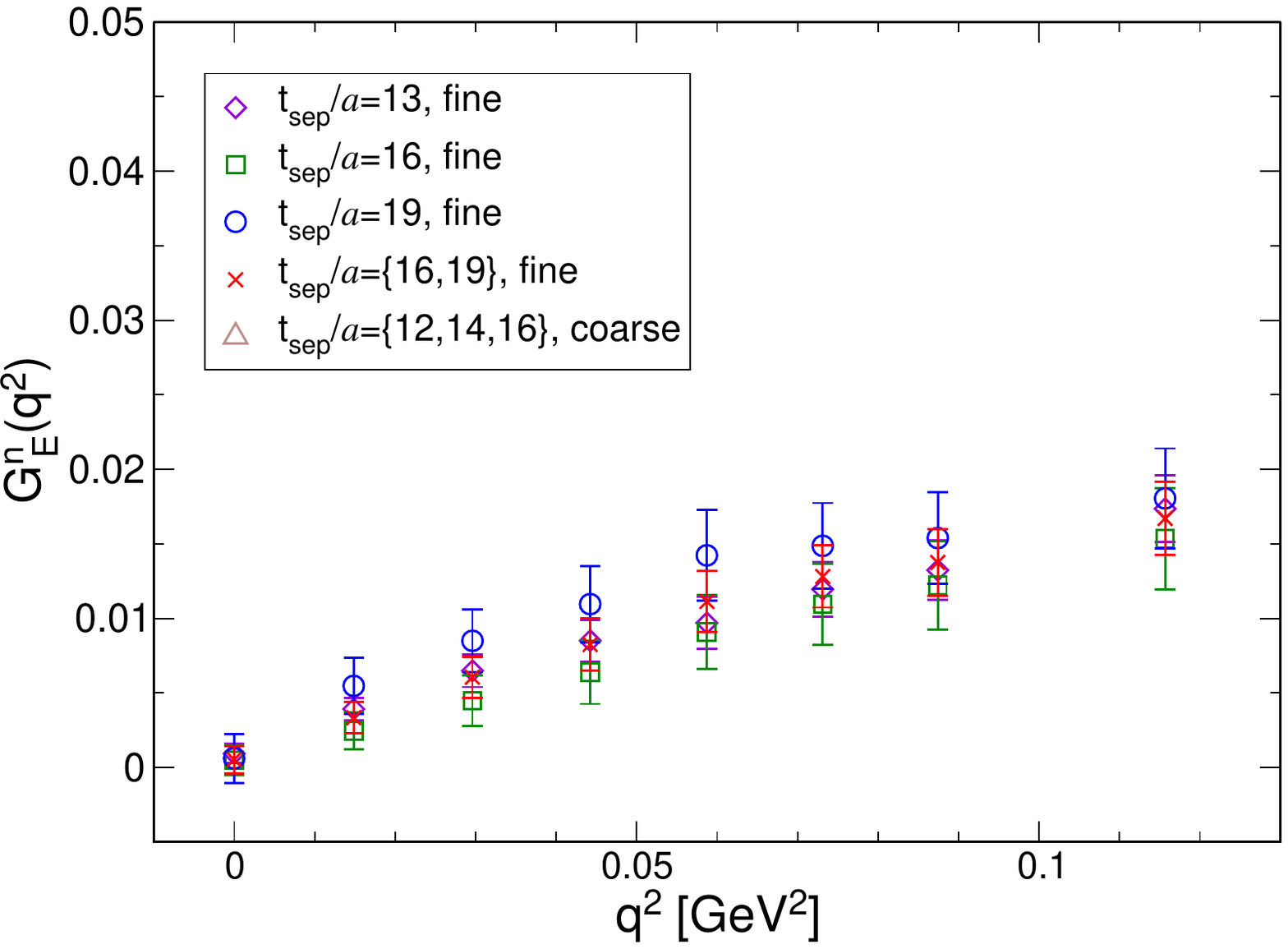}
\caption{
Same as Fig.~\ref{fig:ge_oct22} for the proton (left) and neutron (right).}
\label{fig:gepn_oct22}
\end{figure*}
%
%
\begin{figure*}
\centering
\includegraphics[width=0.49\textwidth,bb=0 0 792 612,clip]{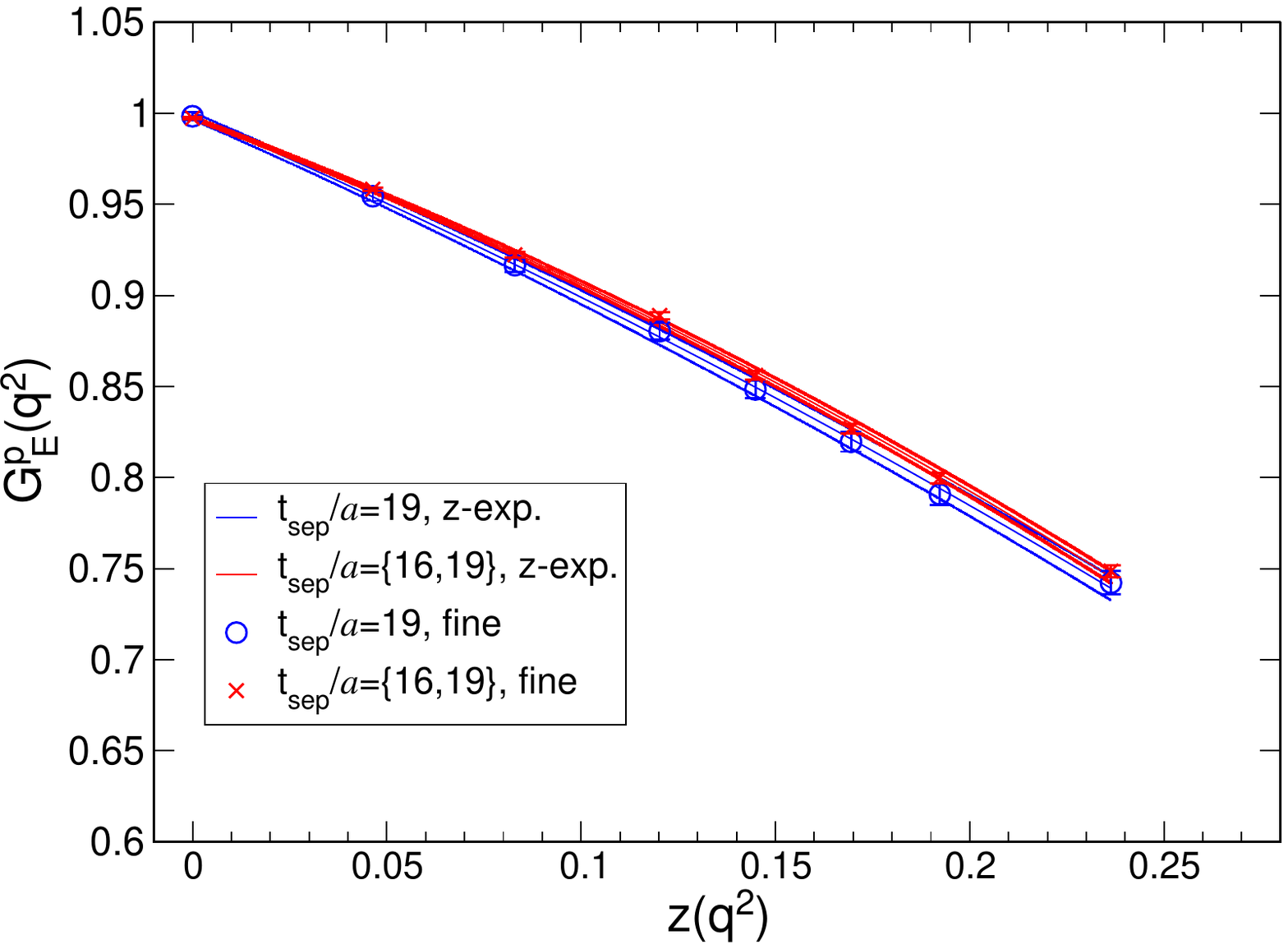}
\includegraphics[width=0.49\textwidth,bb=0 0 792 612,clip]{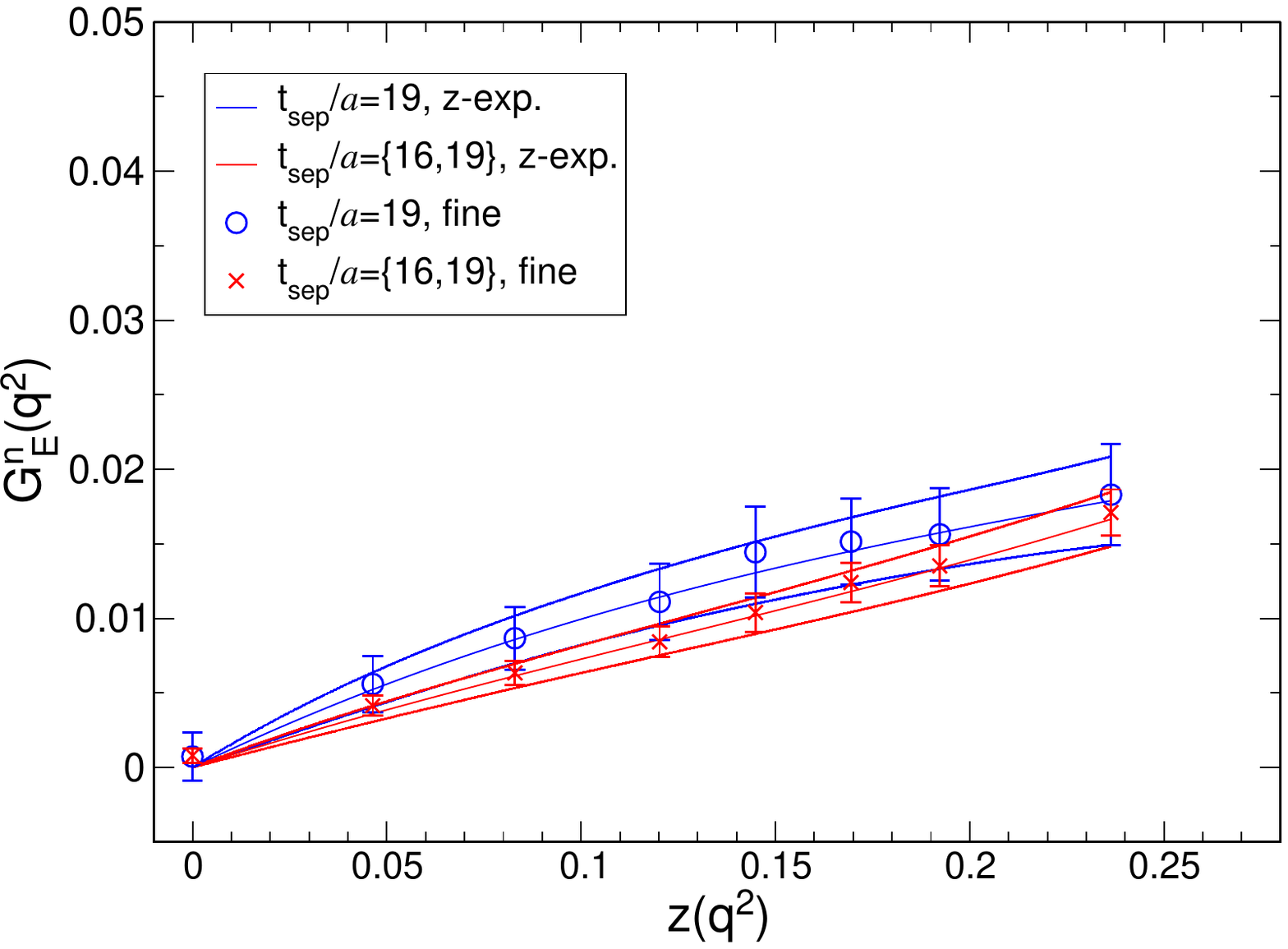}
\caption{
Same as Fig.~\ref{fig:ge_oct22_zexp} for the proton (left) and neutron (right).}
\label{fig:gepn_oct22_zexp}
\end{figure*}
%
%

\clearpage
\subsection{Magnetic form factor and magnetic RMS radius}
\label{sec:gm_rm}
\subsubsection{Isovector sector}
\begin{figure*}
\centering
\includegraphics[width=0.48\textwidth,bb=0 0 864 720,clip]{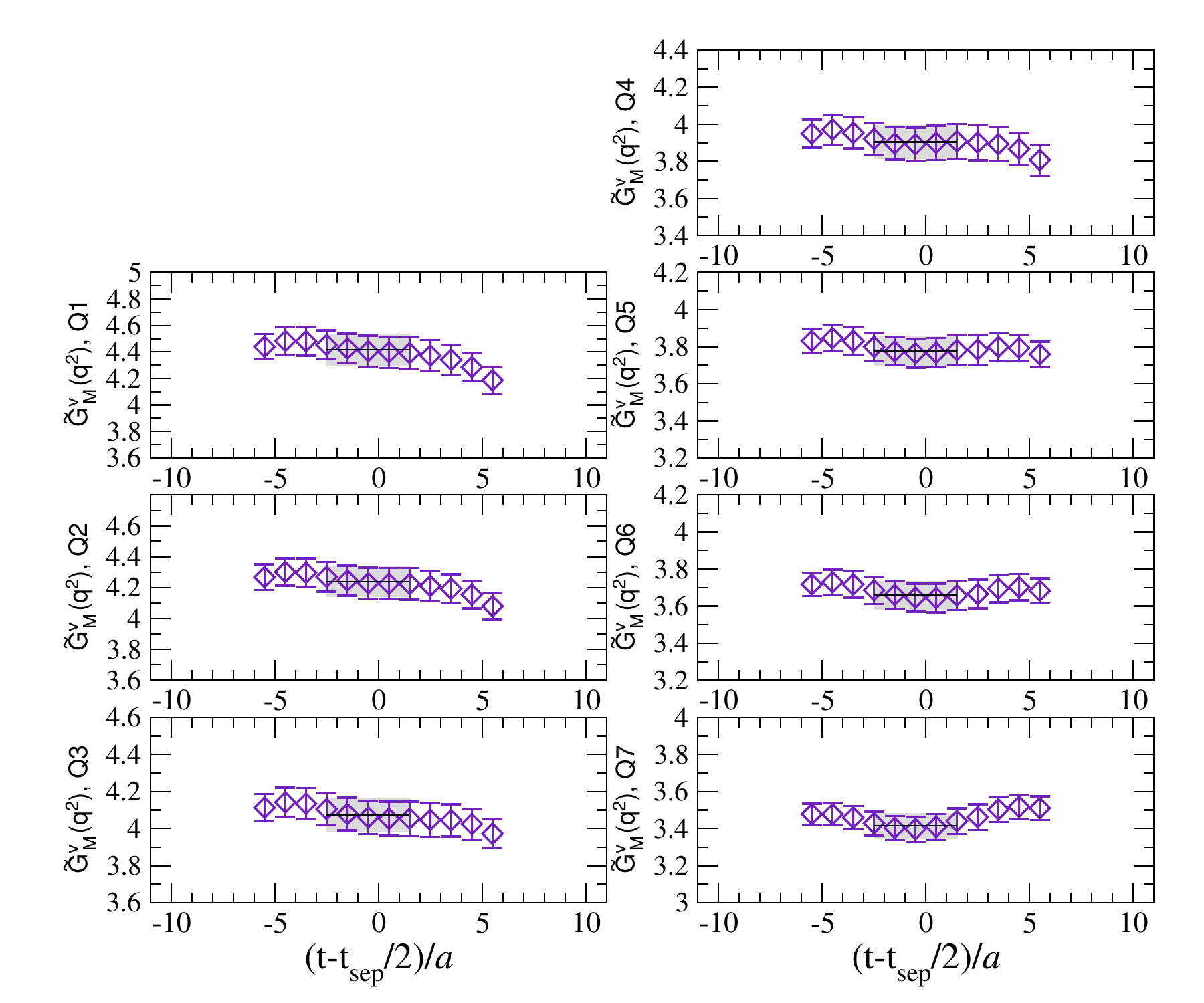}
\includegraphics[width=0.48\textwidth,bb=0 0 864 720,clip]{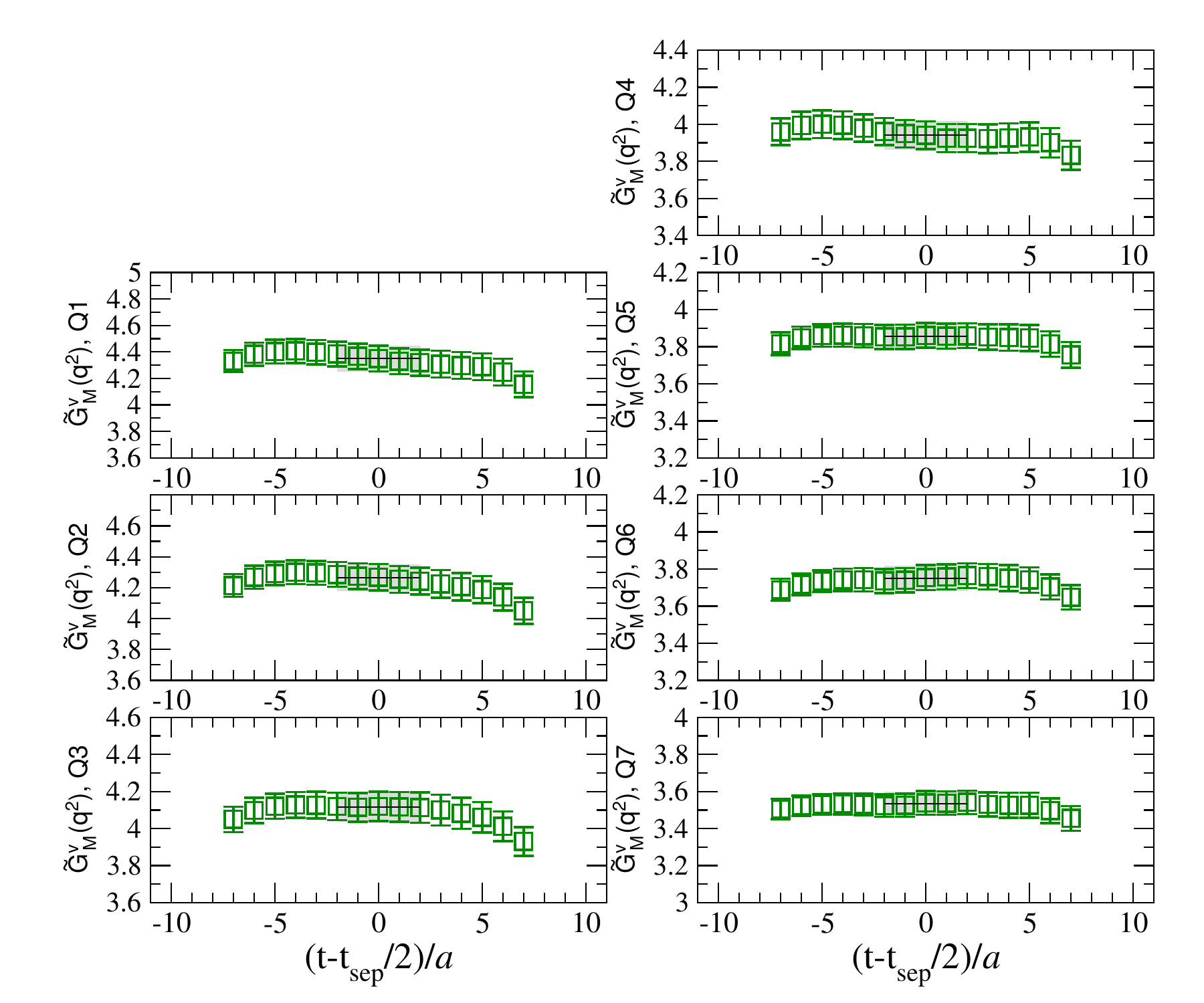}
\includegraphics[width=0.48\textwidth,bb=0 0 864 720,clip]{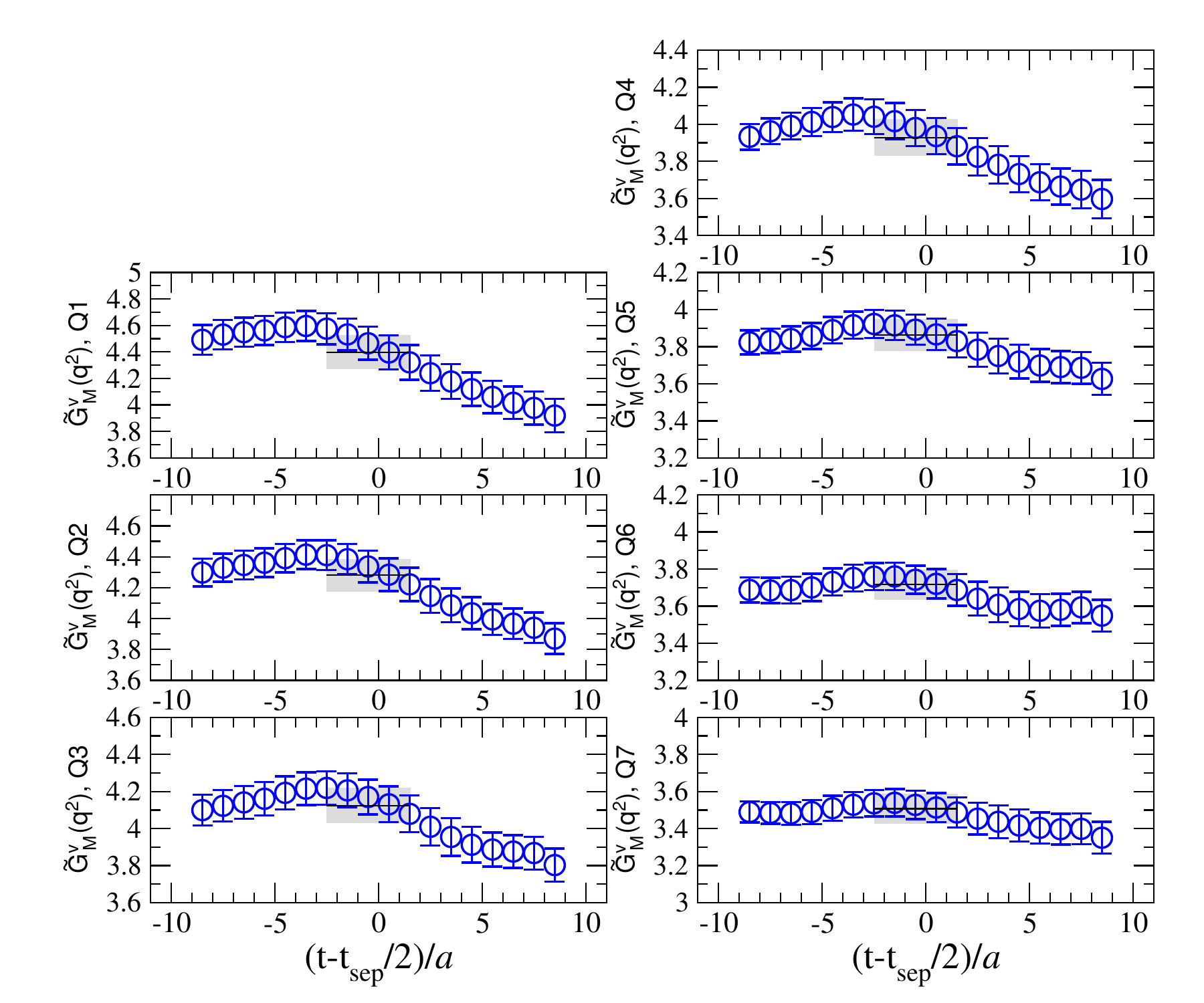}
\caption{Same as Fig.~\ref{fig:ge_qdep_p-n_ts1X} for the isovector magnetic form factor.}
\label{fig:gm_qdep_p-n_ts1X}
\end{figure*}

The magnetic form factor is extracted from the ratio $\mathcal{R}^{5z}_{V_i}(t;\bm{p})$ defined in Eq.~(\ref{eq:gm_def}).
In Fig.~\ref{fig:gm_qdep_p-n_ts1X}, 
the $t$-dependencies of the isovector (bare) magnetic form factor $\widetilde{G}^v_M(q^2)$ for all seven variations of $q^2\neq 0$ with $t_{\mathrm{sep}}/a=\{13,16,19\}$ are displayed.

As is in the case of the electric form factors,
we observe the good plateaus in data sets of $t_{\mathrm{sep}}/a=13$ and $16$ for all variations of $q^2$.
On the other hand, in the case of $t_{\mathrm{sep}}/a=19$,
$t$-dependence of $\widetilde{G}^v_M(q^2)$ for the lower $q^2$ shows a slight wiggle,
which seems to break time reversal between the source and sink points.
However, the time-reversal feature becomes restored for the higher $q^2$.
Indeed, the difference between the top and bottom of the wiggle are at most within the statistical errors and does not affect the analysis with the constant fit. Therefore, 
in all cases of $t_{\mathrm{sep}}$, we extract the values of $\widetilde{G}^v_M(q^2)$ by the standard plateau method with the same fit range for all $q^2$  
as summarized in Table~\ref{tab:measurements}.

\begin{figure*}
\centering
\includegraphics[width=1\textwidth,bb=0 0 864 720,clip]{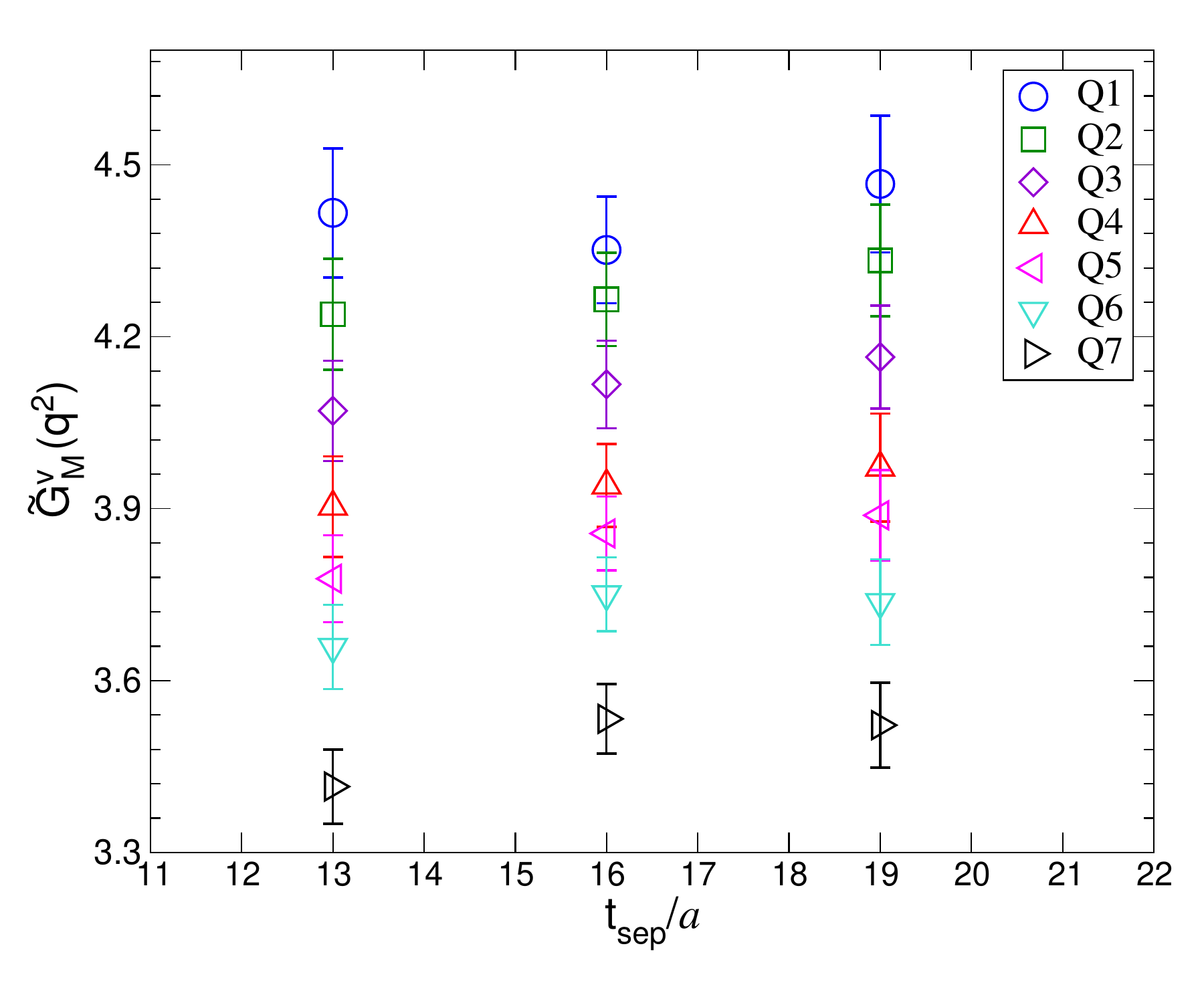}
\caption{Same as Fig.~\ref{fig:ge_tsdep_p-n} for the isovector magnetic form factor.}
\label{fig:gm_tsdep_p-n}
\end{figure*}

In Fig.~\ref{fig:gm_tsdep_p-n},
we show the $t_{\mathrm{sep}}$-dependence of the $\widetilde{G}^v_M(q^2)$,
of which data are summarized in Appendix~\ref{app:table_of_ff}.
Although the values of $\widetilde{G}_M^v(q^2)$ have the larger statistical errors than those of $\widetilde{G}_E^v(q^2)$,
the resultant values of $\widetilde{G}^v_M(q^2)$
show no significant $t_{\mathrm{sep}}$-dependence. 
Therefore, we perform the simultaneous fit with two data sets of $t_{\mathrm{sep}}/a=\{16,19\}$ to evaluate the RMS radius and the magnetic moment as our final estimates.
We also use a single data set
of $t_{\mathrm{sep}}/a=19$ for comparison and quote the difference between two results as the first systematic error.

\begin{figure*}
\centering
\includegraphics[width=1\textwidth,bb=0 0 792 612,clip]{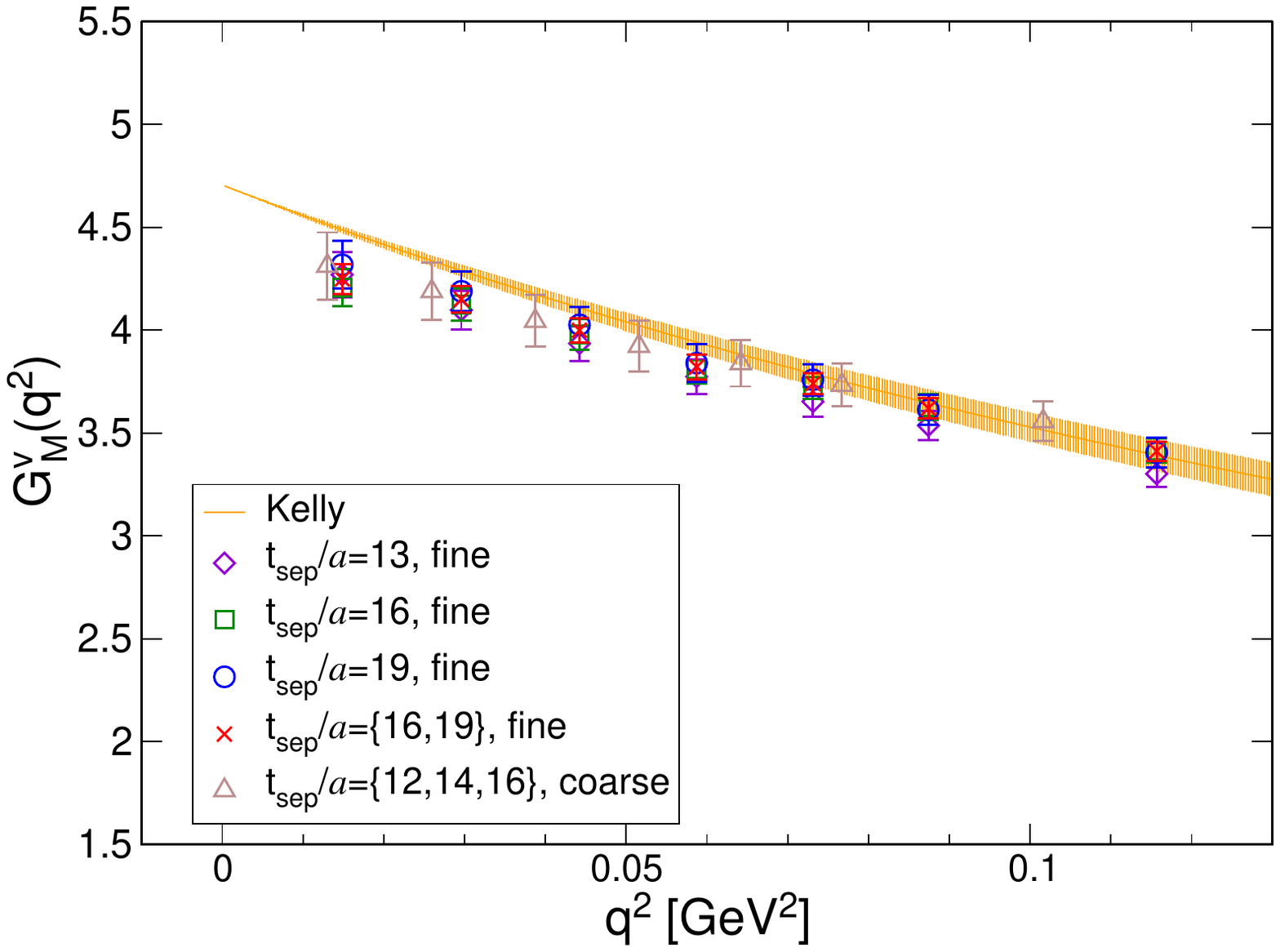}
\caption{
Same as Fig.~\ref{fig:ge_oct22} for the isovector magnetic form factor.}
\label{fig:gm_oct22}
\end{figure*}

Figure~\ref{fig:gm_oct22} shows the $q^2$-dependence of $G^v_M(q^2)=Z_V\widetilde{G}^v_M(q^2)$ with a choice of $q^2_{\mathrm{disp}}$ for the horizontal axis together with the Kelly's fit~\cite{Kelly:2004hm}.
One can see that all our data reproduce the Kelly's fit within their large errors at relatively larger $q^2$,
while they are located slightly below
the Kelly's fit, regardless of
their values of $t_{\mathrm{sep}}$ and lattice spacing $a$.

\begin{figure*}
\centering
\includegraphics[width=1\textwidth,bb=0 0 792 612,clip]{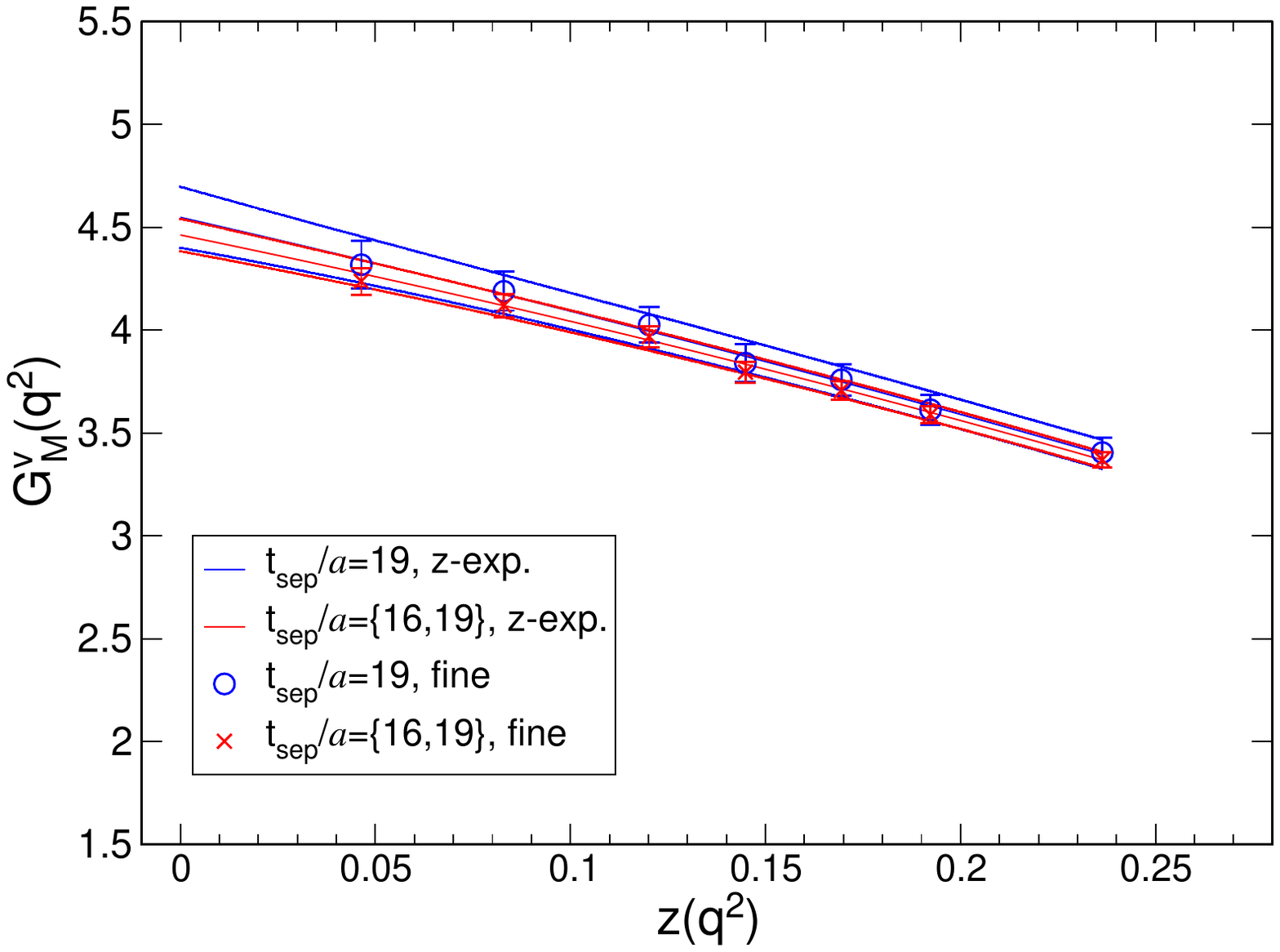}
\caption{
Same as Fig.~\ref{fig:ge_oct22_zexp} for the isovector magnetic form factor.}
\label{fig:gm_oct22_zexp}
\end{figure*}

\begin{figure*}
\centering
\includegraphics[width=0.49\textwidth,bb=0 0 792 612,clip]{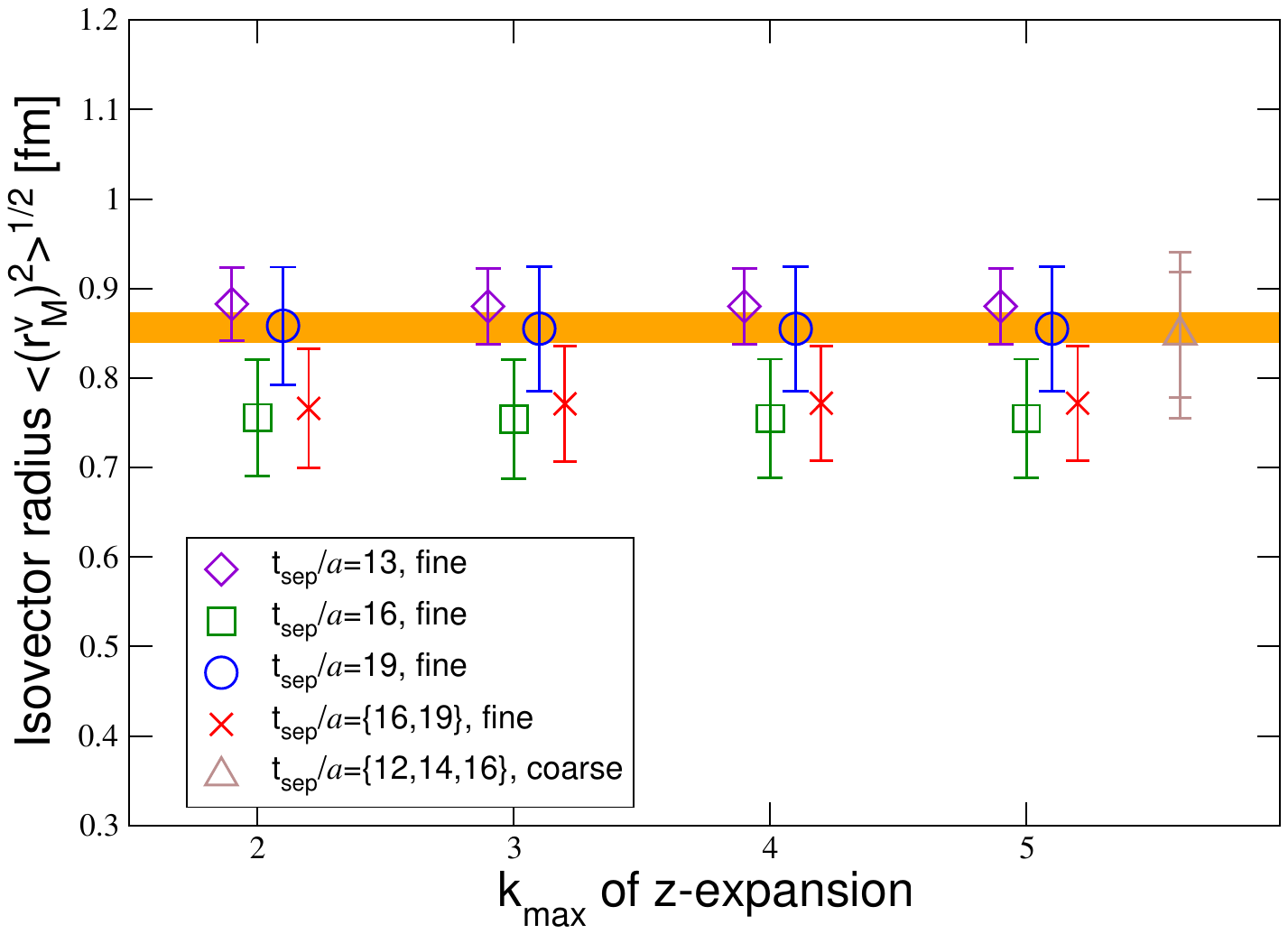}
\includegraphics[width=0.49\textwidth,bb=0 0 792 612,clip]{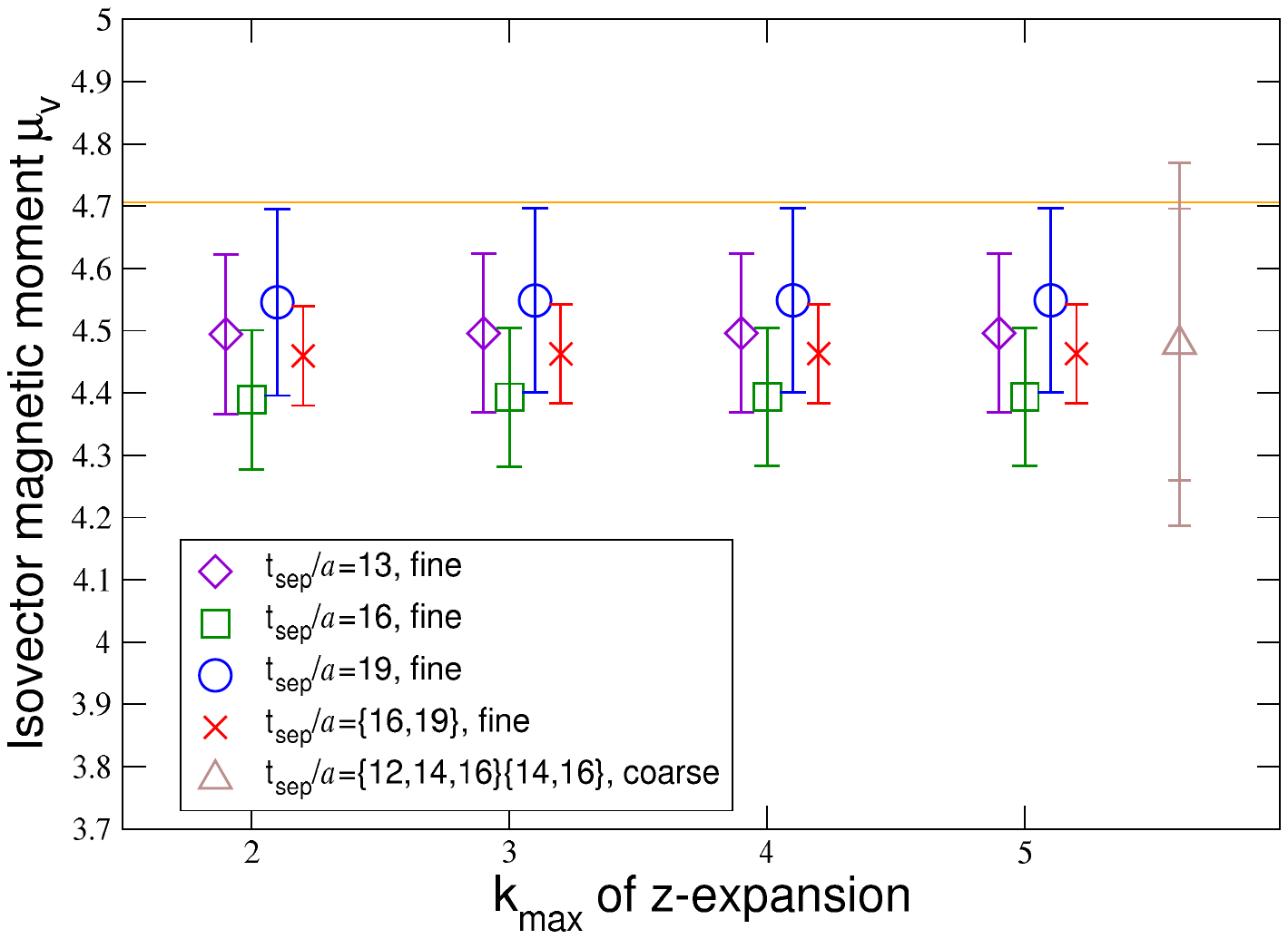}
\caption{
The $k_{\mathrm{max}}$-stability of the isovector magnetic RMS radius (left) and the magnetic moment (right) obtained by $z$-expansion. The horizontal axis represents the $k_{\mathrm{max}}$ for each $z$-expansion, but we slightly shift them for visibility. Results of $t_{\mathrm{sep}}/a=13$ data (violet diamond), $t_{\mathrm{sep}}/a=16$ data (green square), 
$t_{\mathrm{sep}}/a=19$ data (blue circle) and also the result obtained from  the simultaneous fit of $t_{\mathrm{sep}}/a=\{16,19\}$ (red cross mark) are plotted.}
\label{fig:gm_zexp_rmscomp}
\end{figure*}

Next, we evaluate the isovector magnetic RMS radius and magnetic moment by the $z$-expansion method.
Figure~\ref{fig:gm_oct22_zexp} shows the $z(q^2)$-dependence
of $G^v_M(q^2)$ with the fit result from the $z$-expansion method for $t_{\mathrm{sep}}/a=19$ (circle symbols) and a combined data of $t_{\mathrm{sep}}/a=\{16,19\}$ (cross symbols).
In Fig.~\ref{fig:gm_zexp_rmscomp} we show stability of the variation of $k_{\mathrm{max}}$
in extracting both the isovector magnetic RMS radius $\sqrt{\langle (r^v_M)^2 \rangle}$ and the magnetic moment $\mu_v$ for each $t_{\mathrm{sep}}$ data and a combined data of $t_{\mathrm{sep}}/a=\{16,19\}$.
The results of both quantities obtained from the analysis with the $z$-expansion method are 
stable under the variation of $k_{\mathrm{max}}$.
Furthermore, in contrast to the electric one, 
the resultant values for all cases of $t_{\mathrm{sep}}/a=\{13,16,19\}$ are mutually consistent within their statistical errors.

We finally choose the result obtained by the simultaneous fit of a combined data $t_{\mathrm{sep}}/a=\{16,19\}$ with $k_{\mathrm{max}}=4$ for our best estimate, and the systematic errors are quoted in the same way as the electric one.
The analyses with other model-dependent
functional forms are discussed in Appendix~\ref{app:model-dep_anal}.
All results of $\sqrt{\langle (r^v_M)^2 \rangle}$ and $\mu_v$ are summarized in Table~\ref{tab:rm_zexp}.

The statistical uncertainty on the magnetic RMS radius is about three times 
larger than the electric RMS radius. 
Our best estimate of $\sqrt{\langle (r^v_M)^2 \rangle}$ reproduces the experimental value within the statistical error. 
Furthermore, although our result is consistent with our previous result calculated at the coarse lattice albeit with a relatively large error, a discrepancy between
their central values is observed to be about 9.0\%. 
This difference is comparable in the size of the discretization error observed in the electric RMS radius.

On the other hand, the central value of $\mu_v$ has a few standard deviations away from the corresponding experimental value, and is slightly underestimated.
However our best estimate of $\mu_v$ is consistent with the one obtained from the coarse lattice and then does not indicate the presence of discretization errors,
which may resolve the above discrepancy observed in the magnetic moment. 
This issue could be related to the fact that $\mu_v=G^v_M(0)$ is not directly measurable, but
can be accessed by extrapolation of data from regions where $q^2$ is nonzero.
In other words,
the determination of the magnetic moment $\mu_v=G^v_M(0)$ potentially suffers from the systematic uncertainty due to the $q^2$ extrapolation.
In order to avoid such uncertainty, 
Ref.~\cite{Ishikawa:2021eut} advocates a direct calculation method without 
the $q^2$ extrapolation, which will be performed in our future works.

%
%
\begin{table*}[ht!]
     \renewcommand{\arraystretch}{1.1}
    {\tiny
\begin{ruledtabular}
\caption{
Results for the magnetic moments $\mu$ and magnetic RMS radius
$\sqrt{\langle r^2_{M}\rangle}$ for the isovector, proton and neutron channels. 
In the row of ``This work'' we present our best estimates, the first error is statistical one and the second and third are the systematic errors described in the text. Results for the proton and neutron are obtained without the disconnected diagram. 
\label{tab:rm_zexp}}
\begin{tabular}{ccccccccccccc}
  & & & \multicolumn{3}{c}{Isovector}\\
  \cline{4-6}
  Fit type & $q^2$ [GeV$^2$] & $t_{\rm sep}/a$ &
  $\mu_v$ & $\sqrt{\langle (r^v_M)^2\rangle}$ [fm]& $\chi^2$/d.o.f. \\
  \hline
  \multicolumn{2}{ c }{$160^4$ (fine) lattice} \cr
  \multirow{2}{*}{$k_{\rm max}=4$}                      & \multirow{2}{*}{$q^2_{\mathrm{disp}}\le0.116$} & $\{16,19\}$ & { 4.436(89) } & { 0.771(64) } & { 0.21}\\
  &                                  & $19$       & { 4.544(133) } &{ 0.855(70) }  & { 0.16}\\
  \multirow{2}{*}{$k_{\rm max}=3$}                      & \multirow{2}{*}{$q^2_{\mathrm{meas}}\le0.091$} & $\{16,19\}$ & { 4.454(85) } & { 0.781(49) } & { 0.28}\\
  &                                  & $19$       & { 4.565(130) } &{ 0.864(47) }  & { 0.20}\\
  \multicolumn{2}{ c }{This work} & & 
  $4.436(89)(108)(18)$ & $0.771(64)(84)(10)$ & \\
  \multicolumn{2}{ c }{$128^4$ (coarse) lattice} \cr
  \multirow{2}{*}{$k_{\rm max}=4$}                      & \multirow{2}{*}{$q^2_{\mathrm{disp}}\le0.102$} & $\{12,14,16\}$ & { 4.478(218) } & { 0.848(70) } & { 0.9}\\
  &                                  & $\{14,16\}$       & { 4.670(253) } &{ 0.900(65) }  & { 0.9}\\
  \multirow{2}{*}{$k_{\rm max}=4$}                      & \multirow{2}{*}{$q^2_{\mathrm{meas}}\le0.112$} & $\{12,14,16\}$ & { 4.480(221) } & { 0.817(76) } & { 0.9}\\
  &                                  & $\{14,16\}$       & { 4.677(255) } &{ 0.874(65) }  & { 0.9}\\
  \multicolumn{2}{ c }{PACS10 $128^4$ result} & &
  $4.478(218)(192)(2)$ & $0.848(70)(52)(31)$ & \\
  \multicolumn{2}{ c }{Experimental value~\cite{ParticleDataGroup:2022pth}}\\
  & & & 4.70589 & 0.856(16) \\
 \hline\hline
  & & & \multicolumn{3}{c}{Proton} & \multicolumn{3}{c}{Neutron}\\
  \cline{4-6}
  \cline{7-9}
  Fit type & $q^2$ [GeV$^2$] & $t_{\rm sep}/a$ &
  $\mu_p$ & $\sqrt{\langle (r^p_M)^2\rangle}$ [fm]& $\chi^2$/d.o.f. &
  $\mu_n$ & $\sqrt{\langle (r^n_M)^2\rangle}$ [fm]& $\chi^2$/d.o.f. \\
  \hline
  \multicolumn{2}{ c }{$160^4$ (fine) lattice} \cr
  \multirow{2}{*}{$k_{\rm max}=4$}                       & \multirow{2}{*}{$q^2_{\mathrm{disp}}\le0.116$} & $\{16,19\}$ & { 2.702(60) } & { 0.775(74) } & { 0.23}   & { $-$1.695(41) }  & { 0.692(93) } & { 0.32}\\
             &                        & $19$       & { 2.723(92) } & { 0.845(109) } & { 0.14} & { $-$1.722(69) } & { 0.705(157) } & { 0.20}\\
  \multirow{2}{*}{$k_{\rm max}=3$}                       & \multirow{2}{*}{$q^2_{\mathrm{meas}}\le0.091$} & $\{16,19\}$ & { 2.697(56) } & { 0.732(89) } & { 0.23}   & { $-$1.693(40) }  & { 0.652(114) } & { 0.36}\\
             &                        & $19$       & { 2.728(87) } & { 0.831(85) } & { 0.15} & { $-$1.722(68) } & { 0.671(157) } & { 0.22}\\
  \multicolumn{2}{ c }{This work} & &
  $2.702(60)(21)(5)$ & $0.775(74)(70)(43)$ & &
  $-1.695(41)(27)(2)$ & $0.692(93)(13)(40)$\\
  \multicolumn{2}{ c }{$128^4$ (coarse) lattice} \cr
    \multirow{2}{*}{$k_{\rm max}=4$}                       & \multirow{2}{*}{$q^2_{\mathrm{disp}}\le0.102$} & $\{12,14,16\}$ & { 2.741(129) } & { 0.812(98) } & { 1.0}   & { $-$1.718(99) }  & { 0.969(134) } & { 0.08}\\
             &                        & $\{14,16\}$       & { 2.834(163) } & { 0.879(89) } & { 1.6} & { $-$1.842(114) } & { 0.969(134) } & { 0.08}\\
  \multirow{2}{*}{$k_{\rm max}=4$}                       & \multirow{2}{*}{$q^2_{\mathrm{meas}}\le0.112$} & $\{12,14,16\}$ & { 2.742(131) } & { 0.780(104) } & { 1.0}   & { $-$1.718(100) }  & { 0.938(140) } & { 0.08}\\
             &                        & $\{14,16\}$       & { 2.840(165) } & { 0.859(95) } & { 1.6} & { $-$1.845(117) } & { 0.938(140) } & { 0.08}\\
  \multicolumn{2}{ c }{PACS10 $128^4$ result} & &
  $2.741(129)(93)(1)$ & $0.812(98)(67)(32)$ & &
  $-1.718(99)(124)(0)$ & $0.818(134)(151)(33)$\\
  \multicolumn{2}{ c }{Experimental value~\cite{ParticleDataGroup:2022pth}} \cr
  & & & 2.79285 & 0.851(26) & & $-$1.91304 & 0.864(9)\cr  
\end{tabular}
\end{ruledtabular}
}
\end{table*}

%
%
%
%
%
%
%
%
%
%

\subsubsection{Proton and Neutron sector}
\begin{figure*}
\centering
\includegraphics[width=0.48\textwidth,bb=0 0 864 720,clip]{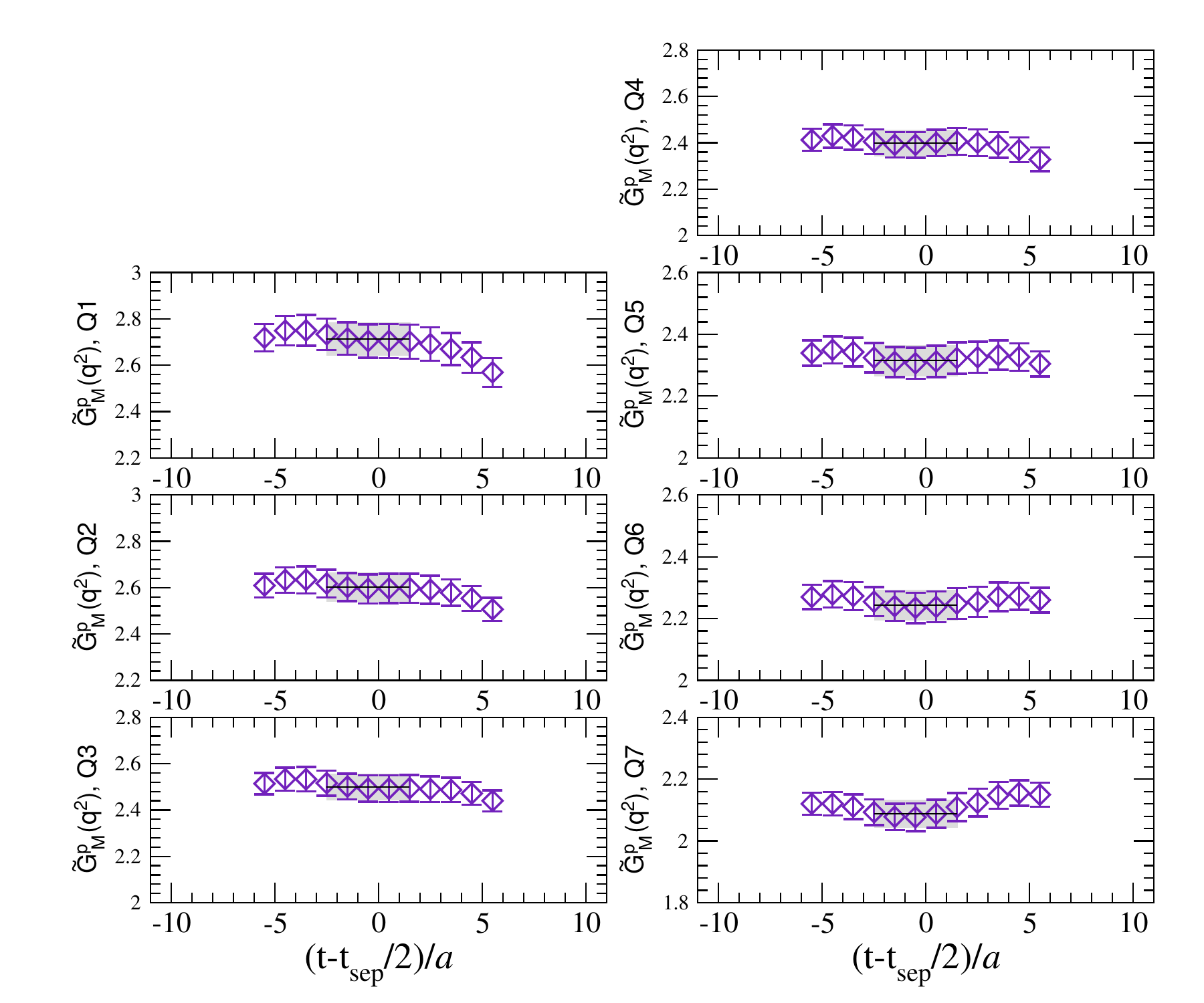}
\includegraphics[width=0.48\textwidth,bb=0 0 864 720,clip]{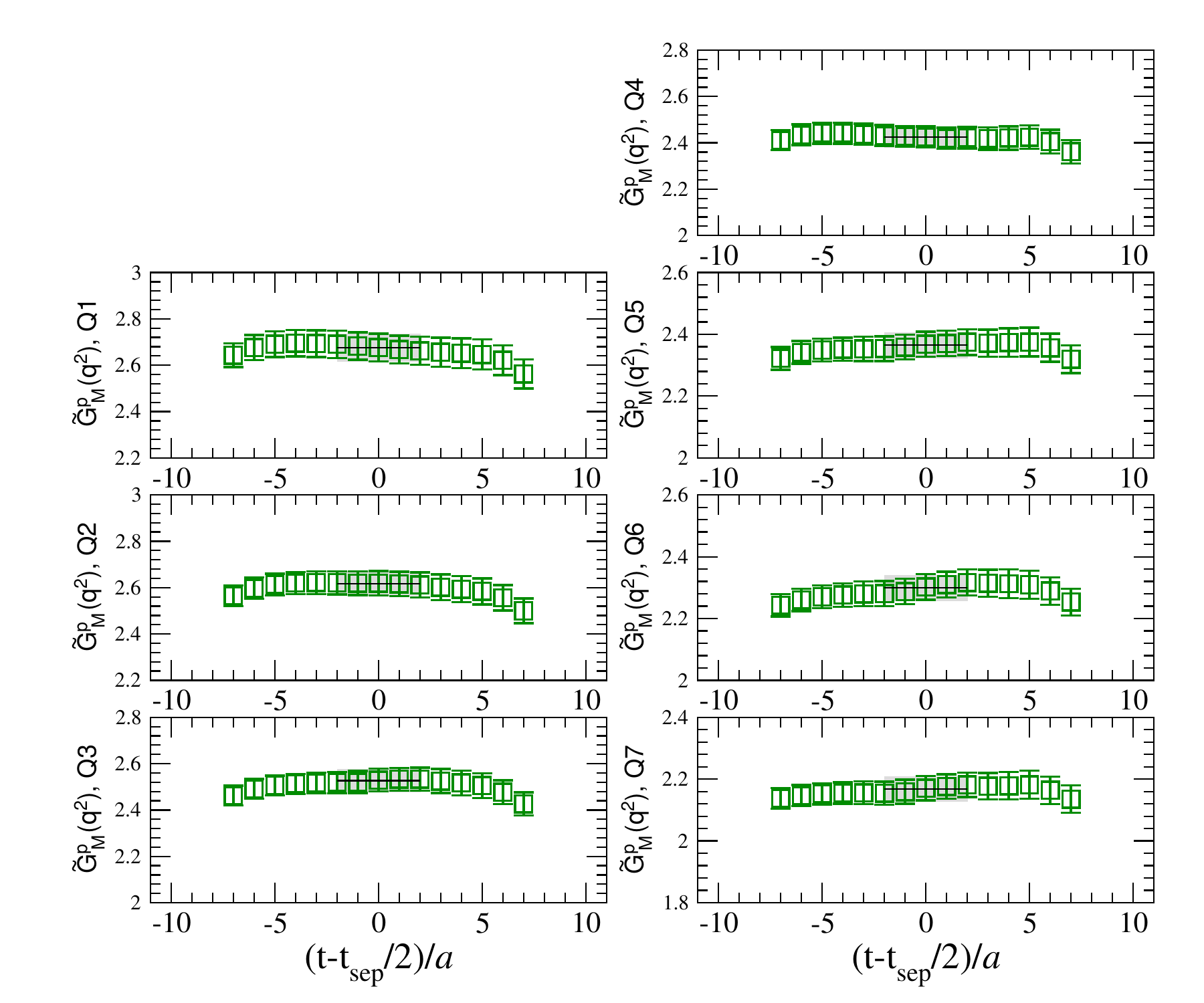}
\includegraphics[width=0.48\textwidth,bb=0 0 864 720,clip]{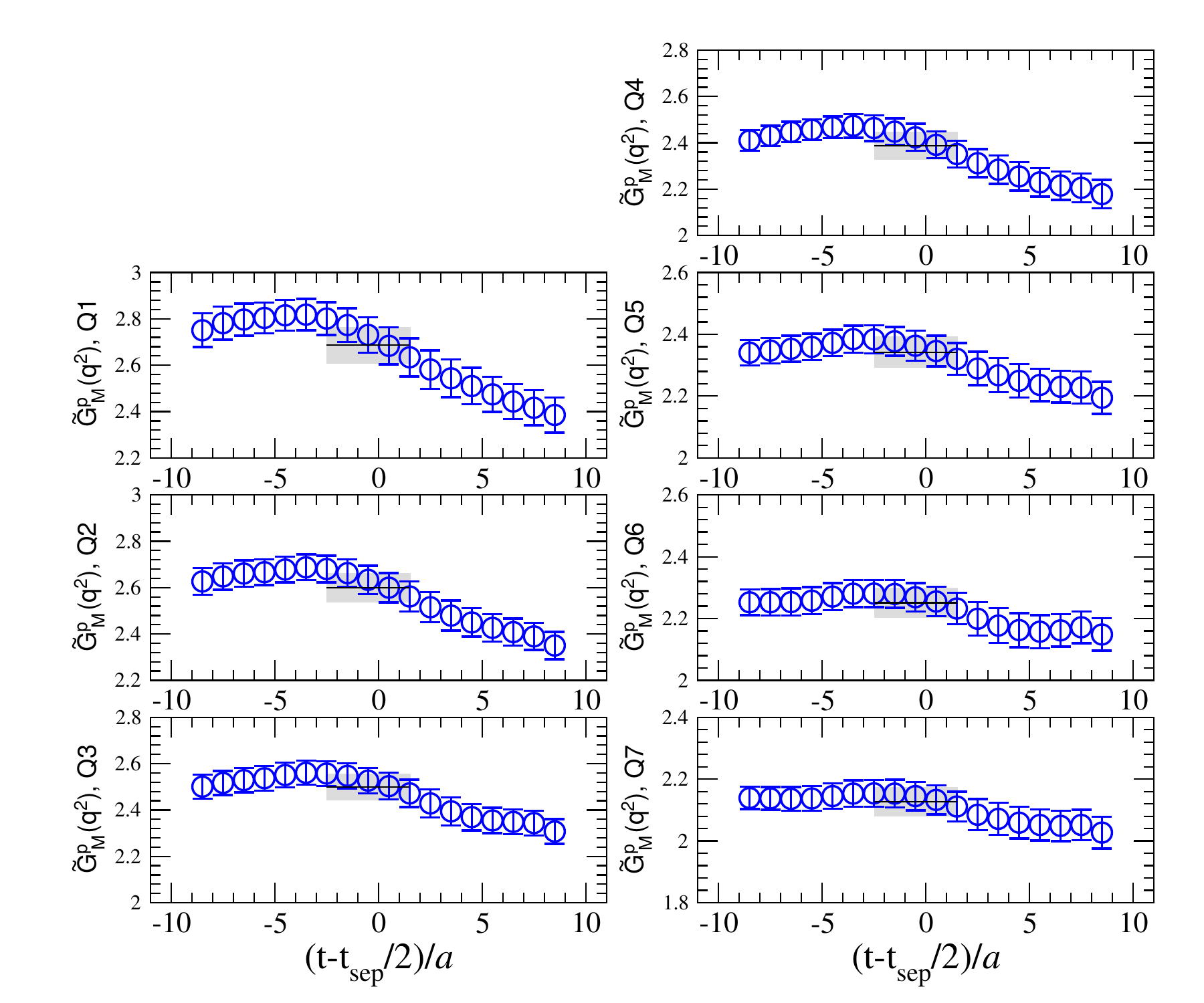}
\caption{Same as Fig.~\ref{fig:gm_qdep_p-n_ts1X} for the proton.}
\label{fig:gmp_qdep_p-n_ts1X}
\end{figure*}

\begin{figure*}
\centering
\includegraphics[width=0.48\textwidth,bb=0 0 864 720,clip]{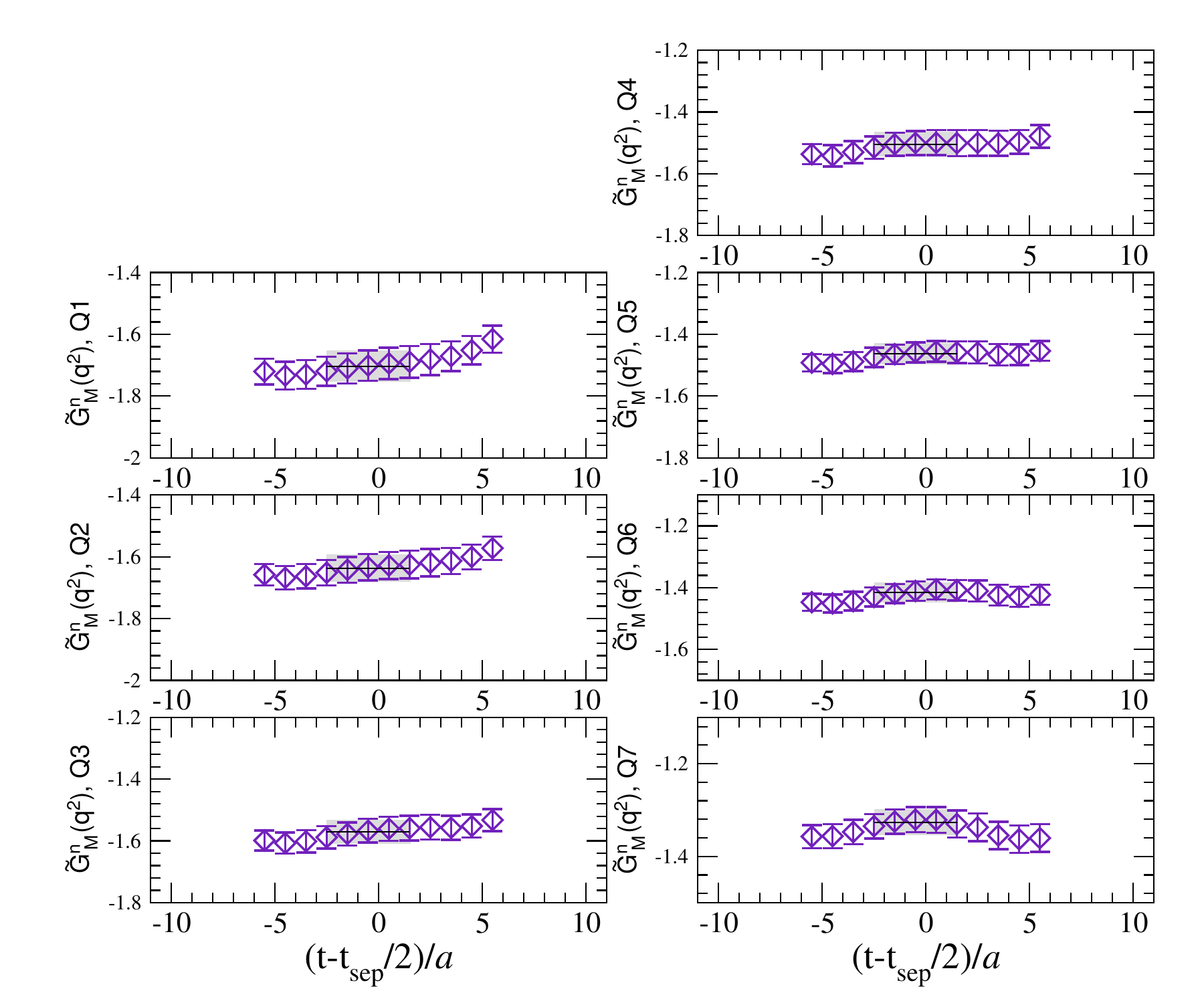}
\includegraphics[width=0.48\textwidth,bb=0 0 864 720,clip]{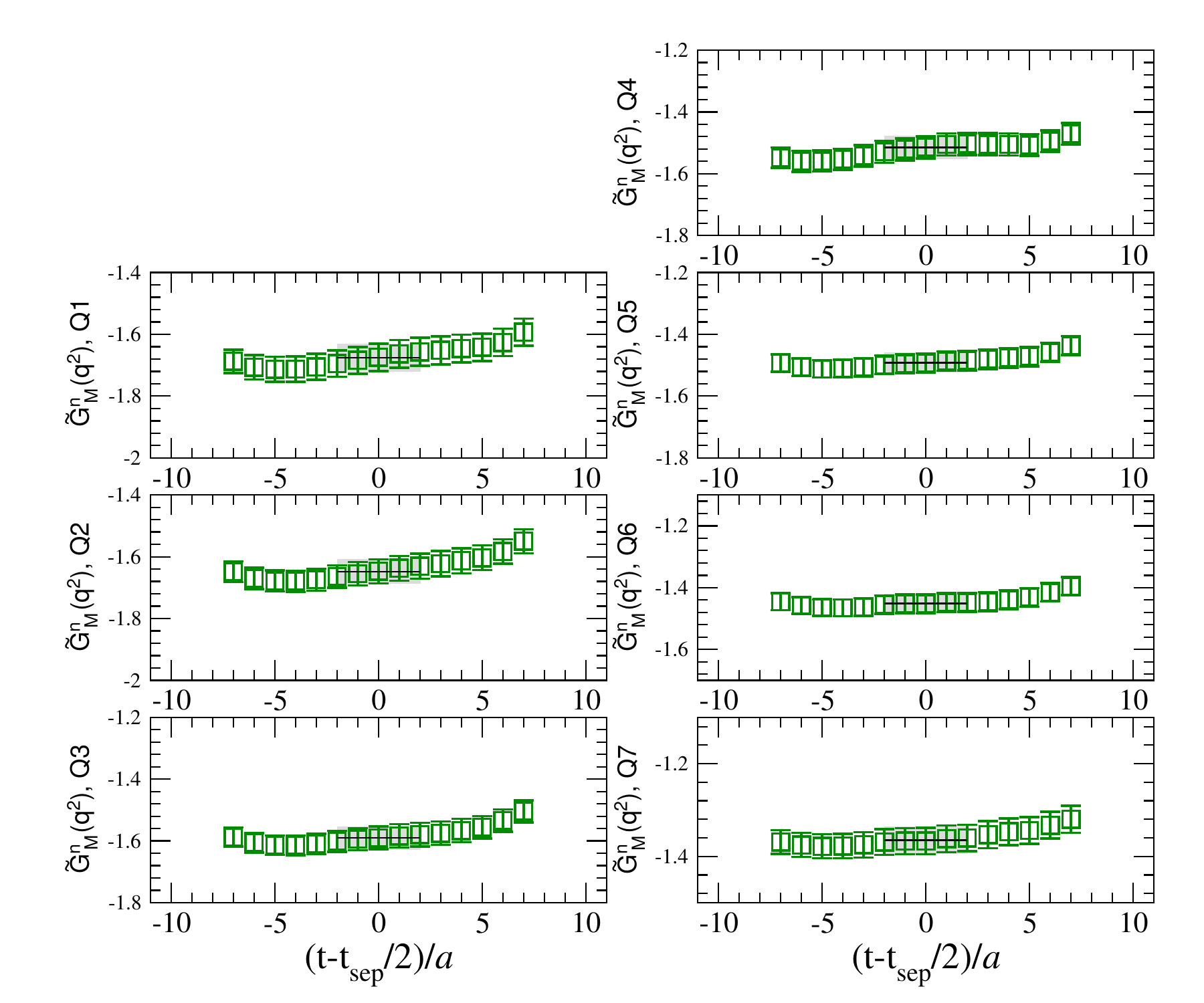}
\includegraphics[width=0.48\textwidth,bb=0 0 864 720,clip]{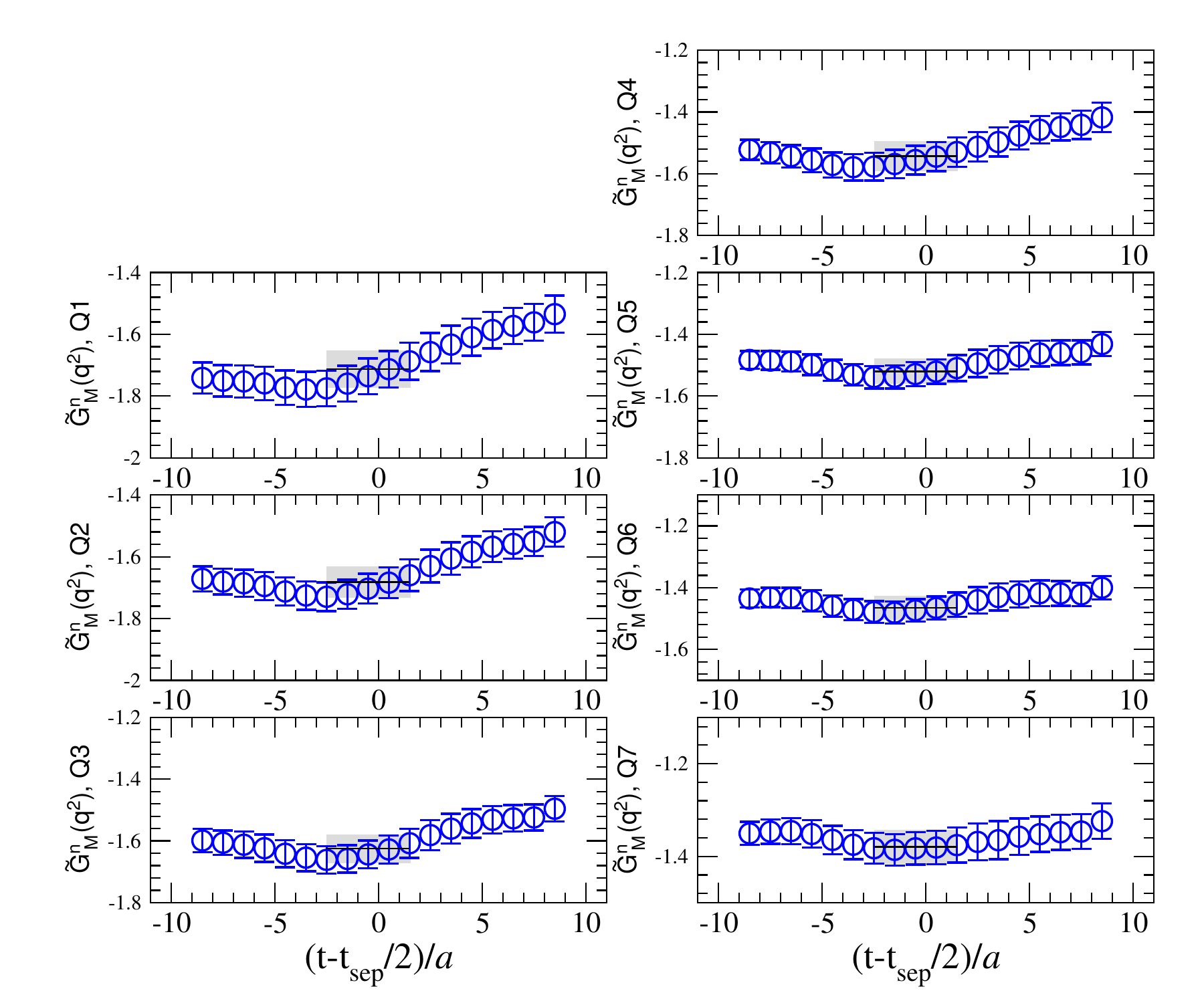}
\caption{Same as Fig.~\ref{fig:gm_qdep_p-n_ts1X} for the neutron.}
\label{fig:gmn_qdep_p-n_ts1X}
\end{figure*}

We evaluate the magnetic form factors, $\widetilde{G}^p_M(q^2)$ and $\widetilde{G}^n_M(q^2)$,
separately from the ratio $\mathcal{R}^{5z}_{V_i}(t;\bm{q})$ for each proton and neutron, where we omit the disconnected contributions.
Similar to the isovector case, Figures~\ref{fig:gmp_qdep_p-n_ts1X} and \ref{fig:gmn_qdep_p-n_ts1X} show the good plateaus in each case of $t_{\mathrm{sep}}$ except for the lower $q^2$ in data set of
$t_{\mathrm{sep}}/a=19$, where the $t$-dependence
is not symmetric between the source and sink points.
As summarized in Table~\ref{tab:measurements}, for all $q^2$,
we simply extract $\widetilde{G}^p_M(q^2)$ and $\widetilde{G}^n_M(q^2)$ by the standard plateau method with the 
same fit range as for the isovector case.

\begin{figure*}
\centering
\includegraphics[width=0.49\textwidth,bb=0 0 864 720,clip]{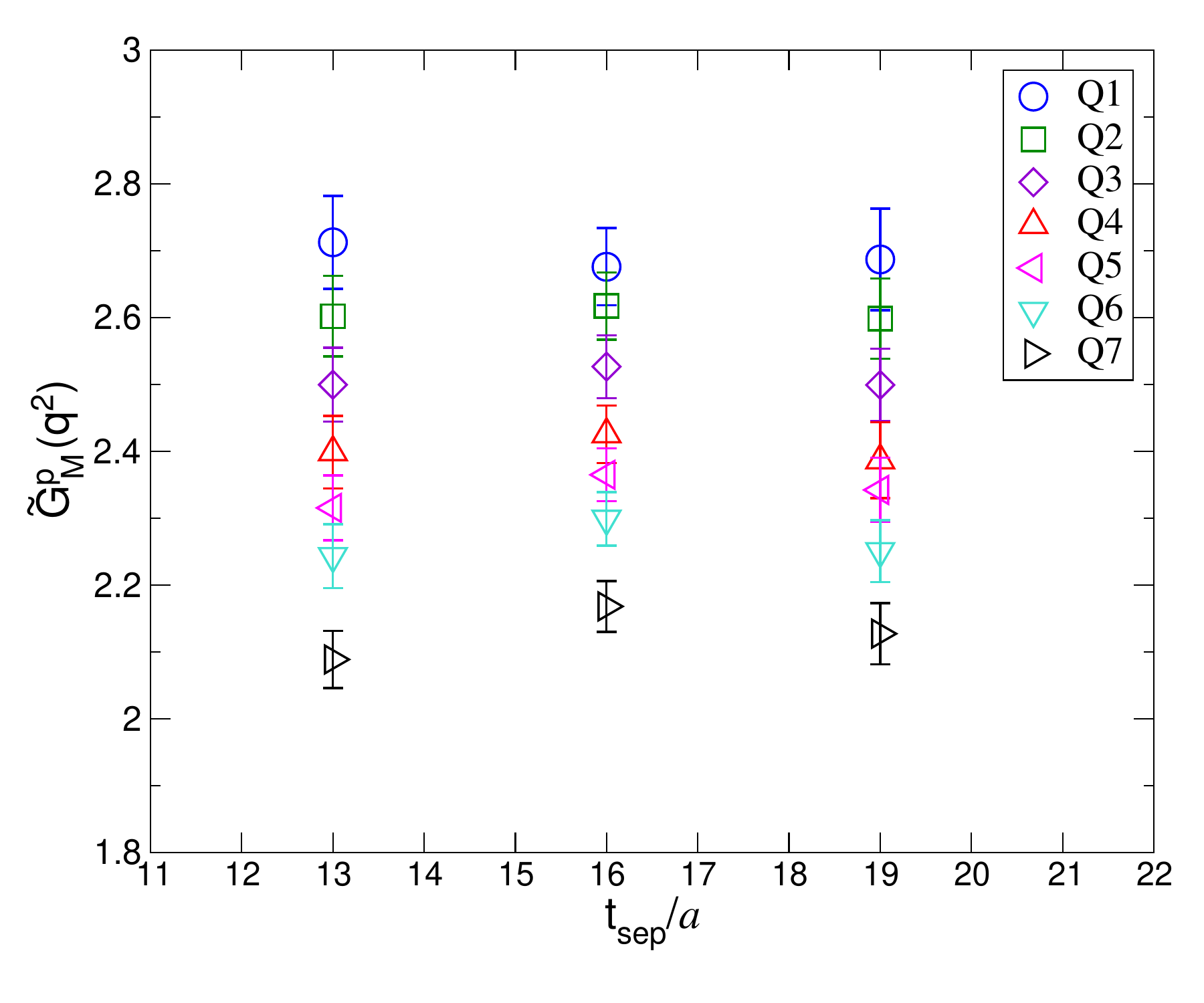}
\includegraphics[width=0.49\textwidth,bb=0 0 864 720,clip]{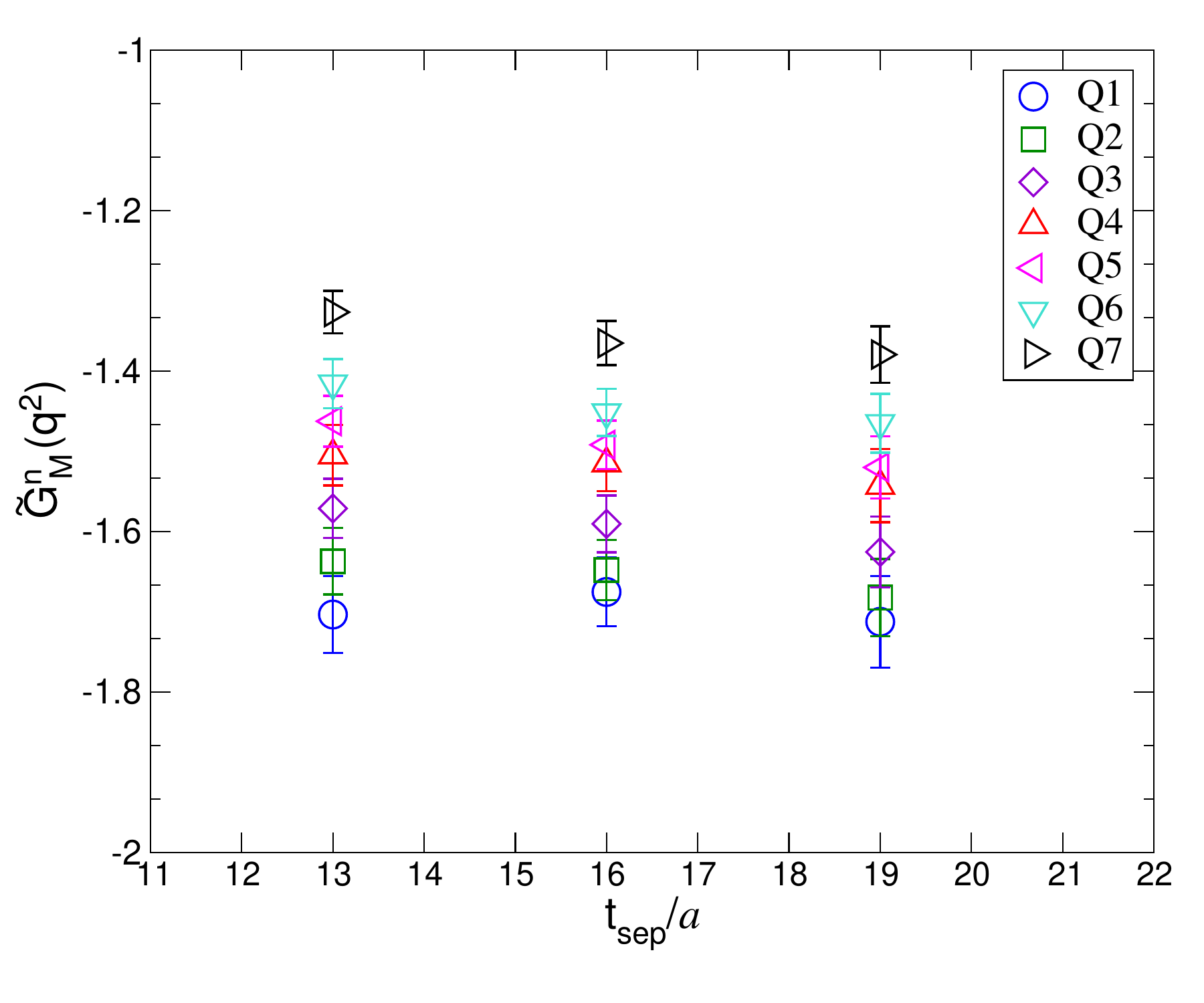}
\caption{Same as Fig.~\ref{fig:gm_tsdep_p-n} for the proton (left) and neutron (right).}
\label{fig:gmpn_tsdep_p-n}
\end{figure*}

\begin{figure*}
\centering
\includegraphics[width=0.49\textwidth,bb=0 0 792 612,clip]{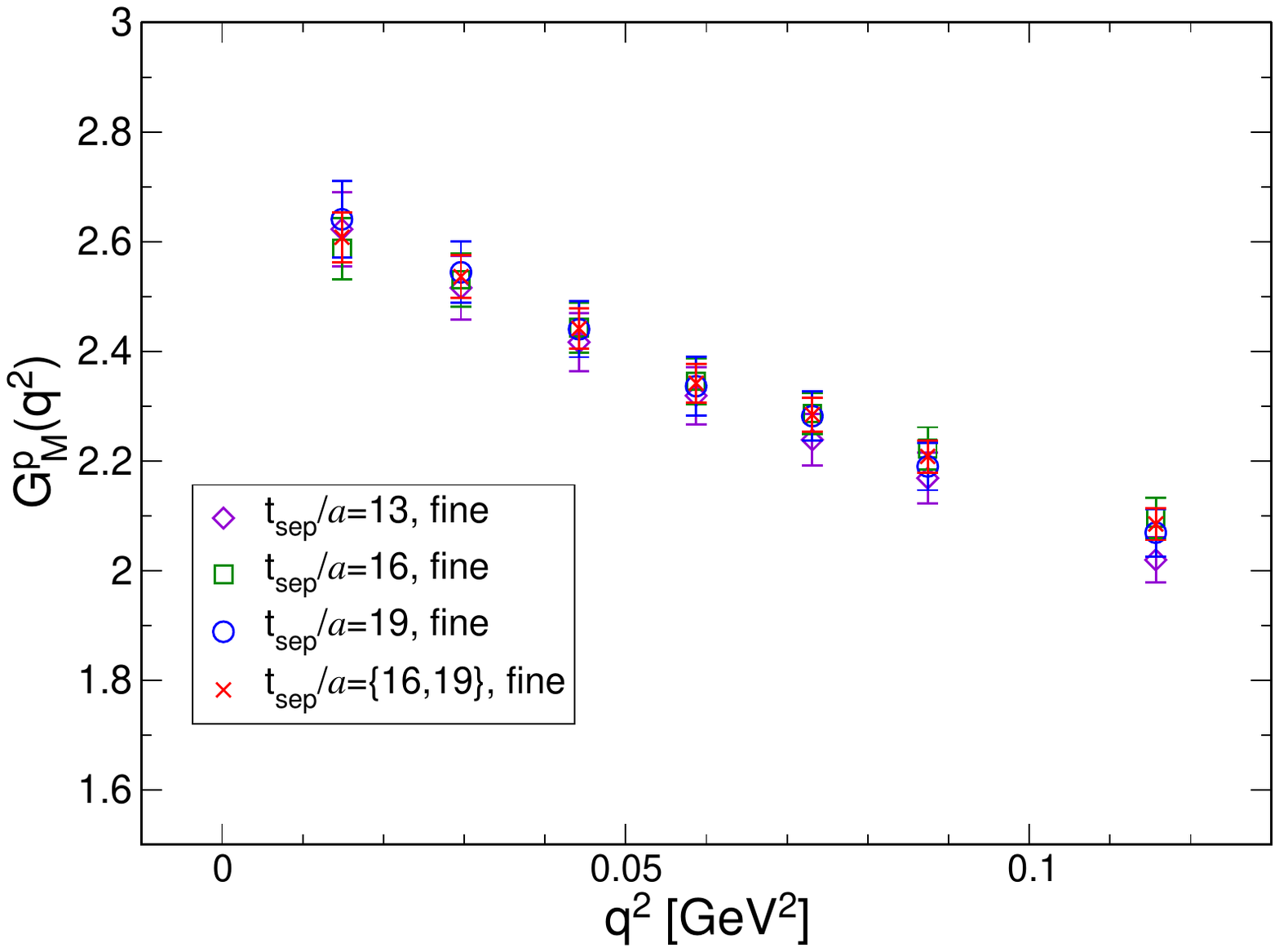}
\includegraphics[width=0.49\textwidth,bb=0 0 792 612,clip]{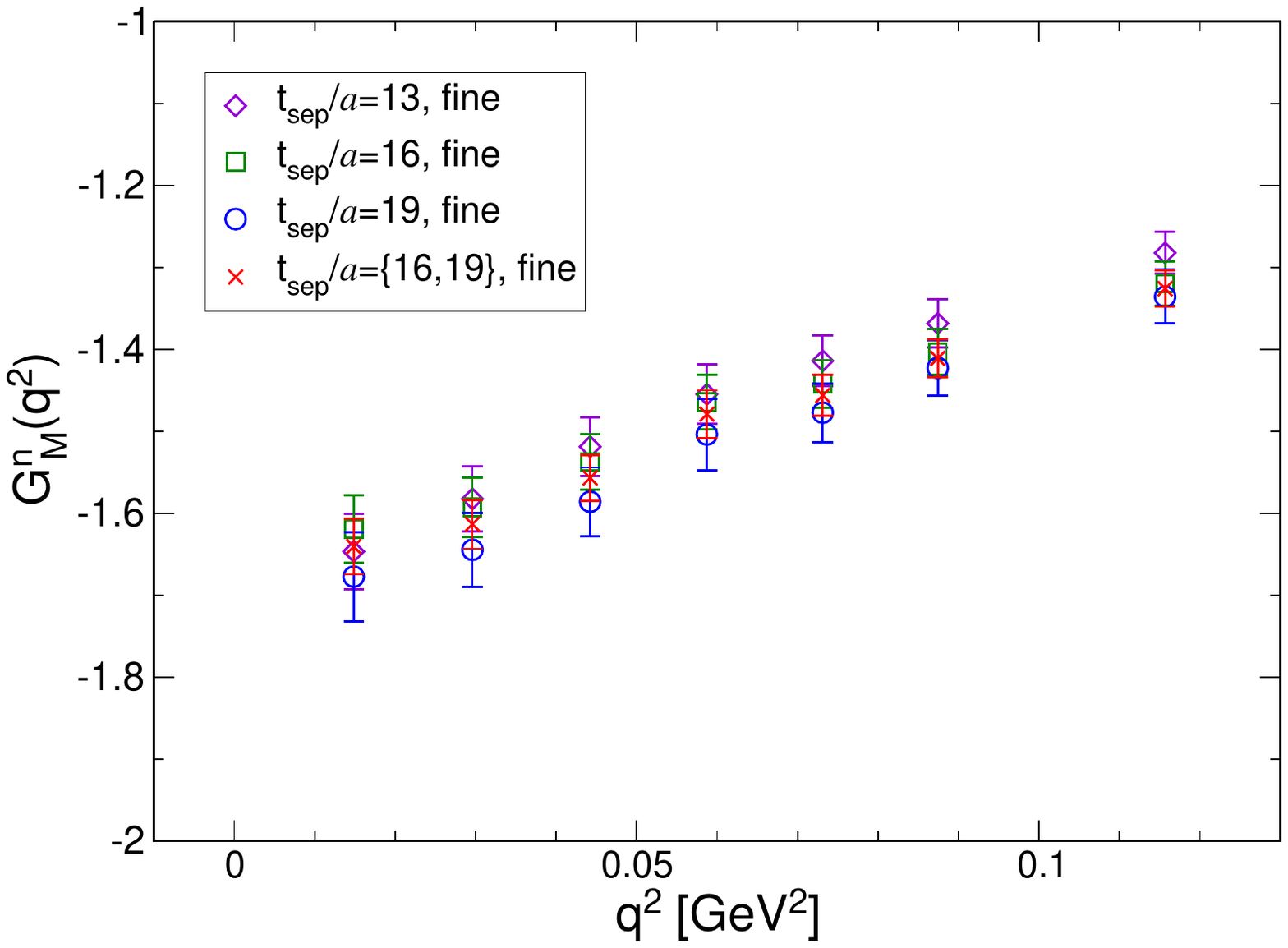}
\caption{
Same as Fig.~\ref{fig:gm_oct22} for the proton (left) and neutron (right).}
\label{fig:gmpn_oct22}
\end{figure*}

In Fig.~\ref{fig:gmpn_tsdep_p-n}, 
we display the $t_{\mathrm{sep}}$-dependence of the extracted $\widetilde{G}^p_M(q^2)$ and $\widetilde{G}^n_M(q^2)$.
It is obvious that there are no significant $t_{\mathrm{sep}}$-dependence for both the proton and neutron.
After employing the simultaneous fit using a combined data of $t_{\mathrm{sep}}/a=\{16,19\}$,
the resultant values of $G^p_M(q^2)=Z_V\widetilde{G}^p_M(q^2)$ and $G^n_M(q^2)=Z_V\widetilde{G}^n_M(q^2)$ at every $q^2$ 
are presented in Fig.~\ref{fig:gmpn_oct22}.

\begin{figure*}
\centering
\includegraphics[width=0.49\textwidth,bb=0 0 792 612,clip]{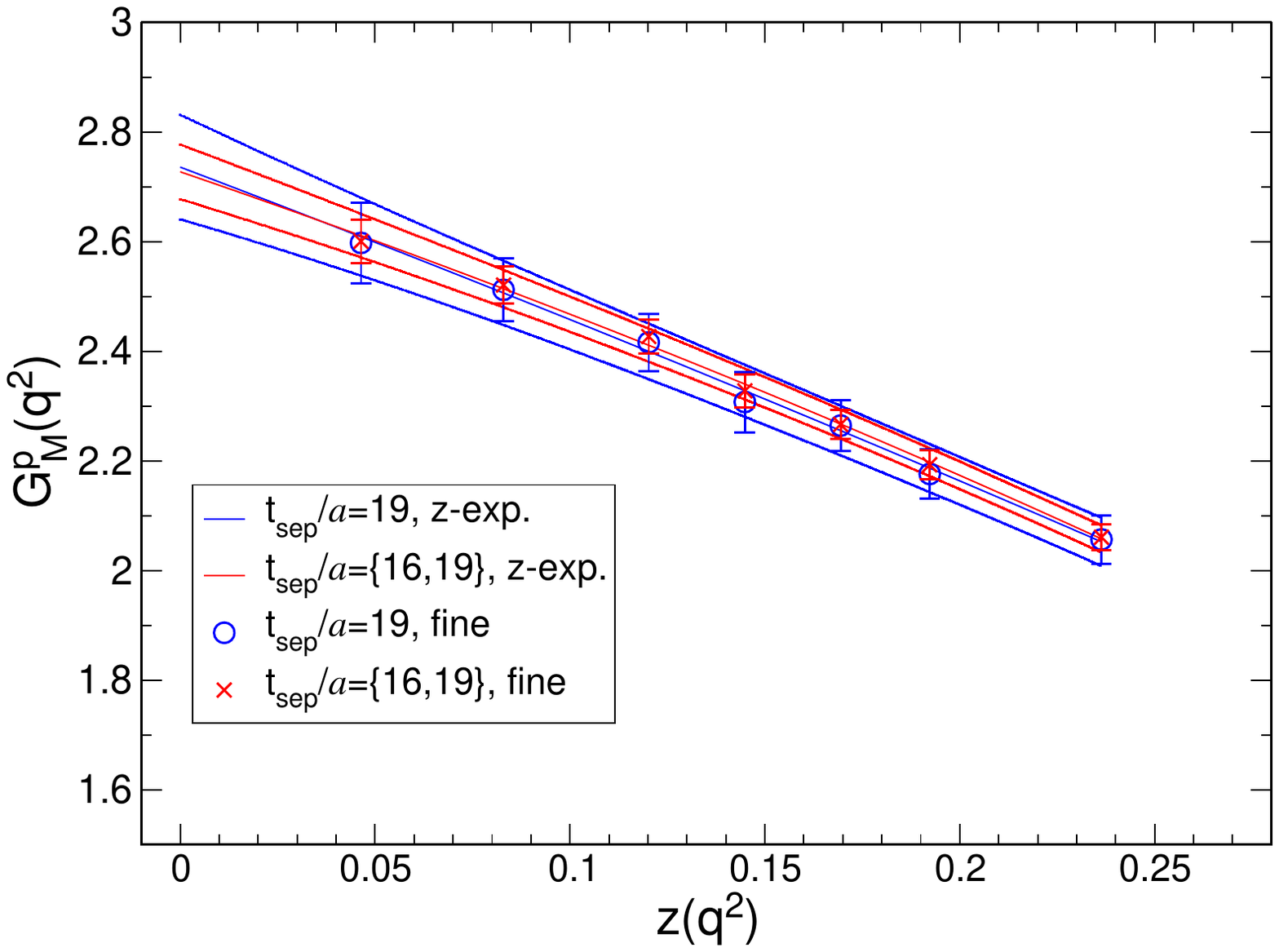}
\includegraphics[width=0.49\textwidth,bb=0 0 792 612,clip]{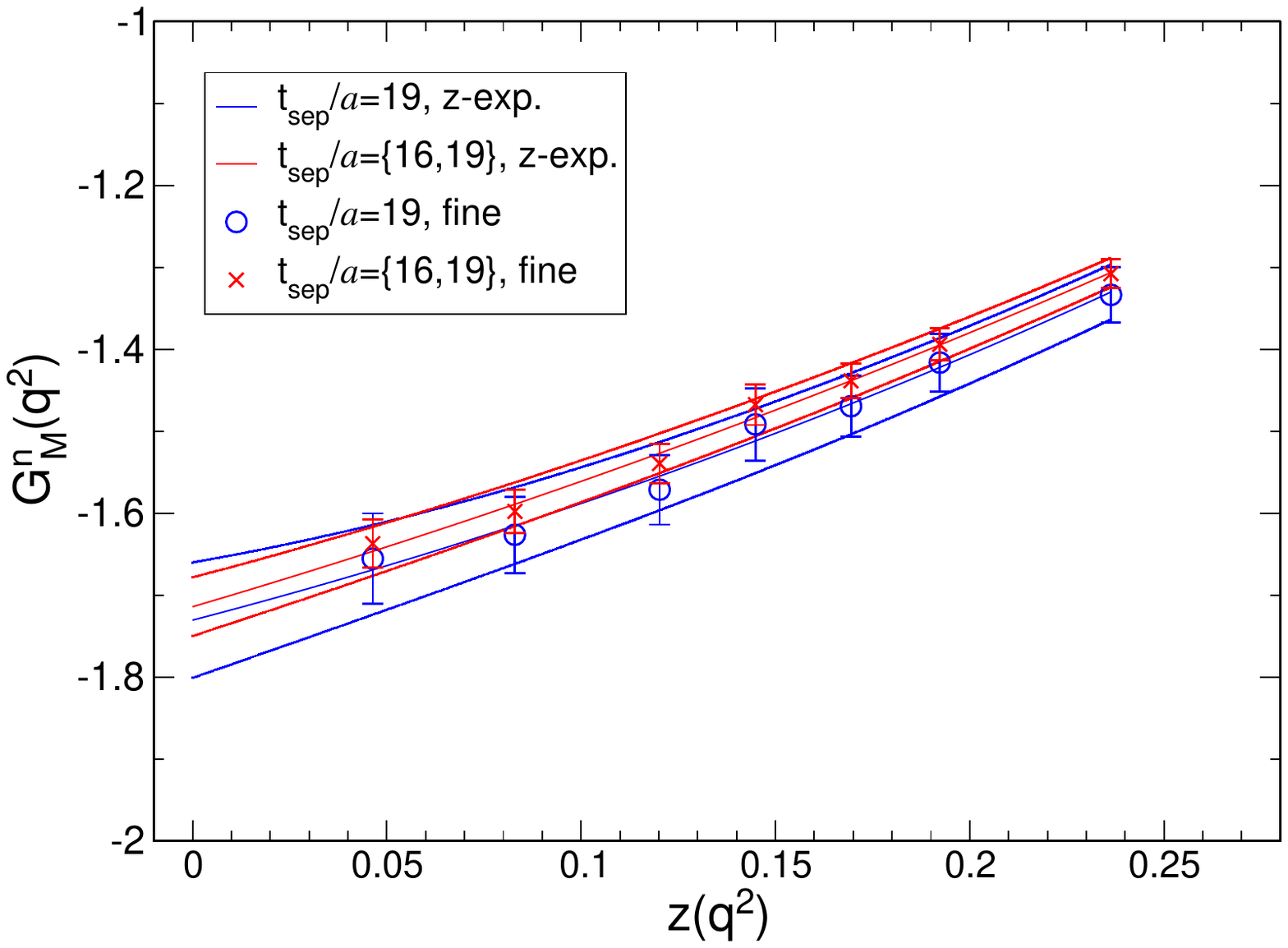}
\caption{
Same as Fig.~\ref{fig:gm_oct22_zexp} for the proton (left) and neutron (right).}
\label{fig:gmpn_oct22_zexp}
\end{figure*}

We obtain both the magnetic RMS radii of the proton 
($\sqrt{\langle (r^p_M)^2 \rangle}$) and neutron ($\sqrt{\langle (r^n_M)^2 \rangle}$),
and the magnetic moments of the proton ($\mu_p$) and neutron ($\mu_n$) by the $z$-expansion method. 
Figure~\ref{fig:gmpn_oct22_zexp} 
shows $G^p_M(q^2)$ and $G^n_M(q^2)$ as a function of $z(q^2)$
together with the fit results obtained from the $z$-expansion method.
The $z$-expansion fitting results are summarized in Table~\ref{tab:rm_zexp}. 
The analyses with other model-dependent functional forms are discussed in Appendix~\ref{app:model-dep_anal}.
Figure~\ref{fig:gmpn_zexp_rmscomp} shows stability of the variation of $k_{\mathrm{max}}$ 
in extracting both the RMS radii and the magnetic moments for each proton and neutron.
For both quantities of the proton and neutron,
we confirm that all results show good stability with respect to the variation of $k_{\mathrm{max}}$
, which is similar to the isovector case.
We employ the simultaneous fit results with $k_{\mathrm{max}}=4$ to get our best estimate of the central value and the statistical error. 
Both results of
$\sqrt{\langle (r^p_M)^2 \rangle}$ and $\sqrt{\langle (r^n_M)^2 \rangle}$
reproduce our previous results from the coarse lattice and the corresponding experimental values within their large statistical errors.
However, it should be too early to conclude that the systematic uncertainties are
well under control at this moment. 
We thus quote two types of the systematic uncertainties, which are similar to those of other RMS radii.

As for the magnetic moments, $\mu_p$ and $\mu_n$ are consistent with the results obtained from the coarse lattice,
while they do not reproduce the experimental values.
The situation is similar to the isovector case, which indicates that these quantities also suffer from the systematic uncertainties due to the $q^2$ extrapolation to $q^2=0$.

%
%
%
%
%
%
%
%
%
%
%
%
\begin{figure*}
\centering
\includegraphics[width=0.49\textwidth,bb=0 0 792 612,clip]{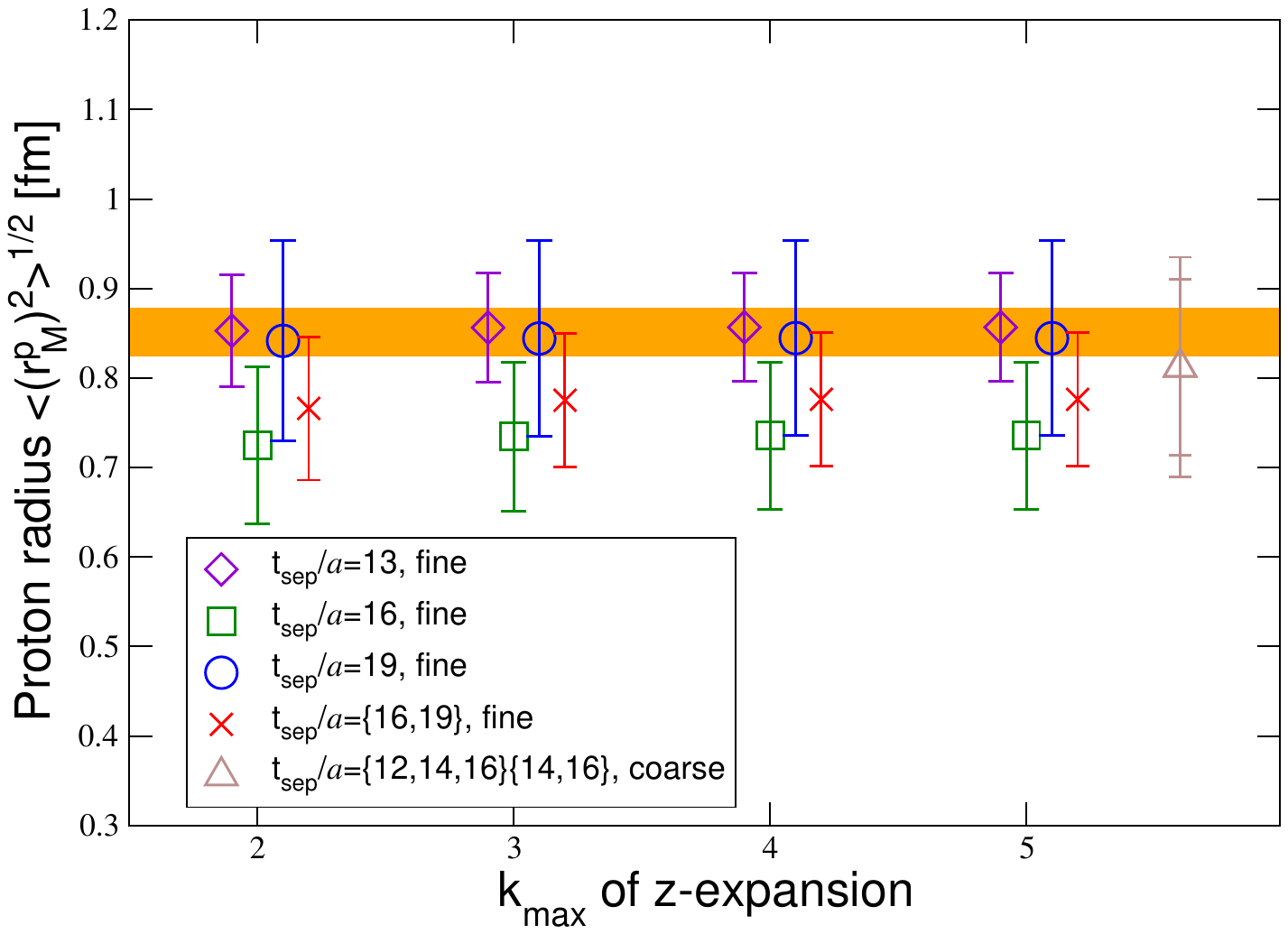}
\includegraphics[width=0.49\textwidth,bb=0 0 792 612,clip]{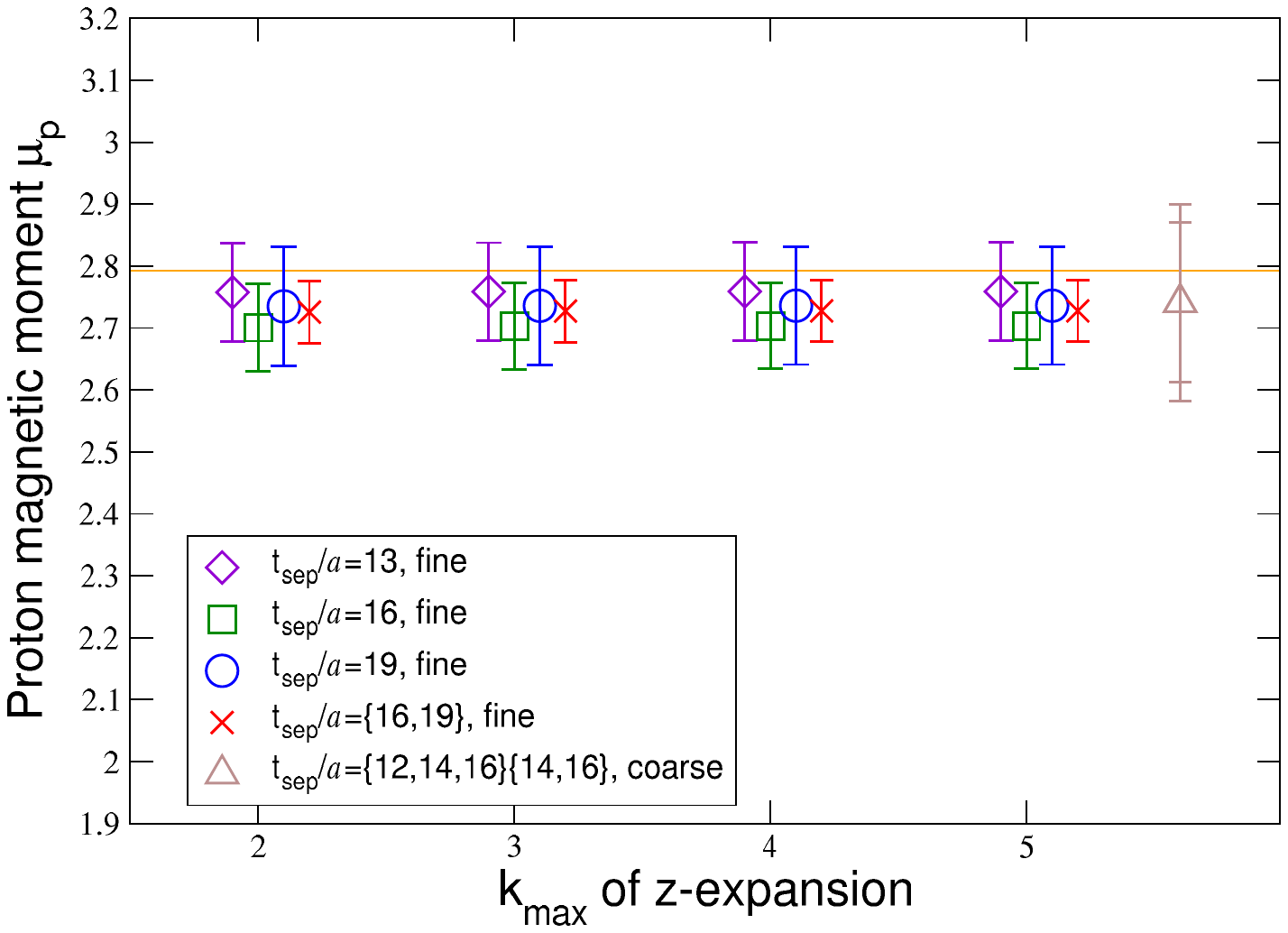}
\includegraphics[width=0.49\textwidth,bb=0 0 792 612,clip]{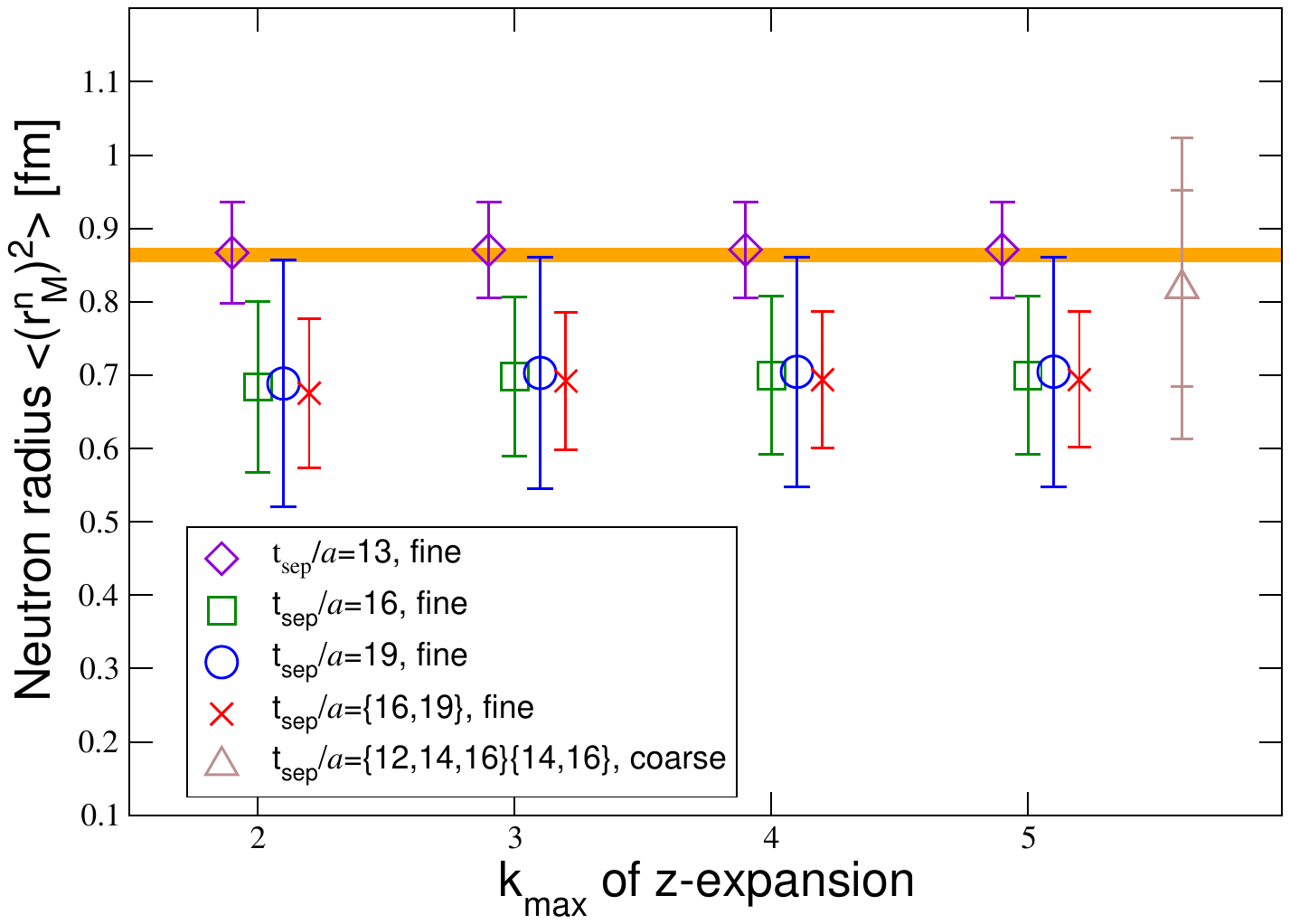}
\includegraphics[width=0.49\textwidth,bb=0 0 792 612,clip]{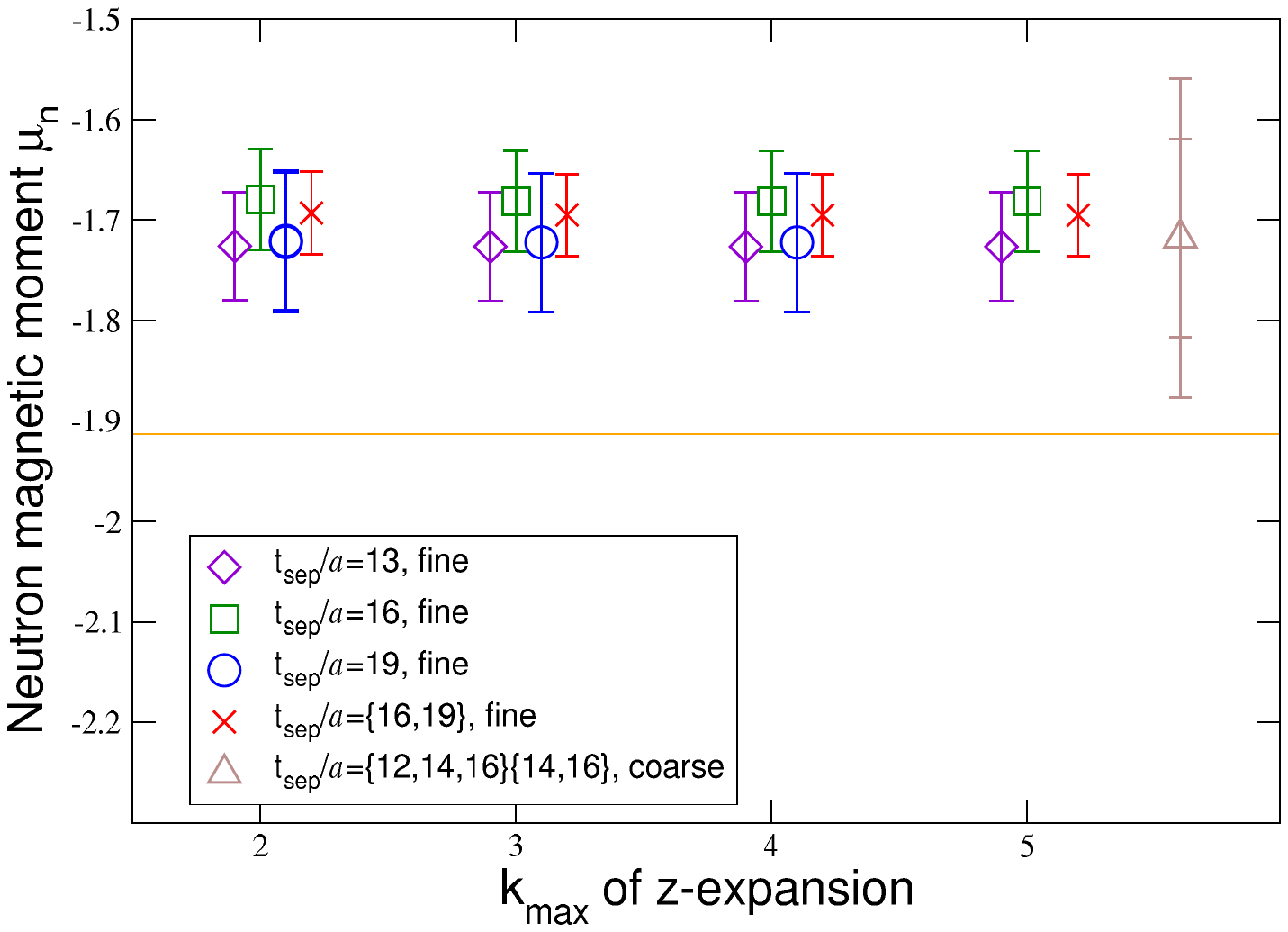}
\caption{
Same as Fig.~\ref{fig:gm_zexp_rmscomp} for the proton (top) and neutron (bottom).}
\label{fig:gmpn_zexp_rmscomp}
\end{figure*}

\clearpage
\subsection{Axial-vector coupling}
\label{sec:ga}
\begin{figure*}
\centering
\includegraphics[width=1\textwidth,bb=0 0 792 612,clip]{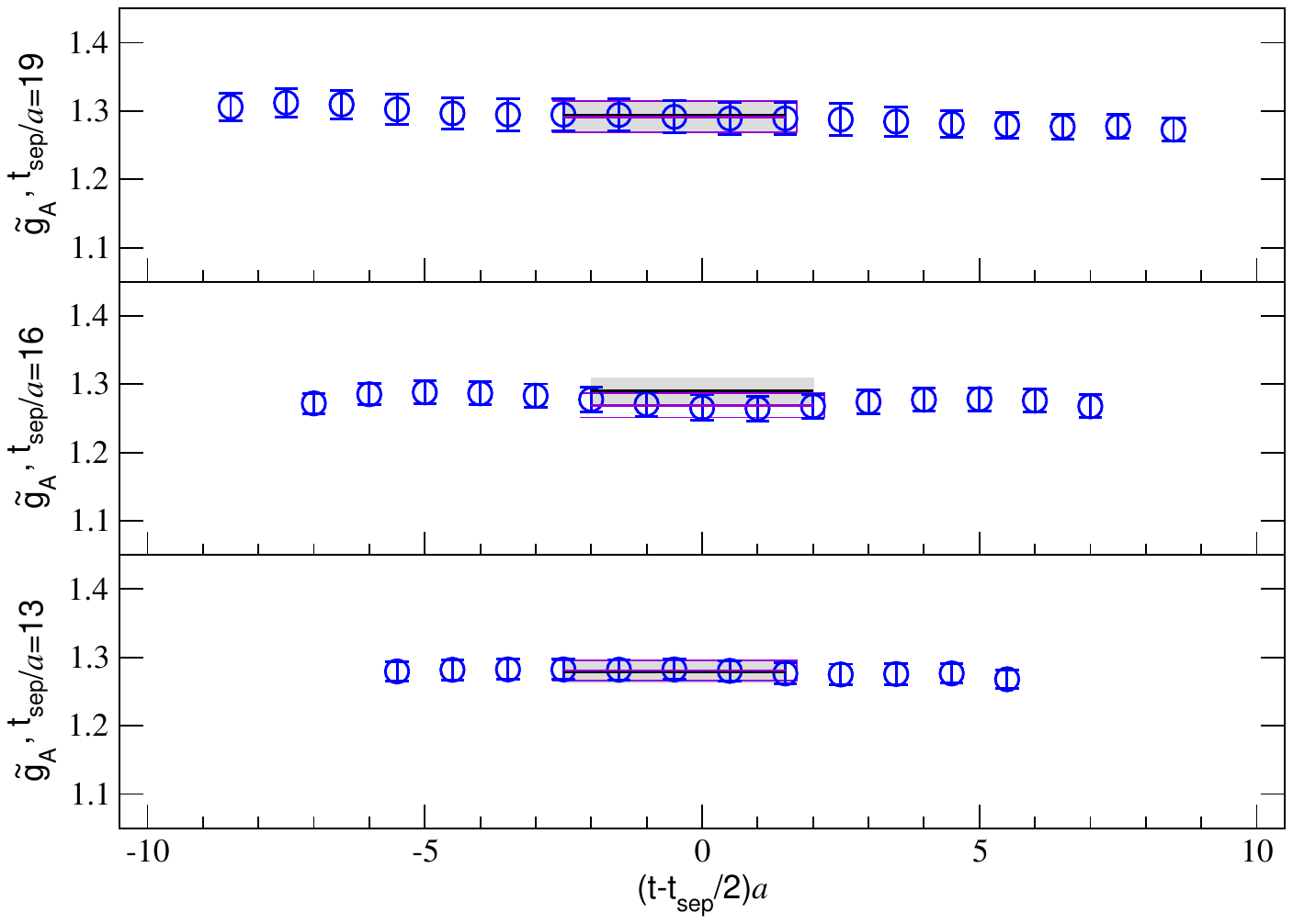}
\caption{The bare value of
the nucleon axial-vector coupling $\widetilde{g}_A$ with $t_{\mathrm{sep}}/a=\{13,16,19\}$ (bottom, center and top respectively). The horizontal axis represents the operator insertion time $t$.
The solid line shows the average value, while
the grey shaded bands and violet boxes display the fit ranges and one standard deviations for the correlated and uncorrelated constant fits.}
\label{fig:ga_tsdep_p-n}
\end{figure*}

The bare value of the nucleon axial-vector coupling $\widetilde{g}_A=\widetilde{F}_A(q^2=0)$ is obtained with the ratio $\mathcal{R}^{5z}_{A_i}(t,\bm{q})$ of Eq.~(\ref{eq:fa_def}) with $\bm{q}=\bm{0}$. 
Figure~\ref{fig:ga_tsdep_p-n} shows the $t$-dependence of the axial-vector coupling $\widetilde{g}_A$.
The good plateau behaviors are equally observed for all choices of $t_{\mathrm{sep}}$.
This indicates that the excited-state contamination is negligible within our statistical precision
due to the optimal choice of the smearing parameters for the nucleon interpolating operator.
Therefore, the bare value of the axial-vector coupling $\widetilde{g}_A$ can be
evaluated by a simple constant fit in the standard plateau method.

The uncorrelated constant fits are employed for extracting the electric and magnetic form factors, $\widetilde{G}_E(q^2)$ and $\widetilde{G}_M(q^2)$,
to keep the same fit range for all $q^2$ as described in Sec.~\ref{sec:ge_re} and \ref{sec:gm_rm}. 
Here as for the analysis of $\widetilde{g}_A$, 
we use both uncorrelated and correlated constant fits and compare the results carefully.
This is simply because a closer look at the $t$-dependence can find a slight undulation appearing in the central region of the data only when $t_{\mathrm{sep}}/a=16$.
In fact, this undulation causes slight systematic difference in the fit results of $t_{\mathrm{sep}}/a=16$
depending on whether one chooses the uncorrelated or correlated method as shown in Fig.~\ref{fig:ga_tsdep_p-n}.
The grey shaded bands and violet boxes in Fig.~\ref{fig:ga_tsdep_p-n} represent the results obtained by the correlated and uncorrelated constant fits and their fit ranges. 
All fit results are summarized in Table~\ref{tab:measurements}.

As for the results of $t_{\mathrm{sep}}/a=13$ and $19$,
there is no difference between the correlated and uncorrelated constant fits.
On the other hand, the difference between the two fits
is certainly observed in the case of $t_{\mathrm{sep}}/a=16$.
The central value given by the correlated constant fit appears slightly
higher than the uncorrelated result and its statistical error is also slightly larger. 
As the fit range is extended, 
the fit results from uncorrelated fits tends to be larger
since the larger values on the near side of the source and sink are incorporated into the fit.
As a result, the difference between the two types of fits becomes smaller.
Therefore, the slight upward shift due to the correlated fit is caused by the strong
data correlation among the high precision data sets of $\widetilde{g}_A$ measured at various time slices of the current operator insertion. 

This particular discrepancy between the correlated and uncorrelated fits is only observed in the determination of $\widetilde{g}_A$ since no other physical quantity is accurate enough to distinguish the difference.
Therefore, we choose the correlated constant fit only to extract $\widetilde{g}_A$ taking into account the data correlation, while the analyses of the form factors use the uncorrelated constant fit in line with our previous analyses on the coarse lattice~\cite{Shintani:2018ozy}.

\begin{figure*}
\centering
\includegraphics[width=1\textwidth,bb=0 0 792 612,clip]{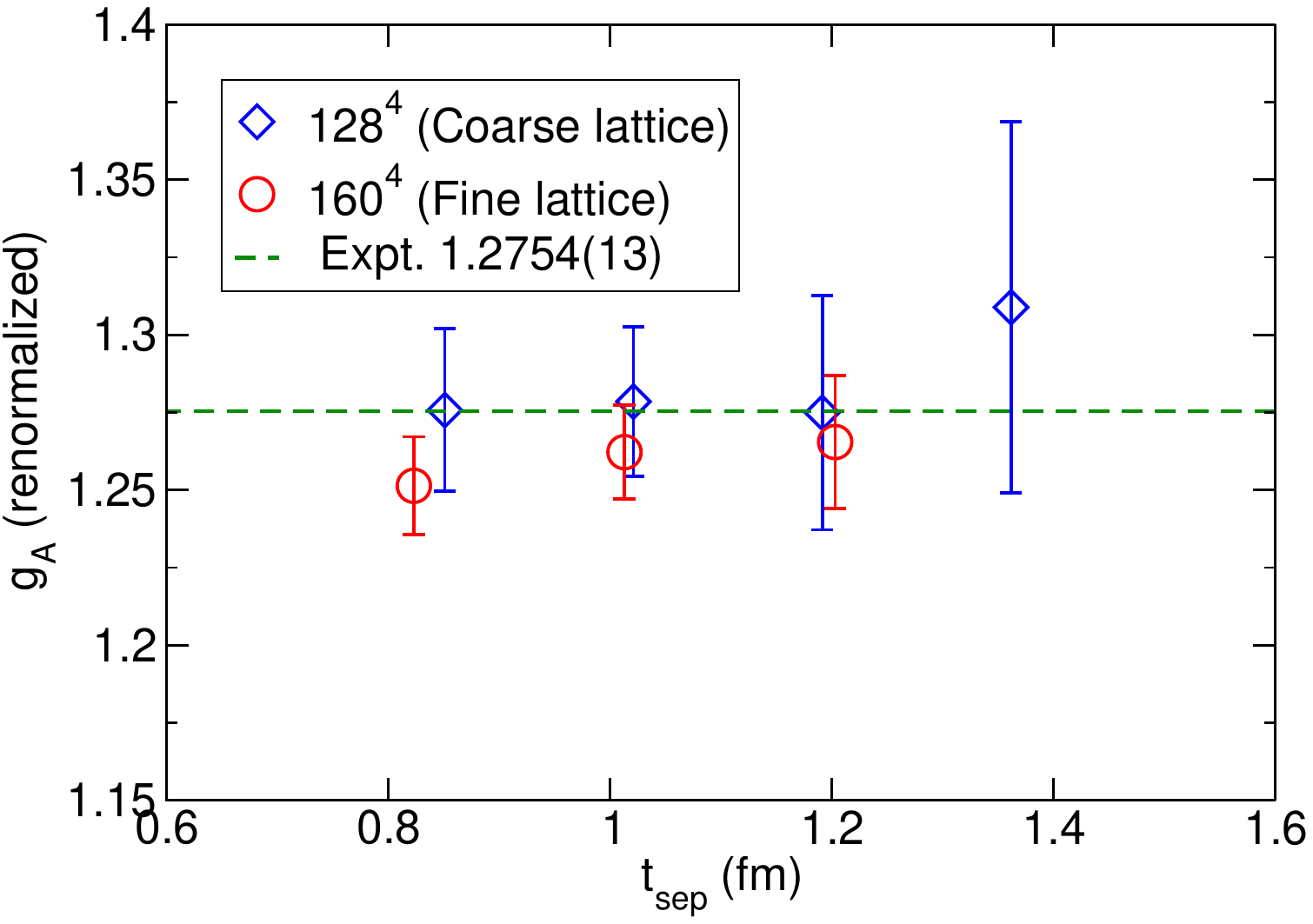}
\caption{
The source-sink separation ($t_{\mathrm{sep}}$) dependence of the renormalized values of $g_A$.
The horizontal axis gives $t_{\mathrm{sep}}$ in physical units. The circle symbols are results in this study, while the diamond symbols are obtained from the coarse lattice~\cite{Tsuji:2022ric}. The horizontal dashed line denotes the experimental value~\cite{ParticleDataGroup:2022pth}.}
\label{fig:ga_tsdep_p-n_1XX}
\end{figure*}

Figure~\ref{fig:ga_tsdep_p-n_1XX} shows the $t_{\mathrm{sep}}$-dependence of the extracted values for the renormalized axial-vector coupling $g_A=Z_A\widetilde{g}_A$, 
comparing with our previous results obtained from the coarse lattice together with the experimental value.
As can be easily seen, the results of $g_A$ obtained from both the fine and coarse lattices, do not show any significant $t_{\mathrm{sep}}$-dependence.
This observation indicates that the systematic uncertainties stemming from the excited-state contamination are well under control at the level of the statistical precision of about 2\% in our calculations.
Combining our previous 
study~\cite{Tsuji:2022ric}
that reveals the finite size effect on the axial-vector coupling is less than 2\% of the statistical precision,
our calculations achieve that all major sources of systematic uncertainties from the chiral extrapolation, finite size effect, excited-state contamination and discretization effect are under control by our statistical precision of 2\%
in \textit{fully dynamical lattice QCD simulations}.

%
%
%
%

\clearpage
\subsection{Axial form factor, axial radius and induced pseudoscalar form factor}
\label{sec:fa_ra_fp}
Since we are only interested in the isovector quantities for the axial form factor and induced pseudoscalar form factor in this study, we simply denote $F_A(q^2)$ and $F_P(q^2)$ for these (renormalized) form factors, hereafter. In the axial-vector channel, the two independent (bare) form factors, namely $\widetilde{F}_A(q^2)$ and $\widetilde{F}_P(q^2)$ can be extracted only from the ratio $\mathcal{R}_{A_{i}}^{5z}(t, \bm{q})$ defined in Eq.~(\ref{eq:fa_def}) with help of
different momentum configurations $\bm{q}=(q_1,q_2,q_3)$ depending on the direction of polarization 
following Ref.~\cite{Sasaki:2007gw}.

In this study, $z$ direction is chosen as the polarized
direction through the definition of the projection operator
${\mathcal P}^{5z}$. 
Indeed, the ratio $\mathcal{R}_{A_{i}}^{5z}(t, \bm{q})$ possesses the following part 
\begin{align}
    \label{eq:fit_fafp}
    C_{i}(\bm{q})
    \equiv
        \widetilde{F}_{A}\left(q^{2}\right)\delta_{i3}
        -
        \frac{q_iq_3}{E_N+M_N}\widetilde{F}_{P}\left(q^{2}\right),
\end{align}
which explicitly depends on the longitudinal momentum
$q_3$ and then makes the difference between the transverse components ($i=1$ or 2) and the longitudinal ($i=3$) component. Furthermore, the dependence of the momentum configuration $\bm{q}=(q_1,q_2,q_3)$ is induced in $C_{i}(\bm{q})$ at fixed ${\bm q}^2$, since the second term in the r.h.s. of Eq.~(\ref{eq:fit_fafp}) also depends on the value of $q_i$.
The value of $C_{i}(\bm{q})$ can be read off from the plateau behavior of the ratio $\mathcal{R}_{A_{i}}^{5z}(t, \bm{q})$ by
multiplying the appropriate factor of $\sqrt{\frac{2 E_{N}}{E_N+M_N}}$ in the standard plateau method.

Taking into account the dependence of the momentum 
configurations, $\widetilde{F}_A(q^2)$ and $\widetilde{F}_P(q^2)$ can be constructed by the 
following combinations of $C_i(\bm{q})$ with the specific momentum configurations of $\bm{q}$. 
We first determine $\widetilde{F}_A(q^2)$ from $C_i(\bm{q})$ 
for either $q_3\neq 0$ or $q_3= 0$
in the following way:
\renewcommand{\arraystretch}{1.2}
\begin{align}
    \label{eq:anal_fa}
    \widetilde{F}_A(q^2) & = \left\{
    \begin{array}{ll}
        C_3 - \frac{q_3}{2}\left[ \frac{C_1}{q_1} + \frac{C_2}{q_2} \right]
        & (q_1 \neq 0\ \mathrm{and}\ q_2\neq 0)\\
        C_3 - \frac{q_3}{q_1}C_1
        & (q_1 \neq 0\ \mathrm{and}\ q_2 = 0),\\
        C_3 - \frac{q_3}{q_2}C_2
        & (q_1 = 0\ \mathrm{and}\ q_2 \neq 0)\\
    \end{array}
    \right.
\end{align}
which depend only on the values of $q_1$ and $q_2$.
Next $\widetilde{F}_P(q^2)$ is determined from $C_i(\bm{q})$
only for the case of $q_3\ne 0$ as
\begin{align}
    \label{eq:anal_fp}
    \widetilde{F}_P(q^2) & = \left\{
    \begin{array}{ll}
        - \frac{(E_N + M_N)}{2}\left[ \frac{C_1}{q_3q_1} + \frac{C_2}{q_3q_2} \right]
        & (q_1 \neq 0\ \mathrm{and}\ q_2\neq 0)\\
        - (E_N + M_N)\frac{C_1}{q_3q_1}
        & (q_1 \neq 0\ \mathrm{and}\ q_2 = 0)\\
        - (E_N + M_N)\frac{C_2}{q_3q_2}
        & (q_1 = 0\ \mathrm{and}\ q_2 \neq 0),\\
        -\frac{E_N+M_N}{q_3^2}
        \left( C_3 - \widetilde{F}_A(q^2)  \right)
        & (q_1 = 0\ \mathrm{and}\ q_2 = 0)
    \end{array}
    \right.
\end{align}
where the last case requires the value of $\widetilde{F}_A(q^2)$
 which should be evaluated in advance. 
\renewcommand{\arraystretch}{1}
Although the above procedure is a bit complicated, 
it can avoid the usage of the ratio $\mathcal{R}_{A_{4}}^{5z}(t, \bm{q})$, which is not applicable in the standard plateau method since the time-reversal odd contributions from the multiparticle states such as the lowest $\pi N$ and $\pi \pi N$
states are inevitable.

\subsubsection{Axial form factor and axial radius}
\label{sec:fa_ra}
\begin{figure*}
\centering
\includegraphics[width=0.48\textwidth,bb=0 0 864 720,clip]{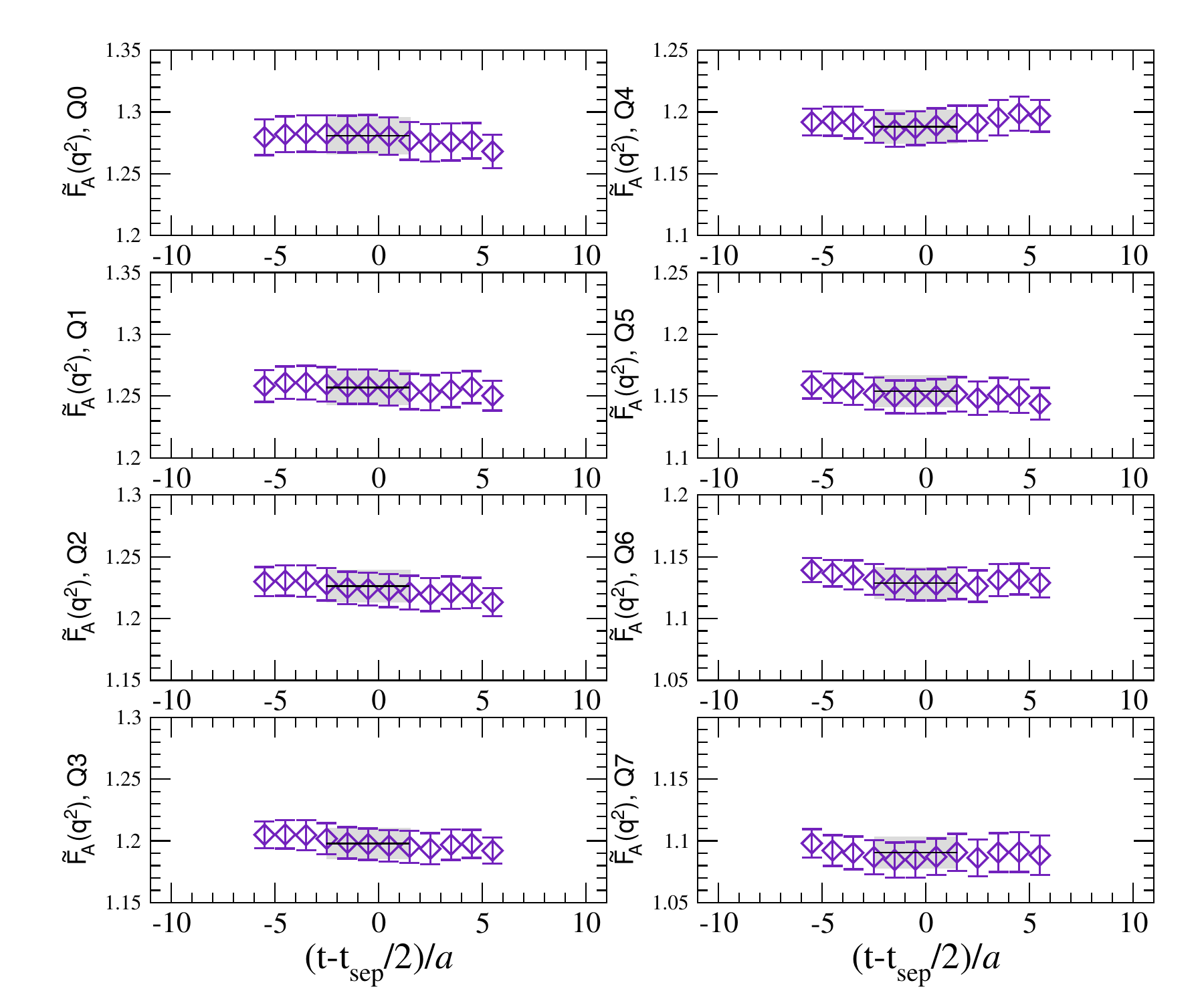}
\includegraphics[width=0.48\textwidth,bb=0 0 864 720,clip]{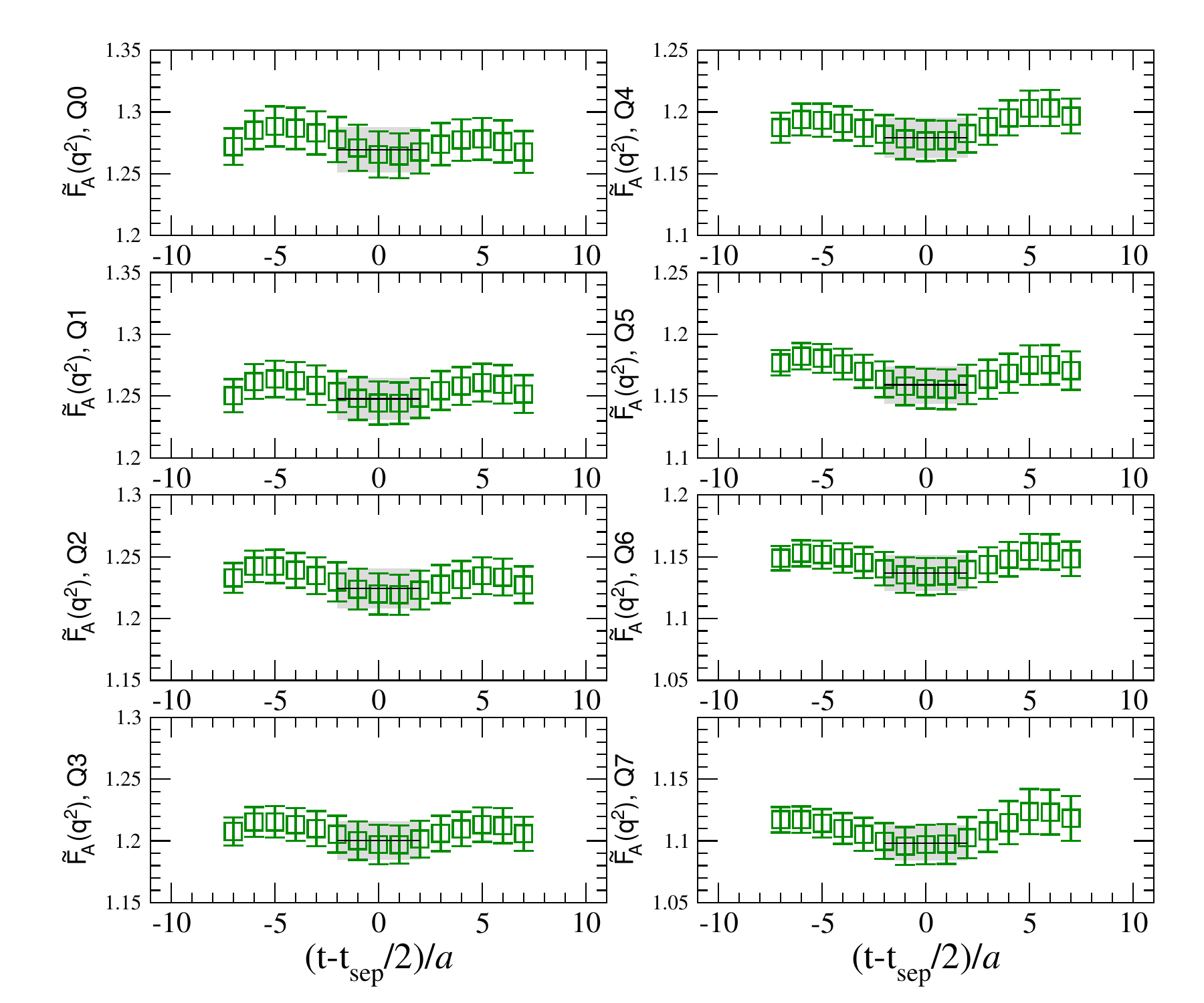}
\includegraphics[width=0.48\textwidth,bb=0 0 864 720,clip]{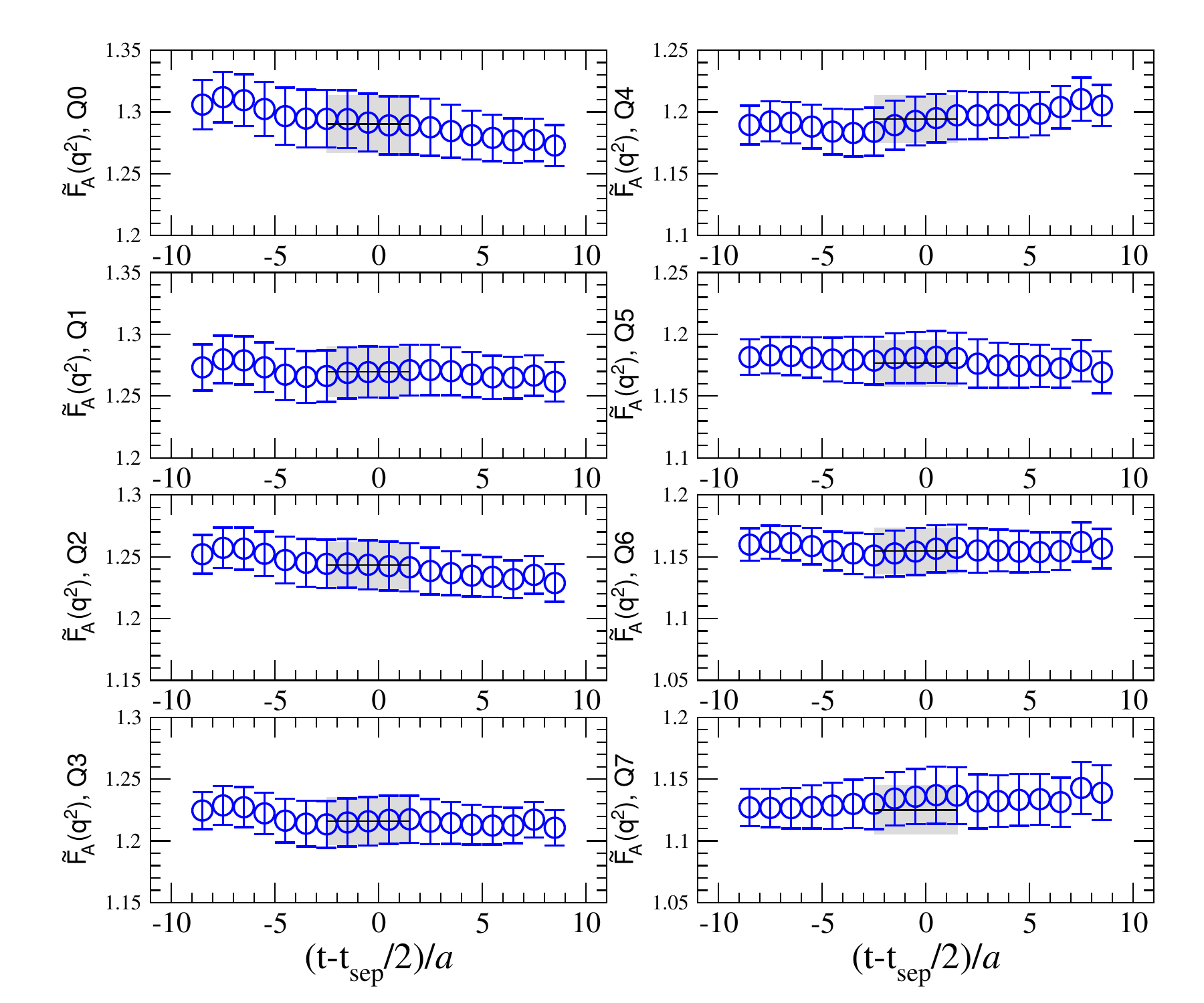}
\caption{Same as Fig.~\ref{fig:ge_qdep_p-n_ts1X} for the axial form factor.}
\label{fig:fa_qdep_p-n_ts1X}
\end{figure*}

\begin{figure*}
\centering
\includegraphics[width=1\textwidth,bb=0 0 864 720,clip]{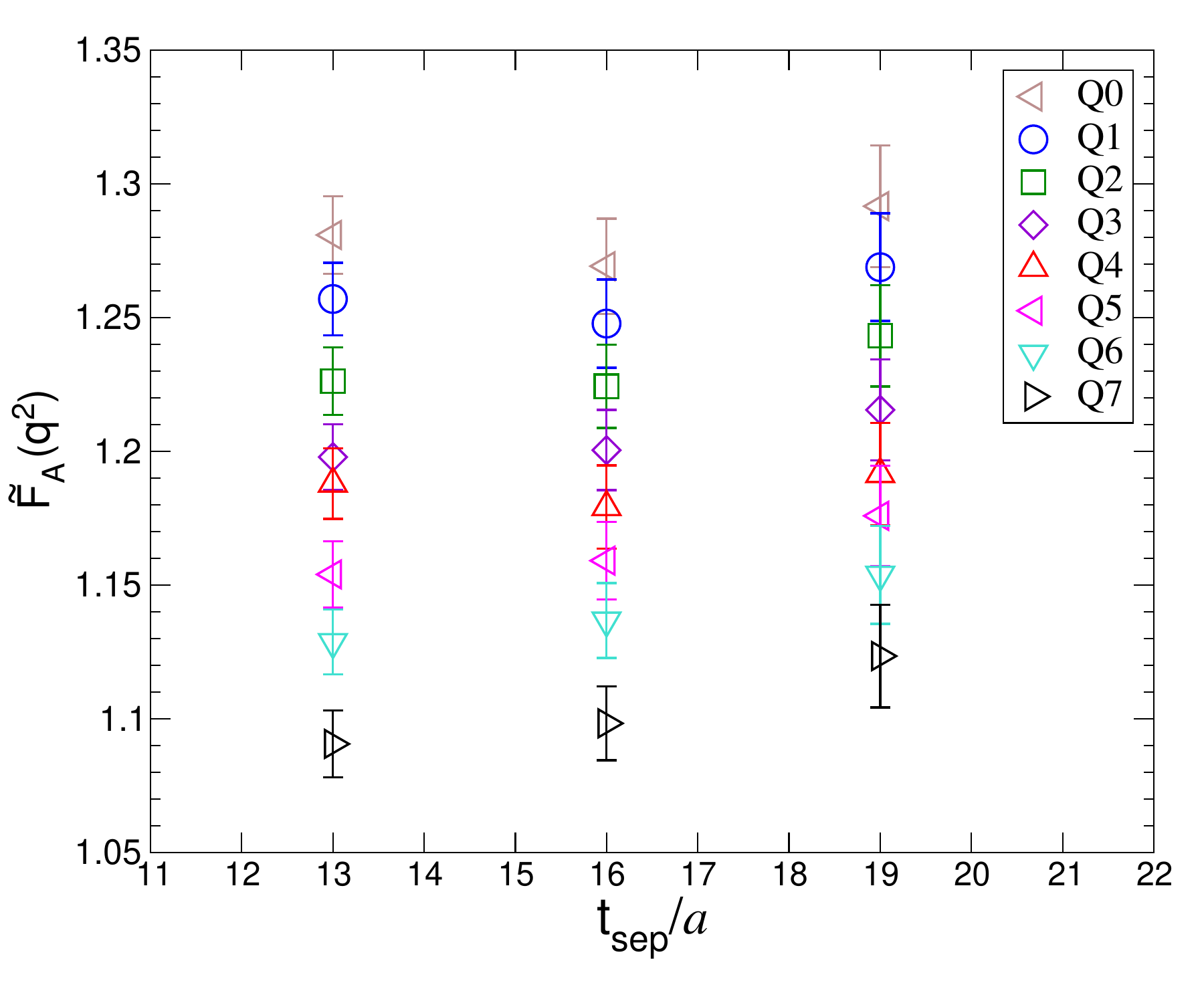}
\caption{Same as Fig.~\ref{fig:ge_tsdep_p-n} for the axial form factor.}
\label{fig:fa_tsdep_p-n}
\end{figure*}

\begin{figure*}
\centering
\includegraphics[width=1\textwidth,bb=0 0 792 612,clip]{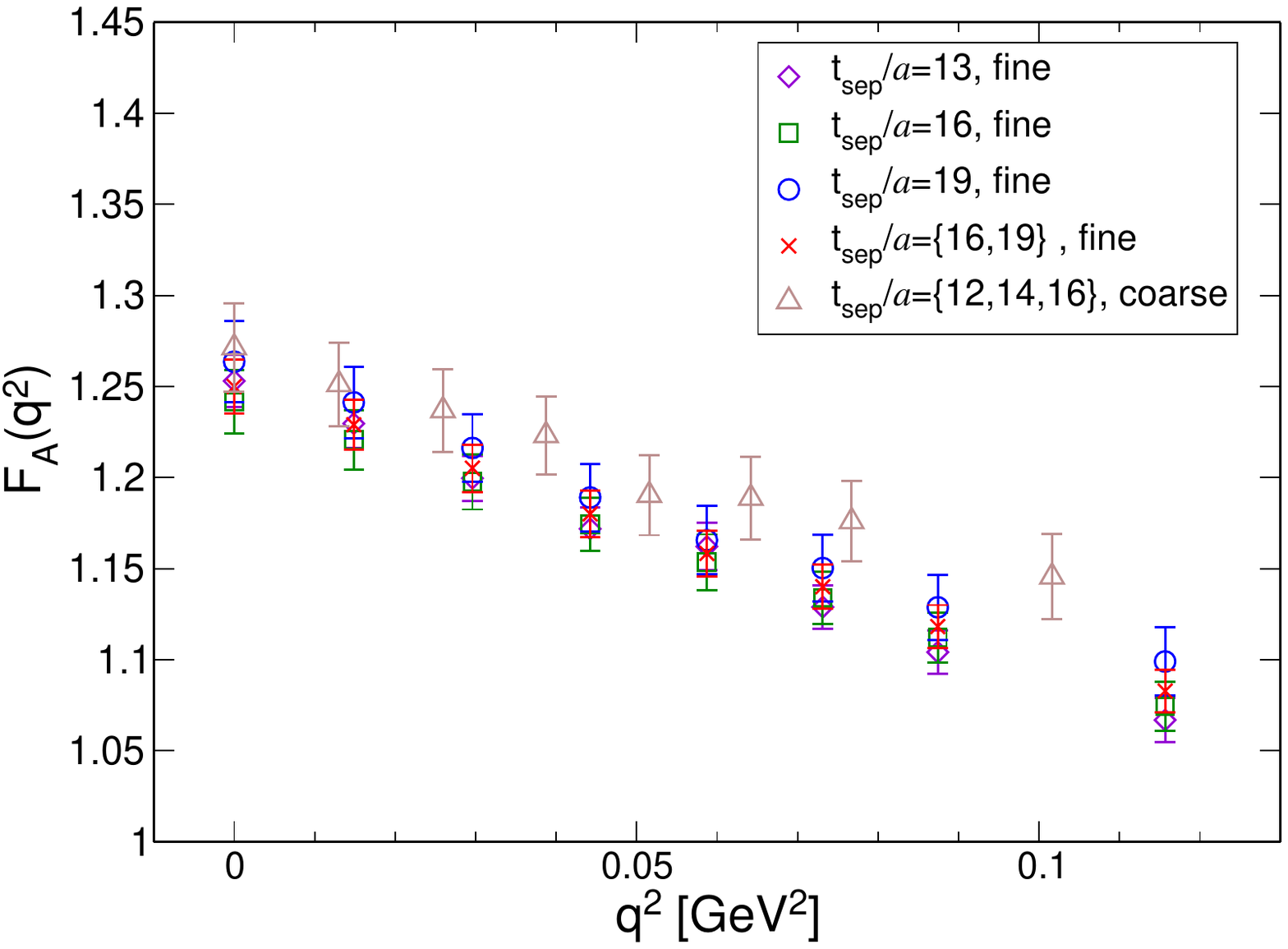}
\caption{
Same as Fig.~\ref{fig:ge_oct22} for the axial form factor.}
\label{fig:fa_oct22}
\end{figure*}

We evaluate the axial form factor $\widetilde{F}_A(q^2)$ as previously described using Eq.~(\ref{eq:anal_fa}). 
As for the $t$-dependence of $\widetilde{F}_A(q^2)$, Figure~\ref{fig:fa_qdep_p-n_ts1X} shows the $t$-dependence of the appropriate combinations of
$\mathcal{R}_{A_{i}}^{5z}(t, \boldsymbol{q})$, which
provides $\widetilde{F}_A(q^2)$ as the asymptotic behavior. 
The value of $\widetilde{F}_A(q^2)$ can be read off from the good plateaus appearing in Fig.~\ref{fig:fa_qdep_p-n_ts1X}.
Next, Figure~\ref{fig:fa_tsdep_p-n} shows the results of $\widetilde{F}_A(q^2)$ which
are evaluated by the uncorrelated constant fits 
in all cases of $t_{\mathrm{sep}}/a=\{13,16,19\}$ 
for all $q^2$, while Figure \ref{fig:fa_oct22} shows the $q^2$-dependence of $F_A(q^2)=Z_A\widetilde{F}_A(q^2)$.
As can be seen, there is no significant $t_{\mathrm{sep}}$-dependence at every $q^2$
within the statistical errors as well as the electric and magnetic form factors.
This indicates that the systematic uncertainties stemming from the excited-state contamination are negligible
within the present statistical precision and
well under control by the optimal choice of the
smearing parameter for the nucleon interpolating operator.

\begin{figure*}
\centering
\includegraphics[width=1\textwidth,bb=0 0 792 612,clip]{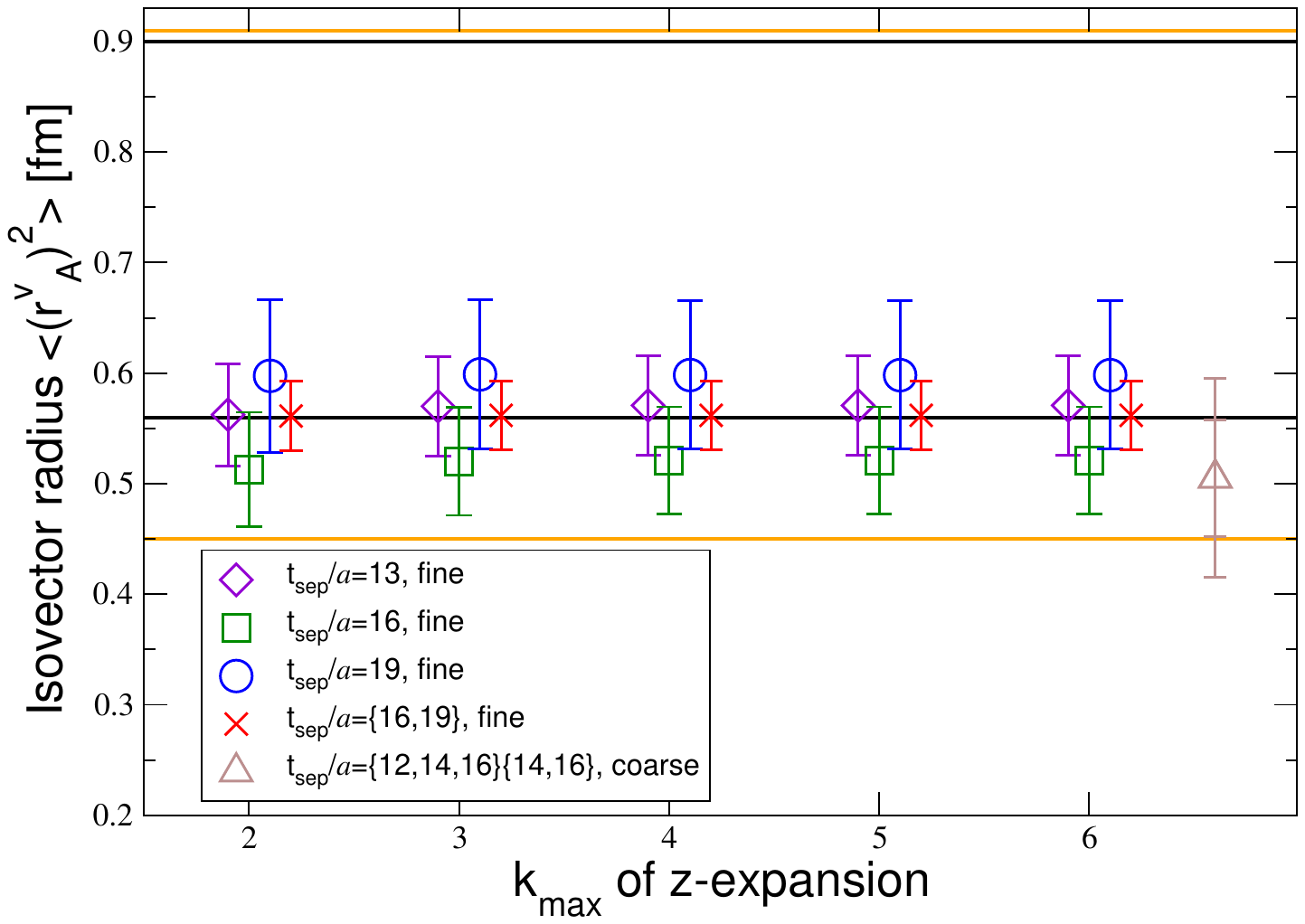}
\caption{
Same as Fig.~\ref{fig:ge_zexp_rmscomp} for the axial form factor. The orange line indicates the phenomenological results using $\nu$-Deutron experiments, the black line is obtained from $\nu$-Nucleon experiments.} 
\label{fig:fa_zexp_rmscomp}
\end{figure*}

Similar to the electric and magnetic radii,
the nucleon axial radius $\sqrt{\langle (r^v_A)^2 \rangle}$ is evaluated by examining the $q^2$-dependence of $F_A(q^2)$ in the model-independent way
using the $z$-expansion method.
The analyses with other model-dependent functional forms are discussed in Appendix~\ref{app:model-dep_anal}.
In Fig.~\ref{fig:fa_oct22_zexp}, we show $z(q^2)$-dependence of $F_A(q^2)$ together with 
the fit results 
obtained from the $z$-expansion method. 
The $z$-expansion fitting results of $\sqrt{\langle (r^v_A)^2 \rangle}$ are summarized in Table~\ref{tab:ra_zexp}.
Figure~\ref{fig:fa_zexp_rmscomp} shows the
stability of the variation of $k_{\mathrm{max}}$ in extracting
$\sqrt{\langle (r^v_A)^2 \rangle}$ for each $t_{\mathrm{sep}}$ data and a combined data of 
$t_{\mathrm{sep}}/a=\{16,19\}$.
The simultaneous fit to the combined data of $t_{\mathrm{sep}}/a=\{16,19\}$ yields
the consistent result with the results for each $t_{\mathrm{sep}}$. 

Considering this observation, the result from the combined data of $t_{\mathrm{sep}}/a=\{16,19\}$ with  $k_{\mathrm{max}}=4$ is quoted for the central value and the statistical error as our final estimate.
Although the final result is consistent with our previous result obtained on the coarse lattice 
within their statistical errors, the discretization error can be estimated as 11.1\% by a difference between the central values of both the coarse and fine lattices. 
The size of the discretization error are comparable to those of the two other RMS radii. 
We will continue discussions on the discretization uncertainties in Sec.~\ref{sec:numerical_results_III}.

%
%
\begin{table*}[ht!]
    {\scriptsize
\begin{ruledtabular}
\caption{Results for the axial-vector coupling $g_A=F_A(0)$ and axial-vector RMS radius $\sqrt{\langle (r^v_A)^2\rangle}$.  In the row of ``This work'' we present our best estimates, where the first error is statistical.
\label{tab:ra_zexp}}
\begin{tabular}{ccccccc}
  Fit type & $q^2$ GeV$^2$ & $t_{\rm sep}/a$ & $g_A$ & $F_A(0)$ & $\sqrt{\langle r^2_A\rangle}$ [fm]& $\chi^2$/d.o.f.   \cr
  \hline
  \multicolumn{2}{ c }{$160^4$ (fine) lattice} \cr
  \multirow{2}{*}{$k_{\rm max}=4$}                       & \multirow{2}{*}{$q^2_{\mathrm{disp}}\le0.116$} & $\{16,19\}$ & & { 1.252(15) } & { 0.562(31) } & { 0.40}\\
             &                        & $19$       & & { 1.267(22) } & { 0.598(67) } & { 0.04}\\
  \multirow{2}{*}{$k_{\rm max}=3$}                       & \multirow{2}{*}{$q^2_{\mathrm{meas.}}\le0.091$} & $\{16,19\}$ & & { 1.251(16) } & { 0.515(41) } & { 0.37}\\
             &                        & $19$       & & { 1.265(22) } & { 0.554(75) } & { 0.03}\\
  \multirow{2}{*}{Correlated}                       & $\cdots$               & $\{16,19\}$       & {1.264(14)} &  &  & 1.1\\
                         & $\cdots$               & $19$       & {1.265(21)} &  &  & 0.03\\
  \multirow{2}{*}{Uncorrelated}                       & $\cdots$               & $\{16,19\}$       & {1.250(14)} &  &  & 0.4\\
                         & $\cdots$               & $19$       & {1.264(22)} &  &  & 0.01\\
  \multicolumn{2}{ c }{This work} & & 1.264(14)(1) & 1.252(15)(15)(1)& 0.562(31)(36)(47)\\
  \multicolumn{2}{ c }{$128^4$ (coarse) lattice} \cr
    \multirow{2}{*}{$k_{\rm max}=4$}                       & \multirow{2}{*}{$q^2_{\mathrm{disp}}\le0.077$} & $\{12,14,16\}$ & & { 1
.279(28) } & { 0.505(53) } & { 0.7}\\
             &                        & $\{14,16\}$       & & { 1.284(42) } & { 0.546(80) } & { 1.3}\\
  \multirow{2}{*}{$k_{\rm max}=3$}                       & \multirow{2}{*}{$q^2_{\mathrm{meas.}}\le0.091$} & $\{12,14,16\}$ & & {1.277(30) } & { 0.445(71) } & { 0.7}\\
             &                        & $\{14,16\}$       & & { 1.279(42) } & { 0.416(131) } & { 1.3}\\
  \multicolumn{2}{ c }{PACS10 $128^4$ result} & & 1.280(24)(4) & & 0.505(53)(41)(60)\\
  \multicolumn{2}{ c }{Experimental value~\cite{ParticleDataGroup:2022pth, Bodek:2007ym}}  &  & & \cr
  & & & 1.2756(13) & & 0.67(1) &\cr
\end{tabular}
\end{ruledtabular}
}
\end{table*}

%
%
%
%
%
%
%
%
\begin{figure*}
\centering
\includegraphics[width=1\textwidth,bb=0 0 792 612,clip]{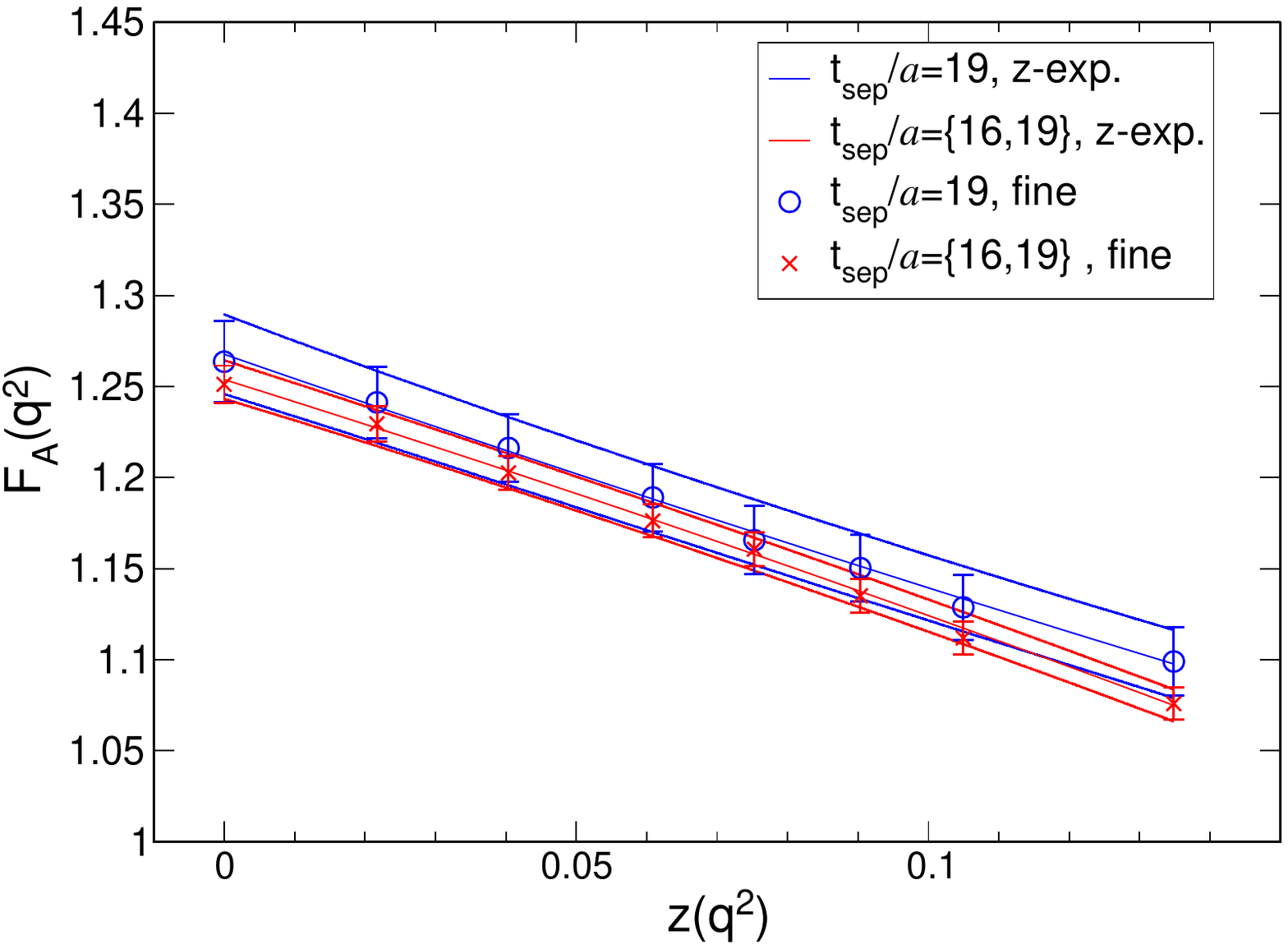}
\caption{
Same as Fig.~\ref{fig:ge_oct22_zexp} for the axial form factor.}
\label{fig:fa_oct22_zexp}
\end{figure*}
%
%

\subsubsection{Induced pseudoscalar form factor}
\label{sec:fp}
\begin{figure*}
\centering
\includegraphics[width=0.48\textwidth,bb=0 0 864 720,clip]{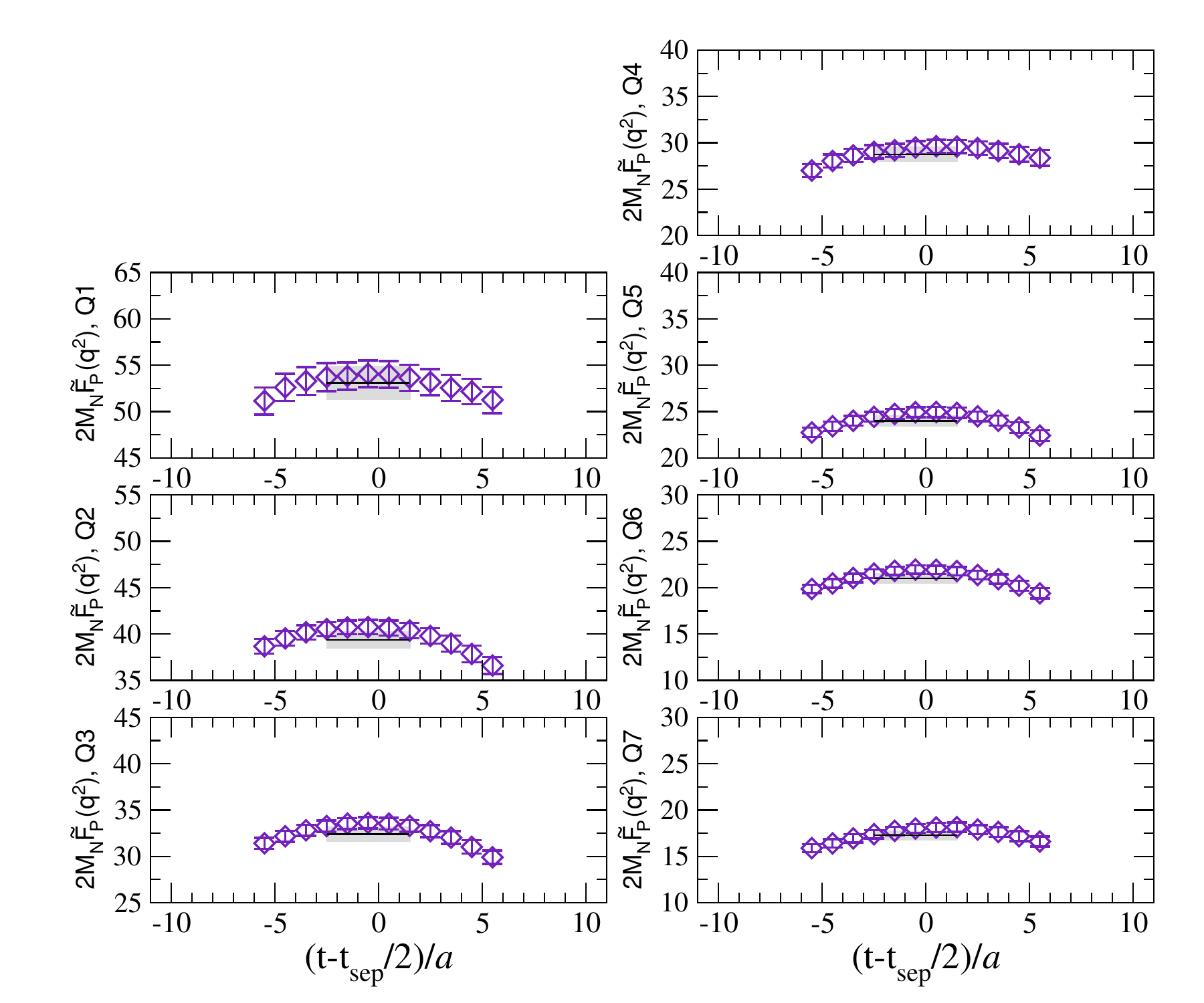}
\includegraphics[width=0.48\textwidth,bb=0 0 864 720,clip]{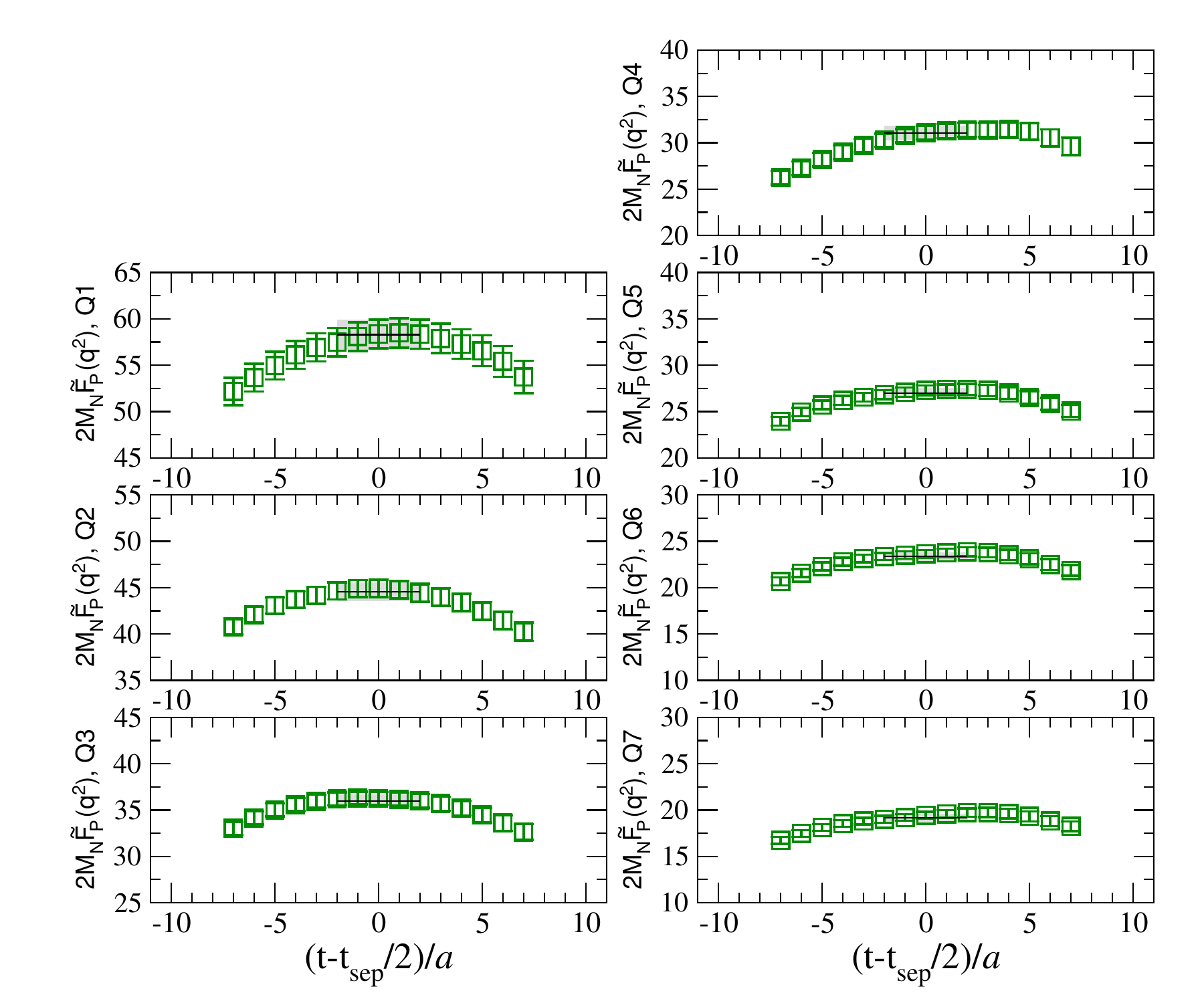}
\includegraphics[width=0.48\textwidth,bb=0 0 864 720,clip]{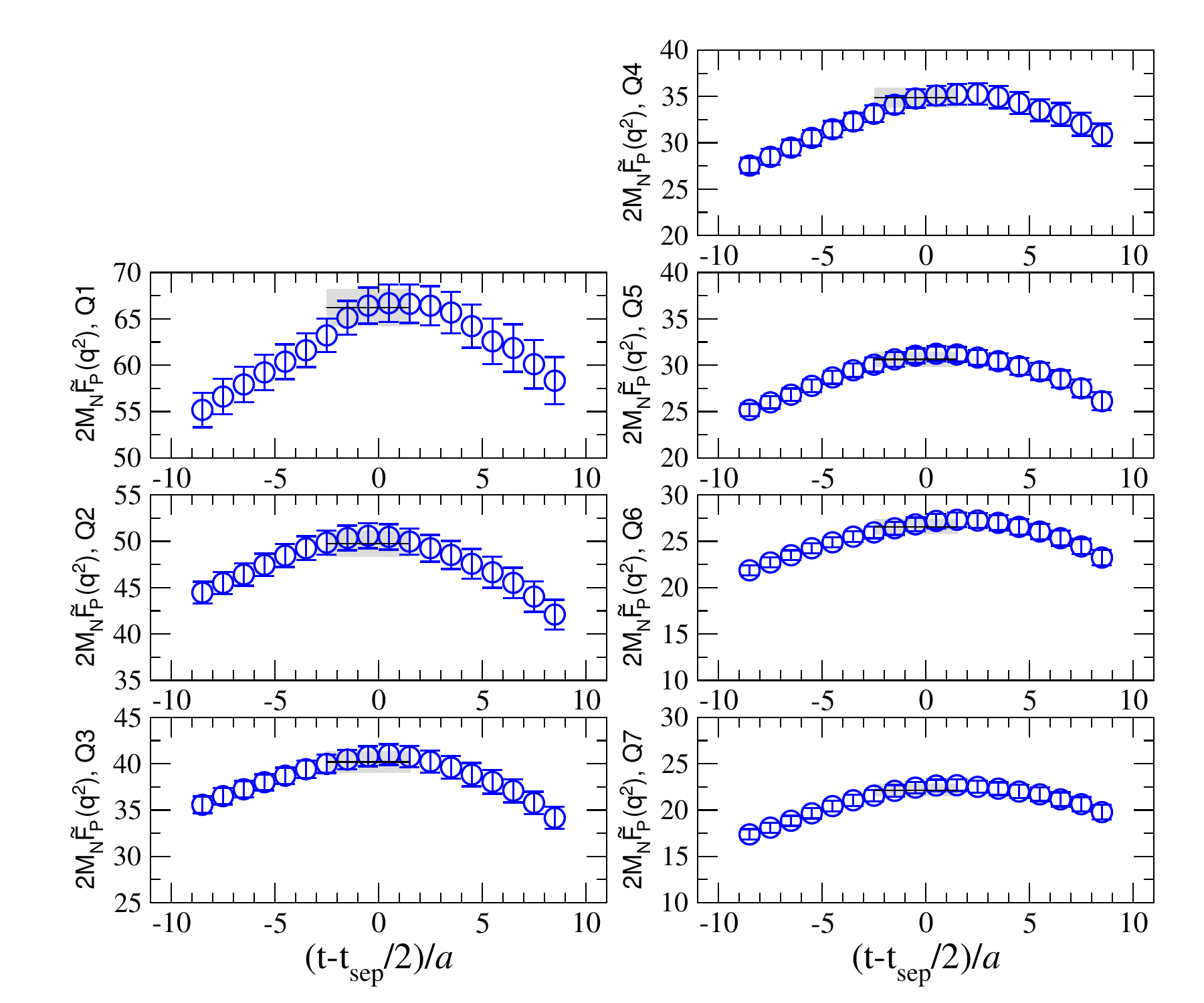}
\caption{Same as Fig.~\ref{fig:ge_qdep_p-n_ts1X} for the induced pseudoscalar form factor multiplying the factor $2M_N$.}
\label{fig:fp_qdep_p-n_ts1X}
\end{figure*}

We evaluate the induced pseudoscalar form factor $\widetilde{F}_P(q^2)$ as previously described using Eq.~(\ref{eq:anal_fp}).
As the $t$-dependence of $\widetilde{F}_P(q^2)$, Figure~\ref{fig:fp_qdep_p-n_ts1X} shows the $t$-dependence of the appropriate combinations of $\mathcal{R}_{A_{i}}^{5z}(t, \boldsymbol{q})$, 
which provides $\widetilde{F}_P(q^2)$ as the asymptotic behavior. 
In contrast of $\widetilde{F}_A(q^2)$ as well as $\widetilde{G}_E(q^2)$ and $\widetilde{G}_M(q^2)$,
the $t$-dependence of $\widetilde{F}_P(q^2)$ has
slight convex shape in all cases of $t_{\mathrm{sep}}/a=\{13,16,19\}$ for all $q^2$.
As reported in our previous works~\cite{{Ishikawa:2018rew},{Shintani:2018ozy}}, 
this convex shape is associated with the excited-state contamination. 
Although in addition to the standard plateau method, we also examine a two-state fitting analysis in this channel,
we can manage to read the value of $\widetilde{F}_P(q^2)$ by the constant fitting in a suitable fit range where the data points overlap within one standard deviation.

\begin{figure*}
\centering
\includegraphics[width=1\textwidth,bb=0 0 864 720,clip]{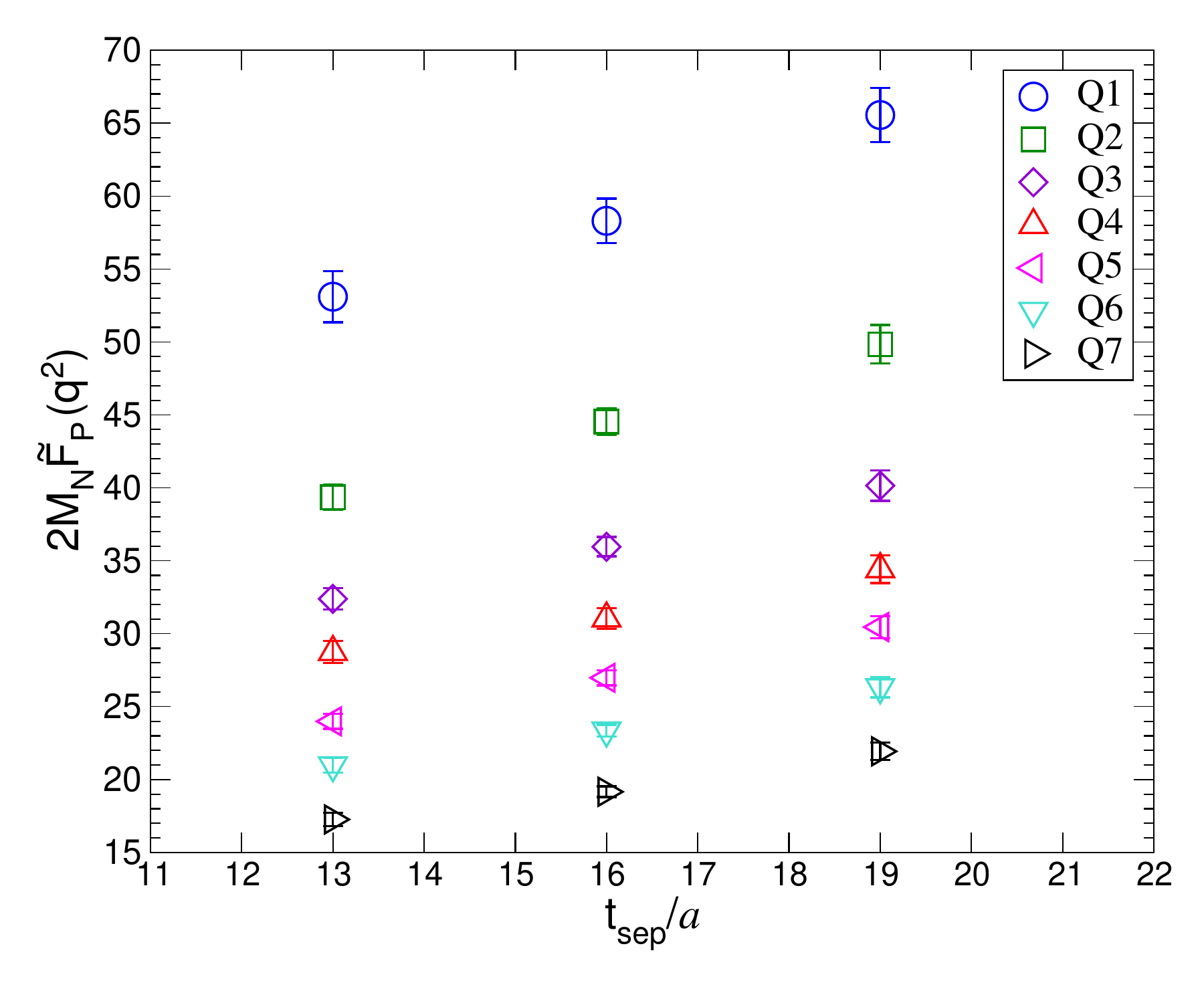}
\caption{Same as Fig.~\ref{fig:ge_tsdep_p-n} for the induced pseudoscalar form factor.}
\label{fig:fp_tsdep_p-n}
\end{figure*}

We plot the $t_{\mathrm{sep}}$-dependence of $\widetilde{F}_P(q^2)$ in Fig.~\ref{fig:fp_tsdep_p-n}. 
The relatively large excited-state 
contamination in $\widetilde{F}_P(q^2)$ is found compared to $\widetilde{F}_A(q^2)$, since the values of $\widetilde{F}_P(q^2)$ systematically increase
as $t_{\mathrm{sep}}$ increases. 
The magnitude of $\widetilde{F}_P(q^2)$ at the lowest $q^2$ for $t_{\mathrm{sep}}/a=19$ becomes about 20\% larger than that of $t_{\mathrm{sep}}/a=13$ at most. 
This observation strongly suggests that the excited-state contributions are not fully eliminated in $\widetilde{F}_P(q^2)$, 
even though there is no significant difference in the evaluation of excited-state contamination based on the two-state analysis as discussed in Appendix~\ref{app:two_state_fit}.

Figure~\ref{fig:fp_oct22} shows that the $q^2$-dependence of the renormalized induced pseudoscalar form factor $F_P(q^2)=Z_A\widetilde{F}_P(q^2)$ for all three cases of $t_{\mathrm{sep}}$ compared to the previous results, which are obtained from the coarse lattice~\cite{Shintani:2018ozy}.
Two experimental results of the muon capture and the pion-electroproduction are marked as blue diamonds and a brown asterisk.
Both of our results from the fine and coarse lattices are significantly underestimated in comparison with both experiments.
The induced pseudoscalar form factor $F_P(q^2)$ is expected to have a pion pole that dominates the behavior near zero momentum transfer.
The colored curves are given by the pion-pole dominance (PPD) 
model~\cite{Nambu:1960xd}, where the induced pseudoscalar form factor is given as
\label{eq:ppd}
\begin{align}
    F^{\mathrm{PPD}}_{P}(q^2)
=
\frac{2M_NF_A(q^2)}{q^2+m_\pi^2}
\end{align}
with the measured values of $m_\pi$, $M_N$ and $F_A(q^2)$. 
The predictions provided by the PPD model with two data sets obtained from the fine and coarse lattices,
successfully describe two experimental results of the muon capture and the pion-electroproduction, 
while they do not agree with our results of $F_{P}(q^2)$.

Recall that in contrast to $F_P(q^2)$, $F_A(q^2)$
has no large $t_{\mathrm{sep}}$-dependence.
Therefore, this discrepancy indicates that in the case of $F_P(q^2)$, the largest choice of $t_{\mathrm{sep}}/a=19$ is not large enough to eliminate the excited-state contributions.

It should be noted that according to baryon chiral perturbation theory, the nucleon matrix element of the axial vector current inevitably has a strong effect due to the contamination of $\pi N$ excited states~\cite{Bar:2018xyi, Bar:2019gfx, Bar:2019igf}.
In order to eliminate such a strong excited-state contamination, several ways of analysis are suggested, 
such as the utilization of the temporal $A_4$ current~\cite{Jang:2019vkm} and proper projections determined by the variational analysis with the explicit $\pi N$ operators~\cite{Barca:2022uhi}.
Further investigation with more sophisticated analyses should be conducted in the future work.

%
%
%
%
%
%
\begin{figure*}
\centering
\includegraphics[width=1\textwidth,bb=0 0 792 612,clip]{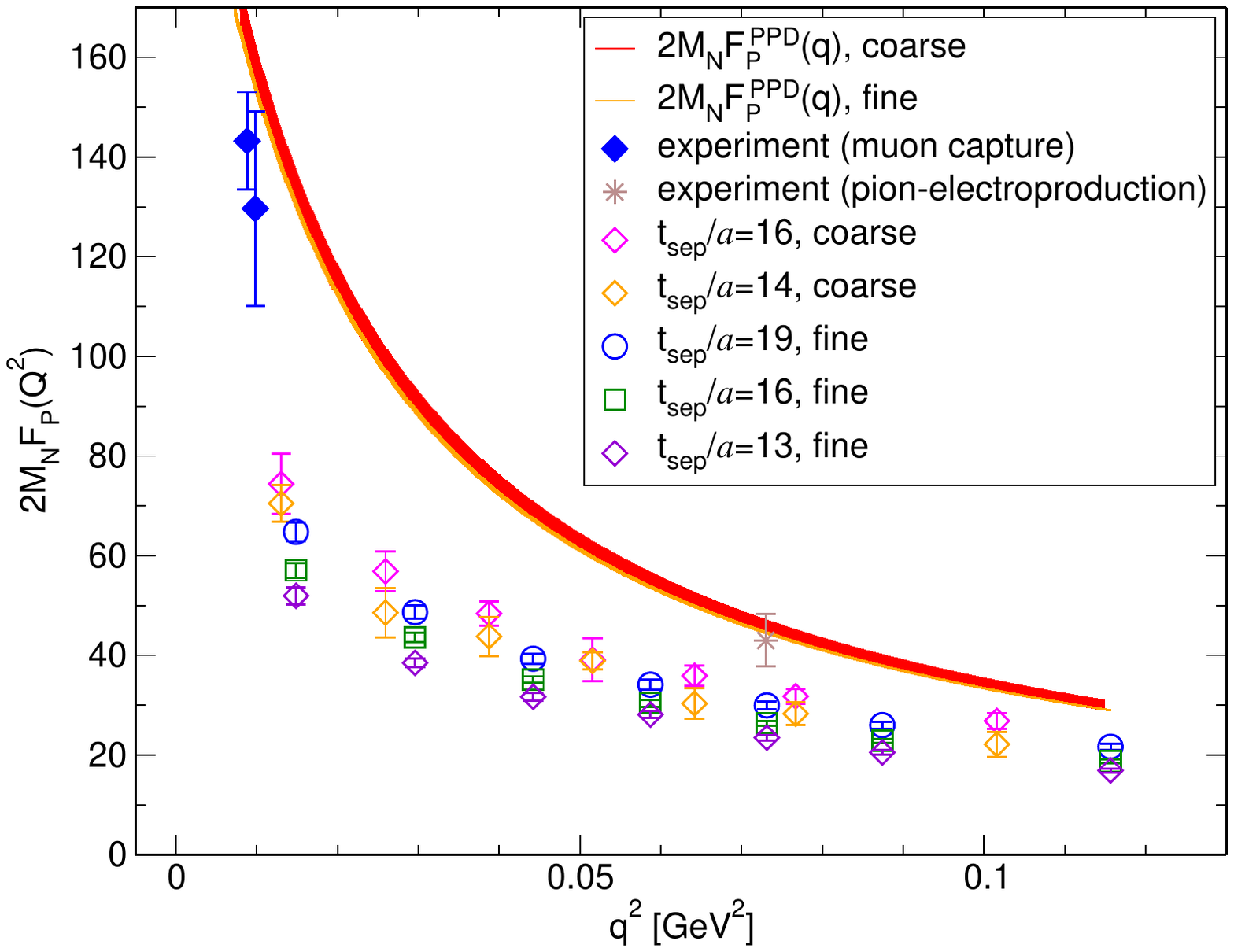}
\caption{
Same as Fig.~\ref{fig:ge_oct22} for the induced pseudoscalar form factor.}
\label{fig:fp_oct22}
\end{figure*}

\subsubsection{Pseudoscalar form factor}
\label{sec:gp}
\begin{figure*}                                 
\centering
\includegraphics[width=0.48\textwidth,bb=0 0 864 720,clip]{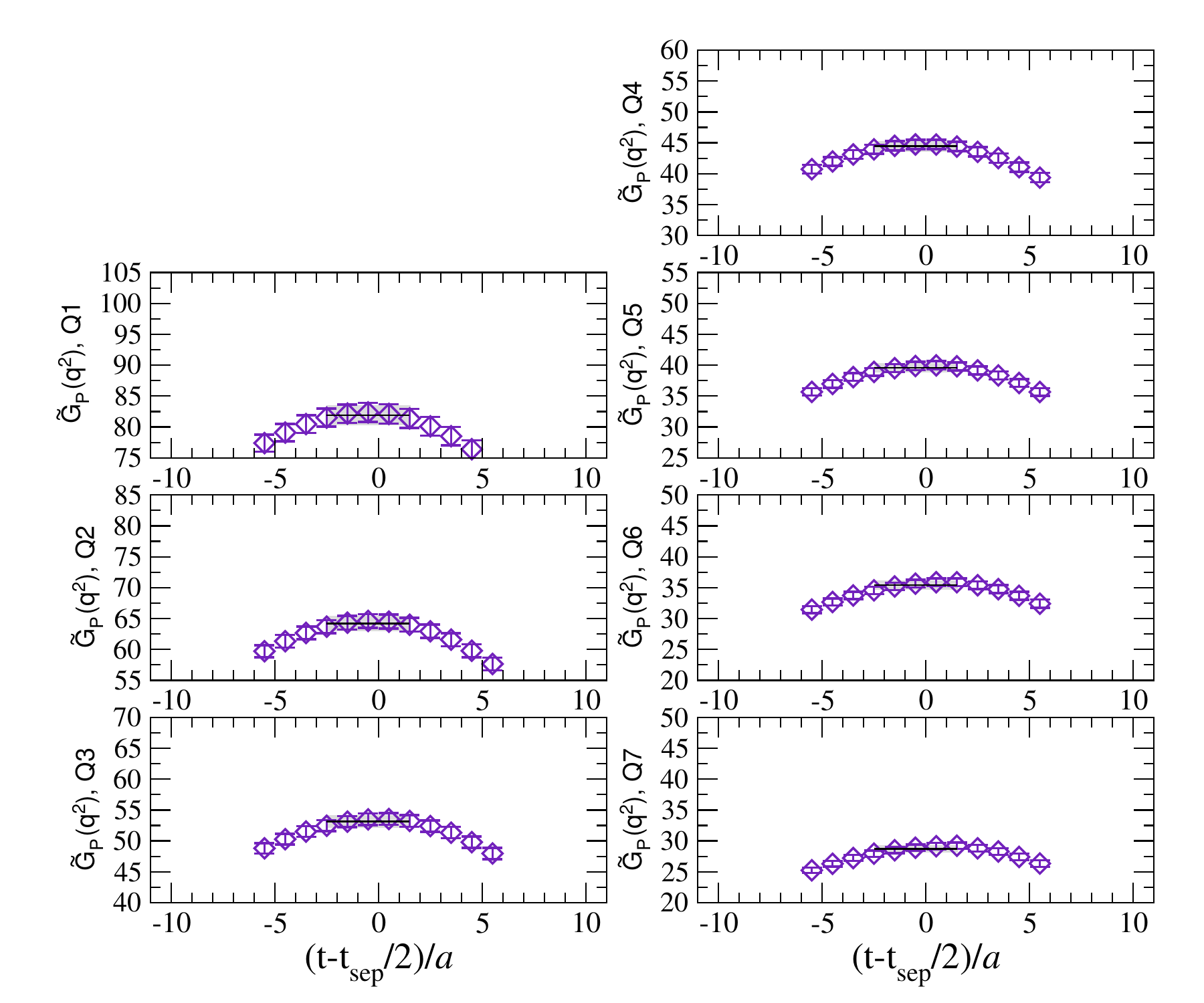}
\includegraphics[width=0.48\textwidth,bb=0 0 864 720,clip]{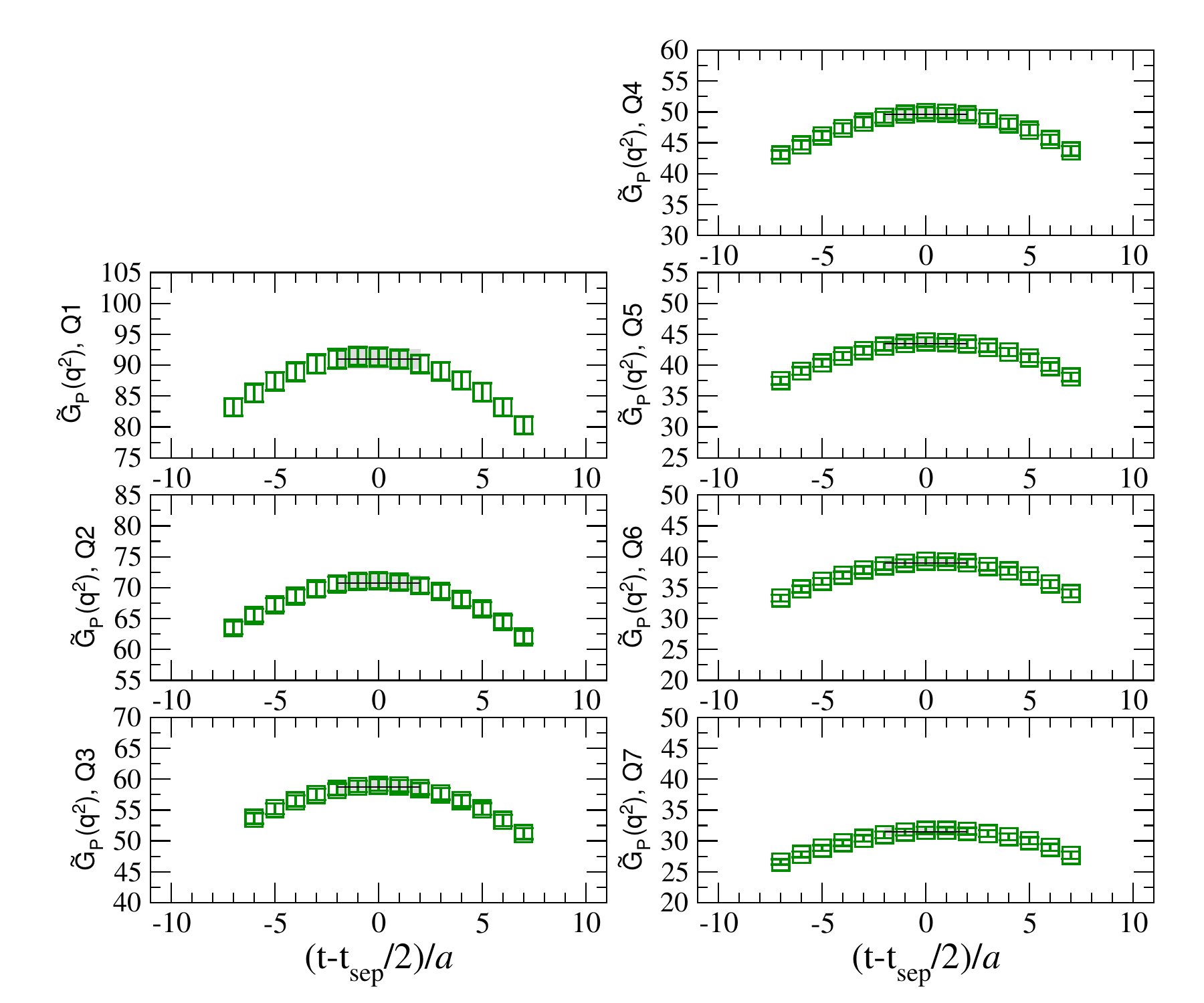}
\includegraphics[width=0.48\textwidth,bb=0 0 864 720,clip]{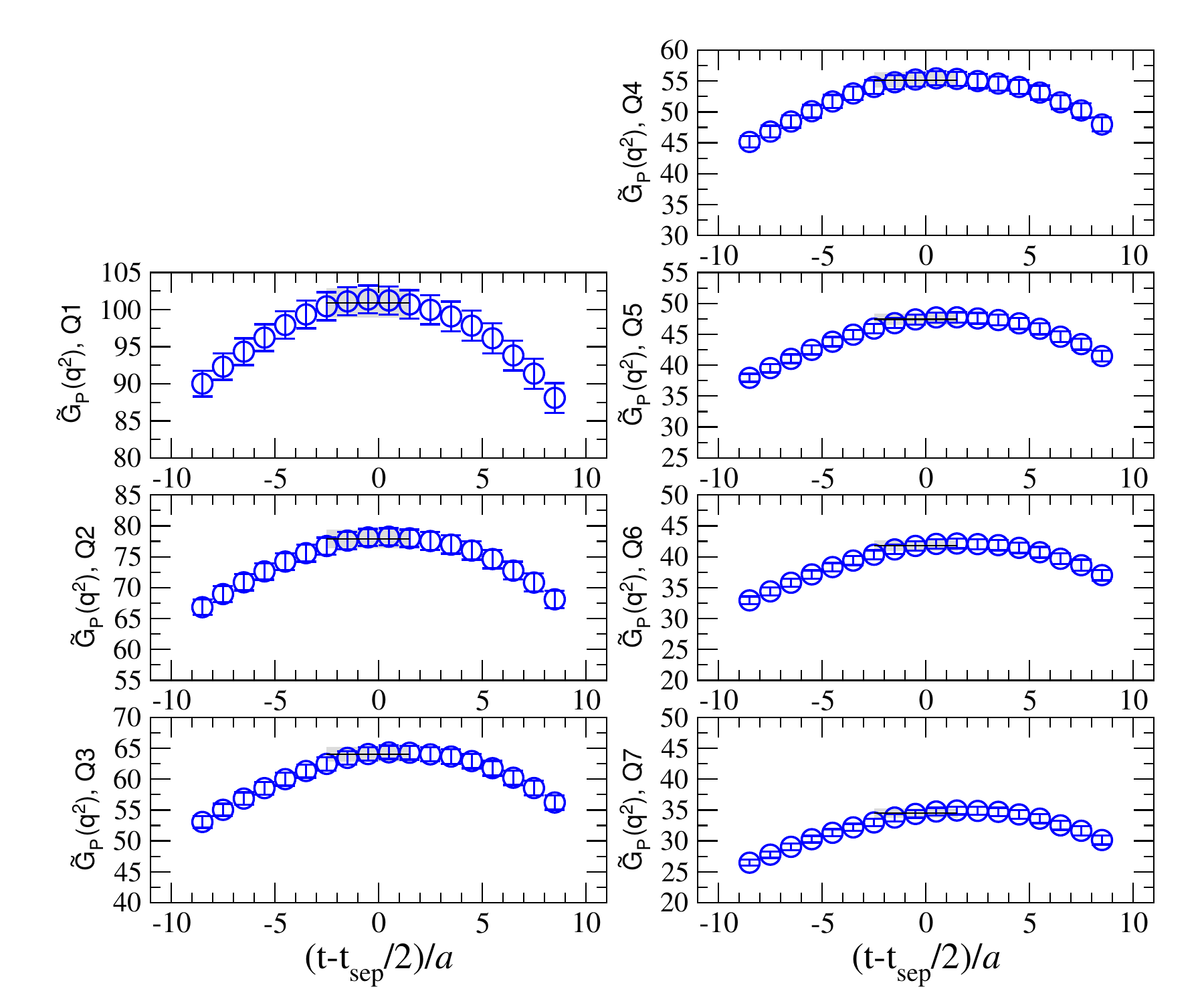}
\caption{Same as Fig.~\ref{fig:ge_qdep_p-n_ts1X} for the pseudoscalar form factor.}
\label{fig:gp_qdep_p-n_ts1X}
\end{figure*}

The pseudoscalar form factor $\widetilde{G}_P(q^2)$ is extracted from the ratio $\mathcal{R}^{5z}_{P}(t; \bm{q})$ defined in Eq.~(\ref{eq:gp_def}).
In Fig.~\ref{fig:gp_qdep_p-n_ts1X}, the $t$-dependencies of $\widetilde{G}_P(q^2)$ for all seven variations of 
$q^2\neq 0$ with $t_{\mathrm{sep}}/a=\{13,16,19\}$ are displayed.
As is in the case of $F_P(q^2)$, 
the slight convex shape, which is associated with the excited-state contributions 
is observed in all cases of $t_{\mathrm{sep}}/a=\{13,16,19\}$ for all $q^2$. 
The data points within the fit range shown
as the gray shaded band in each panel of Fig.~\ref{fig:gp_qdep_p-n_ts1X},
overlap within one standard deviation.
Therefore 
it is adequate to employ a constant fit to estimate the value of $\widetilde{G}_P(q^2)$.

\begin{figure*}
\centering
\includegraphics[width=1\textwidth,bb=0 0 864 720,clip]{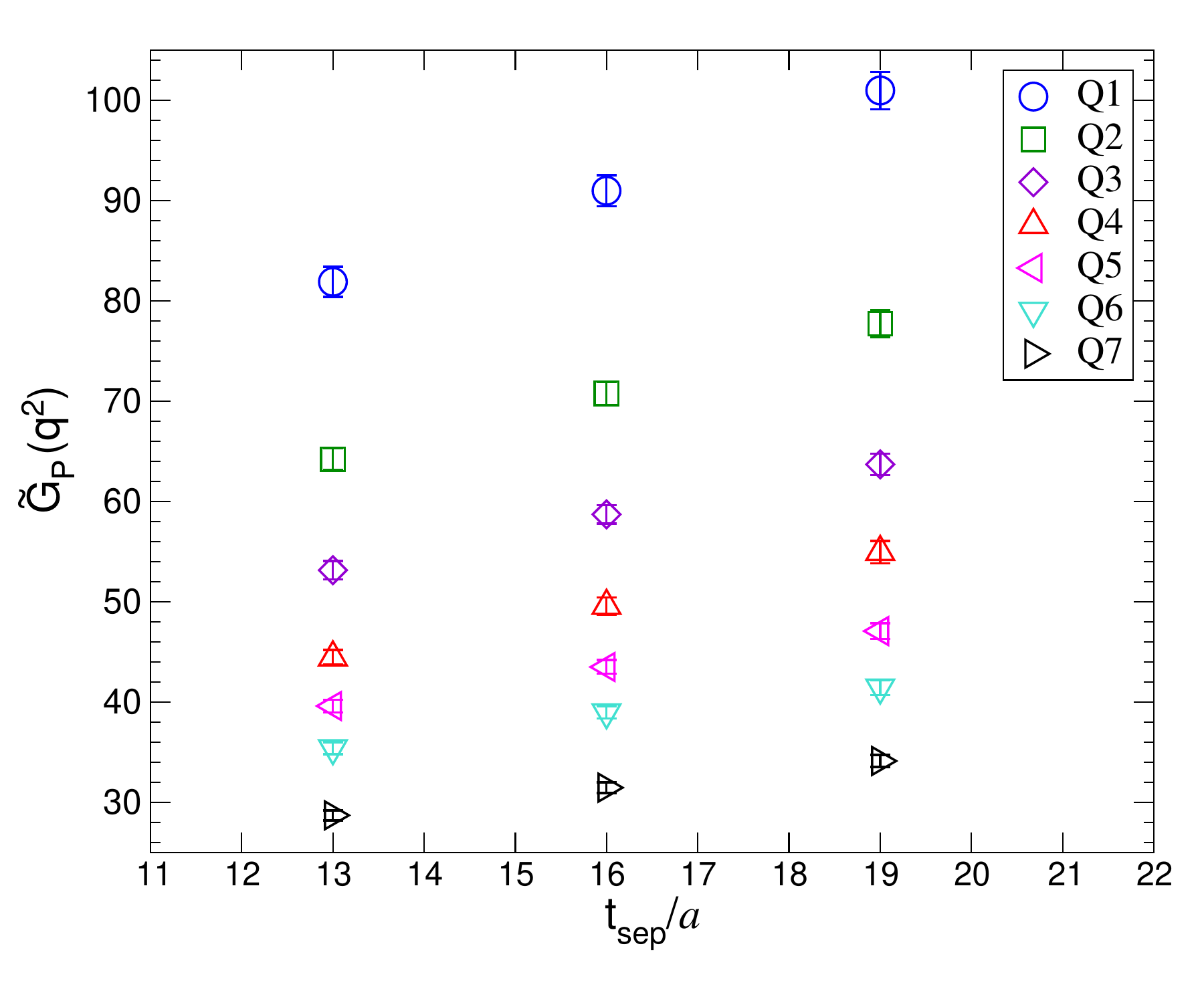}
\caption{Same as Fig.~\ref{fig:ge_tsdep_p-n} for the pseudoscalar form factor.}
\label{fig:gp_tsdep_p-n}
\end{figure*}

Figure~\ref{fig:gp_tsdep_p-n} shows the $t_{\mathrm{sep}}$-dependence of $\widetilde{G}_P(q^2)$ for all $q^2$.
It is clearly observed that $\widetilde{G}_P(q^2)$ shows a large $t_{\mathrm{sep}}$-dependence.
The values of $\widetilde{G}_P(q^2)$ systematically increase as $t_{\mathrm{sep}}$ increases similar to $F_P(q^2)$. 
From $t_{\mathrm{sep}}/a=13$ to $t_{\mathrm{sep}}/a=19$,
the maximum increase observed in the magnitude of $\widetilde{G}_P(q^2)$ at lowest $q^2$ reaches about 20\%. This indicates that $\widetilde{G}_P(q^2)$ involves the significant contribution from
the excited states as well as $F_P(q^2)$. 
However, as discussed in Appendix~\ref{app:two_state_fit},
there is no significant difference in the evaluation of excited-state contamination based on the two-state analysis
in a direct comparison to the standard plateau analysis.
Thus we mainly use the constant fit on the ratio $\mathcal{R}^{5z}_{P}(t; \bm{q})$ to estimate the values of $\widetilde{G}_P(q^2)$ within the standard plateau method in this study.

\begin{figure*}
\centering
\includegraphics[width=1\textwidth,bb=0 0 792 612,clip]{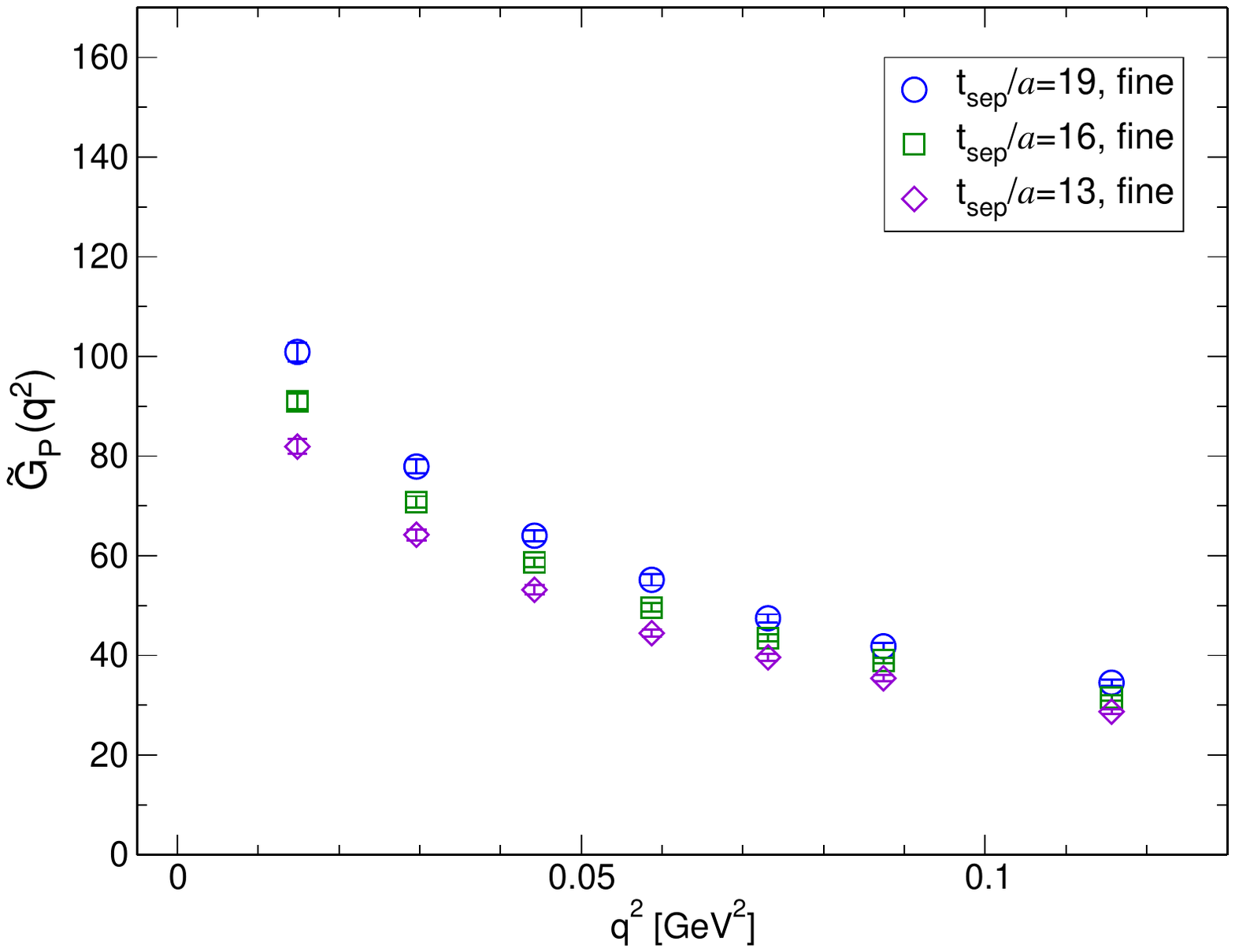}
\caption{
Same as Fig.~\ref{fig:ge_oct22} for the pseudoscalar form factor.}
\label{fig:gp_oct22}
\end{figure*}

We plot the $q^2$-dependence of $\widetilde{G}_P(q^2)$ for all three cases of $t_{\mathrm{sep}}$ in Fig.~\ref{fig:gp_oct22}.
It is obvious that the stronger curvature appears at lower $q^2$ as $t_{\mathrm{sep}}$ increases. 
This particular behavior is shared by both of $F_P(q^2)$ and $\widetilde{G}_P(q^2)$. 
The relatively strong $q^2$-dependence appearing in the lower $q^2$ region can be described by a naive pion-pole dominance form of $\widetilde{G}_P^{\mathrm{PPD}}(q^2)$, which is defined through
the GGT relation with $F_P^{\mathrm{PPD}}(q^2)$ as 
\label{eq:gp_ppd}
\begin{align}
2m
\widetilde{G}_P^{\mathrm{PPD}}(q^2)
=2M_NF_A(q^2)\frac{m_\pi^2}{q^2+m_\pi^2}.
\end{align}
This indicates that the ratio of the PPD forms $\widetilde{G}_P^{\mathrm{PPD}}(q^2)/F_P^{\mathrm{PPD}}(q^2)$ 
yields no dependence on the value of $q^2$ and gives the
low-energy constant $B_0$ as 
\begin{align}
\label{eq:gmor}
\frac{\widetilde{G}_P^{\mathrm{PPD}}(q^2)}{F_P^{\mathrm{PPD}}(q^2)}
=
B_0
\end{align}
with the help of the Gell-Mann--Oakes--Renner relation for the pion mass: $m^2_\pi=2B_0m$.

\begin{figure*}
\centering
\includegraphics[width=1\textwidth,bb=0 0 792 612,clip]{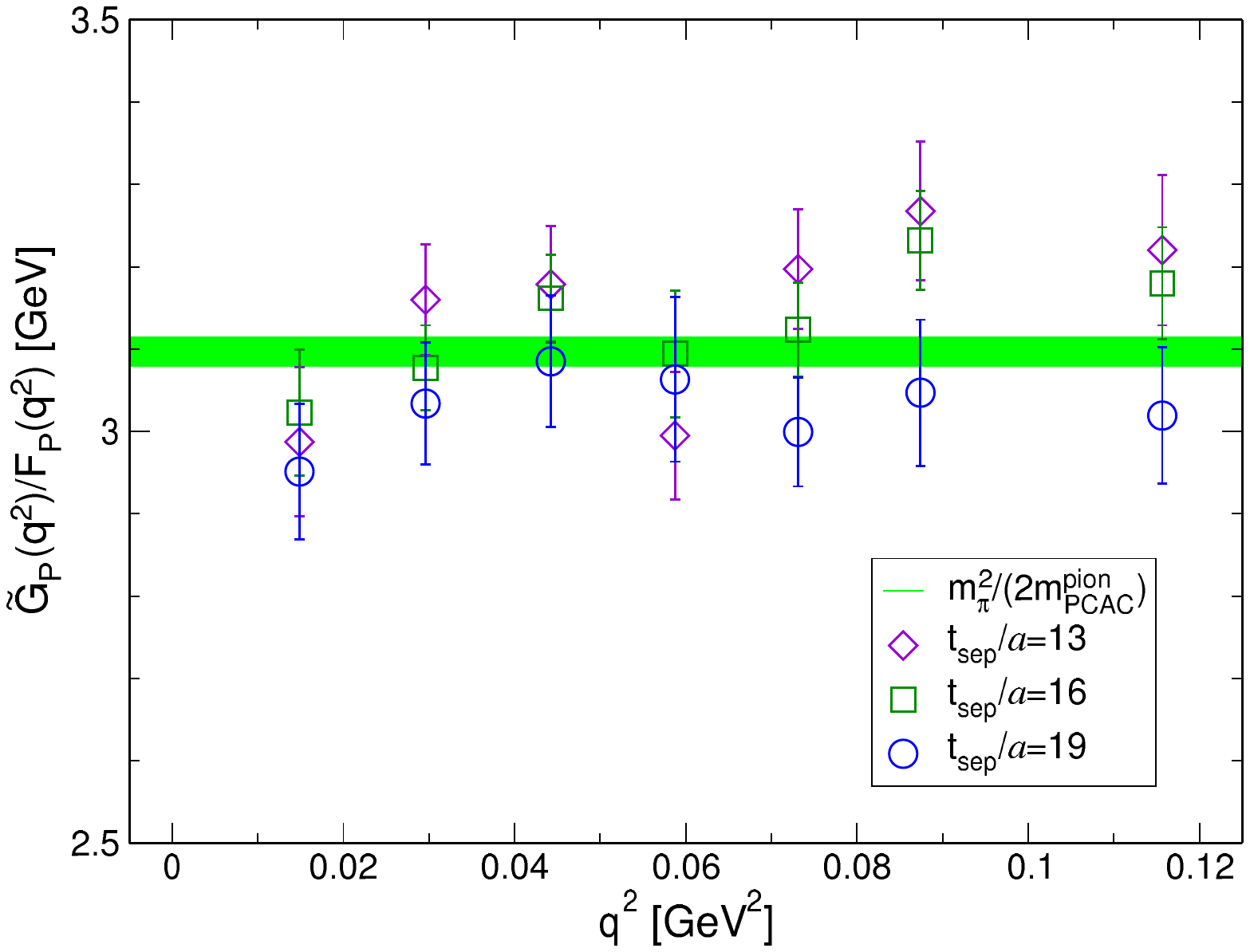}
\caption{Ratio of the pseusoscalar form factor $\widetilde{G}_P(q^2)$ to the induced pseudoscalar form factor $F_P(q^2)$ as a function of $q^2$. The green band represents the low energy constant $B_0$ given by $m_\pi^2/(2m_{\rm PCAC}^{\mathrm{pion}})$. }
\label{fig:frac_gp_fp}
\end{figure*}

As shown in Fig.~\ref{fig:frac_gp_fp}, the corresponding ratio
evaluated with our measured values of the $\widetilde{G}_P(q^2)$ and $F_P(q^2)$,
indeed exhibits a flat $q^2$-dependence for each case of $t_{\mathrm{sep}}$. 
Furthermore, without using knowledge of the PPD model, all ratios are 
in good agreement with the bare value of the low-energy constant, which is evaluated by $m^2_\pi/(2m_{\mathrm{PCAC}}^{\mathrm{pion}})$ with the simulated pion mass $m_\pi$ and the PCAC quark mass $m_{\mathrm{PCAC}}^{\mathrm{pion}}$. 
Here, the PCAC quark mass is determined
by the two-point correlators of the pseudoscalar meson. 

These observations strongly suggest that although our results of $\widetilde{G}_P(q^2)$ and $F_P(q^2)$ suffer from the excited-state contamination, both quantities correctly inherit the low-energy physics associated with the pion-nucleon ($\pi N$) system.
Therefore, once the large excited-state contamination 
is hindered in a certain way, the low-energy constants of the $\pi N$ system, {\it e.g.} $g_{\pi N}$, could be correctly evaluated. 
We do not evaluate the $\pi N$ coupling $g_{\pi N}$ in this paper, since there is no known reasonable way to eliminate 
the excited-state contributions from our results of $\widetilde{G}_P(q^2)$ and $F_P(q^2)$.

%
%
%
%
%
%
%
%

\clearpage
\section{Numerical results II: Test for the axial Ward--Takahashi identity}
\label{sec:numerical_results_II}
\subsection{Quark mass from nucleon correlation functions}
\label{sec:q_mass}
As discussed in Sec.~\ref{sec:fa_ra_fp}, 
both $F_P(q^2)$ and $\widetilde{G}_P(q^2)$ form factors significantly suffer from the excited-state contamination in contrast to $F_A(q^2)$ where
the systematic uncertainties stemming from the 
excited-state contamination are negligible within
the present statistical precision. However, the ratio
of $F_P(q^2)$ and $\widetilde{G}_P(q^2)$ implies 
that both quantities correctly inherit the low-energy
physics which is related to the axial--Ward--Takahashi identity (AWTI).

\begin{figure*}
\centering
\includegraphics[width=1\textwidth,bb=0 0 864 720,clip]{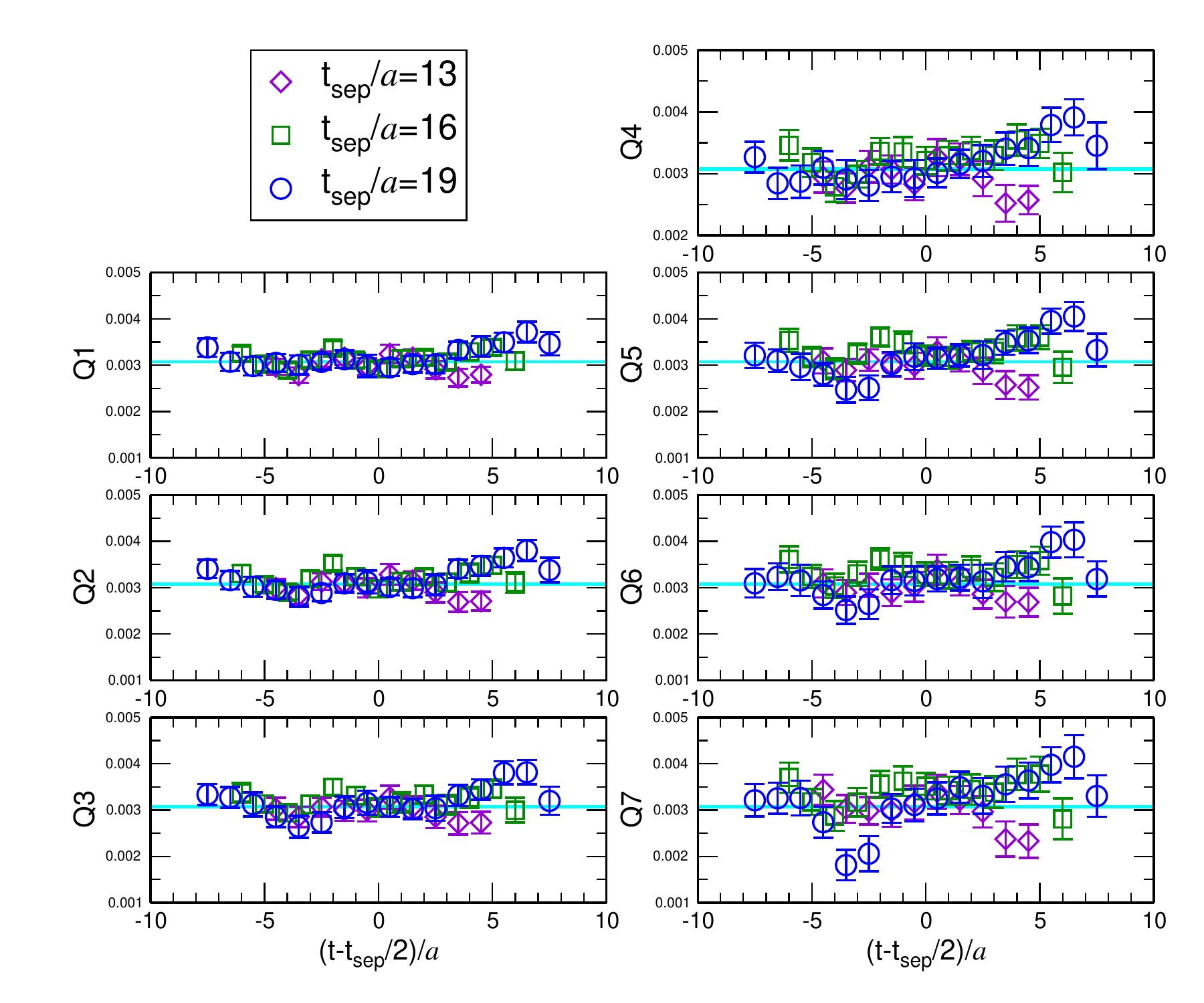}
\caption{
The values of $m_{\mathrm{PCAC}}^{\mathrm{nucl}}$ computed
with $t_{\mathrm{sep}}/a=13$ (diamonds), $16$ (squares) and $19$ (circles) for all momentum transfers as functions of the current insertion time slice $t$. 
In each panel, the value of $m_{\mathrm{PCAC}}^{\mathrm{pion}}$ is presented as a horizontal band.
}
\label{fig:mpcacnu_plateau_p-n}
\end{figure*}

As described in Sec.~\ref{sec:pcac_test}, 
three kinds of the bare quark mass: $m_{\mathrm{PCAC}}^{\mathrm{pion}}$, $m_{\mathrm{PCAC}}^{\mathrm{nucl}}$ and $m_{\mathrm{GGT}}^{\mathrm{nucl}}$ are introduced. 
Let us first consider $m_{\mathrm{PCAC}}^{\mathrm{nucl}}$, which can
verify the PCAC relation using the nucleon three-point
correlation functions, in order to be compared to the value of $m_{\mathrm{PCAC}}^{\mathrm{pion}}$ given by
the pion two-point correlation function.
In Fig.~\ref{fig:mpcacnu_plateau_p-n},
the ratios defined in Eq.~(\ref{eq:m_awti_pcac}) are displayed for all $q^2$ with all three variations of $t_{\mathrm{sep}}/a=\{13,16,19\}$.
All data show good plateaus, which are fairly consistent with $m_{\mathrm{PCAC}}^{\mathrm{pion}}$, regardless of the momentum transfer and $t_{\mathrm{sep}}$.
Thus, the ratio defined in Eq.~(\ref{eq:m_awti_pcac})
can provide an alternative bare quark mass definition
as $m_{\mathrm{PCAC}}^{\mathrm{nucl}}$.

\begin{figure*}
\centering
\includegraphics[width=1\textwidth,bb=0 0 792 612,clip]{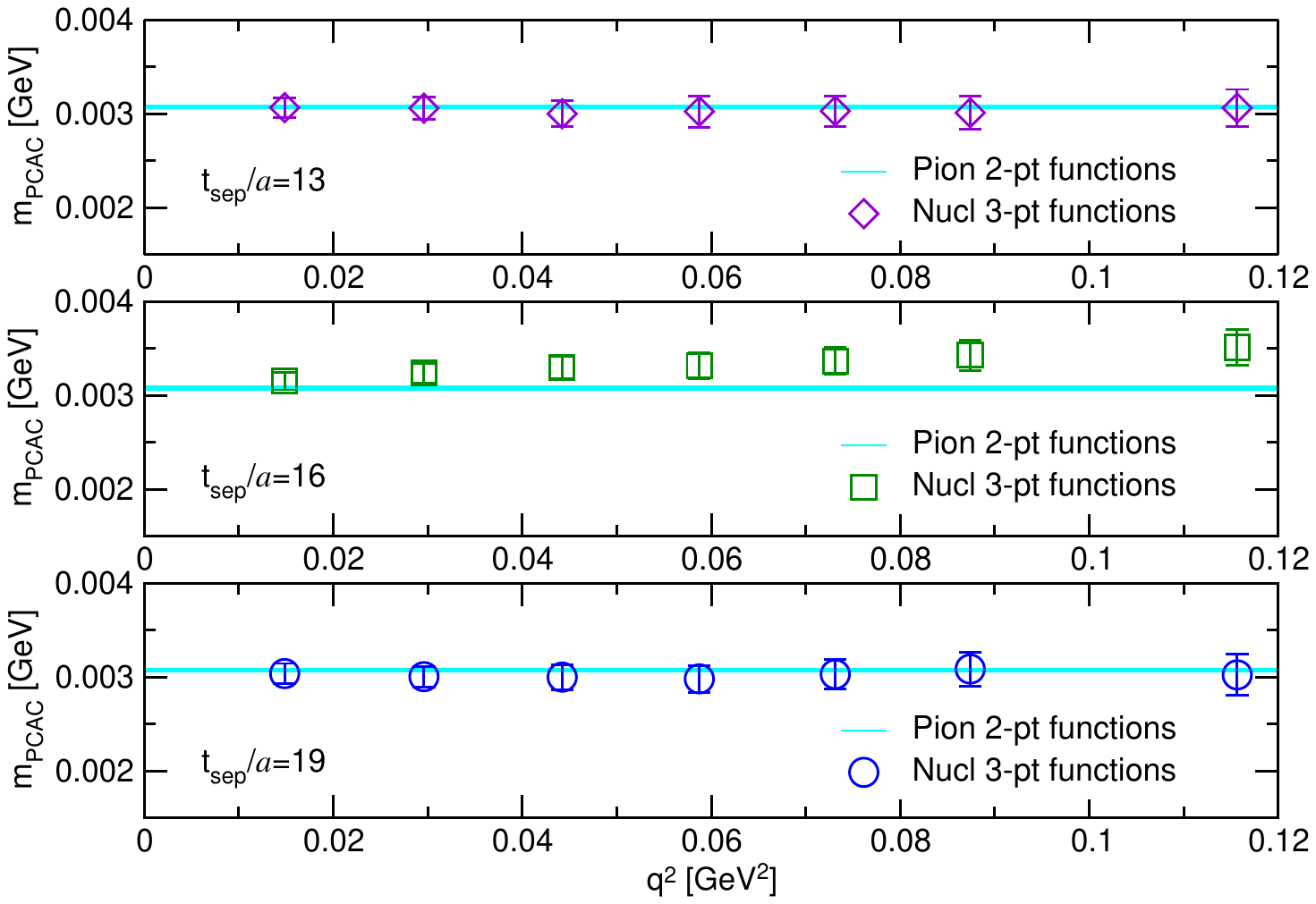}
\caption{
Comparison between $m_{\mathrm{PCAC}}^{\mathrm{pion}}$ (horizontal band)
and $m_{\mathrm{PCAC}}^{\mathrm{nucl}}$ (open symbols) 
in each panel. Results for $t_{\mathrm{sep}}/a=\{13,16,19\}$ are plotted from top to bottom. 
}
\label{fig:mcomp_qdep}
\end{figure*}

For each $q^2$ and $t_{\mathrm{sep}}$, we evaluate the value of $m_{\mathrm{PCAC}}^{\mathrm{nucl}}$ by weighted average using five data points in the central range of $t/a$.
Figure~\ref{fig:mcomp_qdep} shows a direct comparison of $m_{\mathrm{PCAC}}^{\mathrm{pion}}$ (denoted as horizontal line) 
and $m_{\mathrm{PCAC}}^{\mathrm{nucl}}$ (denoted as
diamond symbols) in the case of $t_{\mathrm{sep}}/a=13$ (top panel), $16$ (middle panel), $19$ (lower panel).
As can be easily seen, all data points in $m_{\mathrm{PCAC}}^{\mathrm{nucl}}$ do not show 
strong $q^2$-dependence and reproduce 
the value of $m_{\mathrm{PCAC}}^{\mathrm{pion}}$.
The agreement between $m_{\mathrm{PCAC}}^{\mathrm{pion}}$ and $m_{\mathrm{PCAC}}^{\mathrm{nucl}}$ observed with each finite momentum transfer is highly nontrivial as 
discussed in Sec.~\ref{sec:pcac_test}.

It is worth noting that the definition of $m_{\mathrm{PCAC}}^{\mathrm{nucl}}$
defined in Eq.~(\ref{eq:m_awti_pcac}) does not take into account $O(a)$-improvement of the axial-vector current $\widetilde{A}^{\mathrm{imp}}_\alpha = \widetilde{A}_\alpha+ac_A\partial_\alpha \widetilde{P}$.
The second term in $O(a)$-improvement of the axial-vector current provides the $O(a)$ correction on the value of $m_{\mathrm{PCAC}}^{\mathrm{pion}}$ as
\begin{align}
\label{eq:m_pion_pcac_Oa-imp}
(m_{\mathrm{PCAC}}^{\mathrm{pion}} )^{\mathrm{imp}}
=
m_{\mathrm{PCAC}}^{\mathrm{pion}} - \frac{aZ_Ac_A}{2} m_\pi^2
\end{align}
for the ground state contribution
since the point sink of the pion two-point correlation function is projected onto zero three momentum.
On the other hand, $m_{\mathrm{PCAC}}^{\mathrm{nucl}}$
does not receive the $O(a)$ correction at zero momentum
transfer, though Eq.~(\ref{eq:m_awti_pcac}) can be used 
only when the momentum transfer is finite.
The quark mass is modified by the presence of the 
improvement term as
\begin{align}
\label{eq:m_awti_pcac_Oa-imp}
(m_{\mathrm{PCAC}}^{\mathrm{nucl}} )^{\mathrm{imp}}
&=
Z_A \times 
\frac{\partial_{\alpha}\left(
C^{5z}_{A_{\alpha}}(t;\bm{q}) +
a c_A \partial_{\alpha} C^{53}_{P}(t;\bm{q})
\right)
}{
2 C^{5z}_{P}(t;\bm{q})
}
=
m_{\mathrm{PCAC}}^{\mathrm{nucl}} - \frac{aZ_Ac_A}{2} q^2,
\end{align}
where the $O(a)$ correction is proportional to
the square of momentum transfer $q^2$ and then 
vanishes in the limit of $q^2=0$.
Recall that $m_{\mathrm{PCAC}}^{\mathrm{nucl}}$ and $m_{\mathrm{PCAC}}^{\mathrm{pion}}$ are supposed to be identical in the continuum limit.
In other words,
a difference observed on the lattice 
can be attributed to lattice discretization errors.
However as shown in Fig.~\ref{fig:mcomp_qdep},
$m_{\mathrm{PCAC}}^{\mathrm{nucl}}$
coincides $m_{\mathrm{PCAC}}^{\mathrm{pion}}$ within statistical precision 
without the $O(a)$-improvement over a wide range of $q^2$ $(0.78<q^2/m_{\pi}^2<6.08)$.
Therefore, our finding indicates that the value of $c_A$ is likely to be nearly zero at the $O(10^{-2})$ level
in lattice units.
This suggests that the effect of $O(a)$-improvement of the axial-vector current is negligibly small in our calculations performed at very low $q^2$, and does not change the result of the axial radius.

The second check is made by comparing $m_{\mathrm{GGT}}^{\mathrm{nucl}}$ defined in Eq.~(\ref{eq:m_awti_ggt}) with either $m_{\mathrm{PCAC}}^{\mathrm{pion}}$
or $m_{\mathrm{PCAC}}^{\mathrm{nucl}}$. Recall that
$m_{\mathrm{GGT}}^{\mathrm{nucl}}$ requires 
isolating the ground-state contribution from the
excited-state contributions in determining 
the three form factors $F_A(q^2)$, $F_P(q^2)$ and $\widetilde{G}_P(q^2)$. 
Therefore, if $m_{\mathrm{GGT}}^{\mathrm{nucl}}$ coincides the bare quark mass associated with the PCAC
relation, the ground state dominance is successfully achieved in determination of $F_A(q^2)$, $F_P(q^2)$ 
and $\widetilde{G}_P(q^2)$. This is simply because the GGT relation 
is derived from the axial WT identity in terms of the nucleon matrix elements, not the nucleon three-point 
functions.

\begin{figure*}
\centering
\includegraphics[width=1\textwidth,bb=0 0 792 692,clip]{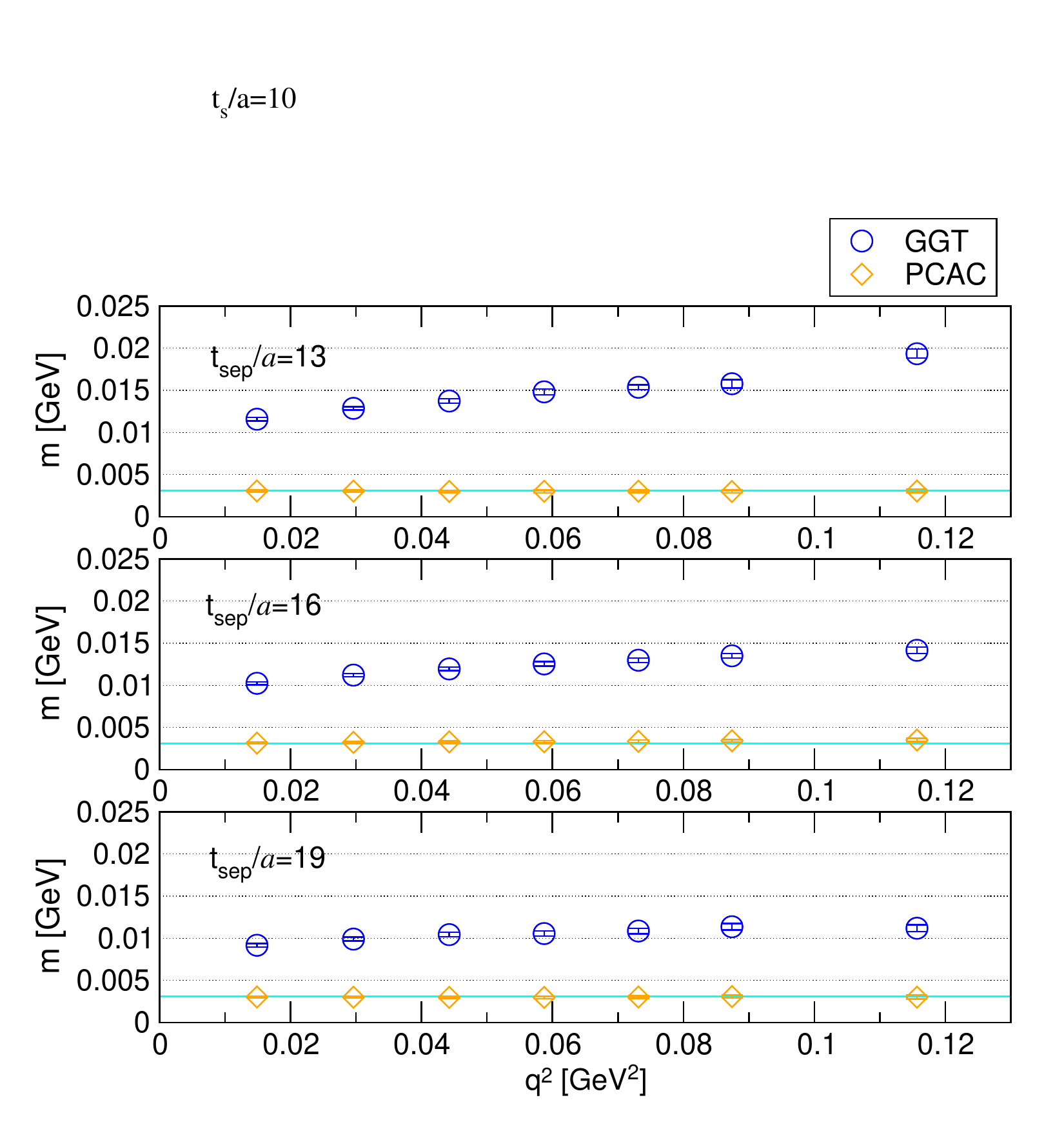}
\caption{
Comparison among three types of the quark mass:
$m_{\mathrm{PCAC}}^{\mathrm{pion}}$ (horizontal band), $m_{\mathrm{PCAC}}^{\mathrm{nucl}}$ (open diamonds) and $m_{\mathrm{GGT}}^{\mathrm{nucl}}$ (open circles) in each panel. Results for $t_{\mathrm{sep}}/a=\{13,16,19\}$ are plotted from top to bottom. 
}
\label{fig:mcomp_qdep2}
\end{figure*}

Figure~\ref{fig:mcomp_qdep2} shows
the results of $m_{\mathrm{GGT}}^{\mathrm{nucl}}$ 
in comparison with the others as a function of $q^2$ for each choice of $t_{\mathrm{sep}}/a=\{13,16,19\}$.
Although
the values of $m_{\mathrm{GGT}}^{\mathrm{nucl}}$ do 
not show strong $q^2$-dependence,
it is obvious that the data points for $m_{\mathrm{GGT}}^{\mathrm{nucl}}$ are deviated from
both of $m_{\mathrm{PCAC}}^{\mathrm{pion}}$ and $m^{\mathrm{nucl}}_{\mathrm{PCAC}}$.
In detail,
the deviation gradually disappears as 
$t_{\mathrm{sep}}$ increases, but
the values of $m_{\mathrm{GGT}}^{\mathrm{nucl}}$ do not reach $m_{\mathrm{PCAC}}^{\mathrm{pion}}$ or $m^{\mathrm{nucl}}_{\mathrm{PCAC}}$ even when $t_{\mathrm{sep}}/a=19$.
This indicates that the form factors used for the construction of $m_{\mathrm{GGT}}^{\mathrm{nucl}}$ 
suffer from the excited-state contamination,
since though $m_{\mathrm{PCAC}}^{\mathrm{nucl}}$ does not require the ground-state dominance, $m_{\mathrm{GGT}}^{\mathrm{nucl}}$ surely does. 

In other words, the maximum $t_{\mathrm{sep}}$ used in our study dose not reach the conditions required in the standard plateau method, where only the ground state is dominant. 
This observation is consistent with 
the strong dependence of $t_{\mathrm{sep}}$ observed in the analyses of $F_P(q^2)$ and $\widetilde{G}_P(q^2)$.

%
%

%
%

%
%

\clearpage
\section{Numerical results III: Discretization error}
\label{sec:numerical_results_III}
Combining our two results from large volume simulations at the 
fine and coarse lattice spacings,
we can discuss the discretization uncertainties appearing in the isovector RMS radii, magnetic moment and also axial-vector coupling.
Recall that the continuum limit results are not yet known in our study,
we only evaluate the differences between two results from different lattice spacing as the lattice discretization uncertainties.
The error budget for the five quantities: $g_A$, $\mu_v$, $\sqrt{\langle (r^v_E)^2 \rangle}$, $\sqrt{\langle (r^v_M)^2 \rangle}$ and$\sqrt{\langle (r^v_A)^2 \rangle}$ are summarized in Table~\ref{tab:err_budget}.

\begin{figure*}
\centering
\includegraphics[width=0.32\textwidth,bb=0 0 792 612,clip]{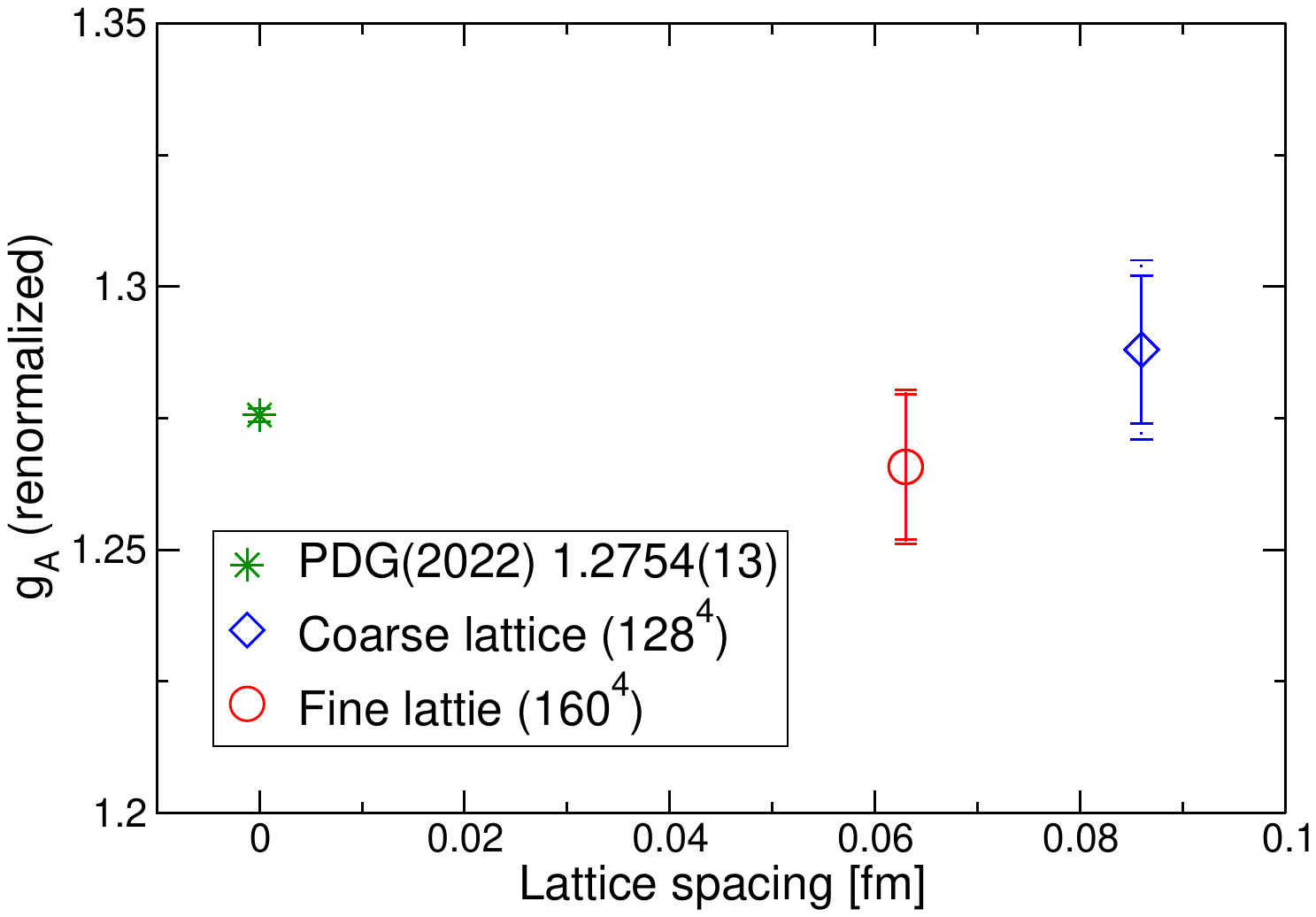}
\includegraphics[width=0.32\textwidth,bb=0 0 792 612,clip]{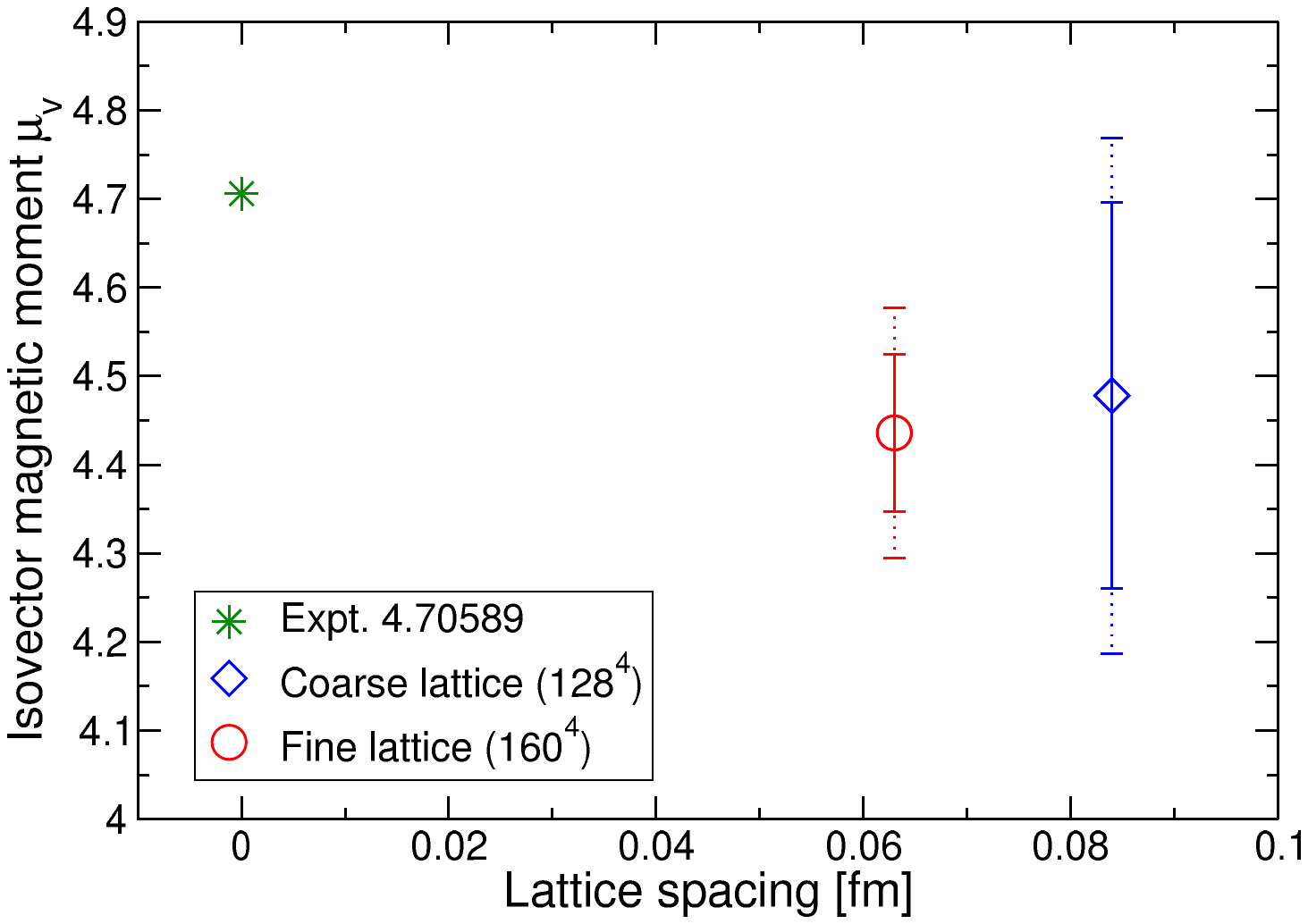}\\
\includegraphics[width=0.32\textwidth,bb=0 0 792 612,clip]{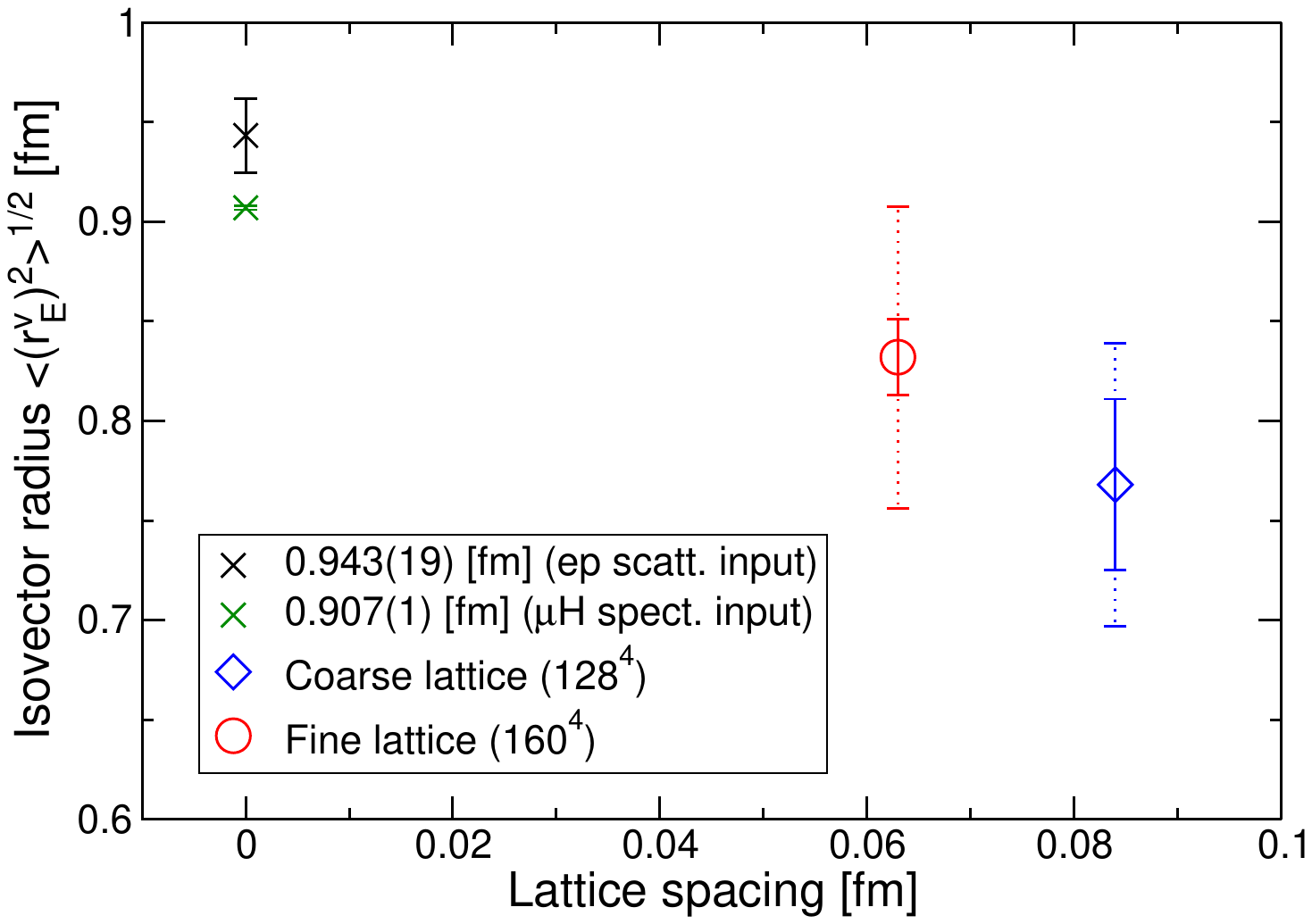}
\includegraphics[width=0.32\textwidth,bb=0 0 792 612,clip]{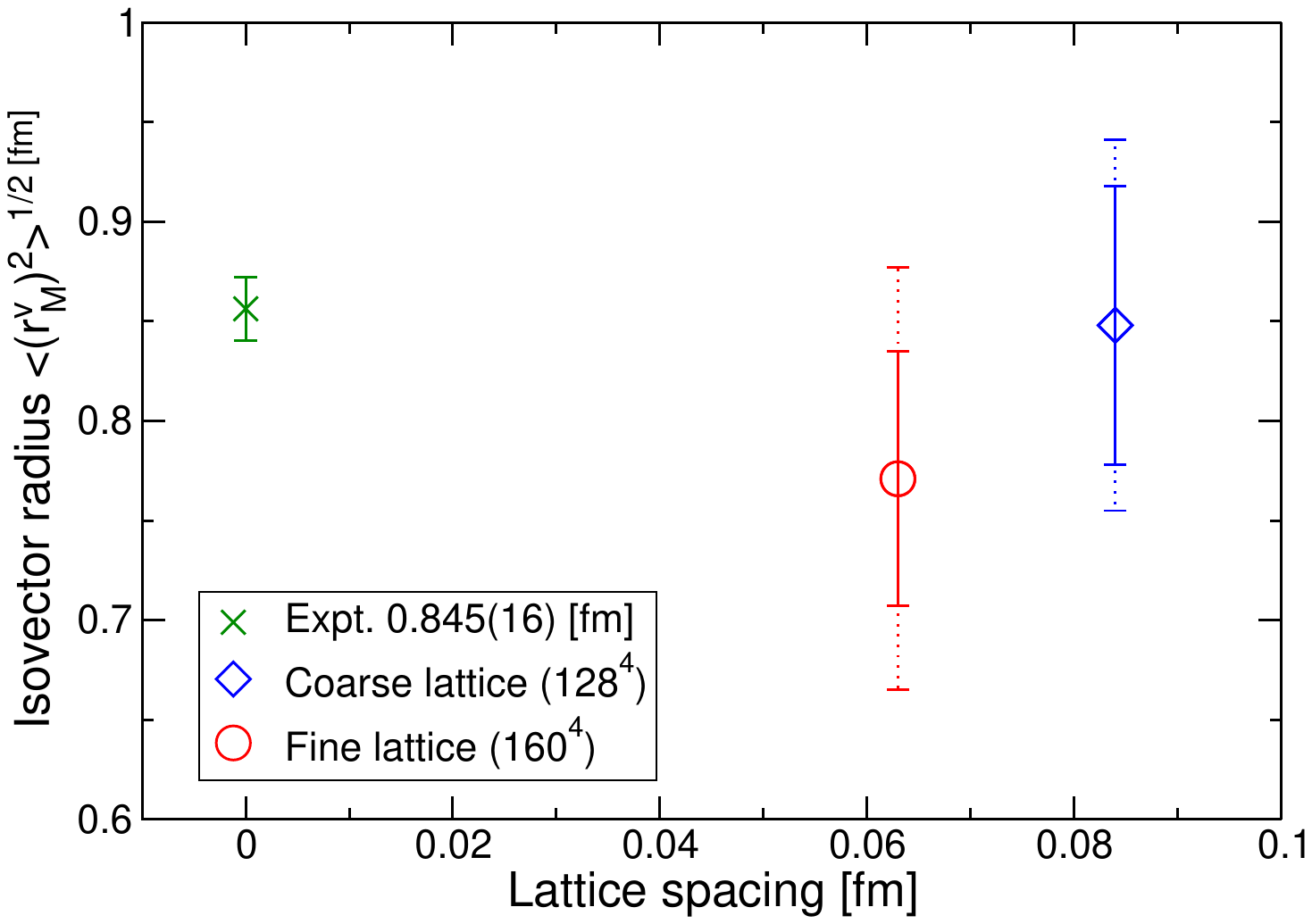}
\includegraphics[width=0.32\textwidth,bb=0 0 792 612,clip]{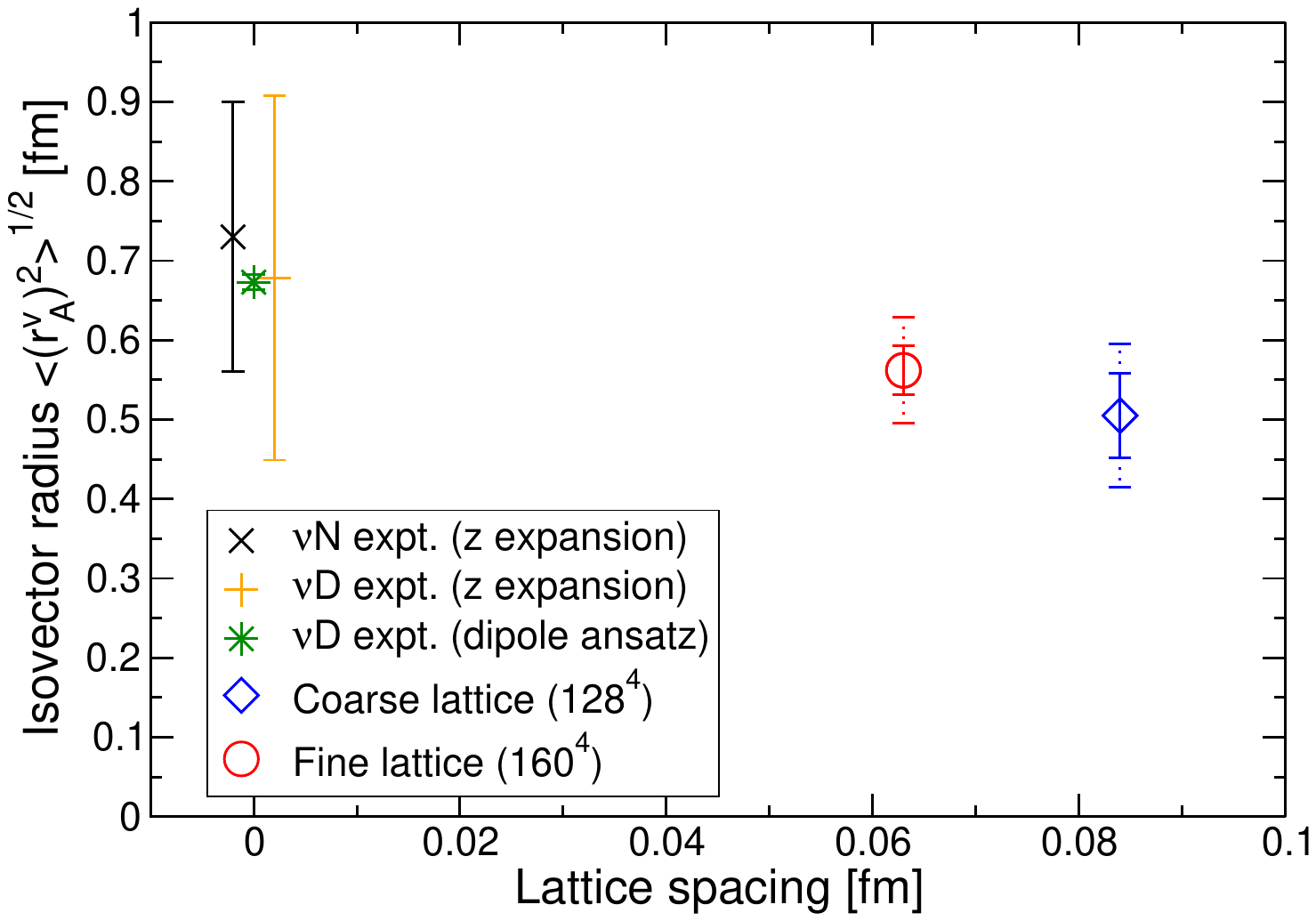}
\caption{
Summary plot for our best estimates and the experimental values of 
the axial-vector coupling (top, left), isovector magnetic moment (top, right) and three kinds of the isovector RMS radius: electric (bottom, left), magnetic (bottom, center) and axial (bottom, right).
 The inner error bars represent the statistical error, while the outer error bars are the total error evaluated by both statistical and systematic errors added in quadrature. Uncertainties associated with the excited-state contamination, the violation of the dispersion relation and other effects are taken into account as the systematic errors.}
\label{fig:continuum_limit}
\end{figure*}

Figure \ref{fig:continuum_limit} shows 
the lattice spacing $a$-dependence for these five quantities.
The inner error bars represent the statistical uncertainties, while the outer error bars represent
the total uncertainties given by adding the statistical
errors and systematic errors in quadrature. The systematic errors take into account
uncertainties stemming from the excited-state contamination and the lattice discretization effects on the dispersion relation.

Let us first discuss the size of the discretization error on the axial-vector coupling $g_A$, 
that is precisely measured by the experiments. 
The axial-vector coupling $g_A=F_A(q^2=0)$ is directly determined from the ratio~(\ref{eq:fa_def})
at zero momentum transfer without the
$q^2$-extrapolation to the zero momentum point.
In the top-left panel of Fig.~\ref{fig:continuum_limit}, 
the two results obtained at different lattice spacing can reproduce the experimental values within statistical precision of at most 2\%.
This implies that the discretization error on the axial-vector coupling is less than 2\%, which is well controlled in our calculations.

The small discretization error, which is less than 1\%,  is also observed for the magnetic moment $\mu_v$
in the top-right panel of Fig.~\ref{fig:continuum_limit}, though the two results for the magnetic moment are both 5-6\% smaller than the experimental value. However, recall that
the magnetic moment is not accessible without the 
$q^2$-extrapolation to the zero momentum point in contrast to the axial-vector coupling. Therefore, the current discrepancy between our lattice result and the experimental value would be caused by the $q^2$-extrapolation, since our data points at the finite
momentum transfer are barely consistent with the Kelly's curve, albeit slightly lower. The obvious disadvantage for magnetic form factor in the standard approach can be overcome by the new method called the derivative of
form factor (DFF) method~\cite{Ishikawa:2021eut}.
In order to fully resolve the current discrepancy,
more comprehensive investigations with the DFF method
are necessary in this particular quantity.

Apart from the question of whether the results are consistent with the experimental values, both quantities, $g_A$ and $\mu_v$, do not seem to be subject to large discretization errors. However, the RMS radii,
which are determined from the form-factor slope at the zero momentum point, may suffer from the $O(qa)$ discretization effects that do not appear in $g_A$ 
and $\mu_v$.
Indeed, as shown in three bottom panels of 
Fig.~\ref{fig:continuum_limit}, the presence of the discretization errors is clearly visible for the isovector RMS radii.
Their sizes can be estimated as 8.1, 9.0 and 11.3\% for $\sqrt{\langle (r^v_E)^2 \rangle}$, $\sqrt{\langle (r^v_M)^2 \rangle}$ and $\sqrt{\langle (r^v_A)^2 \rangle}$, respectively.
These errors are much lager than that of $g_A$.

Especially, in this study, $\sqrt{\langle (r^v_E)^2 \rangle}$
can be evaluated with a statistical error of less than 5\% accuracy, while the magnitude of the discretization uncertainty is much larger than the statistical one.
Therefore, as shown in the bottom-left panel of Fig.~\ref{fig:continuum_limit}, the large discretization uncertainties are clearly observed in $\sqrt{\langle (r^v_E)^2 \rangle}$, which are unexpectedly large. However, this observation may bridge the gap between the lattice results and the two experimental values. 

Similarly, the discretization uncertainties observed in $\sqrt{\langle (r^v_A)^2 \rangle}$ as shown in
the bottom-right panel of Fig.~\ref{fig:continuum_limit}
tend to fill the difference between lattice QCD results and experimental values. It is important to emphasize here that the total errors in the axial radius obtained at two lattice spacings are much smaller than the two estimations obtained from the model-independent $z$-expansion analysis for both $\nu N$ and $\nu D$ scattering data.

As for $\sqrt{\langle (r^v_M)^2 \rangle}$, although the difference between two results obtained for different lattice spacing is comparable to the size of the individual statistical uncertainties, lattice results
agree with the experimental value within 
fairly large total errors. As mentioned earlier, 
the $q^2$-extrapolation without the value of $G_M(0)$
generally leads to large hidden systematic uncertainties,
which can be avoided in the DFF method. 

Finally, it is worth reminding that the discretization error evaluated here depends on the evaluation method and is merely an estimate.
Although the coarse and fine lattice result agree when considering the systematic uncertainties, 
it cannot be excluded that it is due to the finite lattice spacing,
as long as systematic shifts larger than one standard deviation of the statistical error are actually observed.
A detailed discussion should take place only after the continuum limit is properly taken in our future studies.

%
%
\begin{table*}[ht!]
    {\scriptsize
\begin{ruledtabular}
\caption{
The error budgets for $g_A$, $\mu_v$, $\sqrt{\langle (r^v_E)^2 \rangle}$, $\sqrt{\langle (r^v_M)^2 \rangle}$ and $\sqrt{\langle (r^v_A)^2 \rangle}$ obtained 
at the fine lattice spacing.
The discretization errors quoted here are evaluated by differences between the central values of two results  
from two sets of the PACS10 ensemble at the fine and coarse lattice spacings.
\label{tab:err_budget}}
\begin{tabular}{lccccc}
  & 
  $g_A$&
  $\mu_v$&
  $\sqrt{\langle (r^v_E)^2 \rangle}$& 
  $\sqrt{\langle (r^v_M)^2 \rangle}$& 
  $\sqrt{\langle (r^v_A)^2 \rangle}$\\
  \hline
  Statistical:   & 1.9\%& 3.1\%&3.6\%& 14\%& 11\%\\
  Discretization:& 1.6\%& 0.9\%&8.3\%& 9.0\%& 11.3\%\\
\end{tabular}
\end{ruledtabular}
}
\end{table*}

%
%

\clearpage
\section{Summary}
\label{sec:summary}
We have calculated the nucleon form factors 
in the vector, axial-vector and pseudoscalar channels
using the second PACS10 ensemble (lattice spacing of $a=0.063$ fm) that is one of three sets of $2+1$ flavor lattice QCD configurations generated at the physical point on a $(10\;{\mathrm{fm}})^4$ volume.
The PACS10 gauge configurations are generated by the PACS Collaboration with the stout-smeared $O(a)$ improved Wilson quark action and Iwasaki 
gauge action~\cite{Iwasaki:1983iya}. 
In order to achieve the high-precision calculation, the AMA technique that can significantly reduce the statistical 
error is employed.

The axial-vector ($F_A(q^2)$), induced pseudoscalar ($F_P(q^2)$) and pseudoscalar ($G_P(q^2)$) form factors, are calculated for the isovector channel,
while the electric ($G_E(q^2)$) and magnetic ($G_M(q^2)$) form factors are calculated not only for the isovector ones, but also for the individual 
ones of proton and neutron without the disconnected diagram.
Before analyzing the nucleon form factors at finite momentum transfer, we have first examined the nucleon dispersion relation and nucleon axial-vector coupling to demonstrate the validity and reliability of our lattice QCD calculations.
The on-shell $O(a)$ improvement turns out to be effective enough for the momentum range we used, since 
the relativistic continuum dispersion relation
is found to be satisfied less than 1\% for the ground state of the nucleon.
Furthermore, the values of the axial-vector coupling calculated on the fine and coarse lattice reproduce the experimental value with a statistical accuracy of less than 2\%.

In the analyses of the form factors, we have investigated the major systematic uncertainty stemming from the effects of the excited-state contamination in the standard plateau method.
For this purpose, we have calculated appropriate ratios of the nucleon three-point function to the two-point functions by varying $t_{\mathrm{sep}}$ from $0.8$ to $1.2$ fm with $t_{\mathrm{sep}}/a=\{13,16,19\}$.
In the standard plateau method we employed, the form factors can be extracted from the asymptotic plateau of the ratios
between the source and sink points. 
It was found that the condition of $t_{\mathrm{sep}}\ge0.8\ \mathrm{fm}$ is large enough to eliminate the excited-state contamination for $G_E(q^2)$, $G_M(q^2)$ and $F_A(q^2)$ within the present statistical precision, thanks to the elaborated tuning of the sink and source functions.

For our best estimate, we perform the simultaneous fit with two data sets  of $t_{\mathrm{sep}}/a=\{16,19\}$, while we also use a single data set of $t_{\mathrm{sep}}/a=19$ for comparison and quote a difference between two results as the first systematic error. 
In addition, the effect of the lattice discretization error on the dispersion relation is quoted for the second systematic error as follows.
Each form factor is described as a function of $q^2$, which
can be primary evaluated by $q^2_{\mathrm{disp}}=2M_N\left( \sqrt{M_N^2 +(2\pi\bm{n}/(La))^2} - M_N \right)$ on the relativistic continuum dispersion relation with the naive lattice discrete momenta ${\bm{q}}=2\pi{\bm{n}}/(La)$. Alternatively the values of $q^2$ are also
evaluated by $q^2_{\mathrm{meas.}}=2M_N\Delta E$ 
with measured $\Delta E$.
The difference caused by the choice of either $q^2_{\mathrm{disp}}$ or $q^2_{\mathrm{meas}}$ is quoted as 
the systematic error induced in the dispersion relation by the effects of lattice discretization.

To evaluate the RMS radii and the magnetic moment, we have to introduce 
some parameterization of the $q^2$-dependence of the form factors.  
For this purpose, the model-independent $z$-expansion method is 
adopted in this study.
In general, the results of the $z$-expansion method are subject to larger errors than the results of the dipole fit, unless the model dependence of the dipole fit is taken into account.
However, since the results from the $z$-expansion method are model independent, the size of their errors is reasonably small.

Taking account of the systematic uncertainties associated with
the excited-state contamination, lattice discretization effects on the dispersion relation and the uncertainty in the determination of the renormalization factors, our best estimates for the axial-vector coupling and magnetic moments are obtained on the $160^4$ (fine) lattice as follows:
\begin{align}
    \label{eq:ga}
    g_A & = 1.264(14)(1)(-)(3),
\end{align}
\begin{align}
    \label{eq:gm_mm}
    \mu_v &
    = 4.436(89)(108)(18)(2)\quad\mathrm{(isovector)},\nonumber \\
    \mu_p &
    = 2.702(60)(21)(5)(1)\quad\mathrm{(proton)},\\
    \mu_n &
    = -1.695(41)(27)(2)(1)\quad\mathrm{(neutron)},\nonumber 
\end{align}
where 
the first error is statistical one, while the others are systematic ones. The second error is evaluated by the difference between two analyses using either a single data set of $t_{\rm sep}/a=19$ or a combined data set of $t_{\rm sep}/a=\{16,19\}$. The third error is associated with a 
choice of $q^2$ definitions. The fourth error is
associated with the uncertainty in the determination of 
the renormalization factors. For $g_A$, there is no third error, since the value of $g_A$ is directly measurable at $q^2=0$ without $q^2$-extrapolation.
As for the RMS radii, we obtain
\begin{align}
    \label{eq:ge_rms}
    \sqrt{\langle (r_{E}^{v})^2 \rangle} & 
    = 0.832(19)(70)(22)\ [\mathrm{fm}]\quad\mathrm{(isovector)},\nonumber \\
    \sqrt{\langle (r_{E}^{p})^2 \rangle} &
    = 0.804(14)(49)(18)\ [\mathrm{fm}]\quad\mathrm{(proton)},\\
    \langle (r_{E}^{n})^2 \rangle & 
    = -0.054(23)(46)(4)\ [\mathrm{fm}^2]\quad\mathrm{(neutron)}, \nonumber
\end{align}
\begin{align}
    \label{eq:gm_rms}
    \sqrt{\langle (r_{M}^{v})^2 \rangle} &
    = 0.771(64)(84)(10)\ [\mathrm{fm}]\quad\mathrm{(isovector)},\nonumber \\
    \sqrt{\langle (r_{M}^{p})^2 \rangle} &
    = 0.775(74)(70)(43)\ [\mathrm{fm}]\quad\mathrm{(proton)}, \\
    \sqrt{\langle (r_{M}^{n})^2 \rangle} &
    = 0.692(93)(13)(40)\ [\mathrm{fm}]\quad\mathrm{(neutron)}, \nonumber
\end{align}
\begin{align}
    \label{eq:fa_rms}
    \sqrt{\langle r_{A}^{2} \rangle} & 
    = 0.562(31)(36)(47)\ [\mathrm{fm}],
\end{align}
where the first error is statistical, while the second 
and third ones are systematic as explained earlier.
There is no fourth error for the RMS radii.

Since the continuum-limit extrapolation requires results from at least three lattice spacings, we have investigated the systematic uncertainties associated with the finite lattice spacing for 
$g_A$ and isovector RMS radii
from the difference between the current results obtained at two lattice spacings.
It was found that the the finite lattice spacing effect on $g_A$ is kept below the statistical error of less than 2\%, which is currently achieved in our calculations, while both results
of $g_A$ obtained at two lattice spacings
reproduce the experimental value within their statistical precisions.  
Therefore, the lattice discretization 
effect on $g_A$ is negligibly small in our calculations.

On the other hand, 
the systematic errors associated
with the finite lattice spacing
on the isovector RMS radii are
rather large as much as 8-11\%
and cannot be ignored regardless of channel. In particular, in the cases of the electric and axial RMS radii, 
the systematic uncertainties associated with the finite lattice spacing tend to reproduce the experimental values.

One might think that the unexpectedly large systematic errors in the RMS radii are due to not using the 
${\cal O}(a)$-improvement of the vector and axial-vector currents in this study. For the improvement of the 
axial-vector current, $\widetilde{A}^{\mathrm{imp}}_\alpha = \widetilde{A}_\alpha+ac_A\partial_\alpha \widetilde{P}$, 
we have examined the size of $c_A$
by comparing $m^{\mathrm{nucl}}_{\mathrm{PCAC}}$ and $m_{\mathrm{PCAC}}^{\mathrm{pion}}$ based on the PCAC relation,
and found that the value of $c_A$ is likely to be nearly zero at the $O(10^{-2})$ level in lattice units.
This indicates that the effect of $O(a)$-improvement of the axial-vector current 
is not large enough to resolve
the large systematic uncertainties 
observed at very low $q^2$ in our calculations.

Needless to say that additional lattice simulations using the third PACS10 ensemble are required for achieving a comprehensive study of the discretization uncertainties and then taking the continuum limit of our target quantities.
Such planning is now underway.

\clearpage
\begin{acknowledgments}
{
    We would like to thank members of the PACS collaboration for useful discussions.
R.~T. is supported by the RIKEN Junior Research Associate Program.
R.~T. acknowledge the support from Graduate Program on Physics for the Universe (GP-PU) of Tohoku University.
K.-I.~I. is supported in part by MEXT as ``Feasibility studies for the next-generation computing infrastructure".
K.~S. is supported by JST, The Establishment of University Fellowships towards the creation of Science Technology Innovation, Grant Number JPMJFS2106.
We also thank Y. Namekawa for his careful reading of the manuscript.
Numerical calculations in this work were performed on Oakforest-PACS in Joint Center for Advanced High Performance Computing (JCAHPC) and Cygnus  and Pegasus in Center for Computational Sciences at University of Tsukuba under Multidisciplinary Cooperative Research Program of Center for Computational Sciences, University of Tsukuba, and Wisteria/BDEC-01 in the Information Technology Center, the University of Tokyo. 
This research also used computational resources of the K computer (Project ID: hp1810126) and the Supercomputer Fugaku (Project ID: hp20018, hp210088, hp230007) provided by RIKEN Center for Computational Science (R-CCS), as well as Oakforest-PACS (Project ID: hp170022, hp180051, hp180072, hp190025, hp190081, hp200062),  Wisteria/BDEC-01 Odyssey (Project ID: hp220050) provided by the Information Technology Center of the University of Tokyo / JCAHPC.
The  calculation employed OpenQCD system(http://luscher.web.cern.ch/luscher/openQCD/). 
This work is supported by the JLDG constructed over the SINET5 of NII,
This work was also supported in part by Grants-in-Aid for Scientific Research from the Ministry of Education, Culture, Sports, Science and Technology (Nos. 18K03605, 19H01892, 22K03612, 23H01195, 23K03428) and MEXT as ``Program for Promoting Researches on the Supercomputer Fugaku'' (Search for physics beyond the standard model using large-scale lattice QCD simulation and development of AI technology toward next-generation lattice QCD; Grant Number JPMXP1020230409).

}
\end{acknowledgments}

 
\appendix

\clearpage
\section{Two-state fit analysis}
\label{app:two_state_fit}

For an assessment of excited-state contamination, 
the two-state fit analysis is often used in calculation
of the nucleon matrix elements.

The ratio defined in Eq.~(\ref{eq:ratio_3pt_2pt}) can
be described by the following functional form, which includes the forward and backward contributions of the leading excited state 
from source and sink explicitly as below,
\label{eq:twostates_fit}
\begin{align}
    {\cal R}\left(t, t_\mathrm{sep}\right)
    =
    b_0 + b_1 \mathrm{e}^{-b_2\left( t_\mathrm{sep} - t\right)}
    + b_3 \mathrm{e}^{-b_4t},
\end{align}
where $b_0$ is 
the matrix element of the ground state. The parameter $b_1$ ($b_3$) is the amplitude of the overlap between the ground state and the leading excited state, while the parameter $b_2$ ($b_4$) is the energy gap between the excited state and the ground state. The leading exponential terms with four additional parameters $b_1$ through $b_4$ are responsible for the curvature appearing in the ratio as a function of $t$.

For the simplest case if $b_1=b_3$ and $b_2=b_4$~\footnote{This is the case if the kinematics 
is chosen to be $\bm{p}=\bm{p}^\prime=\bm{0}$.}, 
the curvature is represented by $A\cosh\left(
b_2(t-\frac{t_{\mathrm{sep}}}{2})
\right)$ with $A={2b_1}e^{-b_2\frac{t_{\mathrm{sep}}}{2}}$. The excited-state contribution remains at most as the size of $A$ at the center of the source and sink operators $t=\frac{t_{\mathrm{sep}}}{2}$. The value of $A$ corresponds to a typical size of the systematic uncertainty
associated with the excited-state contamination in the standard plateau method. 

We evaluate the amplitude $A$ of $\widetilde{F}_{P}(q^2)$ and $\widetilde{G}_P(q^2)$ at the lowest $q^2$ (Q1) 
for data set of $t_{\mathrm{sep}}/a=13$.
As shown in Fig.~\ref{fig:fp_qdep_p-n_ts1X} for 
$\widetilde{F}_P(q^2)$ and Fig.~\ref{fig:gp_qdep_p-n_ts1X} for $\widetilde{G}_P(q^2)$,
the $t$-dependence of each form factor has a 
slight convex shape which is associated with the excited-state contamination. 
The observed shape is approximately 
symmetric with respect to $t=\frac{t_{\rm sep}}{2}$
within the statistical uncertainties at the lowest $q^2$.
Therefore, for simplicity, we may use the two-state analysis with the symmetric ans\"atz ($b_1=b_3$ and $b_2=b_4$). We then found that the size of $A$ is much smaller than that of $b_0$
, and the resultant $b_0$ is statistically consistent with the result obtained with the standard plateau method.
Indeed, the systematic uncertainty in the standard plateau analysis is at most 1\% even in the most severe case of the excited-state contamination.
This indicates that the two-state fitting analysis is not useful for resolving relatively large excited-state contamination in our data.

\clearpage
\section{Table of nucleon form factors}
\label{app:table_of_ff}

The results for the three isovector form factors
$G_E^v(q^2)$, $G_M^v(q^2)$, and $F_A(q^2)$
obtained with a combined data of $t_{\rm sep}/a=\{16,19\}$ and a single data of $t_{\rm sep}/a=19$
are summarized in Table~\ref{tab:ff_q2_isovector}. 
The electric and magnetic form factors for the
proton and neutron, $G_E^p(q^2)$, $G_M^p(q^2)$,  $G_E^n(q^2)$, and $G_M^n(q^2)$ are complied in
Table~\ref{tab:ff_q2_pn}.

\begin{table*}[hb!]
\caption{
    Results of the three isovector form factors obtained by the standard plateau method using the uncorrelated constant fit with a combined data of $t_{\rm sep}/a=\{16,19\}$ and a single data of $t_{\rm sep}/a=19$. All form factors are renormalized.
\label{tab:ff_q2_isovector}
}
\begin{ruledtabular}
\begin{tabular}{c|cccccc} 
\multirow{2}{*}{$q^2\,[{\rm GeV}^2]$}  & \multicolumn{3}{c}{$t_{\rm sep}/a=\{16,19\}$} & \multicolumn{3}{c}{$t_{\rm sep}/a=19$} \\
  \cline{2-4}\cline{5-7}
  & $G_E^v(q^2)$ & $G_M^v(q^2)$ & $F_A(q^2)$ & $G_E^v(q^2)$ & $G_M^v(q^2)$ & $F_A(q^2)$\\  
  \hline
0.000 & 0.997(1) & ---       & 1.250(15)&  0.998(3) & ---       & 1.264(22)\\
0.015 & 0.954(2) & 4.245(73) & 1.229(14)&  0.949(4) & 4.318(115)& 1.241(20)\\
0.030 & 0.915(3) & 4.150(64) & 1.205(13) &  0.908(6) & 4.189(94) & 1.216(19)\\
0.044 & 0.879(4) & 3.999(60) & 1.118(13)&  0.869(7) & 4.026(87) & 1.189(19)\\
0.059 & 0.843(4) & 3.821(60) & 1.158(13)&  0.834(7) & 3.839(91) & 1.166(19)\\
0.073 & 0.812(5) & 3.709(51) & 1.140(12)&  0.805(7) & 3.759(76) & 1.150(18)\\
0.087 & 0.783(5) & 3.620(49) & 1.118(12)&  0.775(8) & 3.613(72) & 1.129(18)\\
0.116 & 0.730(6) & 3.412(49) & 1.083(12)&  0.724(8) & 3.405(72) & 1.099(19)\\
\end{tabular}
\end{ruledtabular} 
\end{table*}

\begin{table*}[hb!]
\caption{
    Results of the electric and magnetic form factors 
    for the proton and neutron obtained by the standard plateau method using the uncorrelated constant fit with a combined data of $t_{\rm sep}/a=\{16,19\}$ and a single data of $t_{\rm sep}/a=19$. All form factors are renormalized. Our results are determined without the disconnected-type contributions.
\label{tab:ff_q2_pn}
}
\begin{ruledtabular}
  \begin{tabular}{c|cccc cccc}
\multirow{3}{*}{$q^2\,[{\rm GeV}^2]$}  & \multicolumn{4}{c}{$t_{\rm sep}/a=\{16,19\}$} & \multicolumn{4}{c}{$t_{\rm sep}/a=19$} \\
\cline{2-5}\cline{6-9}
  & \multicolumn{2}{c}{Proton} & \multicolumn{2}{c}{Neutron} & \multicolumn{2}{c}{Proton} & \multicolumn{2}{c}{Neutron} \\
  \cline{2-5}\cline{6-9}
  & $G_E^p(q^2)$ & $G_M^p(q^2)$ & $G_E^n(q^2)$ & $G_M^n(q^2)$ & $G_E^p(q^2)$ & $G_M^p(q^2)$ & $G_E^n(q^2)$ & $G_M^n(q^2)$\\  
  \hline
0.000 & 0.997(1) & ---      & 0.0005(9) & ---         & 0.998(2)& ---      & 0.0006(17)& ---\\
0.015 & 0.957(1) & 2.608(45)& 0.0033(11)& $-$1.641(34)& 0.954(3)& 2.608(70)& 0.005(2)  & $-$1.678(54)\\
0.030 & 0.921(2) & 2.536(39)& 0.0061(14)& $-$1.613(30)& 0.916(4)& 2.544(56)& 0.008(2)  & $-$1.649(45)\\
0.044 & 0.887(3) & 2.442(37)& 0.008(2)  & $-$1.557(28)& 0.880(5)& 2.440(51)& 0.011(3)  & $-$1.586(42)\\
0.059 & 0.854(3) & 2.342(35)& 0.011(2)  & $-$1.479(29)& 0.848(5)& 2.336(54)& 0.014(3)  & $-$1.504(43)\\
0.073 & 0.825(3) & 2.284(31)& 0.013(2)  & $-$1.456(25)& 0.819(5)& 2.282(45)& 0.015(3)  & $-$1.477(36)\\
0.087 & 0.797(3) & 2.208(30)& 0.014(2)  & $-$1.411(23)& 0.791(6)& 2.190(43)& 0.015(3)  & $-$1.423(34)\\
0.116 & 0.747(4) & 2.085(29)& 0.017(2)  & $-$1.326(22)& 0.742(6)& 2.069(43)& 0.018(3)  & $-$1.336(33)\\
\end{tabular}
\end{ruledtabular} 
\end{table*}

\clearpage
\section{Model-dependent analyses of RMS radius}
\label{app:model-dep_anal}

In this appendix, a summary of the results obtained in several model-dependent $q^2$ analyses is presented
as supplemental material, though the results obtained by the $z$-expansion method are employed in the main text as model-independent analysis. 

For model-dependent analyses,
we employ three typical models to parameterize the $q^2$-dependence of the form factor $G_l$ in this study:
the linear functional form $G_l(q^2)=d_0+d_1q^2$,
the quadratic functional form $G_l(q^2)=d_0+d_1q^2+d_2q^4$ and
the dipole form $G_l(q^2)=G_l(0)/\left(1+q^2/\Lambda^2_l \right)^2$.
The RMS radius $R_l$ can be determined by
$R_l = \sqrt{-6d_1/d_0}$ (linear fit),
$R_l = \sqrt{-6d_1/d_0}$ (quadratic fit) and
$R_l = \sqrt{12}/\Lambda_l$ (dipole fit)~\footnote{
For the neutron's electric form factor, 
the quadratic fit is only applied to evaluate its mean square radius from $R_l^2 = -6d_1$.}.

All results obtained in these model-dependent analyses are summarized in Table~\ref{tab:re_model}, \ref{tab:murm_model} and \ref{tab:ra_model}.
We employ the uncorrelated fits, where the correlations among data points at different $q^2$ are not considered.


\begin{table*}[ht!]
    {\scriptsize
\begin{ruledtabular}
\caption{
  Results for the electric RMS charge radius $\sqrt{\langle r^2_{E}\rangle}$ in the isovector, proton and neutron channels.
  Results for the proton and neutron are obtained without the disconnected diagram. 
\label{tab:re_model}}
\begin{tabular}{ccccccccccc}
  & & & \multicolumn{2}{c}{Isovector} & \multicolumn{2}{c}{Proton} & \multicolumn{2}{c}{Neutron}\\
  \hline
  Fit type & $q^2$ [GeV$^2$] & $t_{\rm sep}/a$ & $\sqrt{\langle r^2_{E}\rangle}$ [fm] & $\chi^2$/d.o.f. & $\sqrt{\langle r^2_{E}\rangle}$ [fm] & $\chi^2$/d.o.f. & $\langle r^2_{E}\rangle$ [fm$^2$] & $\chi^2$/d.o.f.\cr
  \hline
  \multirow{4}{*}{Linear} & \multirow{2}{*}{$q^2_{\mathrm{disp}}\le0.015$} & $\{16,19\}$&  0.822(14)  &  1.3&  0.793(9)  &  1.1&  $-$&  $-$\\
                                                                           & & $19$&  0.877(21)  &  $-$&  0.832(16)  &  $-$ &  $-$&  $-$ \\
                          & \multirow{2}{*}{$q^2_{\mathrm{meas}}\le0.015$} & $\{16,19\}$&  0.801(14)  &  1.8&  0.770(10)  &  1.0&  $-$&  $-$\\
                                                                            & & $19$&  0.854(21)  &  $-$&  0.810(16)  &  $-$ &  $-$&  $-$ \\
  \hline
  \multirow{4}{*}{Dipole} & \multirow{2}{*}{$q^2_{\mathrm{disp}}\le0.116$} & $\{16,19\}$&  0.827(12)  &  1.1&  0.795(8)  &  1.1&  $-$&  $-$\\
                                                                           & & $19$&  0.847(17)  &  0.4&  0.812(13)  &  0.5&  $-$&  $-$\\
                          & \multirow{2}{*}{$q^2_{\mathrm{meas}}\le0.091$} & $\{16,19\}$&  0.804(14)  &  1.5&  0.774(7)  &  1.4&  $-$&  $-$\\
                                                                           & & $19$&  0.834(18)  &  0.2&  0.799(13)  &  0.2&  $-$&  $-$\\
  \hline
  \multirow{4}{*}{Quadrature} & \multirow{2}{*}{$q^2_{\mathrm{disp}}\le0.116$} & $\{16,19\}$&  0.826(14)  &  1.3&  0.797(10)  &  1.2&  $-$0.050(10)& 0.7\\
                                                                               & & $19$&  0.867(22)  &  0.1&  0.828(16)  &  0.1&  $-$0.067(14)& 0.08\\
                              & \multirow{2}{*}{$q^2_{\mathrm{meas}}\le0.091$} & $\{16,19\}$&  0.856(23)  &  1.5&  0.780  &  1.3(9)&  $-$0.047(10)& 0.9\\
                                                                               & & $19$&  0.853(22)  &  0.08&  0.811(17)  &  0.09&  $-$0.071(15)& 0.04\\
\end{tabular}
\end{ruledtabular}
}
\end{table*}

\begin{table*}[ht!]
    {\scriptsize
\begin{ruledtabular}
\caption{
Results for the magnetic moments $\mu$ and magnetic RMS radius
$\sqrt{\langle r^2_{M}\rangle}$ for the isovector, proton and neutron channels. 
Results for the proton and neutron are obtained without the disconnected diagram. 
\label{tab:murm_model}}
\begin{tabular}{ccccccccccccc}
  & & & \multicolumn{3}{c}{Isovector}\\
  \hline
  Fit type & $q^2_{\rm cut}$ [GeV$^2$] & $t_{\rm sep}/a$ & $\mu_v$ & $\sqrt{\langle r^2_{M}\rangle}$ [fm]& $\chi^2$/d.o.f. \\
  \hline
  \multirow{4}{*}{Linear}     & \multirow{2}{*}{$q^2_{\mathrm{disp}}\le0.030$} & $\{16,19\}$ & { 4.348(90) } & { 0.598(83) } & { 0.4}\\
                              &                                               & $19$    & { 4.448(149) } & { 0.677(115) } & { $-$}\\
                              & \multirow{2}{*}{$q^2_{\mathrm{meas}}\le0.030$} & $\{16,19\}$ & { 4.439(91) } & { 0.585(81) } & { 0.5}\\
                              &                                               & $19$    & { 4.450(149) } & { 0.662(113) } & { $-$}\\
  \hline
  \multirow{4}{*}{Dipole}     & \multirow{2}{*}{$q^2_{\mathrm{disp}}\le0.116$} & $\{16,19\}$ & { 4.422(77) } & { 0.748(22) } & { 0.2}\\
                              &                                               & $19$    & { 4.495(121) } & { 0.779(39) } & { 0.06}\\
                              & \multirow{2}{*}{$q^2_{\mathrm{meas}}\le0.091$} & $\{16,19\}$ & { 4.432(80) } & { 0.739(25) } & { 0.2}\\
                              &                                               & $19$    & { 4.513(127) } & { 0.774(43) } & { 0.08}\\
  \hline
  \multirow{4}{*}{Quadrature} & \multirow{2}{*}{$q^2_{\mathrm{disp}}\le0.116$} & $\{16,19\}$ & { 4.427(84) } & { 0.750(40) } & { 0.2}\\
                              &                                               & $19$    & { 4.511(133) } & { 0.792(66) } & { 0.07}\\
                              & \multirow{2}{*}{$q^2_{\mathrm{meas}}\le0.091$} & $\{16,19\}$ & { 4.426(85) } & { 0.728(58) } & { 0.2}\\
                              &                                               & $19$    & { 4.503(140) } & { 0.757(95) } & { 0.1}\\
 \hline\hline
  & & & \multicolumn{3}{c}{Proton} & \multicolumn{3}{c}{Neutron}\\
  \hline
  Fit type & $q^2_{\rm cut}$ [GeV$^2$] & $t_{\rm sep}/a$ & $\mu_p$ & $\sqrt{\langle r^2_{M}\rangle}$ [fm]& $\chi^2$/d.o.f. & $\mu_n$ & $\sqrt{\langle r^2_{M}\rangle}$ [fm]& $\chi^2$/d.o.f. \\
  \hline
  \multirow{4}{*}{Linear}     & \multirow{2}{*}{$q^2_{\mathrm{disp}}\le0.030$} & $\{16,19\}$ & { 2.660(60) } & { 0.636(82) }  & {0.05}& { $-$1.659(42) }  & { 0.504(108)} & { 0.3}\\
                              &                                               & $19$    & { 2.684(97) } & { 0.710(116) } & { $-$}   & { $-$1.685(69) }  & { 0.523(177)} & { $-$}\\
                              & \multirow{2}{*}{$q^2_{\mathrm{meas}}\le0.030$} & $\{16,19\}$ & { 2.661(59) } & { 0.622(80) }  & {0.05}& { $-$1.660(42) }  & { 0.493(106)} & { 0.3}\\
                              &                                               & $19$    & { 2.685(97) } & { 0.694(114) } & { $-$}   & { $-$1.685(70) }  & { 0.511(173)} & { $-$}\\
  \hline
  \multirow{4}{*}{Dipole}     & \multirow{2}{*}{$q^2_{\mathrm{disp}}\le0.116$} & $\{16,19\}$ & { 2.690(49) } & { 0.636(82) }  & { 0.05}& { $-$1.702(37) }  & { 0.732(26) } & { 0.2}\\
                              &                                               & $19$    & { 2.693(74) } & { 0.768(40) }  & {0.07}& { $-$1.731(59) }  & { 0.748(47) } & { 0.09}\\
                              & \multirow{2}{*}{$q^2_{\mathrm{meas}}\le0.091$} & $\{16,19\}$ & { 2.699(51) } & { 0.740(27) }  & { 0.2} & { $-$1.703(38) }  & { 0.716(38) } & { 0.3}\\
                              &                                               & $19$    & { 2.705(78) } & { 0.766(46) }  & {0.07}& { $-$1.733(61) }  & { 0.733(55) } & { 0.1}\\
  \hline
  \multirow{4}{*}{Quadrature} & \multirow{2}{*}{$q^2_{\mathrm{disp}}\le0.116$} & $\{16,19\}$ & { 2.697(54) } & { 0.756(42) }  & { 0.2} & { $-$1.698(39) }  & { 0.713(51) } & { 0.3}\\
                              &                                               & $19$    & { 2.709(85) } & { 0.795(70) }  & { 0.07}& { $-$1.725(64) }  & { 0.724(88)} & { 0.1}\\
                              & \multirow{2}{*}{$q^2_{\mathrm{meas}}\le0.091$} & $\{16,19\}$ & { 2.697(56) } & { 0.732(59) }  & { 0.2} & { $-$1.700(39) }  & { 0.699(66) } & { 0.3}\\
                              &                                               & $19$    & { 2.710(90) } & { 0.773(97) }  & {0.1}& { $-$1.719(66) }  & { 0.676(120)} & { 0.1}\\
\end{tabular}
\end{ruledtabular}
}
\end{table*}

\begin{table*}[ht!]
    {\scriptsize
\begin{ruledtabular}
\caption{Results for the axial-vector coupling $g_A=F_A(0)$ and axial-vector RMS radius $\sqrt{\langle r^2_A\rangle}$.
\label{tab:ra_model}}
\begin{tabular}{cccccc}
  Fit type & $q^2$ GeV$^2$ & $t_{\rm sep}/a$ & $F_A(0)$ & $\sqrt{\langle r^2_A\rangle}$ [fm]& $\chi^2$/d.o.f. \\
  \hline
  \multirow{4}{*}{Linear} & \multirow{2}{*}{$q^2_{\mathrm{disp}}\le0.015$} & $\{16,19\}$& 1.251(15)& 0.513(39)  &  0.6\\
                                                                           & & $19$& 1.264(22)& 0.526(80)  &  $-$\\
                          & \multirow{2}{*}{$q^2_{\mathrm{meas}}\le0.015$} & $\{16,19\}$&1.250(15)& 0.499(38)  &  0.6\\
                                                                            & & $19$&1.264(22)& 0.512(78)  &  $-$\\
  \hline
  \multirow{4}{*}{Dipole} & \multirow{2}{*}{$q^2_{\mathrm{disp}}\le0.116$} & $\{16,19\}$& 1.251(15)& 0.552(17)  &  0.3\\
                                                                           & & $19$& 1.262(21)& 0.547(28)  &  0.05\\
                          & \multirow{2}{*}{$q^2_{\mathrm{meas}}\le0.091$} & $\{16,19\}$& 1.252(15)& 0.544(17)  &  0.3\\
                                                                           & & $19$& 1.265(21)& 0.549(29)  &  0.02\\
  \hline
  \multirow{4}{*}{Quadrature} & \multirow{2}{*}{$q^2_{\mathrm{disp}}\le0.116$} & $\{16,19\}$& 1.252(15)& 0.559(35)  &  0.4 \\
                                                                               & & $19$& 1.266(22)& 0.590(44)  &  0.02 \\
                              & \multirow{2}{*}{$q^2_{\mathrm{meas}}\le0.091$} & $\{16,19\}$& 1.251(15)& 0.536(29)  &  0.3 \\
                                                                               & & $19$& 1.266(22)& 0.564(54)  &  0.01 \\
\end{tabular}
\end{ruledtabular}
}
\end{table*}

\clearpage
\section{Comparison with the previous lattice QCD calculations}
We discuss a comparison with the results of other recent lattice QCD calculations, which are summarized in Table~\ref{tab:conventional}.

The recent calculation reveals the major sources of uncertainties
(i) statistical noise, (ii) excited-state contamination,
(iii) model dependence of the $q^2$-parameterization
and (iv) extrapolation into the physical point, infinite volume and continuum limit.
Indeed,
Mainz group~\cite{Djukanovic:2021qxp, Djukanovic:2022wru, Djukanovic:2023beb, Djukanovic:2023jag} and the NME Collaboration~\cite{Park:2021ypf}
achieved reducing these uncertainties and reproducing the experimental values of electric radius $\sqrt{\langle (r^v_E)^2 \rangle}$ and magnetic radius $\sqrt{\langle (r^v_M)^2 \rangle}$,
though they are not enough precise to discriminate the proton radius puzzle and the tension about the magnetic form factor.

On the other hand, as for the axial radius $\sqrt{\langle (r^v_A)^2 \rangle}$, the current lattice QCD computation can reproduce the experimental values given by the $z$-expansion method, and has achieved an error accuracy comparable to experiment.
Thus, towards the neutrino oscillation experiment and the physics beyond the standard model, a comprehensive analysis of the scattering data would be possible by combining both the experimental results and the lattice results of the nucleon form factors~\cite{Tomalak:2023pdi}. 

\begin{figure*}
\centering
\includegraphics[width=0.32\textwidth,bb=0 0 532 612,clip]{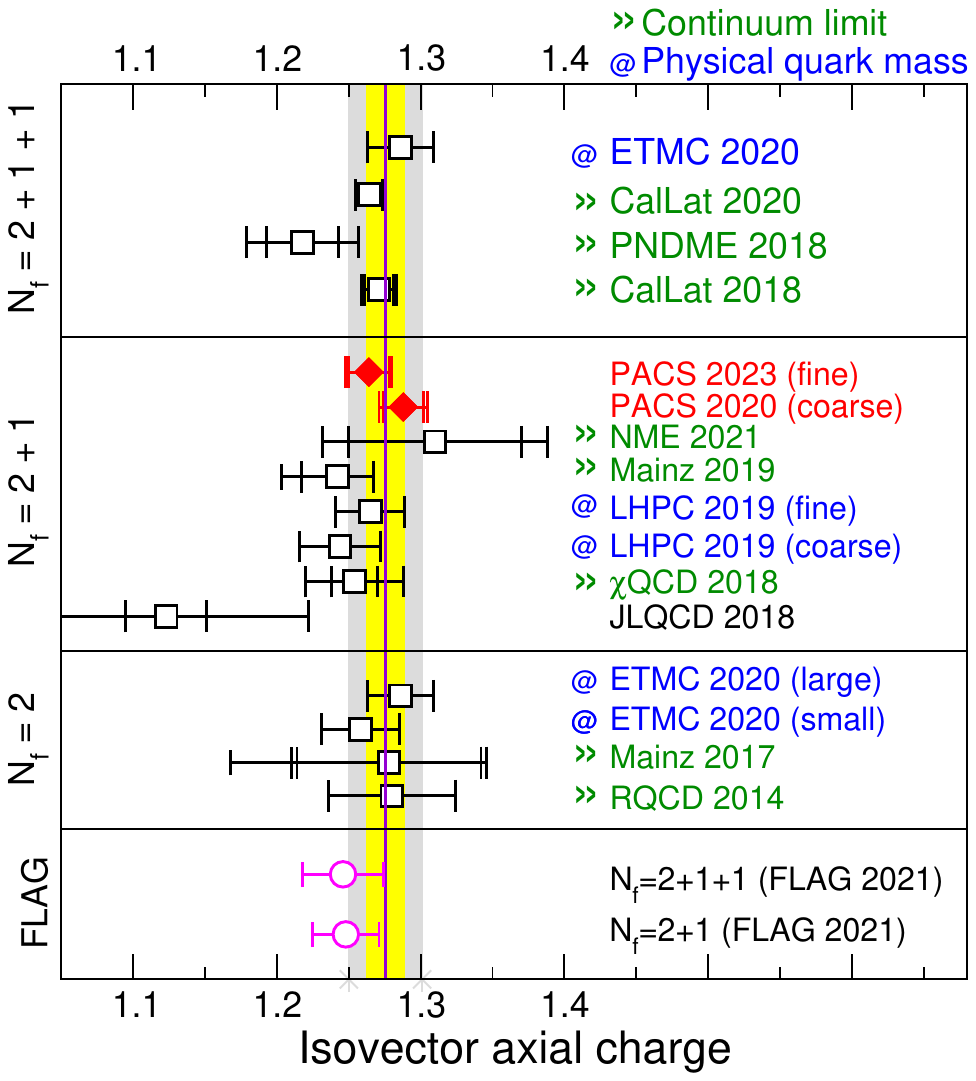}
\includegraphics[width=0.32\textwidth,bb=0 0 532 612,clip]{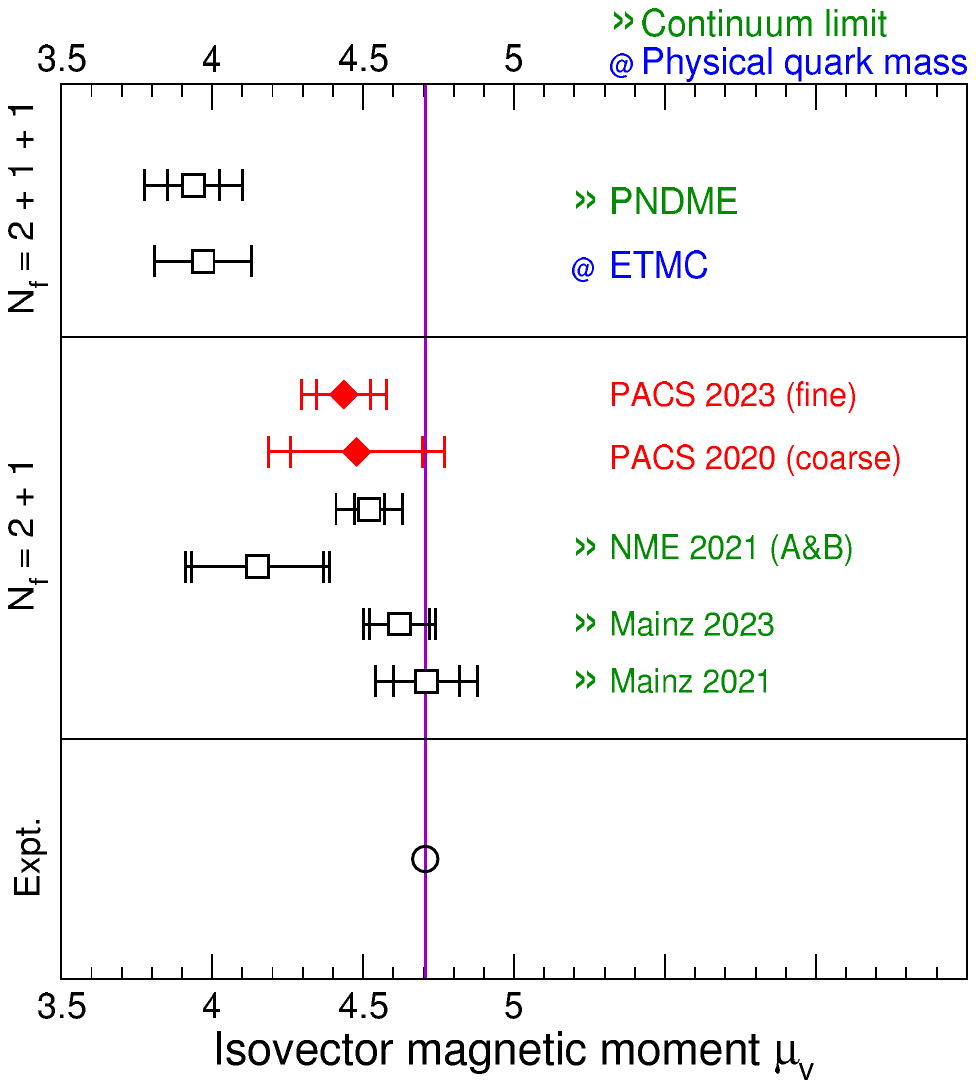}\\
\includegraphics[width=0.32\textwidth,bb=0 0 532 612,clip]{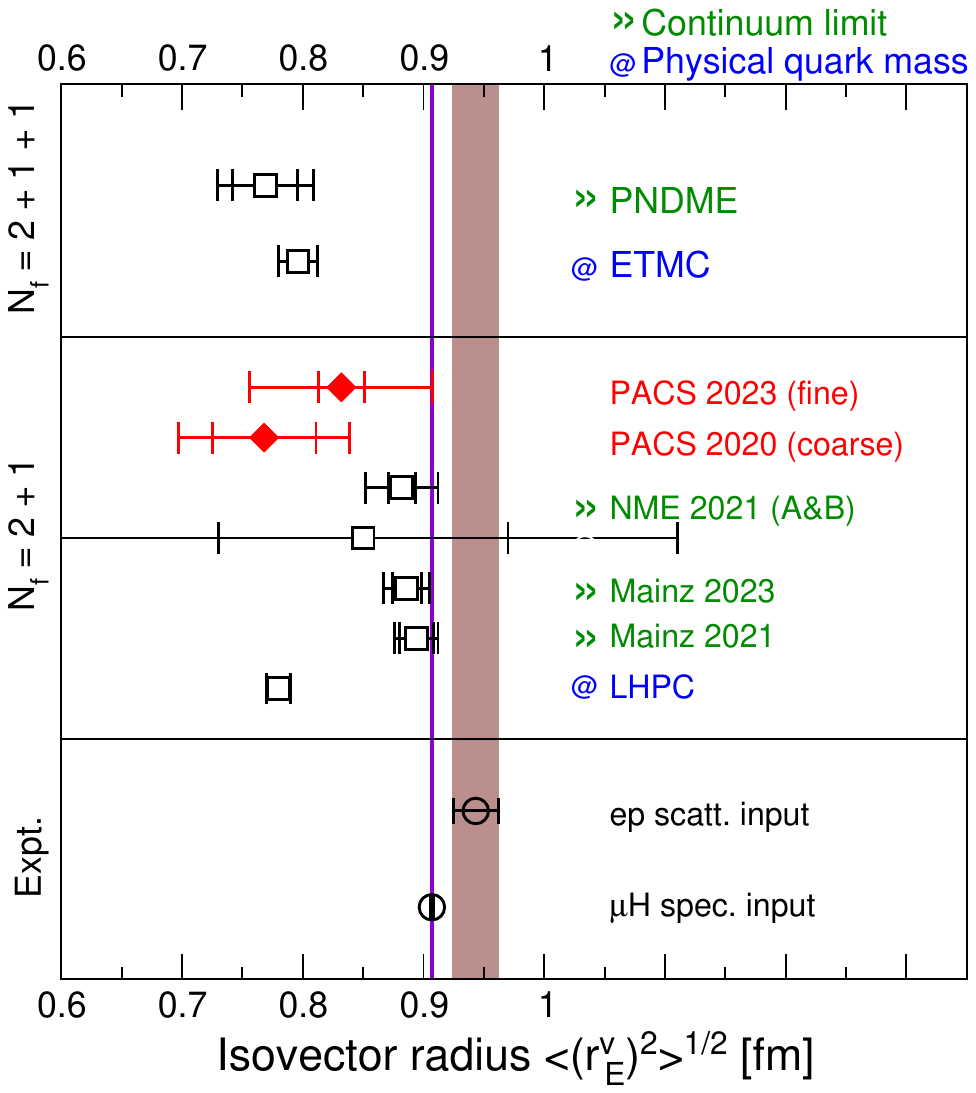}
\includegraphics[width=0.32\textwidth,bb=0 0 532 612,clip]{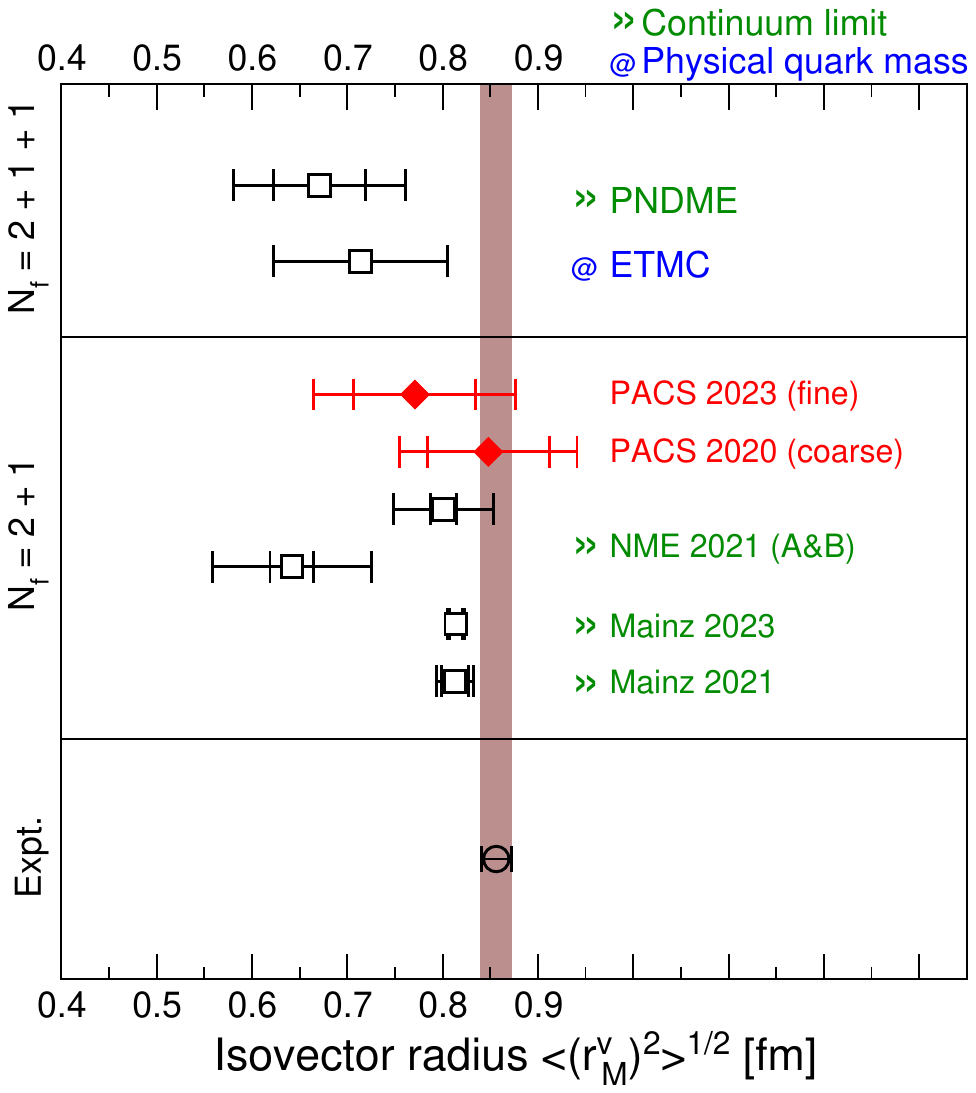}
\includegraphics[width=0.32\textwidth,bb=0 0 532 612,clip]{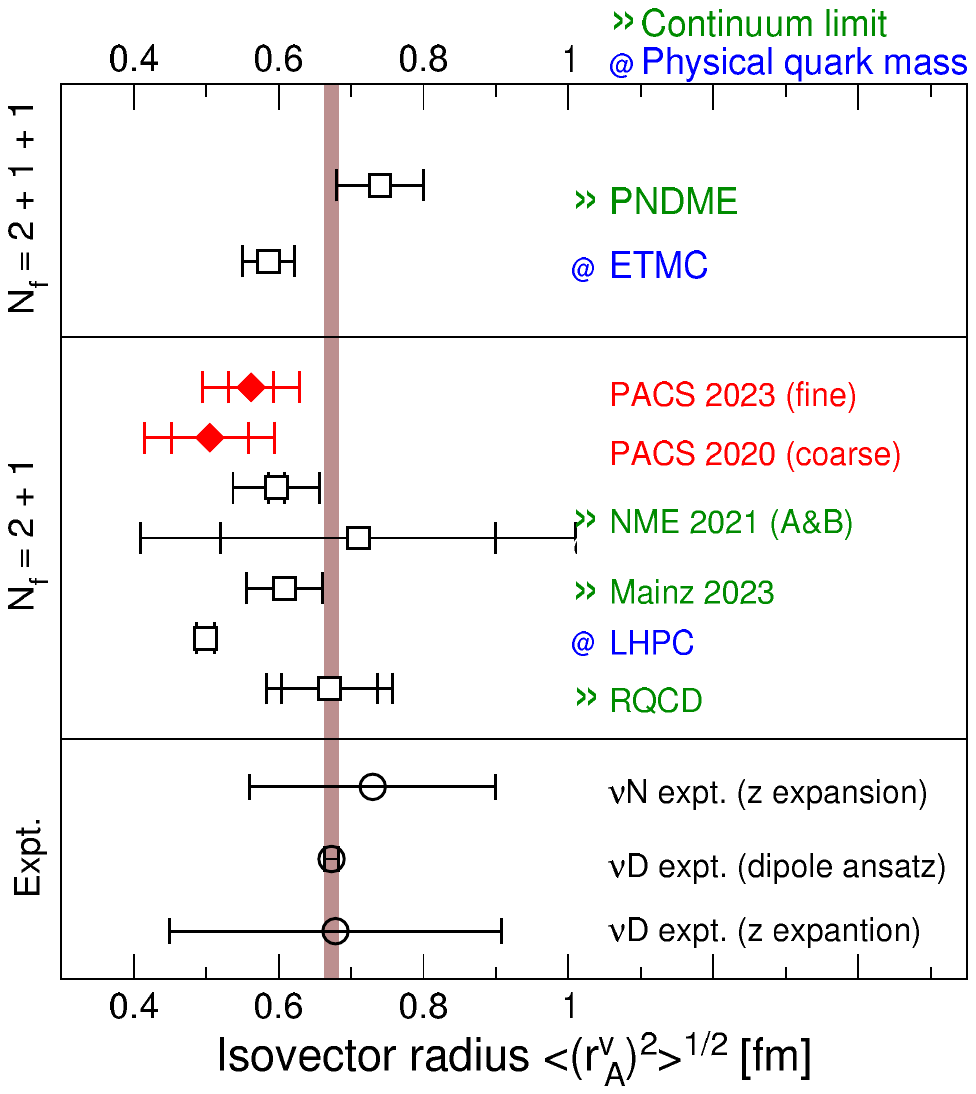}
\caption{
Summary plot for the lattice QCD results and the experimental values of the axial-vector coupling (top, left), isovector magnetic moment (top, right) and three kinds of the isovector RMS radius: electric (bottom, left), magnetic (bottom, center) and axial (bottom, right). The (inner) error bars represent the statistical error, while the outer error bars are the total error evaluated by both the statistical and systematic errors added in quadrature.
Blue labels indicate that the analysis uses the data from lattice QCD simulation near the physical point, while green labels indicate that the continuum extrapolation is achieved. In the top-left panel, yellow and gray bands display 1\% and 2\% deviations from the experimental value.
In addition, violet lines and brown bands appearing in each panel represent the experimental values. 
}
\label{fig:conventional}
\end{figure*}

Although we would like to compare our results with these previous results, 
it should be noted that meaningful and quantitative comparisons are not yet feasible, so no firm conclusions can be drawn at present.
This is simply because the continuum limit was not yet taken in this study.
However, one point we would comment on is the following.
The accuracy of the lattice QCD results for $g_A$ has improved significantly and all results obtained from each lattice study converge within a few percent of the experimental value, while
the lattice QCD results for the RMS radii are not sufficiently consistent with either each other or experiment as shown in Fig.~\ref{fig:conventional}.
This situation may be attributed to the large discretization uncertainties in the RMS radii we observed in this study.

%
%
\begin{table*}[ht!]
    {\scriptsize
\begin{ruledtabular}
\caption{
Summary of lattice QCD recent results of the electric, magnetic and axial RMS radii, magnetic moment and axial-vector coupling obtained from the respective nucleon form factors. The first and second errors represent the statistical and total systematic uncertainties. The latter error is evaluated from all measured systematic errors added in quadrature. The symbol ``$\rightarrow 0$" is used only when the continuum limit is taken.
All data summarized in the table are limited by certain criteria discussed in the text. For the case of $g_A$, see Ref.\cite{FlavourLatticeAveragingGroupFLAG:2021npn} and all the relevant references therein.
\label{tab:conventional}}
\begin{tabular}{lrrcccccccccc}
  \hline
  Publication &
  $a$ [fm]& 
  $m_\pi$ [MeV]&
  $m_\pi L$&
  $\sqrt{\langle (r^v_E)^2 \rangle}$ [fm]&
  $\sqrt{\langle (r^v_M)^2 \rangle}$ [fm]&
  $\mu_v$ &
  $\sqrt{\langle (r^v_A)^2 \rangle}$ [fm]&
  $g_A$ \\
  \hline
  Mainz\cite{Djukanovic:2021qxp, Djukanovic:2022wru, Djukanovic:2023beb, Harris:2019bih}  &$\rightarrow 0$& $\ge130$& $\ge3.05$& $0.886(12)(19)$& $0.814(7)(9)$& $4.62(10)(12)$& $0.608(52)(53)$& $1.242(25)(\genfrac{}{}{0pt}{2}{+0}{-031})$ \\ 
  CalLat\cite{Chang:2018uxx} & 0.12&  $\ge130$&    $3.90$& --& --& --& --& $1.26421(93)$ \\ 
  NME\cite{Park:2021ypf}    &$\rightarrow 0$& $\ge170$& $\ge3.75$& $0.882(11)(28)$& $0.801(14)(50)$& $4.52(5)(10)$& $0.597(11)(59)$& $1.270(11)(22)$ \\ 
  RQCD\cite{RQCD:2019jai}   &$\rightarrow 0$& $\ge128$&  $\ge3.5$& --& --& --& $0.670(66)(57)$& $1.302(86)$ \\ 
  ETMC\cite{Alexandrou:2021jok, Alexandrou:2020okk, Alexandrou:2017ypw, Alexandrou:2017hac}   & 0.08&    $139$&    $3.62$& $0.796(16)$& $0.714(91)$& $3.97(16)$& $0.586(36)$& $1.286(23)$ \\ 
  LHPC\cite{Hasan:2019noy, Hasan:2017wwt}   &0.093&    $135$&       $4$& $0.780(10)$& --& --& $0.499(12)$& $1.27(2)$ \\ 
  PNDME\cite{Yoon:2016dij, Gupta:2017dwj, Gupta:2018qil, Jang:2019jkn, Jang:2019vkm, Jang:2023zts}  &$\rightarrow 0$& $\ge135$&  $\ge3.3$& $0.769(27)(30)$& $0.671(48)(76)$& $3.939(86)(138)$& $0.74(6)$& $1.30(6)$ \\ 
  PACS\cite{Shintani:2018ozy, Ishikawa:2018rew}  &$0.084$& $135$&  $\ge7.6$& $0.776(28)(20)$& $0.748(104)(270)$& $4.468(177)(274)$& $0.532(28)(72)$& $1.273(24)(5)$ \\ 
\end{tabular}
\end{ruledtabular}
}
\end{table*}

%
%

\clearpage
\section{
Vector and Axial current renormalization in the Schr\"{o}dinger functional scheme at $\beta = 2.00$
}
\label{app:sf_scheme}

In this appendix, we explain how we compute the renormalization factors $Z_V$ and $Z_A$ in the main text. 
Our numerical simulation adopts the Schr\"{o}dinger functional (SF) scheme for the RG-improved Iwasaki gauge action with the stout smeared $O(a)$-improved Wilson quark action, which is essentially the same framework reported in Ref.~\cite{Ishikawa:2015fzw}.
See also Refs.~\cite{Luscher:1996jn,DellaMorte:2005xgj,Bulava:2014qgz,Hoffmann:2005cz,PACS-CS:2010gyf,Bulava:2016ktf} 
that determine the renormalization factors for the axial, vector, and pseudoscalar operators with the SF scheme.
In order to define the SF scheme, we consider a finite lattice to temporal and spatial directions, $T$ and $L$.
The Dirichlet boundary condition is imposed in the temporal direction, and hence, the boundary gauge fields at 
$t/a=0$
and $T$ are fixed by it.
The smeared gauge fields in the stout smearing steps are also affected by this boundary condition.

The operators we employ here are the vector and axial current 
\begin{equation}
    V_4^a(x)
    =
    \bar{q}(x)\gamma_4 T^a q(x),
\end{equation}
\begin{equation}
    A_4^a(x)
    =
    \bar{q}(x)\gamma_4 \gamma_5 T^a q(x),
\end{equation}
and the pseudoscalar density
\begin{equation}
    P^a(x) 
    =
    \bar{q}(x) \gamma_5 T^a q(x),
\end{equation}
where $T^a$ is the generator of SU($N_f$).
In our calculation we fix $N_f = 3$.
We assume that the nonperturbative $O(a)$-mixing to the axial current is negligible based on our observation~\cite{Taniguchi:2012gew}, and the unimproved current operator is sufficient to determine the renormalization factor $Z_A$. 

The correlation functions required for the renormalization factors are expressed as
\begin{align}
    f_{XY}(t,s) 
    &=
    -\frac{2}{N_f^2(N_f^2-1)}
    \sum_{\boldsymbol{x},\boldsymbol{y}}
    f^{abc}f^{cde}\left\langle
        O'^d X^a(\boldsymbol{x},t) Y^b(\boldsymbol{y},s) O^e
    \right\rangle,
    \label{eq:fXY}
    \\
    f_{X}(t)
    &=
    -\frac{1}{N_f^2-1}\sum_{\boldsymbol{x}}\left\langle
        X^a(\boldsymbol{x},t)O^a
    \right\rangle,
    \label{eq:fX}
    \\
    f_1
    &=
    -\frac{1}{N_f^2-1}\left\langle
        O'^a O^a
    \right\rangle,
    \label{eq:f1}
    \\
    f_V(t) 
    &=
    \frac{1}{N_f(N_f^2-1)}\sum_{\boldsymbol{x}}if^{abc}
    \left\langle
        O'^a V_4^b(\boldsymbol{x},t) O^c
    \right\rangle,
    \label{eq:fV}
\end{align}
where $f^{abc}$ is the structure constant of SU($N_f$).
The operators $O^a, O'^a$ are defined on the boundary as
\begin{align}
    O^a
    &=
\frac{1}{L^3}
    \sum_{\boldsymbol{y},\boldsymbol{z}}
    \bar{\zeta}(\boldsymbol{y}) \gamma_5 T^a \zeta(\boldsymbol{z}),
    \\
    O'^a
    &=
    \frac{1}{L^3}
    \sum_{\boldsymbol{y},\boldsymbol{z}}
    \bar{\zeta}'(\boldsymbol{y}) \gamma_5 T^a \zeta'(\boldsymbol{z}),
\end{align}
where $\zeta,~\zeta'$ are the boundary quark fields at $t/a=0$
and $T$, respectively.
We substitute $A_4$ and $P$ into $X$ and $Y$.
\par

From the correlation functions in \eqref{eq:fXY}--\eqref{eq:fV}, we define the renormalization factors $Z_V$ and $Z_A$ as 
\begin{equation}
    Z_V 
    =
    \left. 
    \widetilde{Z}_V(T/2) 
    \right|_{\hat{m}_{\rm PCAC}\to 0},
    \quad
    \widetilde{Z}_V(t)
    =
    \frac{f_1}{n_V f_V(t)},
    \label{eq:def_zv}
\end{equation}
and 
\begin{gather}
    Z_A 
    = 
    \left.
    \sqrt{\widetilde{Z}_A(2T/3)}
    \right|_{\hat{m}_{\rm PCAC}\to 0},
    \notag
    \\
    \widetilde{Z}_A(t)
    =
    \frac{f_1}{n_A}
    \left[
        f_{AA}(t,T/3) - 2\hat{m}_{\rm PCAC}f_{PA}(t,T/3)
    \right]^{-1},
\end{gather}
where $n_V, n_A$ are normalization constants such that both $Z_V$ and $Z_A$ become unity at tree level~\cite{Luscher:1996jn,Hoffmann:2005cz}.
The dimensionless PCAC 
mass parameter $\hat{m}_{\rm PCAC}$
is determined by using the average of three points located at the central time slice $t/a=T/2$ as
\begin{equation}
    \hat{m}_{\rm PCAC} 
    =
    \frac{1}{3}\sum_{t/a=T/2-1}^{T/2+1}
        \frac{f_A(t+a) - f_A(t-a)}{4f_P(t)},
    \label{eq:PCACmass}
\end{equation}
which are used to define 
the bare quark mass $m_\mathrm{PCAC}^{\mathrm{pion}}$ in lattice units as
$am_\mathrm{PCAC}^{\mathrm{pion}}=Z_A \hat{m}_{\mathrm{PCAC}}$ that appears in the text.
The massless limit can be taken by adjusting the hopping parameter $\kappa$ so that $\hat{m}_{\rm PCAC}\sim 0$.

As in Ref.~\cite{Ishikawa:2015fzw},
we take $\alpha=0.1$ and $n_{\rm{step}}=6$ for the stout link smearing parameters.
The action parameters are set to be $\beta=2.00$ and $c_{\rm{SW}}=1.02$~\cite{Shintani:2019wai},
and the SF boundary parameters are $c_{t}^{\rm{Plaquette}}=1$, $c_{t}^{\rm{Rectangular}}=3/2$
for the gauge action and $\widetilde{c}_{t}=1$ for the quark action.

\begin{table*}[ht]
    \caption{Simulation parameters and number of trajectories (Traj.) for the calculation of renormalization factors. The acceptance rate in the hybrid Monte-Carlo (HMC) is also presented.}
    \label{tab:parameters}
    \begin{ruledtabular}
    \begin{tabular}{cccccc}
        Run  & $L$, $T$
        & $\theta$ & $\kappa$ & traj. & HMC Acc. \\
        \hline
        (VAS) &  $12 , 30$ & 1/2 & 0.125820 &  10000  & 0.9179   \\
        (VAL) & $16 , 42$ & 1/2 & 0.125820 &  21750  & 0.9109   \\
    \end{tabular}
    \end{ruledtabular}
\end{table*}

\begin{table*}[ht]
    \caption{PCAC masses and renormalization factors $Z_V$ and $Z_A$.}
    \label{tab:main results}
    \begin{ruledtabular}
    \begin{tabular}{cccc}
         Run  & $\hat{m}_{\mathrm{PCAC}}$ & $Z_V$ & $Z_A$  \\ 
        \hline
        (VAS) & $-0.00029(12)$ &  0.96993(108)  & 0.9702(26)  \\
        (VAL) &  0.00028(15) &  0.96677(41) & 0.9783(21)  \\
    \end{tabular}
    \end{ruledtabular}
\end{table*}

The simulation parameters are shown in Table~\ref{tab:parameters} for the two volumes (VAS) and (VAL).
The phase angle $\theta$ is the parameter of the generalized periodic boundary condition for the quark field.
The measurements for the correlation functions are performed at every trajectory, and the statistical errors are estimated with the jackknife method.
In the jackknife method, we built block data with a size of 200 trajectories for (VAS) and 150 trajectories for (VAL).
The hopping parameter $\kappa$ is chosen to be $\hat{m}_{\rm PCAC}\sim 0$ by investigating $\kappa$ dependence of $\hat{m}_{\rm PCAC}$ in the smaller volume (VAS).
The simulation results are tabulated in Table~\ref{tab:main results}, which shows that the PCAC mass is statistically consistent with zero in the (VAS) and (VAL) runs. 

\begin{figure}[]
    \includegraphics[width=130mm]{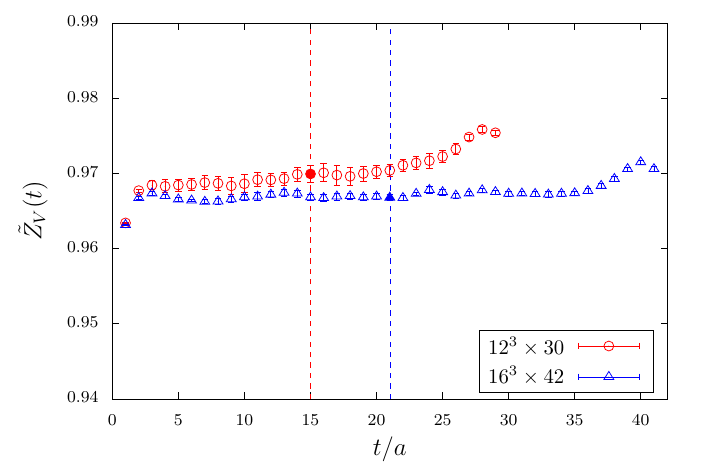}
    \caption{Time dependence of $\widetilde{Z}_V$ for the smaller (red circles) and larger (blue triangles) volumes.
    The filled data expresses the value of $Z_V$.
    }
    \label{fig:tilde{Z}_V}
\end{figure}
The time dependence of $\widetilde{Z}_V$ in (VAS) and (VAL) is shown in Fig.~\ref{fig:tilde{Z}_V}.
In both cases, $\widetilde{Z}_V$ is reasonably independent of time.
As defined in (\ref{eq:def_zv}), the value of $Z_V$ is obtained from the fixed time slice of $\widetilde{Z}_V$ as
\begin{equation}
    Z_V = 0.96677(41)(316),
\end{equation}
where the first and second parentheses represent the statistical and systematic errors, respectively. 
The central value and the statistical error are determined from the results in the larger volume (VAL),
and the systematic error is evaluated by the difference between the central values in the two runs.

\begin{figure}[]
    \includegraphics[width=130mm]{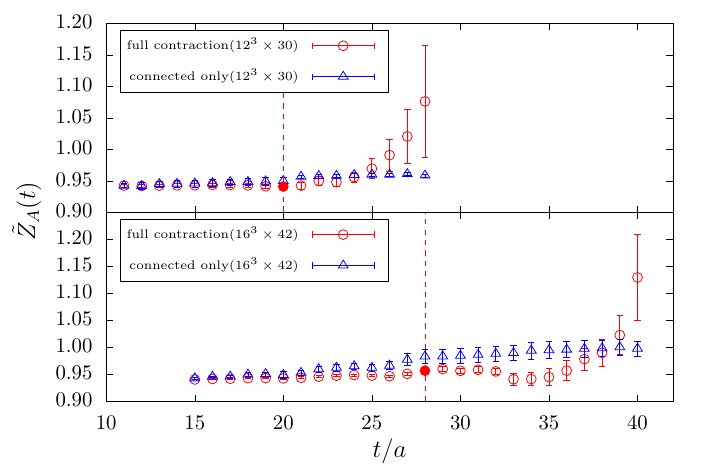}
    \caption{Time dependence of $\widetilde{Z}_A$ for the smaller ((VAS), upper graph) and larger ((VAL), lower graph) volumes.
     The circle and triangle symbols represent the data with and without the disconnected contraction, respectively. The filled data expresses the value of $Z_A$.
    }
    \label{fig:tilde{Z}_A}
\end{figure}
The time dependence of $\widetilde{Z}_A$ in (VAS) and (VAL) is shown in Fig.~\ref{fig:tilde{Z}_A}.
The data of the full contraction are calculated with the connected and disconnected contractions.
Around $t/a=2T/3$,
it is observed that the full contraction has a plateau, while the data of only the connected contraction has a gradual slope.
Therefore, we choose the full contraction result to evaluate $Z_A$, whose value is given by
\begin{equation}
    Z_A = 0.9783(21)(81),
\end{equation}
where the central value and errors are determined in the same manner as for $Z_V$.

\clearpage
\bibliography{main}
\bibliographystyle{man-apsrev}

\end{document}